\documentclass[12pt,a4]{article}

\usepackage{cite}
\usepackage{amsmath}
\usepackage{amssymb}
\usepackage{amsfonts}
\usepackage{array}
\usepackage{graphicx}
\usepackage{tikz}
\usetikzlibrary{calc}
\usepackage{pgfplots}
\pgfplotsset{compat=1.11}
\usepackage{subcaption}
\usepackage{mathrsfs}
\usepackage{multirow}
\usepackage{siunitx}
\usepackage{microtype}
\newcommand{\bea}{\begin{eqnarray*}}
\newcommand{\eea}{\end{eqnarray*}}
\newcommand{\beao}{\begin{eqnarray}}
\newcommand{\eeao}{\end{eqnarray}}

\newcommand{\RR}{{\mathbb R}}
\newcommand{\neos}{\texttt{neos}}

\usepackage{lineno}
\usepackage[hidelinks]{hyperref}
\usepackage{authblk}

\usepackage{xspace}

\newcommand{\tomopt}{\textsc{TomOpt}\xspace}
\newcommand{\mode}{\textsc{Mode}\xspace}
\newcommand{\pytorch}{\textsc{PyTorch}\xspace}
\newcommand{\geant}{\textsc{Geant}~4\xspace}
\newcommand{\xo}{\ensuremath{X_0}\xspace}

\renewcommand{\equationautorefname}{Eq.}

\usepackage{xargs}
\usepackage[colorinlistoftodos,prependcaption,textsize=tiny]{todonotes}
\newcommandx{\gunes}[2][1=]{\todo[linecolor=green,backgroundcolor=green!25,bordercolor=green,inline,#1]{Gunes: #2}}

\newcommand{\eg}{{\em e.g. }}
\newcommand{\egc}{{\em e.g., }}
\newcommand{\ie}{{\em i.e. }}
\newcommand{\iec}{{\em i.e., }}

\begin{document}
\title{Toward the End-to-End Optimization \\ of Particle Physics Instruments \\with Differentiable Programming:\\ a White Paper}


\renewcommand\Authfont{\small}
\renewcommand\Affilfont{\small}

\renewcommand{\Authsep}{ }
\author[{ }]{Tommaso~Dorigo$^{1,2}$, Andrea~Giammanco\footnote{andrea.giammanco@cern.ch}$^{1,3}$, Pietro~Vischia$^{1,3}$ (editors)}
\renewcommand{\Authsep}{, }
\author[4]{\\Max~Aehle}
\author[5]{Mateusz~Bawaj}
\author[1,6]{Alexey~Boldyrev}
\author[1,2]{Pablo~de~Castro~Manzano}
\author[1,6]{Denis~Derkach}
\author[1,7]{Julien~Donini}
\author[8]{Auralee~Edelen}
\author[1,2] {Federica~Fanzago}
\author[4]{Nicolas~R.~Gauger}
\author[1,9]{Christian~Glaser}
\author[1,10]{At\i{}l\i{}m~G.~Baydin}
\author[1,11]{Lukas~Heinrich}
\author[12]{Ralf~Keidel}
\author[1,13]{Jan~Kieseler}
\author[1,14]{Claudius~Krause}
\author[1,3]{Maxime~Lagrange}
\author[1,11]{Max~Lamparth}
\author[1,2,15]{Lukas~Layer}
\author[16]{Gernot~Maier}
\author[1,2,17,7]{Federico~Nardi}
\author[18]{Helge~E.~S.~Pettersen}
\author[19]{Alberto~Ramos}
\author[1,6]{Fedor~Ratnikov}
\author[20]{Dieter~R\"ohrich}
\author[19]{Roberto~Ruiz~de~Austri}
\author[1,21]{Pablo~Mart\'inez~Ruiz~del~\'Arbol}
\author[2,3] {Oleg~Savchenko}
\author[22]{Nathan~Simpson}
\author[1,2]{Giles~C.~Strong}
\author[3]{Angela~Taliercio}
\author[1,2,17]{Mia~Tosi}
\author[1,6]{Andrey~Ustyuzhanin}
\author[1,23]{Haitham~Zaraket}

\affil[1]{MODE Collaboration, \url{https://mode-collaboration.github.io/}}
\affil[2]{Istituto Nazionale di Fisica Nucleare, Sezione di Padova, Italy}
\affil[3]{Centre for Cosmology, Particle Physics and Phenomenology (CP3), Universit\'e catholique de Louvain, Belgium}
\affil[4]{Chair for Scientific Computing, Technische Universit\"at Kaiserslautern, Germany}
\affil[5]{Universit\`a di Perugia and INFN, Sezione di Perugia, Italy}
\affil[6]{HSE University, Russia}
\affil[7]{Universit\'e Clermont Auvergne, Laboratoire de Physique de Clermont, CNRS/IN2P3, France}
\affil[8]{SLAC National Accelerator Laboratory, USA}
\affil[9]{Department of Physics and Astronomy, Uppsala University, Sweden}
\affil[10]{Department of Computer Science, University of Oxford, UK}
\affil[11]{Physik-Department, Technische Universit\"at M\"unchen, Germany}
\affil[12]{Center for Technology and Transfer, University of Applied Sciences Worms, Germany}
\affil[13]{CERN, Switzerland}
\affil[14]{NHETC, Dept. of Physics and Astronomy, Rutgers University, USA}
\affil[15]{Università di Napoli ``Federico II", Italy}
\affil[16]{Deutsches Elektronen-Synchrotron (DESY), Germany}
\affil[17]{Universit\`a degli Studi di Padova, Italy}
\affil[18]{Department of Oncology and Medical Physics, Haukeland University Hospital, Norway}
\affil[19]{Instituto de Física Corpuscular, UV-CSIC, Spain}
\affil[20]{Department of Physics and Technology, University of Bergen, Norway}
\affil[21]{Instituto de Física de Cantabria, UC-CSIC, Spain}
\affil[22]{Lund University, Sweden}
\affil[23]{Multi-Disciplinary Physics Laboratory, Optics and Fiber Optics Group, Faculty of Sciences, Lebanese University, Lebanon}

\maketitle


\begin{abstract}

The full optimization of the design and operation of instruments whose functioning relies on the interaction of radiation with matter is a super-human task, given the large dimensionality of the space of possible choices for geometry, detection technology, materials, data-acquisition, and information-extraction techniques, and the interdependence of the related parameters. On the other hand, massive potential gains in performance over standard, ``experience-driven'' layouts are in principle within our reach if an objective function fully aligned with the final goals of the instrument is maximized by means of a systematic search of the configuration space.
The stochastic nature of the involved quantum processes make the modeling of these systems an intractable problem from a classical statistics point of view, yet the construction of a fully differentiable pipeline and the use of deep learning techniques may allow the simultaneous optimization of all design parameters. 

In this document we lay down our plans for the design of a modular and versatile modeling tool for the end-to-end optimization of complex instruments for particle physics experiments as well as industrial and medical applications that share the detection of radiation as their basic ingredient. We consider a selected set of use cases to highlight the specific needs of different applications.
\end{abstract}

\clearpage
\tableofcontents
\clearpage

\section{Introduction}
\label{sec:intro}

The optimal choice of layout, characteristics, materials, and information-extraction procedures of a measuring instrument constitutes a loosely constrained problem, with a very large number of free parameters related by non-obvious correlations. Although typically quite complex, similar problems may sometimes still be tractable by standard means, in the sense that a parameterized model of the system allows the definition of a likelihood function $L = p(x|\theta)$, given simulated data $x$, and a solution by minimization of $-\ln L$ with respect to the modeling parameters $\theta$. If, however, the instrument bases its functioning on quantum phenomena, such as the interaction of radiation with matter, the optimization problem is intractable: the probability $p(x|\theta)$ of observing data $x$ given underlying parameters $\theta$ may not be written explicitly. In such circumstances, one has access at best to the generating function of the observed data through forward simulation, a setting commonly referred to as likelihood-free or simulation-based inference~\cite{cranmer2020frontier}.

Over the course of the past eighty years, the intractability of the design optimization problems commonly encountered in particle physics has not prevented physicists from successfully conceiving, commissioning, and operating detectors of huge complexity. The development of new, high-performance instruments followed a robust strategy that, while systematically leveraging technological advancements in electronics and material science, duly exploited well-tested paradigms proven to work by previously acquired experience. For example, a long-standing paradigm for the detection of particles in collider physics experiments is the need to measure the momentum of all electrically charged particles by magnetic bending in gaseous or light materials before exploiting the electromagnetic and hadronic showers produced by both charged and neutral particles in dense matter. Another paradigm is the requirement of significant redundancy in the detection systems, to enable cross-calibration of the different components and offer robustness of the resulting inference. A further typical default is the choice of a symmetric layout of the detection components, such as the equal spacing of scintillating and passive elements along the depth of a calorimeter. While these paradigms have a strong motivation if we look at the past of particle detection practice, the same cannot be said if we look into the future of our field.

\begin{figure}[h!]
\begin{center}
    \includegraphics[width=15cm]{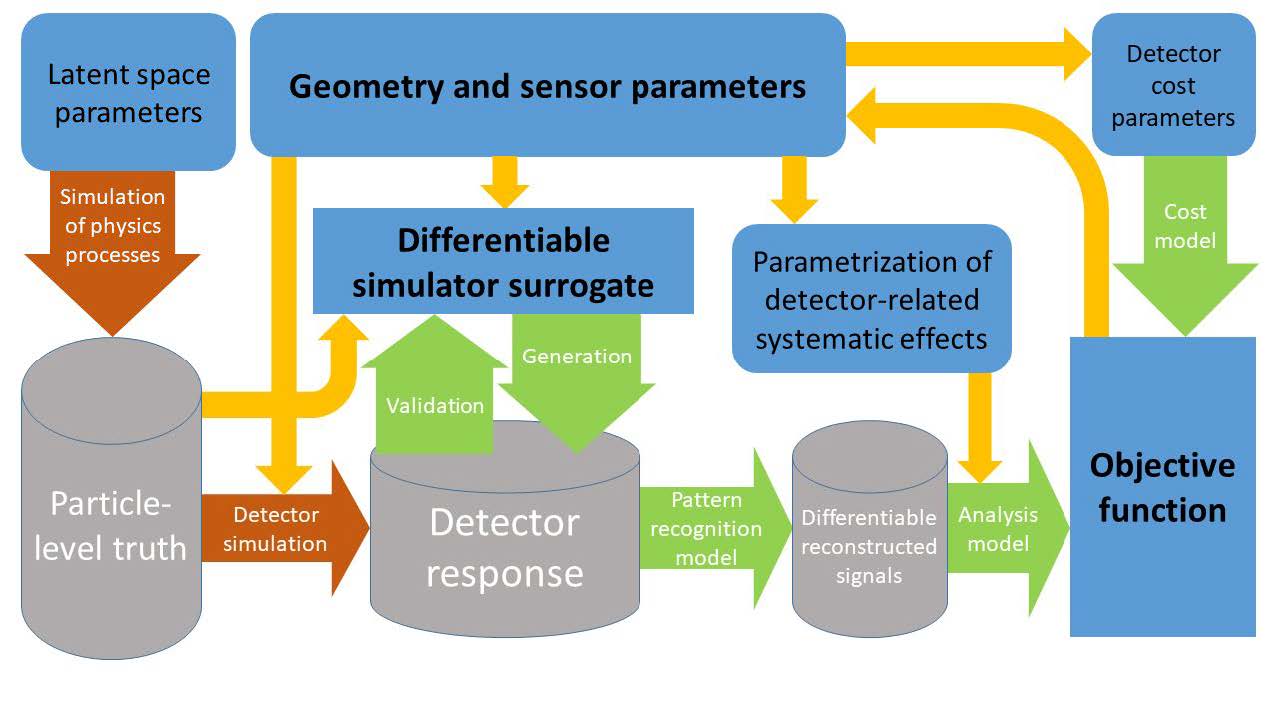}
    \caption {\em Block diagram for the optimization of a generic detector. Data from a simulator (left, cans labeled ``Particle-level truth'' and ``Detector response'') are used to train and validate a differentiable model (``Differentiable simulator surrogate'') of the relevant physical processes. Models of pattern recognition, inference extraction, cost of components, and a loss function may then become a function of detector geometry and construction layout parameters. A back-propagation loop of loss derivatives through the functional elements of the system allows their optimization. The figure is adapted from Ref.~\cite{npnipaper}.
    }
    \label{f:pipeline}
\end{center}
\end{figure}

The fast progress of computer science in the past twenty years, together with the development of deep neural networks and optimization software based on differentiable programming, offers us an unprecedented opportunity to rethink the foundations of our design strategies, and to identify and investigate novel, possibly revolutionary solutions we have been unable to figure out by ourselves. The typical design problems we face involve the choice of hundreds, if not thousands of parameters defining the placement and geometry of materials and detection layers, their specifications and performance, and their monetary cost. The full exploration of this high-dimensional space of design solutions is a wholly super-human task: to move forward, we must turn to the differentiable programming tools that make this exploration possible.

It must be noted that the breadth of the space of design solutions has also been increasing with our technological advancements. Nowadays we can 3D print scintillation detectors~\cite{Mishnayot:2014nya}, as well as design more complex detection elements with thin layers of AC-coupled resistive silicon sensors~\cite{Giacomini:2019kqz}. These advancements can best be exploited if we endow ourselves with the capability of performing continuous scans of the geometry space of the devices we wish to construct: this is something we achieve by developing differentiable programming pipelines.

Another reason for revisiting our detector design paradigms while accounting for the availability and development of new computer science tools is the evolution of the pattern recognition and inference procedures we have been adopting in the extraction of information from raw detector readouts. The demands posed to our instruments are continuously increasing, as we move, \egc toward the high-luminosity (HL) phase of the Large Hadron Collider (LHC)~\cite{Evans_2008}, or toward larger and larger detection volumes in cosmic ray and neutrino physics. At the HL-LHC, in a few years we will be reconstructing high-energy particle collisions within ${\cal{O}}(200)$ pileup interactions taking place during the same bunch crossing; the performance of standard reconstruction algorithms for charged tracks will be strongly reduced in the presence of an exponential increase of the combinatorial background. If deep learning methods will be employed for those pattern recognition tasks (such as those described in Refs.~\cite{Farrell:DLPS2017,Farrell:2018cjr,Amrouche:2019wmx,Ju:2020xty,Akar:2020jti,Shlomi:2020ufi,Choma:2020cry,Siviero:2020tim,Fox:2020hfm,Amrouche:2021tlm,goto2021development,Biscarat:2021dlj,Akar:2021gns,Thais:2021qcb,Ju:2021ayy,Dezoort:2021kfk,Edmonds:2021lzd,Lavrik:2021zgt,Huth:2021zcm,Goncharov:2021wvd,Lazar:2022ixi}), the question arises of whether the detectors have been conceived to be optimal for those tools. Such a potential misalignment between design and exploitation is even more evident if we look further into the future, to the construction of colliders, such as the proposed Future Circular Collider (FCC)~\cite{Benedikt:2020ejr}, characterized by higher center-of-mass energies than the current machines: since we are currently sitting on a rapidly growing curve of performance of artificial-intelligence-powered methods~\cite{2020arXiv200504305H}, in order for our future detectors to be most effective we need to consider their design as an optimization problem that includes a model of the pattern recognition and inference extraction procedures available at operation time, however hard it may be to envision their power today.

The above considerations motivate us to pursue a wide-ranging plan of investigations that has the primary purpose of educating ourselves and our community on how to best integrate all the elements of a detector design problem---from the modeling of the stochastic quantum phenomena to the description of detector layout, geometry, and performance; from the pattern recognition to the inference extraction procedures; and from the interplay of geometry and systematic uncertainties to the physical and economic constraints---into a single optimization problem, as exemplified in Fig.~\ref{f:pipeline}. We believe that the capability to compute derivatives of the objective function with respect to any one of the parameters of the system, provided by implementing the whole pipeline using differentiable programming, will be key to enable the successful exploration of the large space of design choices, and the discovery of innovative solutions. 

What we are facing is an extremely tall order if we consider a detector of the scale of collider experiments such as ATLAS or CMS. In fact, it is doubtful that we have today the resources, expertise, and skills required to attack a problem of that complexity. Hence we must proceeds in steps, by considering less ambitious, but achievable, goals. In this document we propose, and discuss in some detail, a series of design optimization tasks that are interesting in their own right, and whose solution via the above plan may enable us to build a framework of methods and software tools that together may constitute the building blocks for solving harder problems. While the specificity of a detector leaves little room for reuse of the differentiable surrogate models of particle interaction with active and passive components that may have been developed to study them, there is instead significant device-independence in recently developed reconstruction algorithms empowered by deep learning~\cite{Kieseler:2020wcq}, and a clear possibility of reusing the models we may develop for the monetary cost of the components, for the interaction between geometry- and detector-related systematic uncertainties, and for inference extraction. 

The core of all optimization procedures is a carefully defined objective function, which should encode as closely as possible the explicit goals of the instrument we are designing. For a large scientific endeavor, specifying this function may at first sight appear an impossible task, given the multi-purpose nature of the detectors, the breadth of physics studies they enable, and the arbitrariness of the relative value of different scientific objectives of the experiment. However, we argue that the exercise of appraising those goals and proposing an evaluation metric {\em can} be beneficially carried out, and an objective function---or a family of objective functions that address different points of view---can proficuously be specified. Indeed, such an exercise is not altogether different from the one of defining a trigger menu for a collider physics experiment, which produces a list of triggers with relative selection strategies, bandwidths, and prescaling factors. We stress that the resulting optimization study cannot be expected to produce a final answer, but rather that it may indicate advantageous combinations of design choices and ``sweet spots'' in the space of design parameters, guiding our hand toward robust and effective decisions.

The present document, which builds on the ideas succinctly described in Ref.~\cite{npnipaper}, is structured as follows.
In Sec.~\ref{sec:state-of-the-art} we provide an overview of the state of the art of the computer science ingredients that can be used to construct the software pipeline for an end-to-end optimization study, and we provide a brief survey of today's solutions to optimization problems in other fields of research. In Sec.~\ref{sec:problem-description} we outline the concrete way of defining a detector optimization problem and the way of assembling a set of modules to construct a closed-loop pipeline employing differentiable programming techniques. In Sec.~\ref{sec:exemplary_use_cases} we provide a discussion of example applications and the specific needs of each, and assess their feasibility and requirements. In Sec.~\ref{sec:hardware} we discuss the hardware and software requirements for the solution of the typical problems we have considered. We offer some conclusions in Sec.~\ref{sec:conclusions}.

\clearpage
\section {The State of the Art in Design Optimization and Differentiable Programming}
\label{sec:state-of-the-art}
Scientists and engineers have leveraged the steady growth of available computing power over decades to continuously improve the accuracy of their numerical simulations. Nowadays, in many technical disciplines and for setups too complicated for traditional theoretical approaches, simulations can  answer the question ``what happens when we do $X$?'' The oftentimes excellent agreement of simulation with physical experiments at a tiny fraction of their cost makes it natural to pose the next question, ``which design $X$ gives the \emph{optimal} outcome?''

The ``model pupil'' among the technical disciplines to deal with this kind of question  is certainly computational fluid dynamics (CFD). While first attempts to numerically improve the shape of airfoils~\cite{hicks1974assessment,hicks_wing_1978} were based on finite differences, the analytic derivation of sensitivity equations~\cite{pironneau_optimum_1974,pironneau_optimum_1973,jameson_aerodynamic_1988} proved much more efficient when the number of design parameters is large. During the last two decades, automatic differentiation\footnote{Also called algorithmic differentiation, or simply autodiff.} (AD) has been integrated into various CFD codes~\cite{towara_discrete_2013,Albring_etal2016b,luers_adjoint-based_2018} and enabled a large number of design optimization studies~\cite{nemili_accurate_2017,luers_adjoint-based_2018,zhou_discrete_2018,bombardieri_aerostructural_2021}.

``Optimality'' can relate to much more than the technical performance of a design. In a wider sense, parameter and model fitting optimize the predictive quality of a theoretical model given empirical data. AD has enabled parameter fitting studies in fields as diverse as ice sheet modeling~\cite{MGDLS2021}, optimal control~\cite{andersson2012casadi}, and quantitative finance~\cite{achdou_computational_2005}. Deep learning~\cite{goodfellow2016deep,lecun2015deep} can be understood as a special case of parameter fitting with an emphasis on representation learning and specialized model architectures for AI-related tasks such as computer vision~\cite{janai2020computer}, natural language processing~\cite{goldberg2017neural}, and reinforcement learning~\cite{sutton2018reinforcement}. AD capability in machine learning (ML) frameworks such as PyTorch~\cite{pytorch} and TensorFlow~\cite{abadi2016tensorflow} is at the core of many recent advances in the machine learning community~\cite{baydin_automatic_2017}.

The relatively new term ``differentiable programming''\footnote{A term initially proposed by Christopher Olah~\cite{olahfuncprog},
David Dalrymple~\cite{dalrymple}, and Yann LeCun~\cite{lecunfacebookpost} from a deep learning point of view.} is used to capture the essence of deep learning practice, where one constructs differentiable computer code---via AD---in order to solve various tasks, and optimizes them via gradient-based optimization of an objective based on training data. The term emphasizes the general-purpose programming aspect of recent machine learning approaches, in contrast with conventional neural networks. In this context, neural networks---viewed as compositions of a series of non-linear transformations---are only a member in the more general family of differentiable programs, which covers the space of all differentiable algorithms with, \egc control flow, loops, and recursion, and crucially includes differentiable numerical simulators in science and engineering disciplines.

In particle physics, measuring instruments rely on the extraction of inference from the data they collect, and thus base their operation on the aforementioned functionalities: therefore, their design optimization is closely connected to the precision of the models and the effectiveness of those parameter estimate procedures. However, instruments that rely on the interaction of radiation with matter in their data-collection mechanisms, such as particle detectors, add more complexity to the task, in that they introduce an element of intrinsic stochasticity from the quantum nature of the involved physical processes, and thus necessitate the deployment of special solutions, such as those described in Sec.~\ref{sec:problem-description}. Hence, the full optimization of particle detectors and accelerators is a frontier topic examined by only a few examples in the literature; see \eg Refs.~\cite{Shirobokov:2020tjt,dorigo2020geometry,ratnikov2020,cisbani2020,npnipaper,edelen:ml2,Koser:ML1,van_der_veken:ml1,meyer_optimization_2020}.

There are two main ways to make a simulation differentiable. The first one consists of using AD directly in the simulation code, making use of the AD tools based on operator overloading or source code transformation in the programming language in which the simulator is implemented. As we will detail below in Secs.~\ref{sec:ad-implementation} and~\ref{sec:ad-adaptations}, the complexity of such a task varies widely from case to case, sometimes making it not viable. The second way to have a differentiable simulation consists of using deep learning techniques to produce a differentiable surrogate model of the simulator, using supervised training data sampled by running the original simulator. We detail the advantages of such an approach in Sec.~\ref{sec:ad-surrogates}.

The remainder of this section is organized as follows. Section~\ref{sec:state-of-the-art-intro} introduces the notation for a general optimization problem. The optimization algorithms listed in Sec.~\ref{sec:gradient-based} rely on the gradient of the objective function. Sec.~\ref{sec:derivatives-overview} discusses multiple ways to compute derivatives, and Secs.~\ref{sec:ad-forward} and~\ref{sec:ad-reverse} illustrate the mathematical background of the forward and reverse mode of AD, respectively. Section~\ref{sec:ad-implementation} summarizes the general implementation aspects, while Sec.~\ref{sec:ad-adaptations} lists various considerations that might make program-specific adaptations necessary. Finally, Sec.~\ref{sec:ad-surrogates} introduces the notion of surrogate models.

\subsection{Optimization}
\label{sec:state-of-the-art-intro}

\begin{figure}
\begin{subfigure}[t]{0.9\textwidth}
%
%
\includegraphics[width=\textwidth]{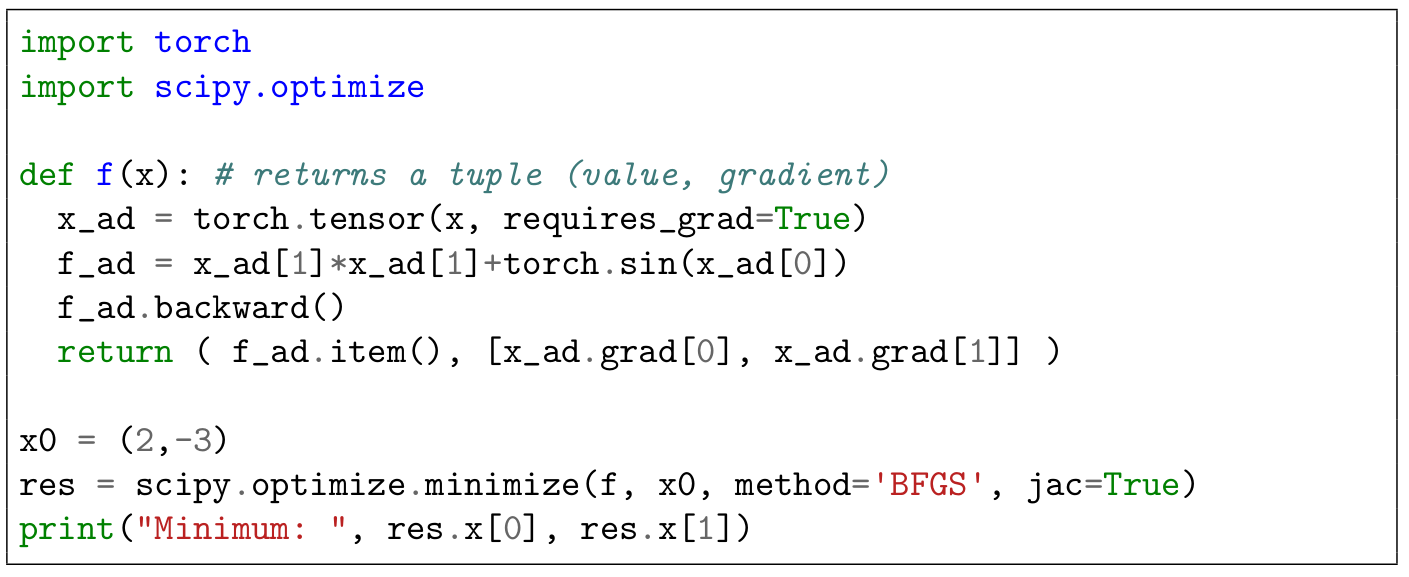}
\caption{Using SciPy~\cite{2020SciPy-NMeth} for optimization. See
Fig.~\ref{fig:ad-codes-pytorch} for how the gradient is obtained.}
\label{fig:ad-codes-optimization}
\end{subfigure}
\\[0.5cm]
\begin{subfigure}[t]{0.45\textwidth}
%
%
%
%
  \includegraphics[width=\textwidth]{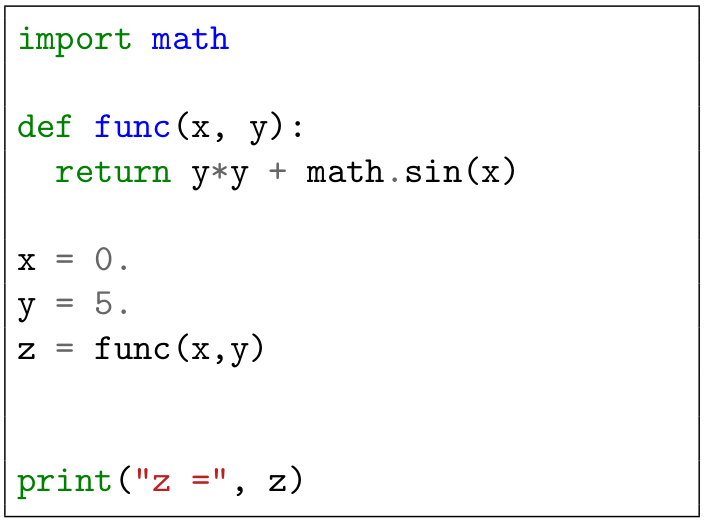}
\caption{\emph{Primal}, \ie  undifferentiated, program.}
\label{fig:ad-codes-primal}
\end{subfigure}
\qquad
\begin{subfigure}[t]{0.45\textwidth}
%
%
%
  \includegraphics[width=\textwidth]{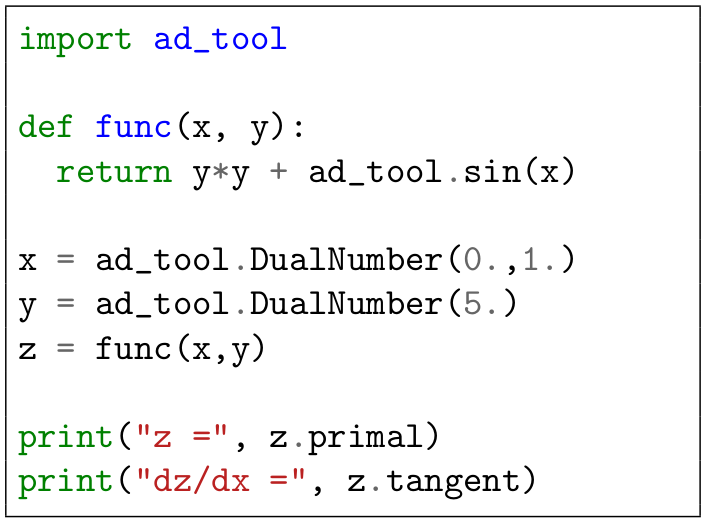}
  \caption{Program differentiated in \emph{forward mode} AD using the {\em ad-hoc} tool in Fig.~\ref{fig:ad-codes-forwardtool}.}
\label{fig:ad-codes-forward}
\end{subfigure}
\\[0.5cm]
\begin{subfigure}[t]{0.45\textwidth}
%
%
%
  \includegraphics[width=\textwidth]{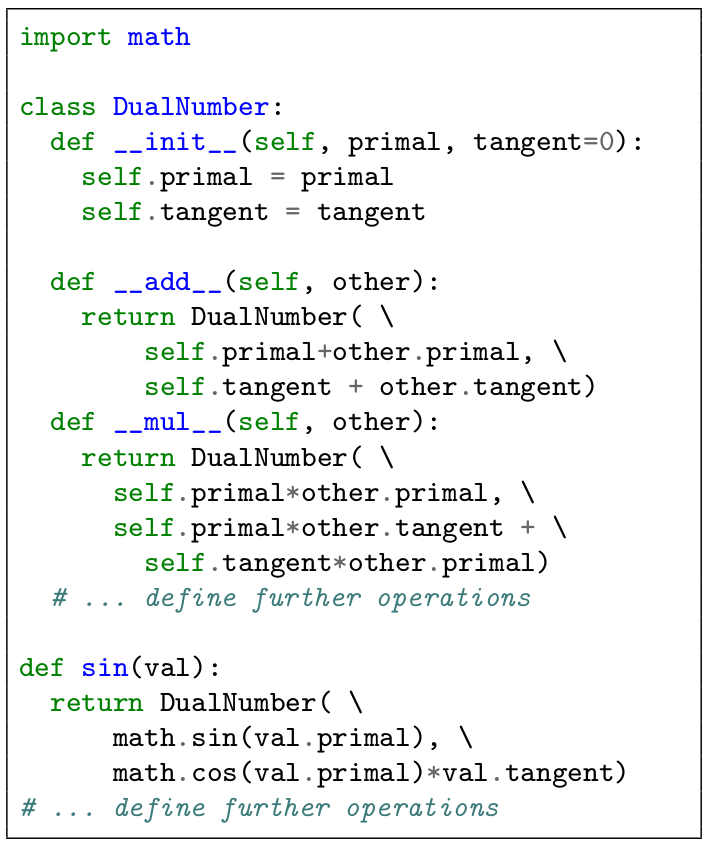}
\caption{{\em Ad-hoc} tool for forward AD.}
\label{fig:ad-codes-forwardtool}
\end{subfigure}
\qquad
\begin{subfigure}[t]{0.45\textwidth}
%
%
%
  \includegraphics[width=\textwidth]{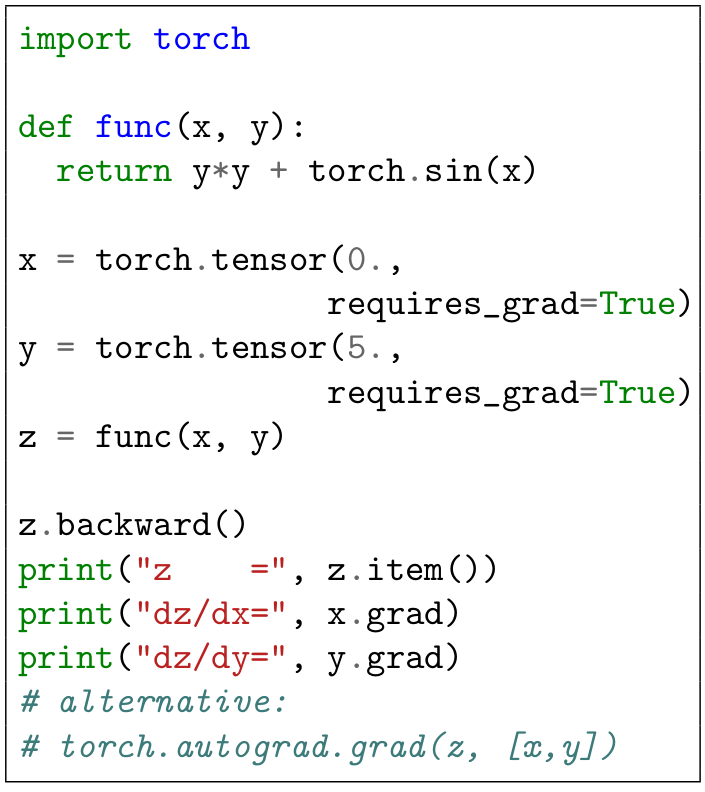}
\caption{Program differentiated in \emph{reverse mode} AD with PyTorch~\cite{pytorch}.
}
\label{fig:ad-codes-pytorch}
\end{subfigure}
    \caption{Applying AD and optimization to a simple function $f(x) = y^2 + \sin(x)$. See the text for more detail.}
    \label{fig:ad-codes}
\end{figure}

The most generic formulation of a mathematical optimization problem is
\begin{equation}\label{eq:general-optimization-problem}
\min_{x \in \mathcal{X}} f(x)
\end{equation}
where 
\begin{itemize}
\item $\mathcal{X}$ is a space of possible choices, usually a subset of some $\RR^d$, and
\item $f: \mathcal{X}\to\RR$ is an \emph{objective}, \emph{cost}, or \emph{loss} function, quantifying the idea of an optimal choice.
\end{itemize}
The terms ``loss'' and ``cost'' usually refer strictly to a function defined on a data point; \egc measuring a distance between a prediction and a target value. ``Objective'' is the most general term for any function we would like to optimize, and can cover cases where there is no explicit notion of a loss between a prediction and a target; it can also include regularization terms and other constraints. Note that these terms are sometimes used interchangeably by some authors.
Both $\mathcal{X}$ and $f$ are problem-specific and need to be determined by domain experts. Regarding the end-to-end optimization of detectors, $\mathcal{X}$ could represent the space in which each dimension corresponds to a numerical parameter affecting the detector hardware and software design.

The function $f$ encodes the accuracy and technical and financial constraints. Evaluating those involves a whole pipeline of software modules that simulate the physical processes inside the detector, generate the detector response, post-process it, and finally assess the objective. We describe these physics-specific aspects in Sec.~\ref{sec:problem-description} and present concrete examples of detector optimization projects in Sec.~\ref{sec:exemplary_use_cases}. This section deals with the theory for solving a general minimization problem as defined in \equationautorefname~\eqref{eq:general-optimization-problem}.

\subsection{Gradient-Based Optimization}
\label{sec:gradient-based}

Most of the design variables $x_i$ of the detector systems of our interest (\eg those considered {\em infra}, Sec.~\ref{sec:exemplary_use_cases}, as well as others of similar scope) can be chosen continuously and influence the objective function $f$ in a smooth way, making them subject to differentiable programming and gradient-based optimization. 

Gradient descent algorithms have a general form where a differentiable function $f$ is minimized over a number of iterations by starting from an initial parameter $x^{(0)}$ and generating a sequence of updated parameters $x^{(1)}$, $x^{(2)}$, etc. by using an update rule 
\begin{equation}\label{eq:gradient-descent}
x^{(k+1)} = x^{(k)} - \eta_k \cdot \nabla_x f(x^{(k)})\;.
\end{equation} 
A small step into the direction of the negative gradient decreases the value of $f$, as $-\nabla_x f(x^{(k)})$ points into the direction of steepest descent of $f$ around $x^{(k)}$. The step size $\eta_k>0$ has to be chosen in advance, or adjusted in each step, to make sure that the gradient descent algorithm does not step ``beyond'' the minimum. 

If $f$ is twice differentiable, the Hessian matrix $H_x f$ can be used in the (damped) Newton's method:
\begin{equation}\label{eq:newton-method}
x^{(k+1)}  = x^{(k)} - \eta_k \cdot H_x f(x^{(k)})^{-1} \cdot \nabla_x f(x^{(k)})\;.
\end{equation}
Quasi-Newton methods such as BFGS~\cite{shanno_conditioning_1970,goldfarb_family_1970,fletcher_new_1970} and L-BFGS-B~\cite{byrd_limited_1995,zhu_algorithm_1997} only require the knowledge of $\nabla_x f$ and form a internal approximation of the Hessian.

\emph{Stopping criteria} indicate that the present solution $x^{(k)}$
is close to a \emph{local minimum}, \ie a point that minimizes $f$
within some neighbourhood. For example, the norm of the gradient $\nabla_x f(x^{(k)})$ can be tested to fall below a user-defined
threshold. In this case, no further improvement is expected and the
algorithm terminates.

A general function $f$ can have many local minima. The one selected by an optimization algorithm not necessarily is the \emph{global minimum} as well. A finite set of iterates $x^{(1)}, \dots, x^{(k)}$ cannot
comprehensively explore the full design space $\mathcal{X}$. Therefore in theory, better local minima might remain undiscovered no matter how
ingeniously the optimization algorithm uses the information on the value and derivatives of $f$ at the iterates. For applications, solutions that do not perfectly or provably reach the global minimum may still present a valuable improvement over previous designs.

Many popular optimization algorithms have been implemented in open-source packages such as SciPy~\cite{2020SciPy-NMeth}; Figure~\ref{fig:ad-codes-optimization} shows an example calling a SciPy optimizer with the gradient of the objective supplied by the PyTorch's AD module. Essentially, the user must provide an initial solution and code to evaluate both $f$ and its derivatives automatically. Three ways to to obtain the derivative are listed in the next subsection. 

\subsection{Computing Derivatives: An Overview}
\label{sec:derivatives-overview}

\emph{Numerical differentiation} can approximate the components of a gradient $\tfrac{\partial f}{\partial x_i}(x^{(k)})$ by evaluating $f$ at several points around $x^{(k)}$. The most commonly used formulas are the forward and central finite difference quotients,
\begin{equation}\label{eq:difference-quotients}
\frac{f(x^{(k)}+h\cdot \mathrm e^{(i)}) - f(x^{(k)})}{h} \quad \text{and} \quad \frac{f(x^{(k)}+h\cdot \mathrm e^{(i)}) - f( x^{(k)}-h\cdot \mathrm e^{(i)})}{2h},
\end{equation}
where $\mathrm e^{(i)}$ is the $i$-th unit vector, and $h>0$ is a small real number. The approximations in \equationautorefname~\eqref{eq:difference-quotients} converge to the true derivative of $f$ at the limit $h\to 0$. In numeric code $h$ cannot be chosen to be arbitrarily small and the error in the approximation can never be eliminated because of truncation and round-off errors in floating-point operations~\cite{burden2015numerical,baydin_automatic_2017}. Hence a suitable value for $h$ must be selected whenever computing a finite-difference quotient to minimize this error.
Numerical differentiation is usually easy to implement, but its time complexity scales linearly with the number of input variables. 

\emph{Analytic differentiation} by hand, and \emph{symbolic differentiation} by a computer algebra system such as Mathematica, provide exact derivatives as mathematical (symbolic) expressions. These are however usually only applicable to conventional closed-form expressions that \eg cannot easily describe programming-language concepts such as loops and control flow. Symbolic differentiation also involves expression swell, where the derivative expression obtained can be significantly more costly to compute than the original expression in terms of computational complexity.

\emph{Automatic-} or {algorithmic-differentiation} (AD), extends the computer code implementing a given objective function $f$ by additional arithmetics to compute specific partial derivatives of the involved variables (input, intermediate, and output). AD is exact up to floating-point accuracy, and its so-called \emph{reverse mode} beats numeric differentiation also with respect to (asymptotic) time complexity. AD and gradient-based optimization are the two main ingredients of \emph{differentiable programming}, where solutions to optimization problems are implemented as computer code, differentiated via AD, and optimized using gradient-based algorithms.

We present a summary of the two \emph{modes} of AD in Secs.~\ref{sec:ad-forward} and~\ref{sec:ad-reverse}, and implementation aspects in Sec.~\ref{sec:ad-implementation}. For a more complete introduction to AD, see the textbook by Griewank and Walther~\cite{griewank_evaluating_2008} as a classical resource from the numerical simulation community, or a recent survey by Baydin {\em et al.}~\cite{baydin_automatic_2017} from a machine learning perspective.

\subsection{The Forward Mode of AD}\label{sec:ad-forward}
The forward mode of AD~\cite{wengert1964simple} extends each variable $a$ by a variable $\dot a$ for its partial derivative in some direction, also called a ``tangent''. This is typically achieved by introducing a new dual data type, as exemplified with the {\em ad-hoc} AD tool sketched in Fig.~\ref{fig:ad-codes-forwardtool}, which defines a new class that has a member \texttt{tangent} for $\dot a$ besides the member \texttt{primal} (the ordinary ordered computation or primal program) for $a$.  Whenever some $a$ is evaluated via an operation like $a = b_1 \cdot b_2$, $\dot a$ can be evaluated alongside according to the analytic rules of differentiation, like $\dot a = \dot b_1 \cdot b_2 + b_1 \cdot \dot b_2$. Therefore the elementary operations like \texttt{\_\_mul\_\_} have been overloaded in Fig.~\ref{fig:ad-codes-forwardtool}. Figure~\ref{fig:ad-codes-forward} shows how to use such a tool: to compute $\tfrac{\partial f}{\partial x_i}(x^{(k)})$, all $\dot x_j$ are initialized with $0$ except for $\dot x_i = 1$. Then $y=f(x^{(k)})$ is evaluated using the extended arithmetics and $\tfrac{\partial f}{\partial x_i}(x^{(k)})$ can be read from $\dot y$. The time complexity to compute the full gradient vector $\nabla_x f(x^{(k)})$ is proportional to the time complexity to evaluate $f$ times the number of input variables. Any particular directional derivative $\nabla_x f(x^{(k)})^T \cdot v$ can be computed within a time proportional to the evaluation of $f$ by initializing $\dot x_j = v_j$ for all $j$.

\subsection{The Reverse Mode of AD}\label{sec:ad-reverse}
The reverse mode of AD~\cite{linnainmaa1970representation,speelpenning1980compiling,rumelhart1985learning} consists of two phases: first, the program is executed using the ordinary arithmetic operations, but all statements are recorded, usually in a computational graph or a stack-like data structure called the \emph{tape}. After this \emph{primal} run, each primal variable $a$ is extended by an \emph{adjoint variable} $\bar a := \tfrac{\partial f}{\partial a}$. The adjoint variables are successively updated while revisiting the statements in reverse, starting from the output $y=f(x)$ and going towards the inputs $x_i$. For instance, a primal statement $a = b_1 \cdot b_2$ that forms an intermediate step of the computation of $f$ in the recording phase translates to the following updates in the reversal phase:
\begin{equation}\label{eq:ad-adjoint-update-example}
\bar b_1 \,+\!\!= \bar a \cdot b_2, \qquad \bar b_2 \,+\!\!=\bar a \cdot  b_1\,.
\end{equation}
The update of adjoints $\bar b_1$ and $\bar b_2$ in \equationautorefname~\eqref{eq:ad-adjoint-update-example} reflect the difference between the derivatives of the output $f$ with respect to the values of $b_1$ and $b_2$ prior to and after $a=b_1\cdot b_2$. The adjoint $\bar a$ accounts for the dependency of the output $f$ on the value of $a$, and is itself computed as a result of reverse propagation of adjoints from $f$ to $a$. Through the combination of forward and reverse propagations for $a = b_1 \cdot b_2$, this two-phase algorithm computes the partial derivatives of $f$ through the chain rule $\frac{\partial f}{\partial b_1} = \frac{\partial f}{\partial a} \frac{\partial a}{\partial b_1} = \bar a \frac{\partial a}{\partial b_1}$ and $\frac{\partial f}{\partial b_2} = \frac{\partial f}{\partial a} \frac{\partial a}{\partial b_2} = \bar a \frac{\partial a}{\partial b_2}$ and accumulates them in the adjoints $\bar b_1$ and $\bar b_2$ respectively.

All the adjoint variables are to be initialized with 0, except for the output variable $y=f(x)$, which has $\bar y = 1$. After the reverse pass, the adjoint input variables $\bar x_i$ contain the derivatives $\tfrac{\partial f}{\partial x_i}$. The time complexity of the reverse mode relative to the primal computation is independent of the number of input variables, making it faster than forward-mode AD or numerical differentiation when computing the gradient of a scalar-valued function with a large number of inputs variables. However, recording a tape requires a significant amount of memory. The reverse mode of AD is also significantly more difficult to implement. In Fig.~\ref{fig:ad-codes-pytorch} we differentiate a simple function using the machine learning framework PyTorch~\cite{pytorch} using the reverse mode.

\subsection{Implementation Aspects of AD}
\label{sec:ad-implementation}
Depending on the programming language, the primal program can be extended by AD arithmetics in different ways.

The most straightforward strategy is certainly to substitute any arithmetic operations by calls to an AD library that implements the additional AD arithmetics. Just like the {\em ad-hoc} forward AD tool of Fig.~\ref{fig:ad-codes-forwardtool}, many AD tools make use of polymorphism and \emph{operator overloading}~\cite{Walther2012Gsw,Hogan2014FRM,Lotz2016Hat,SaAlGauTOMS2019}. This is a feature of many contemporary programming languages, by which the compiler or interpreter automatically dispatches any calls to arithmetic operators or functions to custom implementations if one of the involved variables is of a custom type. It makes adopting AD as easy as replacing the floating-point datatype (\egc {\ttfamily double} or {\ttfamily float}) by a type from the AD library. 

Alternatively, AD arithmetics can be added by a modified or special compiler~\cite{Vassilev_Clad,NEURIPS2020_9332c513} or through \emph{source-to-source transformation} tools~\cite{Bischof1997AAE,Bischof1996AAD,utke_openad_2008,hascoet_tapenade_2013}. An extensive overview of AD tools can be consulted online~\cite{autodiffwebsite}. In some cases, the entire simulation code needs to be rewritten in the AD framework, which easily makes this approach prohibitive.

In Ref.~\cite{madjax}, AD was added to the general purpose matrix element generator MadGraph~\cite{Alwall:2014hca} by using \texttt{JAX}~\cite{jax2018github}. However, some modifications of the original code were necessary for the implementation to work;
this is very typical for complex codes, as we outline {\em infra}.

\subsection{Adaptations to the Primal Program}
\label{sec:ad-adaptations}

\begin{figure}[h!]
\centering
\begin{subfigure}[t]{0.32\textwidth}
\begin{tikzpicture}
\begin{axis}[xlabel={$x$},ylabel={$y$},ticks=none,height=4.5cm,width=4.5cm,xmin=-1.2,xmax=1.2,ymin=-0.2,ymax=1.2]
\addplot[domain=-1:1,samples=200,black,very thick]  {x*x};
\addplot[blue,domain=-1:1,samples=200]  {x*x - 0.1*x*x*x + 0.1*sin(200*x) - 0.1}; 
\end{axis}
\end{tikzpicture}
\caption{Good.}
\label{fig:badapproxderiv-a}
\end{subfigure}
\begin{subfigure}[t]{0.32\textwidth}
\begin{tikzpicture}
\begin{axis}[xlabel={$x$},ylabel={$y$},ticks=none,height=4.5cm,width=4.5cm,xmin=-1.2,xmax=1.2,ymin=-0.2,ymax=1.2]
\addplot[domain=-1:1,samples=200,black,very thick]  {x*x};
\addplot[blue,domain=-1:1,samples=200]  {0.05+floor(x*x*10)/10 }; 
\end{axis}
\end{tikzpicture}
\caption{Bad: The derivative\\ is 0, if it exists at all.}
\label{fig:badapproxderiv-b}
\end{subfigure}
\begin{subfigure}[t]{0.32\textwidth}
\begin{tikzpicture}
\begin{axis}[xlabel={$x$},ylabel={$y$},ticks=none,height=4.5cm,width=4.5cm,xmin=-1.2,xmax=1.2,ymin=-0.2,ymax=1.2]
\addplot[domain=-1:1,samples=200,black,very thick]  {x*x};
\addplot[blue,domain=-1:1,samples=600]  {x*x  + 0.05*sin(20000*x) + 0.05*cos(31400*x*x)}; 
\end{axis}
\end{tikzpicture}
\caption{Bad: The derivative is dominated by high-frequency noise. }
\label{fig:badapproxderiv-c}
\end{subfigure}
\caption{Value-wise good approximations can, but do not need to be, good derivative-wise as well. Reproduced from Ref.~\cite{aehle-derivatives-2022}.}
\label{fig:badapproxderiv}
\end{figure}
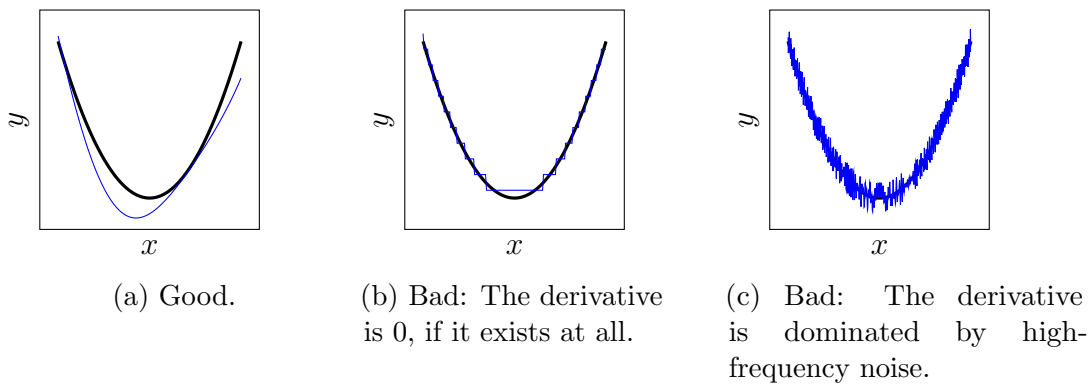

One common goal in the development of AD tools is to make their integration into an existing primal program as automatic as possible. Problem-specific adaptations of the primal program and the AD workflow can however be necessary, for various reasons such as the following:
\begin{itemize}
\item The input/output of the program must be extended to initialize and output the tangent or adjoint variables;
\item The primal program might call \emph{external functions} from \eg numerical libraries that are only available in compiled form. Here an analytic, numerical, or surrogate derivative must be provided;
\item Recording every statement might violate memory limits in reverse mode. \emph{Checkpointing}~\cite{hutchison_data-flow_2006} allows the program to re-execute parts of the program so they can be taped just before this chunk of the tape is needed. \emph{Preaccumulation} consists in immediately finding the derivative of blocks with few input and output variables, instead of recording them on the tape. Some numerical algorithms use many operations to solve a mathematically simple problem, and the knowledge of their analytic derivative can be used via external functions;
\item When differentiating shared-memory parallel code in reverse mode, shared reads in the forward run translate to concurrent writes in the adjoint updates, \equationautorefname~\eqref{eq:ad-adjoint-update-example}. While the AD tool can  use atomic updates as a general solution to prevent race conditions, the programmer may manually disable them  where the data access patterns allow~\cite{bluhdorn_event-based_2021};
\item If the primal program only approximates the objective function, there is no guarantee that the (accurate) derivative of the primal program is related to the derivative of the actual objective, as illustrated in Fig.~\ref{fig:badapproxderiv}. Manual adaptations might be necessary to ensure that we are in the case of Fig.~\ref{fig:badapproxderiv-a}.

\item Related with this caveat and independent from AD, the objective function itself can have properties that make it not amenable to gradient-based optimization, such as having many discontinuities or a very large number of local minima. In such a case, the objective function might be replaced by a surrogate model as discussed in the next section.

\end{itemize}

\subsection{Surrogate Models}
\label{sec:ad-surrogates}
Algorithmic differentiation can also be applied to a \emph{surrogate model} instead of the original objective function~\cite{Shirobokov:2020tjt,Adelmann:2022ozp}. While the latter can be a very complex code dealing with the very specific design problem at hand, a surrogate model is usually taken from a simple and generic class of functions, such as various neural network architectures commonly used in deep learning. The set of parameters (or weights) specifying the surrogate is determined by means of a fitting (or training) procedure that makes it imitate the original objective.

With a surrogate based on a deep-learning architecture, AD is immediately available within the machine learning framework used to train the surrogate. Note that the surrogate can be differentiable even if the original function is not. In addition, evaluation of the surrogate (and also its derivatives) is usually several orders of magnitude faster~\cite{kasim2020up} than the evaluation of the ``true'' model, mainly due to vectorization and access to hardware parallelism of GPUs and TPUs available in machine learning libraries. However, training the surrogate requires a substantial number of evaluations of the original function, which does scale at least with the number of design parameters. Also, a poorly trained surrogate that does not reproduce the original function well enough can introduce a bias in the subsequent analysis. We discuss several kinds of suitable neural network architectures in Sec.~\ref{sec:gen-model}.

\clearpage
\section{Problem Description and Possible Solution}
\label{sec:problem-description}

In this section we consider the problem of optimizing a customizable objective function for an instrument that employs the interaction of radiation with matter as part of its data-generating processes. The above abstract definition embraces a variety of detectors and instruments and a wealth of use cases in fundamental physics research (including high-energy particle physics, astro-particle physics, high-energy nuclear physics, and neutrino physics) as well as industrial applications ranging from hadron therapy or irradiation facilities, to muon tomography scanners for border control, geological monitoring, and archaeological prospections. 

Besides the stochastic element provided by particle interactions with matter, which is common to all, the above applications share some specific tasks and lack others. In the following we discuss separately the most important and critical of those tasks, partitioning them in such a way that their interplay may be expected to be the loosest: they may therefore lend themselves to be studied separately both in a modeling phase and in their separate initial optimization on a reduced set of parameters while fixing (``freezing'') all the others, before a global optimization loop can be proficuously carried out on the full unfrozen system of parameters, together with the other ingredients of the problem.

\subsection{Problem Statement}
\label{sec:problem-statement}

An end-to-end detector design optimization task can be briefly formalized in the following way. We start with a simulation of the physics processes of relevance for the considered application, which generates a multi-dimensional, stochastic input variable $x$, distributed with a probability density function (PDF) $f(x)$. The input is turned by the simulation of the detection apparatus into sensor readouts $z$ distributed with a PDF $p(z|x,\theta)$, which constitute the observed features of the physical process; readouts $z$ depend through $p(\cdot)$ on parameters $\theta$ that describe the physical properties of the detector and its geometry. Note that in general $f(x)$, and consequently $z$, also depend on other latent features---parameters that characterize the underlying physics process and that may not be known precisely. These latent features constitute an additional potential source of systematic uncertainty in the measurement task in addition to the detector-related uncertainties we discuss here; we ignore their existence in the following simplified treatment, noting that the inclusion of their effect in the problem is comparatively straightforward in most cases. 

The observations $z$ are used by a reconstruction model $R(\cdot)$ that produces high-level features,
\begin{equation}
\zeta(\theta) = R[z,\theta,\nu(\theta)]
\end{equation}
(\eg particle four-momenta, in a collider application), by employing knowledge of the detector parameters as well as a model of the detector-driven nuisance parameters $\nu(\theta)$ that affect the pattern recognition task. In turn, high-level features $\zeta(\theta)$ constitute the input of the data analysis step: this is a further dimensionality-reduction task, typically performed by a classifier or regressor $A(\cdot)$ powered by a neural network (NN). Once properly trained for the task at hand, the NN produces a low-dimensional summary statistic $s = A[\zeta(\theta)]$ with which inference can finally be carried out to produce the desired goal of the experiment. Suitable optimization metrics derivation follows from that final step: \egc the power $1-\beta(s)$ of a hypothesis test on the presence of smuggled material in a container, if we are discussing muon tomography for border control (see {\em infra}, Sec.~\ref{sec:muography-fom}); or the total uncertainty in the cross section of production of a new particle, as a simplified proxy to the experimental goals of a collider detector use case.

In general, one may formally specify the problem of identifying optimal detector parameters as that of finding estimators $\hat{\theta}$  that satisfy
\begin{equation}
\hat{\theta} = arg~min_{\theta} \int L[A(\zeta),c(\theta)] p(z | x,\theta) f(x) dx dz\;,
\end{equation}
where we omitted for simplicity to specify the dependence of the high-level features $\zeta$ on $z$ and the nuisances $\nu$. Moreover, in the solution above $c(\theta)$ is a function modeling the cost of the considered detector layout of parameters $\theta$, and the loss function $L[A,c]$ is constructed to appropriately weight the result of the measurement in terms of its desirable goals, as well as to obey cost constraints and other use-case-specific limitations. For example, for the aforementioned search of high-atomic-number material smuggled in a container, one might write:
\begin{equation}
  L=(1+e^{(k(c-c_0))} ) \sum_Z[ u(Z) w_{50} [s(Z)]]\;,
\end{equation}
where $k$ is an external parameter describing the importance of preventing cost $c$ from exceeding a given budget $c_0$, $u(Z)$ is defined {\em a priori} to weigh the relative importance of successfully detecting material of atomic mass $Z$ in a given benchmark search case involving a set of possible $Z$ values, and $w_{50}[s(Z)]$ is the mass of concealed material for which the power $1-\beta(s)$ to accept the alternative hypothesis (that there is concealed material) in a test at a fixed type-I error rate $\alpha$ (\egc 0.05) is 50\%,
\begin{equation}
    w_{50} [s(Z)] = \beta^{-1} (0.5)\;.
\end{equation}
Since in the cases of interest the PDF $p(z|x,\theta)$ is not available in closed form (as the considered models are implicit), we must rely on forward simulation to sample from it. The problem is solved by approximating $\hat{\theta}$ with
\begin{equation}
\hat {\theta}_a = arg~min_{\theta} ~\frac{1}{n}\sum_{i=1}^n L[A(R(z_i )),c(\theta)]\;,
\end{equation}
where $z_i$ is distributed as $F(x_i,\theta)$ to emulate $p(z|x,\theta)$, as $x_i$ is sampled from its PDF $f(\cdot)$ by the simulator. One may thus obtain an estimate of the loss function and the detector parameters that minimize it. 

In order to cast the problem formulated above in a differentiable framework, which makes it possible to search for optimal solutions by gradient descent, it has been demonstrated how it is viable to approximate the non-differentiable stochastic simulator $F(\cdot)$ with a local surrogate model, $z = S(y,x,\theta)$, that depends on a parameter $y$ describing the stochastic variation of the approximated distribution~\cite{Shirobokov:2020tjt}. This allows to descend to the minimum of the approximated loss $\hat{L(z)}$  by following its surrogate gradient,
\begin{equation}
\nabla_\theta \hat{(L(z))} = \frac{1}{n} \sum_{i=1}^n \nabla_\theta L[A(R(S(y_i,x_i,\theta))),c(\theta)]\;.
\end{equation}
The above recipe requires one to learn the differentiable surrogate $S(\cdot)$: this task can be carried out independently from the optimization procedure.

It should transpire from the above succinct description that the components of the final optimization goal are sufficiently decoupled from one another to allow for modular solutions. So, \egc the development of a detailed model of event reconstruction that closely matches present-day capabilities in advanced pattern recognition tools, and that allows to obtain high-level features $\zeta = R[z,\theta,\nu(\theta)]$ from the detector outputs $z$ (or approximations thereof, learned by the generative model), can be performed by a separate learning task and then incorporated in the architecture. A similar note concerns the triggering and data acquisition parts of the detection apparatus, most relevant to HEP applications: while the online identification and triggering of processes of interest constitute a valid subject for an independent optimization task, they may be incorporated in a simplified way as an independent modeling block with similar techniques to those describing pattern recognition and offline event reconstruction; the output may then be encoded into a set of efficiency maps $\epsilon(z,\theta)$, and the latter included as collected events weights in the global optimization task. 

\subsection{Modeling of Particle Detectors}
\label{sec:modeling}

Several tasks of widely varying complexity fall within the scope of the modeling of particle interactions with matter. These range from the simple propagation of individual particles subjected to multiple scattering, which is relevant {\em e.g.} to muon tomography applications (as discussed in Sec.~\ref{sec:muography}), or the similarly straightforward modeling of charge deposition and collection in silicon strip sensors (which was done in a study of MUonE, see Sec.~\ref{sec:muone}), to the enormously complex description of hadronic showers in a calorimeter for a collider detector (Sec.~\ref{sec:calo-lhcb}), or the modeling of beam-induced backgrounds in a detector for a high-intensity muon collider (Sec.~\ref{sec:calo-muoncoll}).

\begin{center}
\begin{figure}[!ht]
\begin{minipage}{0.62\linewidth}
\includegraphics[width=8.5cm]{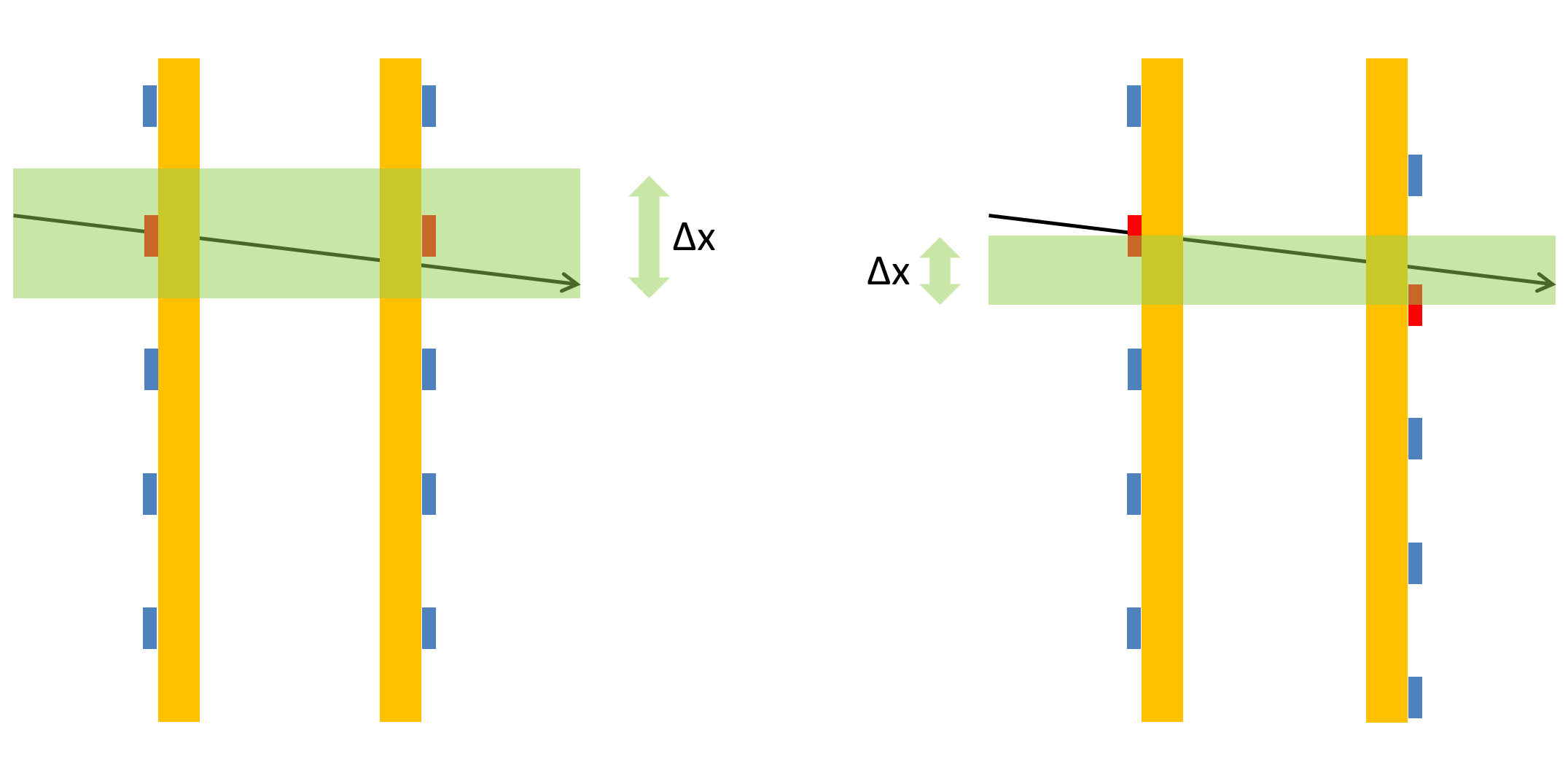}
\end{minipage}
\begin{minipage}{0.37\linewidth}
\includegraphics[width=6.5cm]{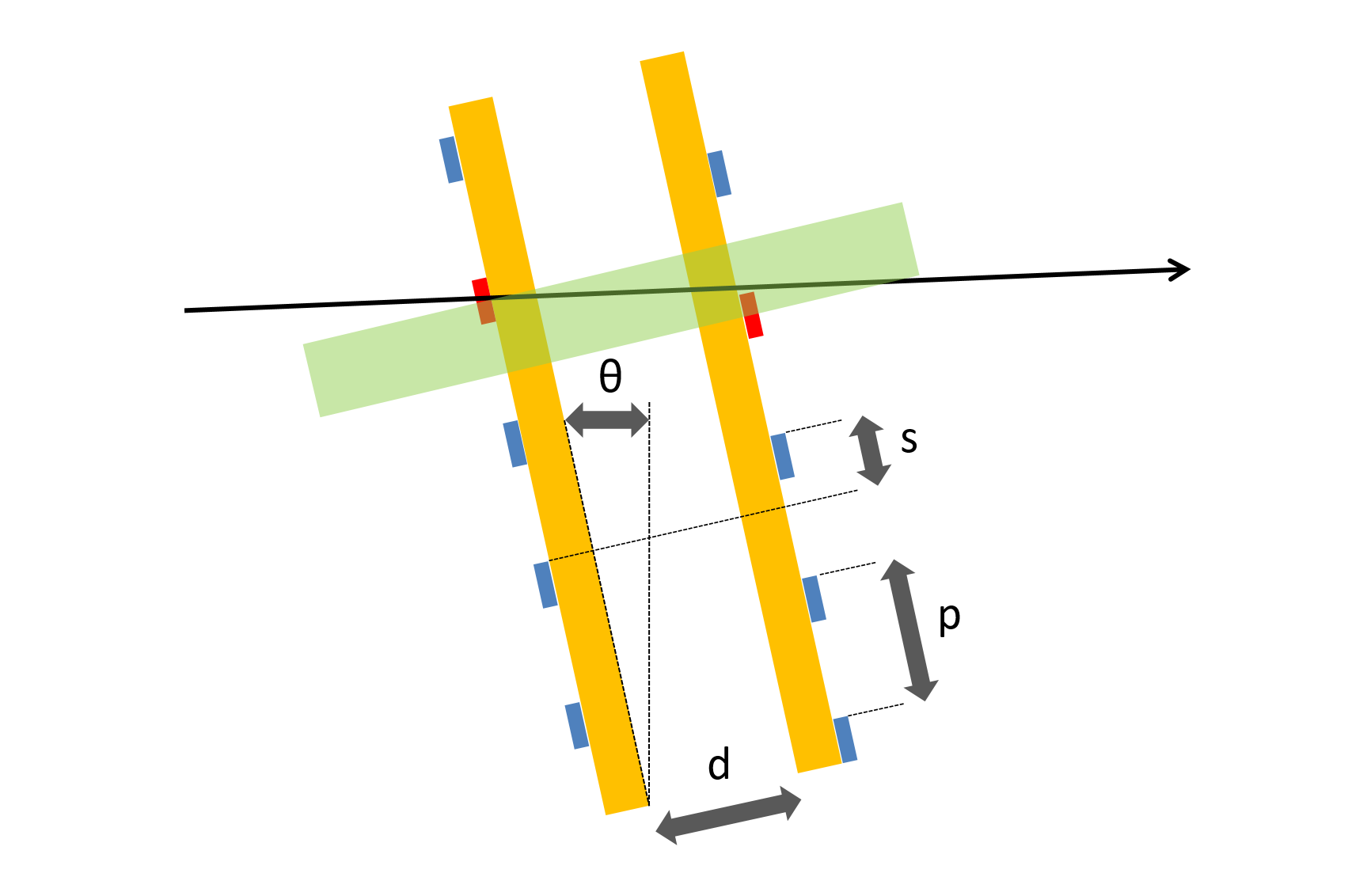}
\end{minipage}
\caption{Left and center: a double-sided silicon strip sensor produces twice smaller resolution $\Delta x$ on single-strip hit position for an orthogonally incident particle if strips on the two sides are staggered by half the strip pitch. Right: the four parameters affecting single-strip hit position resolution (tilt angle $\theta$, strip pitch $p$, sensor distance $d$, staggering $s$).} 
\label{f:twosidedsensors}
\end{figure}
\end{center}

\subsubsection{Modeling the Geometry of Detection Elements}

A continuous model of the layout of detection elements is straightforward to construct for most applications, and the methods to do so are rather general. Indeed, for most of the use cases of our interest, a selection of which is provided in Sec.~\ref{sec:exemplary_use_cases} {\em infra}, we may employ solutions similar to those of existing implementations developed for full~\cite{GEANT,allison_geant4_2006,allison_recent_2016,FLUKA} or fast~\cite{Delphes} detector simulations. The challenge is however to identify what other parts of the global pipeline are most affected by variations of the geometry of the instrument, such that we may choose parameterizations that allow a simpler functional mapping of the geometry parameters into those dependent features. An example will clarify this point.

Consider a set of silicon tracking layers made up of two strip sensors glued back-to-back, such as those that constitute the MUonE detector (described in Sec.~\ref{sec:muone}). Almost any relevant figure of merit for such a detector is heavily dependent on the resolution of the track parameters, which may be extracted from a fit to the reconstructed hit positions. The latter will in turn be subjected to systematic uncertainties due to positioning and alignment of the sensors: there is thus a clear path through which these systematic uncertainties affect the utility function. For silicon strip sensors, the root-mean-square hit resolution on the coordinate orthogonal to the strips equals $p/\sqrt{12}$, where $p$ is the pitch, if the particle ionization causes the system to detect a single-strip hit~\footnote{ Estimated single-strip hit positions distribute with a uniform density function, which in a differentiable study can easily be parameterized by using a constant connected to sigmoid functions at the extrema.}; however, a significant resolution gain is obtained if particles create multiple-strip hits, as charge sharing then provides added information on the track intercept. For particles moving along the beam axis (here taken as the $z$ coordinate) or very close to it (the vast majority in MUonE, and also those that are most relevant to extract the final measurement of high-$q^2$ scattering cross section, which is the goal of the experiment), the fraction of multiple-strip hits has a functional dependence on four geometry parameters: pitch, $z$-spacing of the glued sensors, staggering of strips on one sensor relative to the other, and (if allowed) tilt angle of the sensors with respect to the $z$ axis, as illustrated in Fig.~\ref{f:twosidedsensors}. An optimization study of the placement of those detection elements which varied those four variables independently would waste significant resources to investigate their functional dependence at resolution minimum, which may be instead determined in isolation from the rest of the problem (and potentially more accurately) through a dedicated full simulation, and thus inform a constrained parameterization that keeps the four parameters bounded within the subspace of optimal charge-sharing conditions. The smaller-dimensional description would also allow a simpler tracing of position and alignment-related systematic uncertainties. This example shows how domain knowledge---in this case, insights in the operation of a detector along with the inference extraction procedures applied to its output---may help find the most suitable data representations for an optimization task, and how this may be beneficial on the whole.

Surrogates based on deep generative models learn the distribution of data (either explicitly or implicitly, see below) and sample new events from this distribution. Data either come from experiment or from dedicated simulations, for example using GEANT4~\cite{GEANT,allison_geant4_2006,allison_recent_2016} or other alternatives~\cite{Sinha:2022ogd}, depending on the specific case at hand. Energy depositions in the continuous space of the detector are then discretized into voxels representing read-out channels, forming the feature space $z$ in which training is carried out. The number of these voxels, usually of the order $\mathcal{O}$(100-1000), should be chosen to be as large as possible to have fine-grained description of the detector module at hand. However, deep generative models do not work with an arbitrarily high number of dimensions, so the number of voxels cannot be too large. Current state-of-the-art calorimeter shower simulations uses at most $\mathcal{O}(10^4)$ voxel: Ref.~\cite{Belayneh:2019vyx} uses 65k voxels and Ref.~\cite{Buhmann:2020pmy} uses 27k voxels. 

In order to be differentiable in the additional parameters $\theta$ that describe the physical properties of the detector, its geometry, or other features of interest for the optimization task, the models have to be trained with $\theta$ as a conditional label~\cite{Baldi:2016fzo}. 
Examples for $\theta$ are discussed in Sec.~\ref{sec:calo-lhcb} and include thickness, position, angle, or size of detector modules (as long as the number of voxels remains constant); tuning  parameters of GEANT4; or atomic number $Z$ of the detector material. Conversely, there are a few parameters that cannot be studied. Since the dimensionality of the feature space is hardcoded into the models and given by the training data, the number of voxels is fixed.  
The resulting augmented dataset $(z, \theta)$ is then used to train one of the deep generative models described in the next subsection.

\subsubsection{Learning the Simulated Particle Interactions with Matter through Surrogates}
\label{sec:gen-model}

The models which we will consider for learning the datasets mentioned {\em supra} strongly depend on the situation at hand. As an illustration, we focus on modeling a particle shower produced by an energetic particle in a calorimeter for a collider detector with deep generative models. Based on the similarity of voxelized calorimeter showers and pixelated digital images, all deep generative models used to generate artificial images can also be used to learn calorimeter showers, albeit some modifications improve the quality of the samples. Note that the task is always to reproduce the shower of a single incoming particle in a given submodule of a larger detector. Such showers are statistically independent from each other and several of them can be combined into the simulation of a more complicated, single event.

Current, state-of-the-art deep generative models are based on one of the following three main architectures: Variational AutoEncoders (VAEs)~\cite{2013arXiv1312.6114K,2014arXiv1401.4082J,2019arXiv190602691K}, Generative Adversarial Networks (GANs)~\cite{2014arXiv1406.2661G,2020arXiv200106937G}, and Normalizing Flows (NFs)~\cite{2015arXiv150505770J,2019arXiv190809257K,2019arXiv191202762P}. Each of them has its own advantages and disadvantages, ranging from memory footprint, to training time and stability, to sampling time, and sampling quality. The best surrogate model is one that is both fast and faithful, however, in practice there is a trade-off between these metrics.

VAEs~\cite{2013arXiv1312.6114K,2014arXiv1401.4082J,2019arXiv190602691K} build upon simple autoencoder (AE) structures. AEs consist of two neural networks, an encoder $e$ and a decoder $d$. The encoder maps the $n$-dimensional data to an $m$-dimensional latent space, where usually $m\ll n$. The decoder maps the latent vector back into data space, making the entire AE architecture (the combination of encoder and decoder) a bottleneck. Training of an AE is done by minimizing the reconstruction loss between the input data $x$ and the encoder-decoder-transformed data: $d(e(x))$. By randomly sampling points in the latent space and passing them through the decoder, the architecture becomes a generative model. In a perfect AE, the NN learns a lossless compression of data to latent space. This, however, can become problematic if the NN overfits and simply remembers a one-to-one mapping between data and latent space. In this case, the AE performs poorly as generative model, as the latent space only remembers the data and newly-sampled points confuse the decoder. A VAE avoids these issues by giving structure to the latent space, thereby regularizing it. The decoder now maps the input to $2m$ numbers, that are understood as mean and standard deviation of an $m$-dimensional, multivariate Gaussian. A pass through the decoder-encoder chain now involves a sampling step in the latent space. The loss function of the VAE consists of two terms: the first one is the reconstruction error of the AE architecture, the second one is a Kullback--Leibler (KL) divergence~\cite{kullback1951information} that compares the latent space distribution to an $m$-dimensional standard normal distribution. A relative weight between the two terms emphasizes a more regular latent space, \ie a smooth morphing of one data point into another through a trajectory in the latent space, or a higher sample quality. Sampling is usually fast, as generation only involves a single pass through the decoder network. Without further modifications or post-processing, samples from VAEs usually have comparatively low quality.
This setup, including modifications and post-processing, has been considered for calorimeter simulations in high-energy physics~\cite{ATL-SOFT-PUB-2018-001,Buhmann:2020pmy,2020arXiv200606704D,Buhmann:2021lxj,Howard:2021pos,Buhmann:2021caf,Hariri:2021clz}.

GANs~\cite{2014arXiv1406.2661G,2020arXiv200106937G} train two NNs, a generator and a discriminator, in an adversarial objective. The generator tries to generate realistically-looking images while the discriminator tries to  separate them from real ones. The resulting saddle-point optimization objective is harder to train, resulting sometimes in mode-collapse and artifacts in samples. A significant improvement in performance can be obtained by using a Wasserstein GAN (WGAN)~\cite{2017arXiv170107875A,2017arXiv170400028G}. In this case, the discriminator is replaced by a critic that evaluates the Wasserstein distance of a given generated sample from the training data. Model selection and evaluation of GANs is a difficult task~\cite{2018arXiv180203446B}, as the loss value of the critic in training does not always correlate with the sampling quality. Once trained well, GANs generate realistically-looking samples across various domains. Hence, there have been several applications of GANs to calorimeter shower simulation~\cite{deOliveira:2017pjk,Paganini:2017hrr,Paganini:2017dwg,Bellagente:2019uyp,Vallecorsa:2019ked,SHiP:2019gcl,Chekalina:2018hxi,ATL-SOFT-PUB-2018-001,Carminati:2018khv,Vallecorsa:2018zco,Musella:2018rdi,Erdmann:2018kuh,Deja:2019vcv,Derkach:2019qfk,Erdmann:2018jxd,Oliveira:DLPS2017,deOliveira:2017rwa,Hooberman:DLPS2017,Belayneh:2019vyx,Buhmann:2020pmy,2009.03796,Maevskiy:2020ank,Rehm:2021zoz,Rehm:2021qwm,Kansal:2021cqp,Khattak:2021ndw,Anderlini:2021qpm,Buhmann:2021caf}.

NFs~\cite{2015arXiv150505770J,2019arXiv190809257K,2019arXiv191202762P} learn a bijective mapping between two distributions, which are usually the (complicated) data distribution on one side and a (computationally easy) base distribution---such as a standard normal distribution---on the other side. The bijector does not only provide the coordinate transformation between the two distributions, but also the Jacobian of the transformation, making NFs suitable for many different applications: when points from the data-space are mapped to the base distribution, the Jacobian of the transformation, together with the probabilities of the points under the base distribution, give a probability of the original points, thus making NFs density estimators. In the inverse direction, noise generated under the base distribution can be mapped to data-space: the NF then acts as a generative model. Since the log-likelihood of the data points is available by construction, NFs can be trained by minimizing the negative log-likelihood. The resulting training is usually very stable and gives a reliable density estimator, if the NF is expressive enough. Such a model, when used as a generative model, does not suffer from mode collapse. In addition, the log-likelihood does provide a good metric for model selection that directly correlates with the quality of the fit to the data distribution. Requiring the transformations to have an analytic inverse and tractable Jacobian puts constraints on the NN architecture realizing them. State-of-the-art bijectors consist of a series of spline-based transformations~\cite{2019arXiv190604032D}, the parameters or which are predicted by NNs. To further ensure that the Jacobian can be evaluated in linear time, the architectures must be either bipartite~\cite{2016arXiv160508803D} or autoregressive~\cite{2017arXiv170507057P,2016arXiv160604934K}. The former run equally fast in both directions (density estimation for training and sampling generation for application), the latter strongly favor one direction over the other. Masked Autoregressive Flows (MAFs)~\cite{2017arXiv170507057P} are fast in density estimation, but slower in sampling by a factor $d$, given by the dimension of the space to be learned. Inverse Autoregressive Flows (IAFs)~\cite{2016arXiv160604934K} are fast in sampling, but a factor $d$ slower in estimating the density of data points. Only recently NFs (based on MAF and IAF architectures) have been applied to calorimeter shower simulation~\cite{Krause:2021ilc,Krause:2021wez}, surpassing the quality of showers generated by an older GAN trained on the same dataset~\cite{Paganini:2017hrr,Paganini:2017dwg}. 

Modeling calorimeter showers with deep generative models is a very active area of research with considerable interest of the community, as can be seen at the ``Fast Calorimeter Simulation Challenge 2022''~\cite{CaloChallenge}.

\subsection{Modeling of Pattern Recognition and Event Reconstruction Procedures}
While the simulation of multiple particles with matter can be factorised on a particle-by-particle basis and the corresponding deposits can be superimposed, the subsequent reconstruction of their patterns can be highly affected by correlations between particles, \eg through spatial overlap. These correlations can in principle even span over whole detectors or sub-detectors. 
This poses more stringent requirements on the resources to model this step.

While typical detectors or sub-detectors consist of hundreds to millions of individual sensors to achieve the resolution needed to measure the quantities of primary interest, the latter usually belong to a set orders of magnitude smaller than the number of detector inputs. Therefore, the deposits in sensors left by particles traversing the detector or being stopped by detector elements (hits) are used to reconstruct those very same particles as so-called physics objects, which relate more directly to the quantity of primary interest, such as tracks in tracking detectors, or full particle candidates in particle-flow algorithms~\cite{Arbor,Pandora_1,CLICPF_1,Pandora,Pandora_kit,Sefkow_2016,Tran_2017,Evans_2008,CMSPFPaper,ATLASPF}. This concept of pattern recognition has proven to be very powerful, however most algorithms suffer from inherent limitations with respect to their differentiability, as they rely heavily on seeding mechanisms that select a certain area or point of interest as starting point for a reconstruction algorithm based on a certain particle type assumption. Then, the reconstruction is subsequently refined in steps, each individually optimized and coming with its own selection thresholds. As such, this reconstruction procedure contains many non differentiable steps, and furthermore does not generalize easily if the detector geometry or the individual sensor properties are changed. Therefore, these algorithms can introduce biases towards those detector designs they were originally developed for, which makes them not applicable for a generic differentiable detector design optimization.

For a truly generic reduction of dimensionality from hits to physics objects, the algorithm needs to be geometry agnostic, be differentiable, encode generic physics considerations, and adhere to the given computing resources. Machine-learning algorithms can offer this flexibility as they are typically differentiable by construction, and can also easily adapt to changing conditions. However, neither those algorithms that rely on a regular geometry (such as convolutional neural network based approaches~\cite{lecun1998gradient,Bols:2020bkb,TopTaggers,FCChh-CDR,aleksa2019calorimeters}), nor algorithms based on dense neural networks alone that make no assumptions at all on the structure of the problem, are applicable: the former cannot generalize to irregular geometries, and the latter cannot fit resource constraints. In addition, recurrent algorithms cannot be made really generic, as conceptually they rely on a certain ordering of the inputs. 

The only viable option to date to handle this high-dimensional and irregular input space are graph neural networks~\cite{scarselli2009graph} (GNN). Compared to other classes of neural networks, GNNs are a quite recent development, but have already proven to be very powerful, also in the physics domain~\cite{GravNet,JEDINet,ParticleNet,HEPTrkX2,Qasim:2021hex,Shlomi:2020gdn}. GNNs do not require a certain type of input ordering or regularity; they solely rely on a set of input points that can represent the detector hits, and connections between them. These connections can simply be chosen as nearest neighbours in a physical or a latent space. This procedure has two advantages: it encodes a notion of locality directly into the network architecture, and therefore maps the physics of locally propagating particles through the detector, and it also comes with advantages in terms of computing resources, since the number of operations to be performed does not grow quadratically with the number of inputs anymore, as it is the case for a dense neural network approach. The exception are fully connected graph neural networks, which might be applicable for a small number of inputs, but cannot conceptually work for more complex detectors. Such neural network architectures have already shown their potential for tracking~\cite{HEPTrkX2} and calorimetry~\cite{GravNet,Qasim:2021hex}.

However, none of the neural network architectures themselves perform a reduction of dimensionality; the dimensionality reduction is performed in a second step. First approaches use edge classifiers, which learn a score that determines whether or not a connection between two neighbouring hits corresponds to a connection between hits of the same object, and then the objects get segmented by following edges with scores above a certain threshold~\cite{HEPTrkX2}. The object properties are then derived in a second step based on the selected hits. This is a natural way to approach a tracking problem, with clearly separable objects, but the paradigm breaks with larger overlaps, and when object sizes are similar to the spatial detector resolution, \eg in calorimeters. The object condensation approach has been recently proposed as a way to overcome this limitation~\cite{Kieseler:2020wcq}. It is being used already for reconstruction in the CMS HGCAL~\cite{HGCAL-TDR,Qasim:2021hex} and machine-learning driven particle flow in CMS~\cite{Pata:2021oez}. Here, the object properties are directly accumulated in representative so-called condensation points, and a simple clustering in a learned clustering space resolves ambiguities and performs the dimensionality reduction, without the need for full segmentation. This is done by collecting hits around points with a large condensation score in the learned clustering space.\footnote{In a fully differentiable pipeline one could omit the ambiguity, by resolving clustering and feeding the points with high condensation score directly to subsequent steps.} 

Neither the graph neural networks nor the training procedures discussed above make any conceptual distinction between hits in different detector systems. Therefore, these approaches also provide a basis for investigating designs that break with existing paradigms such as hybrid calorimeters as discussed in Sec.~\ref{sec:calo-fcc}. 

Once the patterns of individual particles are identified using information from the whole detector, and the information is reduced in dimensionality by orders of magnitude, it can be either passed on directly to information-extraction procedures, or a more classical approach can be taken \eg by clustering jets and deriving higher-level quantities such as missing (transverse) momentum in high-energy physics experiments. The latter being a simple 4-vector sum can trivially be part of a fully differentiable pipeline, while the jet clustering incorporates non-differentiable assignments of a particle to either one jet or the other by introducing cut-off parameters. However, while the assignment is not differentiable, gradients for all particle properties can be passed through the calculation of the final jet properties. Moreover, by introducing weights in the clustering, it is also possible to equip the cut-off parameters off the jet clustering themselves with effective gradients. Therefore, the full pipeline from hits to higher-level inputs to the information-extraction procedures used to determine the final quantities of interest can be made differentiable.

\subsection{Modeling of Information-Extraction Procedures and Detector-Related Biases}
A crucial ingredient in the global optimization task for a measuring instrument is a precise model of the conversion of high-level information (produced by the pattern recognition and event reconstruction tasks described {\em supra}) into the summary statistics which either are by themselves the final product of the measurement procedure, or constitute the direct input for its extraction. Here the problem
displays a new layer of complexity, as the design choices of a measuring instrument have implications on the existence of biases and imprecision in the measurements, which can be only partly be corrected by calibration procedures or detailed simulation studies. The resulting uncertainties propagate directly into a worsening of the final performance, and must therefore be included in an optimization pipeline. In this section we discuss how those effects can be tamed by methods that themselves employ differentiable programming solutions. The inclusion of these inference extraction procedures in the optimization task can thus not only properly account for the impact of design choices on the final metrics, but also allow for an alignment of the optimization of the whole system with the most performing inference strategies.

\subsubsection{Systematic-Aware Summary Statistics}

In the last decade, classification and regression models have become very popular in High Energy Physics (HEP) to construct powerful summary statistics that are used for inference. This is largely due to the presence of high-fidelity simulators that provide us with accurate models of underlying physical processes, which we can use as training data. However, standard machine learning-driven loss functions become misaligned with respect to physics goals when the simulated observations come with a notion of systematic uncertainty. Given this, it is then desirable to seek an objective function that is ``systematics-aware", which can optimize any set of free analysis parameters, including learnable components like neural networks.

This issue has been first addressed by the \textsc{INFERNO} algorithm~\cite{DeCastro:2018psv}, that aims at directly minimizing the expected variance of the parameter of interest (POI), accounting for the effect of relevant nuisance parameters.
\begin{figure}[ht]
  \centerline{\includegraphics[width=1.\textwidth]{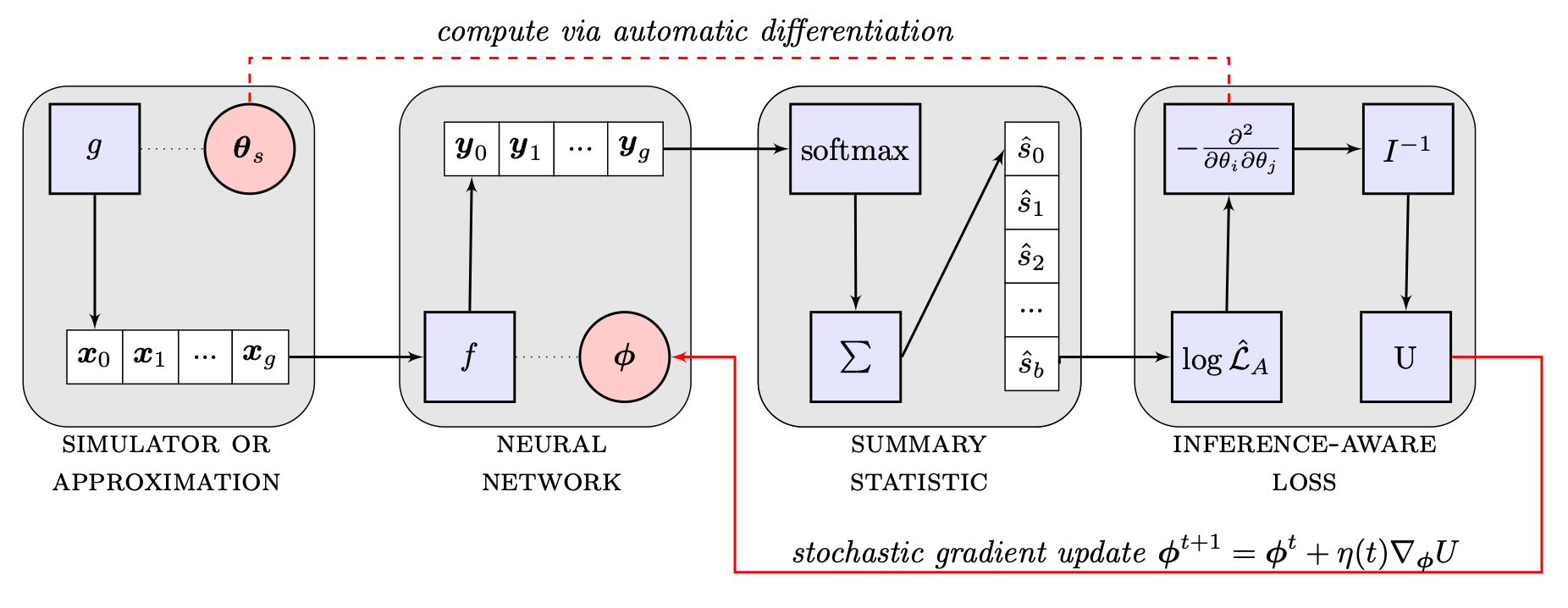}}
  \caption{Sketch of the INFERNO algorithm. Batches from a simulator are passed through a neural network and a differentiable summary statistic is constructed that allows to calculate the variance of the POI. The parameters of the network are then updated by stochastic gradient descent. The figure is reproduced from Ref.~\cite{DeCastro:2018psv}.}
  \label{inferno_sketch}
\end{figure}
The parameters of a neural network are optimized by stochastic gradient descent via automatic differentiation. 
A sketch of the \textsc{INFERNO} algorithm is shown in Fig.~\ref{inferno_sketch}.
An inference-aware summary statistic is learnt by optimizing the parameters $\boldsymbol{\phi}$ of a neural network $f$ in order to reduce the dimensionality of the input data $\boldsymbol{x}$:
\begin{equation}
f(\boldsymbol{x} ; \boldsymbol{\phi}):  \mathbb{R}^{d} \rightarrow \mathbb{R}^{b} ~.
\end{equation}
The network is trained with batches of simulated samples $G_{s}$ obtained from a simulator $g$ with parameters $\boldsymbol{\theta}_{s}$. The number of nodes in the last layer of the network determines the dimension $b$ of the summary statistic. Since histograms are not differentiable, the original algorithm uses a softmax function as a differentiable approximation for the neural network output $y$:
\begin{equation}
\hat{s}_{i}(\boldsymbol{x} ; \boldsymbol{\phi})=\sum_{x} \frac{e^{f_{i}(\boldsymbol{x} ; \phi) / \tau}}{\sum_{j=0}^{b} e^{f_{j}(\boldsymbol{x} ; \boldsymbol{\phi}) / \tau}}
\end{equation}
where the temperature hyper-parameter $\tau$ regulates the softness of the operator. In the limit of $\tau \rightarrow 0^{+}$, the probability of the largest component will tend to $1$ while others to $0$. With this approximation it is possible to construct a summary statistic for each batch by computing the Asimov Poisson-count likelihood $\hat{\mathcal{L}}_{A}$:
\begin{equation}
\hat{\mathcal{L}}_{A}(\boldsymbol{\theta} ; \boldsymbol{\phi})=\prod_{i=0}^{b} \operatorname{Pois}\left(\hat{s}_{i}\left(G_{s} ; \boldsymbol{\phi}\right) \mid \hat{s}_{i}\left(G_{s} ; \boldsymbol{\phi}\right)\right)~.
\end{equation}
where the HEP-jargon term ``Asimov" means that the value of $\hat{\mathcal{L}}_A$ is computed with the expected values based on the simulated samples $G_{s}$, such that the maximum likelihood estimator for the Asimov likelihood is the parameter vector $\boldsymbol{\theta}_{s}$ used to generate the simulated dataset $G_{s}$. In Statistics this is also referred to as a saturated model.
From the Asimov likelihood the Fisher information matrix is then calculated via automatic differentiation according to:
\begin{equation}
\boldsymbol{I}(\boldsymbol{\theta})_{i j}=\frac{\partial^{2}}{\partial \theta_{i} \partial \theta_{j}}\left(-\log \hat{\mathcal{L}}_{A}(\boldsymbol{\theta} ; \boldsymbol{\phi})\right)~.
\end{equation}
The covariance matrix can be estimated from the inverse of the Fisher information matrix if $\hat{\boldsymbol{\theta}}$ is an unbiased estimator of the values of $\boldsymbol{\theta}$:
\begin{equation}
\operatorname{cov}_{\boldsymbol{\theta}}(\hat{\boldsymbol{\theta}}) \geq I(\boldsymbol{\theta})^{-1} ~.
\end{equation}
It is also possible to include auxiliary measurements that constrain the nuisance parameters, characterized by likelihoods $\left\{\mathcal{L}_{C}^{0}(\boldsymbol{\theta}), \ldots, \mathcal{L}_{C}^{c}(\boldsymbol{\theta})\right\}$,
by considering the augmented likelihood $\hat{\mathcal{L}}_{A}^{\prime}$:
\begin{equation}
\hat{\mathcal{L}}_{A}^{\prime}(\boldsymbol{\theta} ; \boldsymbol{\phi})=\hat{\mathcal{L}}_{A}(\boldsymbol{\theta} ; \boldsymbol{\phi}) \prod_{i=0}^{c} \mathcal{L}_{C}^{i}(\boldsymbol{\theta}) ~.
\end{equation}
The loss function used for the optimization of the neural network parameters $\boldsymbol{\phi}$ can be any function of the inverse of the Fisher information matrix at $\boldsymbol{\theta}_{s}$, depending on the concrete inference problem. 
The diagonal elements $I_{i i}^{-1}\left(\boldsymbol{\theta}_{s}\right)$ correspond to the expected variance for the parameter $\theta_{i}$. Thus, if the aim is efficient inference about one of the parameters $\omega_{0}=\theta_{k}$ a possible loss function is:
\begin{equation}
U=I_{k k}^{-1}\left(\boldsymbol{\theta}_{s}\right)
\end{equation}
which corresponds to the expected width of the confidence interval for  
$\omega_{0}$ accounting also for the effect of the other nuisance parameters in $\boldsymbol{\theta}$.
The algorithm performance was originally studied with a synthetic example inspired by a typical cross section measurement. In that setup it was shown that the confidence intervals obtained using \textsc{INFERNO}-based summary statistics result narrower than those using binary classification and tend to be closer to those expected when using the true model likelihood for inference. The improvement over binary classification increases when more nuisance parameters are considered. Recently INFERNO has been re-implemented in \texttt{pytorch}, allowing for its use in a drop-in fashion~\cite{gilesstrong_2021_5040810} and enabling the use of approximately differentiable histograms with sigmoid functions. By making use of interpolation techniques, and a suitable preprocessing of the systematic variations, the INFERNO algorithm has been extended to run with realistic HEP-like systematics. First studies of the algorithm with real data based on CMS Open Data indicate that the algorithm is able to mitigate the impact of systematic uncertainties also in realistic LHC analysis.

More recently, several methods have been developed that build upon the ideas behind INFERNO. A promising approach is taken by the authors of \neos{} that aims at directly optimizing the expected sensitivity of an analysis. This also serves to minimize the uncertainty on the maximum likelihood estimate in a different fashion compared to INFERNO, since the sensitivity is calculated from the distribution of a test statistic that is monotonically related to the MLE, as shown in Ref.~\cite{asymptotics}. To approximate histograms, \neos{} uses a binned version of a kernel density estimate (KDE), where the smoothness is determined by the bandwidth parameter of the KDE. To get a differentiable analysis sensitivity as targeted by \neos{}, one can leverage fixed-point \textit{implicit differentiation} in order to calculate expressions for the gradients of maximum likelihood estimates, as found in the profile likelihood. In particular, these gradients only involve one update step of the optimization procedure, and do not require the costly unrolling of the entire optimization loop during back-propagation.  \neos{} takes advantage of recent progress in the \texttt{pyhf} package, which implements full HistFactory likelihoods and their inference using many different automatic differentiation backends, \eg \texttt{tensorflow}, \texttt{pytorch}, and \texttt{JAX}~\cite{pyhf, pyhf2}. 

Despite their close similarity, divisions clearly exist in the implementations of INFERNO and \neos{} from a software standpoint. To this end, work is being done on a library that serves as a toolbox of drop-in differentiable operations that are designed for use in this kind of workflow, called \texttt{relaxed}~\cite{relaxed}. This will facilitate design of a pipeline with easily exchangeable components, allowing for more flexibility in algorithm design, and easier replicability through a unified implementation. Moreover, both INFERNO and \neos{} have been separately implemented with this library, and one can trivially combine elements of each approach to reach a potentially more optimal analysis pipeline.

In the context of a complete detector optimization task, the integration of a differentiable analysis workflow in a complete pipeline as above may be tricky. A way to perform this in practice is to simplify the problem, by considering the inference extraction task as one to first order independent on the system's parameters, and to independently optimize the inference step assuming frozen values for those geometry and detector-related parameters (such as, \egc calibration errors, imperfect efficiency maps, alignment and positioning accuracy) which introduce potential imperfections in the model. The trained dimensionality reduction model may then be integrated in the global pipeline, and only updated when significant changes occur to the value of parameters to which the model is most sensitive. Future studies are needed to test the most advantageous ways to include the inference extraction step in an end-to-end optimization task. 

\subsection {Modeling the Cost of Detectors}
\label{sec:costs}


Monetary cost plays a key role in the conception of any detector and acts as a major constraint in terms of technology and design choices. In this context, detector optimization cannot rely exclusively on physics performance features such as resolution or efficiency. Along with case-specific technical constraints, construction cost has to be implemented in the loss function (see Sec.~\ref{sec:problem-statement}) to fix boundaries to the parameter space and guarantee the feasibility of the project. In order to preserve the adaptability of the optimization to any experiments, one can compute the effect of construction costs on the loss function in two main steps each of them depending on different sets of parameters:

\begin{equation}
    L_{cost} = c(\theta,\phi)
\end{equation}

\begin{itemize}
    \item Local cost parameters $\theta$ are specific to the technology used: \eg active components material, photo-detection and light transport techniques.
    \item Global cost $c(\theta, \phi)$ can be expressed as a function of local cost parameters $\theta$ and a set of parameters $\phi$ describing the overall detector conception such as number and size of detector modules and their respective positions.
\end{itemize}

Modeling the dynamics of the global cost with respect to local costs may seems unfeasible for large scale detectors such as experiments at the LHC or even larger future colliders (see Sec.~\ref{sec:colliders}), but can surely be done for setups of moderate complexity such as cosmic-muon trackers for muon radiography (see Sec.~\ref{sec:muography}), and it is indeed one of the features being included in the TomOpt package described {\em infra} (Sec.~\ref{sec:tomopt}).\\

Most of the time, the same detection task can be achieved by several different technologies. For each of them one must establish relations between their local cost parameters $\theta$ and their physics performance $\gamma$. In a first approximation, it can be done by considering a detector as two separate modules, active detection system and electronics:

\begin{itemize}
\item The choice of an active detector module fixes most of the detector performance parameters $\gamma$, such as spatial and time resolution. Its cost $C$ should be expressed as a function these performance proprieties and normalized to its active area/volume and number of readout channels.
\begin{equation}
    C_{technology} = C(\gamma)\ [\ m^{-2} .\ readout^{-1}\ ]
\end{equation}
\item Electronics: In many detectors, front-end electronics along with data acquisition systems account for the largest share of detector cost, which makes their cost estimation critical. To ensure compatibility with active detector cost, it must be normalized to the number of channels.
\end{itemize}

Such a splitting of the cost is done under the assumption that it scales linearly with surface or volume and the number of readout channels, which is likely to be a fair approximation for simple setups but becomes biased for large scale detectors.

Of course the total monetary cost of a detector is not only a function of specification and number of its components: other fixed expenses such as infrastructure, laboratory equipment, and maintenance can occasionally dominate the overall cost of a detector. Nevertheless, modeling such costs and including them in the optimization process is not relevant since their dependency with detector performance parameters is limited. Instead, they are rather related to the global scale of the detector set up, a parameter whose order of magnitude is known before optimization. Besides, infrastructure costs might vary from one technology to another, hence it would be more pertinent to evaluate them aside from the optimization phase. These fixed costs can then be added to the variable cost modeled in the loss function. 

\subsection {Modeling of External Constraints}

\begin{figure}[!ht]
\begin{center}
    \includegraphics[width=12cm]{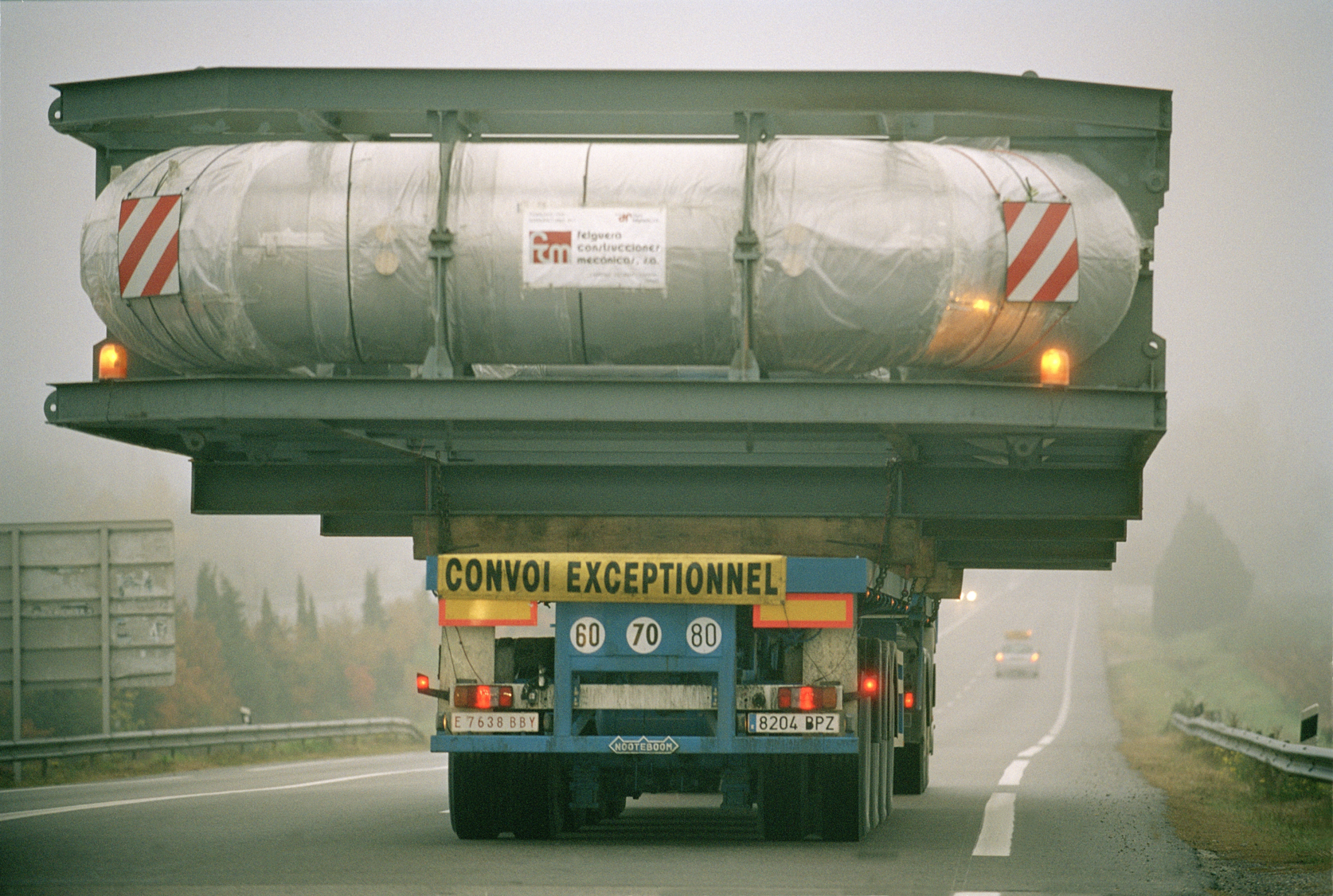}
    \caption{Road transport of a structure for the ATLAS air toroids. Photo reproduced from Ref.~\cite{Ginter:737473}.}
    \label{f:transport}
\end{center}
\end{figure}

Often a detector design task needs to consider, in addition to performance and cost of the instrument, external constraints coming from a number of specific conditions that characterize the project. For example, the construction of the detectors of the LHC was conditioned by engineering constraints connected with the placement and operation of the instruments in underground caverns; these, \egc affected the largest size and weight of elements that could be assembled on the surface before lowering them down in the caverns; pieces built externally by participating laboratories or contractor industries also posed considerable logistic challenges connected to their transportation to CERN (Fig.~\ref{f:transport}). Other physical constraints are often connected with power consumption and payload (factors of high relevance for detectors to be operated in space), operation temperature and cooling infrastructure, online computing and data acquisition limits. Further, for large-scale projects such as the LHC, even the sheer availability of construction materials may play a role, and require their insertion in a global optimization task.

A different kind of external constraints often comes from the timeline of the projects. At the end of the nineties, when upgrades to the CDF and DZERO detectors were being planned prior to the start of Tevatron Run 2, the construction schedule of the LHC was a very important ingredient affecting the decisions that the laboratory took on how much to invest in upgrades which could grant Fermilab a fighting chance to discover the Higgs boson before the European machine would take over with its larger energy and luminosity. A careful modeling of commissioning timelines may in similar situations heavily affect the absolute value of attainable goals depending on their expected time to completion, and cannot therefore be ignored in a serious optimization study. 

While the above examples could fuel criticism toward the naivety of the idea that an automated scanning of detector design configurations can provide significant help to the hand of the expert detector builder, we argue that in fact everything which adds to the complexity of the task only strengthens the value of complete modeling approaches. In fact, all the mentioned constraints are relatively straightforward  to insert in a differentiable pipeline, and in most cases they are also only weakly coupled to the other ingredients, making their inclusion less impacting and simpler. At any rate, as we already argued in Sec.~\ref{sec:costs}, we believe that whenever some external factor is impossible or inconvenient to include in an optimization pipeline, one needs not worry about it too much. It is already very helpful to optimize the system under design for various scenarios, to pin down the most optimal set of parameters in each of them and reduce the complexity of the decision to a comparison between a few discrete options, where non-quantitative and even political or sociological arguments can find their place.

\clearpage

\section{Example Use Cases}
\label{sec:exemplary_use_cases}

In this section we consider the specific modeling needs of instruments designed for a variety of different goals, ranging from pure research to industrial and medical applications. Our choice of illustrative use cases, which is far from being exhaustive, is driven by the need to clarify, to ourselves and to the reader, how a modular optimization pipeline such as the one described in Sec.~\ref{sec:problem-description} may be customized and adapted to very different problems, with only a minor reconfiguration of its basic ingredients and minimal changes in the optimization methodology. These problems all constitute interesting benchmarks to us because of our specific focus on those research areas.

We start in Sec.~\ref{sec:colliders} with a discussion of a representative selection of use cases for AD-powered design optimization taken from accelerator experiments. 
In Sec.~\ref{sec:astro} we discuss some use cases from astro-particle and neutrino physics, which offer a variety of additional complications and intriguing problems to solve. In Sec.~\ref{sec:muography} we consider in detail the use case of muon tomography, which is an excellent test-bed for the development of a full optimization pipeline, given the relatively simple physics involved and the well-defined nature of the optimization target. Section~\ref{sec:protonct} describes the special optimization challenges of instruments designed for proton-computed tomography. In Sec.~\ref{sec:neutronpp} we offer an example drawn from research in low-energy particle physics, where one tries to optimize the transport of cold neutrons. We complete our pot-pourri of potential applications of differentiable programming to design optimization in Sec.~\ref{sec:erran} with the consideration of how to optimize the calculations in lattice QCD.

\subsection{Experiments at Accelerators}
\label{sec:colliders}
Fundamental research has driven the need of accelerating particles and collide them at ever-increasing energies, to study the products of the resulting collisions with the hope of understanding the fundamental structure of nature and detect the presence of physics processes never observed before, if they exist.
Physicists have used the data from these collisions to identify elementary particles, directly or indirectly observable, and understand their interactions~\cite{thomson_2013,Zyla:2020zbs}. To do so, they have built particle detectors of ever-increasing size, which detect final-state particles by exploiting their interaction with matter. The particle detectors needed to collect data taken from collisions at the energies reached by the LHC~\cite{Evans_2008} have reached an unprecedented level of complexity; a full optimization of the next generation of particle detectors with a differentiable pipeline is the holy grail of the MODE Collaboration~\cite{npnipaper}.
Particle accelerators and detectors have been originally designed for fundamental research, but the technologies that drive their functioning have been however promptly adapted to a vast range of other applications in scientific, medical, and industrial fields.

The optimization of an entire accelerator or detector is a daunting task that is probably still beyond our present-day capabilities. Nevertheless, differentiable-programming-based optimization has been successfully applied to the optimization of portions of these machines and to the their automated control. In Sec.~\ref{sec:accelerators} we outline existing work and future perspectives for the design and control of particle accelerators; in Sec.~\ref{sec:calo-lhcb} we consider the optimization of the electromagnetic calorimeter of the LHCb experiment; in Sec.~\ref{sec:calo-fcc} we describe how one could approach the task of optimizing a hybrid calorimeter design for a future collider; and in Sec.~\ref{sec:calo-muoncoll} we discuss the optimization of an electromagnetic calorimeter for a future muon collider experiment. We conclude our survey with Sec.~\ref{sec:muone}, where we describe the optimization of MUonE, a proposed detector with a relatively simple tracking and calorimetry geometry, and Sec.~\ref{sec:exotics}, where we describe the perspectives for a cost-effective optimization of the MilliQan detector.

\subsubsection{Particle Accelerator Design and Control}
\label{sec:accelerators}
Particle accelerators are some of the most complex instruments in existence; their development drives numerous types of scientific, medical, and industrial applications, including high-energy particle physics experiments. Optimization and control of particle accelerators is challenging: these instruments have many interconnected sub-systems (\eg low-level dynamic control of radio-frequency cavities, high-level optimization of focusing/steering magnets and cavity settings), numerous adjustable settings often consisting of hundreds-to-thousands of variables, and in many cases have highly nonlinear beam responses to different combinations of input settings. In addition, there are time-varying inputs and responses that must be taken into account, both due to unintended drift (\eg due to temperature changes) and deliberate changes in state (\eg to achieve different beam parameters). The challenge of optimizing these systems both in design phase and during operation increases as we push toward the energy and intensity frontiers of beam physics, where the beam responses become increasingly nonlinear and sensitive to machine settings, noise, and other sources of uncertainty.

\begin{figure}[b!]
            \includegraphics[width=0.90\linewidth]{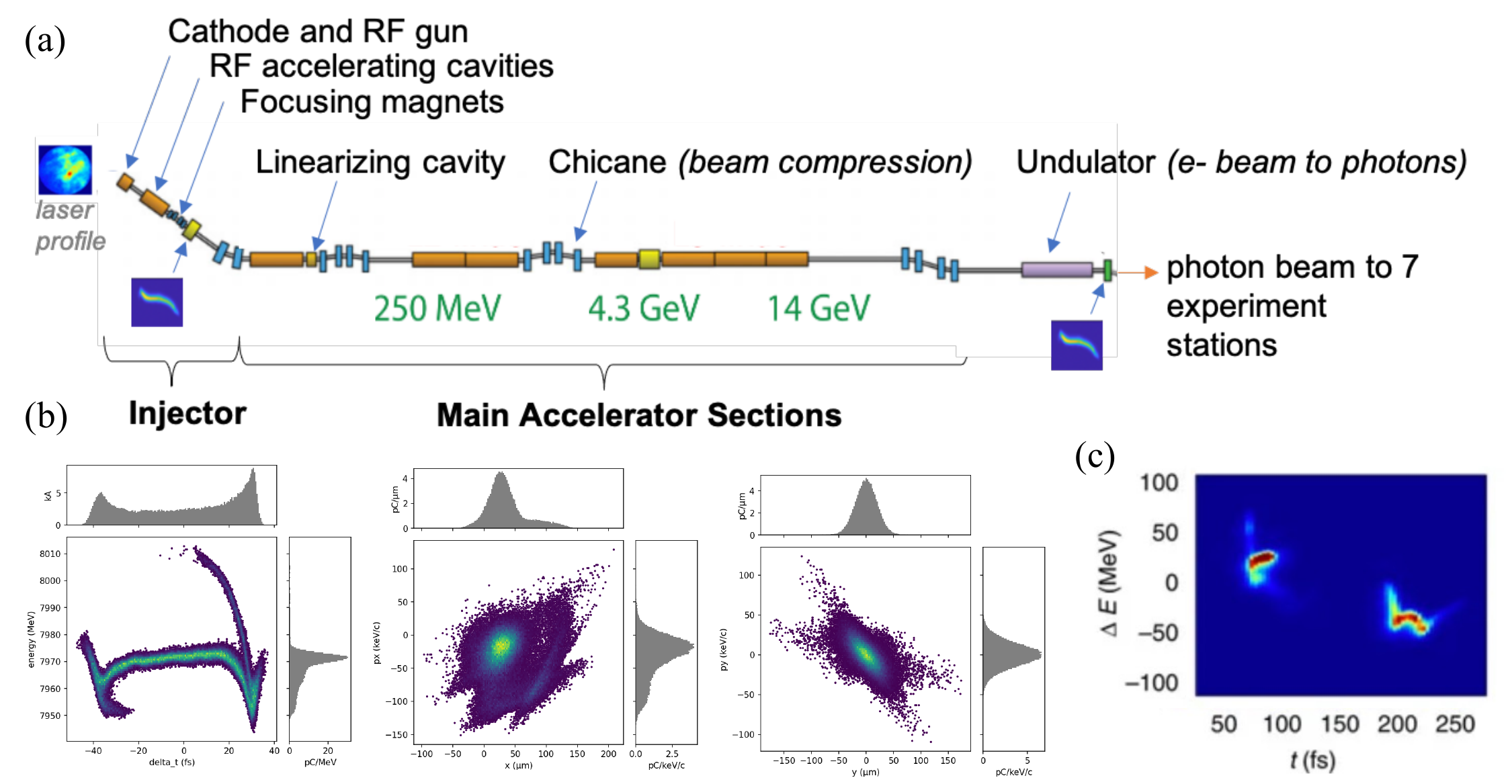}
    \begin{centering}
    \vspace*{-2mm} 
    \caption{Example of a linear accelerator and major components (in this case the Linac Coherent Light Source) (a). Examples of 2D projections of 6D beam phase space, from physics simulations in {\fontfamily{qcr}\selectfont{Bmad}}~\cite{SAGAN2006356} (b). Example of measured longitudinal (duration vs. energy) phase space in an operating mode with two electron bunches (c, reproduced from Ref.~\cite{twopulse}).}
    \label{fig:lcls-schematic}
    \end{centering}
\end{figure}

The beam itself is typically represented in physics simulations as a cloud of particles in 6D position-momentum phase space. In the past, using bulk statistical scalar metrics of the beam distribution ( such as energy spread or bunch duration) was sufficient for optimization in many applications. However, increasingly the full information about the phase space distribution is needed to meet the tolerances of applications at the energy and intensity frontiers. For example, in high-energy particle beams, the beam ``halo'' at the edges of the beam distribution can contain enough energy to damage accelerator components. Advanced phase space manipulation techniques to produce, for example, beams that are highly elongated in one dimension is often challenging due to the sensitivity of the many correlated responses of the beam to different accelerator components. The need to precisely optimize (in design phases) and then control the beam phase space distribution in detail is common across a wide variety of accelerator types and applications, ranging from light source user facilities, to isotope production facilities, to colliders. An example of an accelerator and particle beam phase space distribution is shown in Fig.~\ref{fig:lcls-schematic}.

\paragraph{Challenges for accelerator design and online optimization.} The initial design of accelerators is driven by detailed physics simulations of the accelerator components (such as the type of RF cavities) and, to varying degrees of detail, simulations of the nominal range of accelerator settings and placement of components as a whole. Automated multi-objective optimization is an essential component of this process and has traditionally used heuristic methods, such as genetic algorithms~\cite{deb2002afast} and particle swarm optimization, coupled with high performance computing. Accelerator physicists then examine beam parameter tradeoffs at different working points when making design decisions (for example, examining the achievable beam intensity at different beam energies).

While this design paradigm has a long history of success, the computational expense of most simulations that include the majority of expected nonlinear collective beam effects makes this process a slow and resource-heavy endeavor.  Simulations that include nonlinear collective beam effects typically use computations on hundreds-of-thousands of individual particles in a particle beam to predict its evolution along the accelerator. Increased execution speed can be achieved by making simplifications to simulations (\eg not including all expected beam effects, or including only the most significant accelerator variables in optimizations) at the cost of accuracy and comprehensiveness. 

The design process and online optimization are also segmented at present: many accelerator design optimizations examine the tradeoff between only two-to-three scalar beam properties at a time, online optimization often examines only a few output parameters as objectives at a time, and sub-systems are often optimized sequentially or semi-independently (\eg injector systems are optimized separately from down-stream sections). This approach is known from simulation studies to produce sub-optimal results, in contrast to simultaneously optimizing settings across the entire accelerator~\cite{s2eopt}. Emerging accelerator designs that have more stringent tolerances on the beam parameters and involve operation at the energy and intensity frontiers cannot rely on such simplifications to the same degree as older or more conventional accelerator systems where the sensitivity of the beam dynamics and impacts of nonlinear beam collective effects is lower. For novel acceleration schemes such as plasma-based acceleration, the computational burden and level of complexity of the physics effects is even more substantial. 

The computation burden of detailed accelerator physics simulations has numerous impacts that suggest where future improvements can be made: (1) it drives a need for more sample-efficient and high-dimensional optimization methods, (2) it limits in practice the extent of detailed analyses, such as comprehensive treatment of uncertainties due to component misalignments and noise, or inclusion of the majority of available variables in optimization, and (3) it limits the extent to which simulations are used online with operational accelerators, including the extent to which physics simulations are updated to match the as-built accelerator system. Many of these challenges are starting to be addressed with advanced computational methods, including the use of AI/ML and differentiable simulators. 

In addition, at present there is a need for truly ``end-to-end'' accelerator system modeling that enables more comprehensive design and control. Looking to the future, this should include co-design of the accelerator and its controls along with the experimental equipment (such as particle detectors) and their associated analyses. New computational techniques for accelerator simulations and optimization (including both machine learning approaches and differentiable programming) coupled with corresponding developments in particle detector design and physics analysis provide promising avenues toward bringing this type of integrated approach to fruition. Much emphasis in recent years has been on ML and AI applications that can directly aid accelerator operation (\eg see Ref.~\cite{van_der_veken:ml1} for some applications to the LHC), but substantial opportunities also exist for design optimization. In addition to the brief introductions below, more detail on accelerator modeling challenges and emerging opportunities is given in Ref.~\cite{Snowmass_WP_AM}. 

\paragraph{New computational techniques for accelerator simulations and optimization.}  Recently, a variety of computational techniques have begun to open up new capabilities in accelerator simulation and design. Many of these techniques are in their infancy with respect to application to accelerators and represent a significant opportunity for future development. 

\begin{figure}[b!]
   \includegraphics[width=0.99\linewidth]{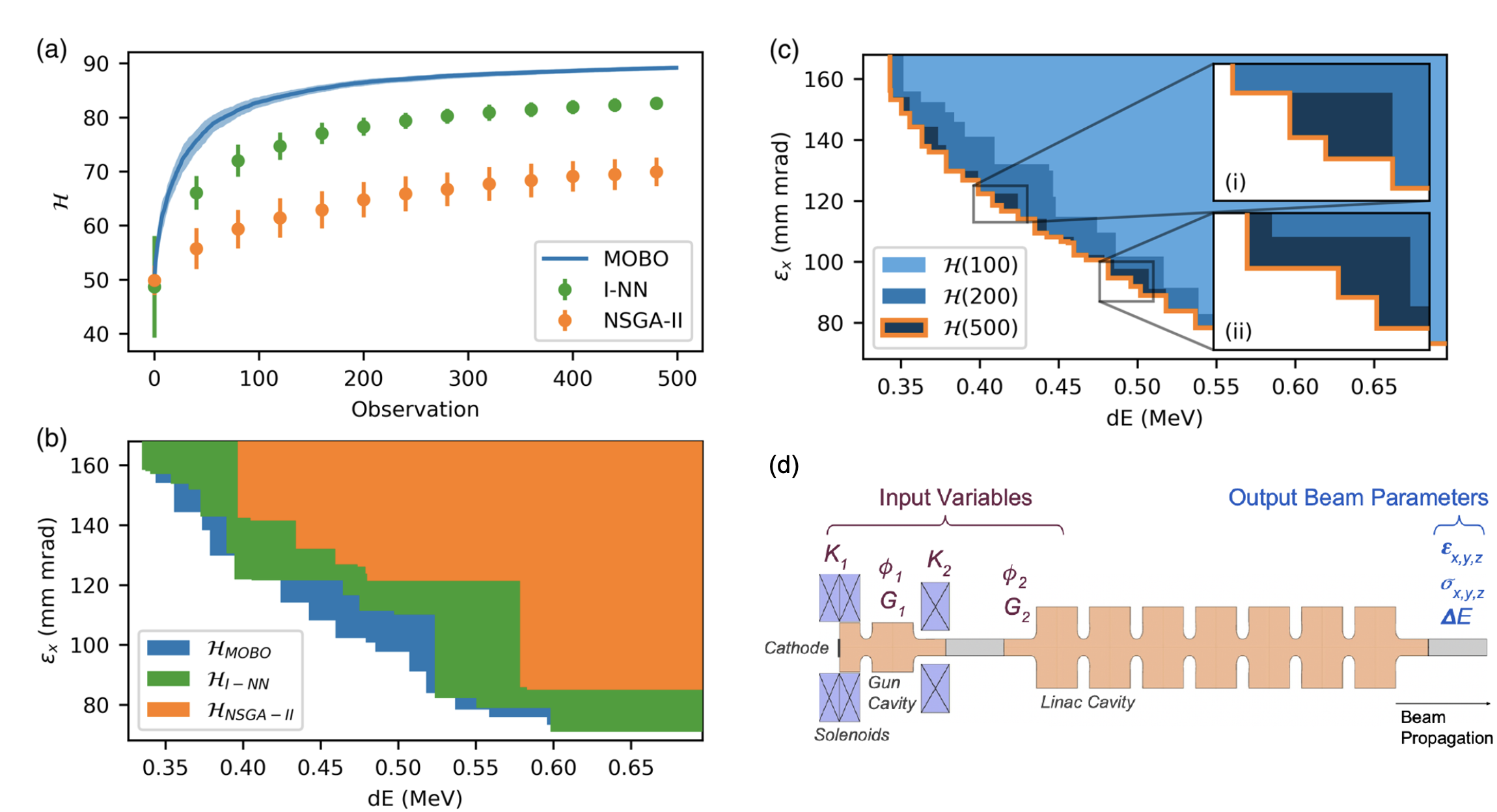}
    \begin{centering}
    \vspace*{-2mm} 
    \caption{ Multi-objective Bayesian optimization of the Argonne Wakefield Accelerator (reproduced from Ref.~\cite{roussel2021multiobj}). (a) Multi-objective Bayesian optimization is more sample-efficient in converging to the Pareto-optimal front in a 7D objective space consisting of desired beam parameters (here denoted by the convergence in the hypervolume) than classical genetic algorithms~\cite{deb2002afast} and neural network boosted genetic algorithms~\cite{edelen:ml2}. (b) The optimal front is outlined with higher resolution due to the uncertainty-aware sampling (an example for one of the 2D projections of the front is shown). (c)  Running multi-objective Bayesian optimization for more iterations results in strategic refinement of the Pareto front resolution. (d) Schematic of the Argonne Wakefield Accelerator injector, along with input settings and output beam objectives used in this problem. For more detail, see Ref.~\cite{roussel2021multiobj}.}
    \label{fig:mobo}
    \end{centering}
\end{figure}

\textit{ML-based optimization} algorithms have demonstrated increases in sample-efficiency and ability to extend to higher dimensionality, resulting in higher-quality solutions and reduced computational burden. For example, multi-objective Bayesian optimization~\cite{roussel2021multiobj} and Bayesian parameter space exploration~\cite{roussel2021turnkey} enable more sample-efficient optimization and characterization of accelerators (see, \egc Fig.~\ref{fig:mobo}), which is important both in design optimization and online optimization. These techniques can be used even when no previous data is available, making them appealing for commissioning of new systems. They can also be combined with learning where constraint violations on output parameters are likely to occur, to prevent undesirable conditions (such as losing the beam online, or wasting computational resources on poor simulation runs)~\cite{roussel2021turnkey,KirschnerSafeBO2019}. Bayesian optimization can also leverage expected correlations from physics models to help increase convergence speed and improve sample-efficiency in model learning~\cite{duris2020bayesian,hanuka2021physics}. Feed-forward corrections with ML system models also have been used to speed up optimization convergence, both online and in the context of experiment design~\cite{edelen2016neural,edelen:ml2,scheinker2018demonstration}.

\textit{ML-based surrogate models} trained on physics simulations have been shown to increase execution speed of accelerators system models by orders of magnitude~\cite{edelen:ml2,edelen2016first,ogren2021surrogate}. These fast-executing models can be used in conjunction with standard optimization and feedback algorithms to speed up design. They can also be used online to provide fast predictions of unobserved beam behavior (\ie ``virtual diagnostics'')~\cite{emma2018machine,EdelenDiss,Gupta_2021,scheinker2021adaptiveML,scheinker2018demonstration}. This is of great interest in many accelerator applications where diagnostics are invasive to the beam and cannot be used during downstream delivery.  Recent lines of inquiry have also examined the use of Bayesian neural networks~\cite{mishra2021uncertainty} and ensembling~\cite{convery2021uncertainty} to obtain uncertainty estimates from these models. Such uncertainty estimates are essential when making design or control decisions, and when using model predictions to inform physics analysis or convey information about the beam to downstream user experiments. On a more granular level, numerous individual simulation calculations involved in accelerator physics codes can also be sped up and made differentiable with machine learning. For example, calculation of the  impact of coherent synchrotron radiation (CSR) on each particle in the beam at each simulation timestep can be very computationally intensive; preliminary studies have shown that computation of the CSR wakefield can be signficantly sped up using a neural network~\cite{NN_CSR}.

\textit{Differentiable physics simulations} have not yet been explored extensively in accelerators. 
While automatic differentiation has a long history in accelerator physics~\cite{Berz:1988aj}, the major simulation codes do not support arbitrary computation of gradients. Making these codes differentiable will be a challenge, especially in instances where simpler analytic or transport matrix based representations do not suffice and more detailed simulations (such as those based on particle-in-cell calculations) are needed. However, some of the potential benefits of differentiable physics models have been explored with differentiable ML models that are trained on detailed physics simulations. The differentiability of ML models enables them to be readily calibrated to match measurements from operational accelerator systems, for example to increase prediction accuracy and identify sources of systematic errors (such as unknown initial beam distributions, \eg in Ref.~\cite{Koser:ML1,scheinker2021adaptive}). This is important for understanding beam dynamics effects in the observed system. The differentiability of ML models also enables them to be combined with other ML algorithms directly. For example, in Ref.~\cite{EdelenNeurIPS2017} back-propagation through a forward ML model trained on simulations of an injector system was used to train a neural network controller to quickly switch between beam energies while maintaining optimal beam size and divergence into an undulator. In that case, the forward model had to be updated occasionally with new data as the simulation entered new regions of the parameter space (where the forward model was no longer accurate). In contrast, having a differentiable physics simulation for this type of problem would enable one to avoid this intermediate step, and also would enable more flexibility in adjusting the problem to tackle other combinations of input and output variables. There are numerous accelerator control problems that could benefit from this type of approach. As another example, in Ref.~\cite{hanuka2021physics} the Hessian of a neural network system model and physics simulation (numerically estimated) is used to easily incorporate expected correlations between input variables into a Gaussian process model, resulting in faster Bayesian optimization. Having differentiable physics simulations of accelerators would greatly enhance the ease and speed with which this type of approach could be applied to arbitrary accelerator optimization tasks.

Recent work has also highlighted the utility of using differentiable accelerator physics models by exploring this approach with simpler analytic representations. In Ref.~\cite{roussel2022differentiable} a physics-based model of hysteresis is made differentiable and used to describe accelerator focusing magnets. The differentiable physics model is then used to learn a magnet's hysteresis response based on measured data, enabling rapid solving of the hysterion density. This procedure can even be done indirectly by using, for example, beam size measurements; this capability is essential for applying this technique to existing accelerator beamlines where the individual magnetic field responses cannot be directly measured.  It is then shown that this approach can be combined directly with Bayesian optimization for improved precision in optimization of an accelerator quadrupole triplet that exhibits hysteresis. Hysteresis is a major concern for accelerator optimization in practice, especially for high-energy systems, such as final focus systems for future colliders. Similarly, many high-level accelerator beam dynamics problems can be modeled in bulk analytically with transport matrices and Taylor matrices. These idealized representations can be cast into a differentiable form and used with gradient-based optimization of free parameters to make the model more closely match measured data, as was done in Ref.~\cite{Ivanov2020} for a Taylor map representation of beam dynamics in a ring.

Looking forward, differentiable physics simulations  would help enable (1) tighter integration of physics simulations of accelerators with ML-based optimization routines, (2) gradient-based calibrations of predictions and uncertainty estimates when comparing simulations to measured data, (3) identification of unknown input parameters and sources of error, and (4) fine gradient-based optimization of beam phase space distributions at the particle level. Having differentiability for simple transport matrix-based accelerator codes and codes used for detailed predictions of beam phase space that include nonlinear collective effects (often requiring expensive particle-in-cell computations) would open up a host of new capabilities in this area. This could substantially benefit higher-precision modeling, characterization, and tuning of accelerators. Linking these modeling and optimization tasks directly with end user experiments (such as particle detectors) would further enhance the overall precision with which we can design, characterize, and control the entire end-to-end experiment.

\paragraph{Optimization and simulation infrastructure for accelerators and associated experiments.} Finally, analogous to efforts for particle detector experiments, there is much work in the accelerator physics community to make software tools and standards that allow interoperability as ease-of-use when running simulations for different types of accelerator sub-systems, and exchanging data between them. This is essential for running end-to-end simulations and optimizations, as different parts of accelerators typically are simulated with different codes depending on the relevant physics effects for each section. Examples include frameworks such as LUME~\cite{lume} to run different simulations and stitch them together, and Xopt~\cite{christopher_mayes_2021_5559141} to drive high-level optimization and save datasets. This also includes standards for describing the beam distribution (such as OpenPMD~\cite{openPMDapi,openPMDprojects,openPMD}) and other aspects of the simulation and data sets (such as those used in LUME~\cite{lume}). Efforts to develop standards for exchanging information with user experiments is also underway, for example in doing end-to-end simulations of accelerator-based light source user experiments. This provides a solid foundation for beginning to integrate accelerator simulations with particle physics detector simulations and analyses.

\subsubsection{Calorimeter Optimization}
\label{sec:calo-lhcb}
Calorimeters are a crucial component for most detectors mounted on modern colliders. Their tasks include identifying and measuring the energy of photons and neutral hadrons, recording energetic hadronic jets, and contributing to the identification of electrons, muons, and charged hadrons. To fulfill these many tasks while keeping costs reasonable, the calorimeter construction requires good and thoughtful balancing with other components of the detector.

In practice, the construction of a calorimeter depends on the choice of many options, which affects drastically both the physics performance of the resulting detector and its overall cost.
Such options include:
\begin{itemize}
\item granularity of the detector;
\item materials for absorber and active component;
\item mechanical construction of absorber and active component;
\item light collection and transport techniques;
\item photo-detecting techniques;
\item photodetector signal acquisition and processing.
\end{itemize}

The performance of the calorimeter as a detector may be described in terms of its:
\begin{itemize}
\item energy resolution;
\item spatial resolution;
\item time resolution;
\item sustainability to in-time and out-of-time backgrounds;
\item radiation hardness contributions.
\end{itemize}

In the following we present a calorimeter optimization task for the upgrade~\cite{LHCb:2018roe} of the electromagnetic calorimeter~\cite{LHCb:2000vji} of the LHCb detector at the LHC~\cite{LHCb:2008vvz}. The upgraded calorimeter is expected to operate in the challenging conditions of LHC's high luminosity Run 5 and beyond~\cite{Apollinari:2017lan}.

\paragraph{Problem statement and current optimization approach}

The ultimate goal of the optimization of a calorimeter is to achieve the best performance to fulfil the physics program of the experiments, while fitting within the available budget. 
Taking into account
the different tasks for the detector as mentioned above, the list of 
requirements is a trade-off between different properties, including:
\begin{itemize}
\item radiation hardness to sustain the expected lifetime span;
\item energy and spatial resolution for good photon reconstruction and
  electron identification;
\item high granularity to facilitate better precision, both spatial
  and in energy, which in turn improves reconstruction algorithms;
\item good timing resolution to facilitate pileup suppression in high-occupancy areas
as well as better matching of separate signal components.
\end{itemize}

Thus, the optimization problem statement is to find a technology for the construction of calorimeter modules, \eg homogeneous, Shashlik~\cite{Barsuk:2000vna}, or SpaCal~\cite{Jenni:1987cz,Lucchini:2013loa} (see Ref.~\cite{fabjan2003calorimetry} for a review on calorimetry technologies for particle physics), and to choose the construction details for the chosen technology, such as used materials, granularities, geometries, etc.
    
To evaluate the physics performance of the different possible configurations,
a detailed GEANT4-driven~\cite{GEANT,allison_geant4_2006,allison_recent_2016} simulation of the detector modules is needed. Thus the first task in an optimization pipeline consists in evaluating necessary target low-level performance metrics like energy, spatial, and timing resolutions from physics-based first principles, and validating with simulation if a particular
configuration fulfils the requirements.
The challenge of this approach is that, although the individual detector
channel response may be simulated, achieving even a low-level performance
requires the development of a reconstruction algorithm tuned to
the particular configuration under study. 
Moreover, this approach decouples the local optimization of the
calorimeter low-level performance metrics from the global optimization of the physics performance of
the entire detector. To include the ultimate physics performance into
consideration, a more comprehensive optimization loop is necessary~\cite{fedor:chef2019}.

\begin{figure}[htbp]
\centering 
\includegraphics[width=0.8\textwidth]{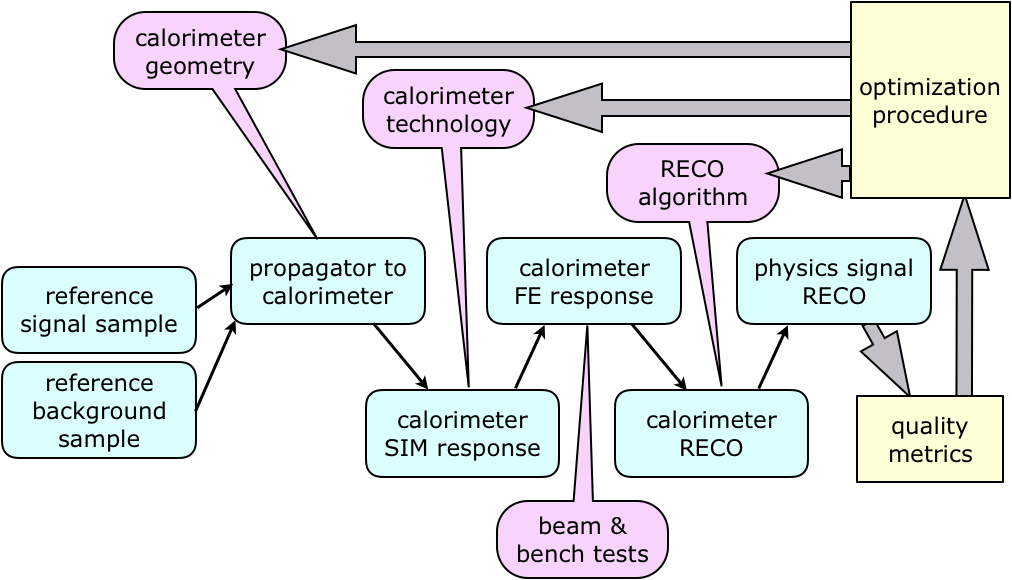}
\caption{\label{fig:pipeline} A general pipeline for the calorimeter
  optimization includes several steps. Blue blocks indicate data processing pipeline steps; pink bubbles represent configurations and conditions for pipeline steps; yellow blocks close the optimization loop.}
\end{figure}

The typical workflow for the optimization of calorimeter components is
sketched in Fig.~\ref{fig:pipeline} and proceeds as follows:
\begin{itemize}
\item Selected event samples, both signal and background, are used to
  initiate an optimization cycle for comparing the performance in terms of signal recovery and background suppression;
\item the calorimeter is usually installed downstream of the detector, so a propagation of events from their origin to the calorimeter is necessary. This step is dependent on the properties of the elements of the detector between origin and calorimeter. Additionally, if the calorimeter detector has a non-homogeneous configuration, the details of the global geometry are to be accounted for in this step for the optimization to be based on physics performance quality metrics;
\item the construction technology used for the individual calorimeter modules is a central point for the detector R\&D. To evaluate the impact of a choice of construction technology, we need to simulate its effect on observable event characteristics. This is done using response simulation models, typically based on GEANT4. The details of the calorimeter technology used drive such simulation; 
\item the behaviour of the front-end electronics is another important contribution to the physics quality of the detector. Although the properties driving the behaviour are hard to simulate, good data samples may be obtained from beam or bench tests;
\item a reconstruction algorithm is absolutely necessary to evaluate the quality of converting the detector response into the physics objects;
\item physics quality metrics may be calculated using reconstructed objects and it can be used as a target function for the optimization procedure;
\item all aspects of the calorimeter may be optimized: the details of the calorimeter technology, the geometrical layout, and the possible reconstruction algorithms.
\end{itemize}

Such an optimization cycle, built on top of an event-processing pipeline, makes it possible to obtain physics-motivated optimal values for the detector parameters.

To evaluate the physics performance of a particular configuration for a possible future calorimeter detector, one needs to run the optimization cycle described above. A good fine-tuning of the individual blocks is important to properly propagate the properties of the configuration under study to the ultimate physics performance.
For the regular operation of a stable detector, these blocks are carefully tuned based on the actual detector configuration. In contrast, for the R\&D of a new detector, many different possible configurations are studied simultaneously during the optimization stage.
Nevertheless, reasonable representations of the simulation and reconstruction steps, which are tuned for each of the configurations studied, are necessary for inferring consistent conclusions about the physics performance of these configurations. This is a time consuming work, if done manually. Fortunately, these studies use well-labelled data sets either from MC simulation or from test beam measurements. Thus surrogate models may be built and trained on labelled data using regular ML approaches. This makes it possible to speed up model building for different pipeline steps. Importantly, such training may be automated and requires minor expert supervision.

The big slowdown factor for running an optimization cycle is the necessity of fine tuning reconstruction algorithms for every new calorimeter technology and geometry configuration. ML may help tuning the reconstruction in an automatic way. Indeed, as soon as a sample of calorimeter responses is available, the corresponding regressor may be trained to extract physics information from the raw response.

\begin{figure}[htbp]
\centering 
\includegraphics[width=0.8\textwidth]{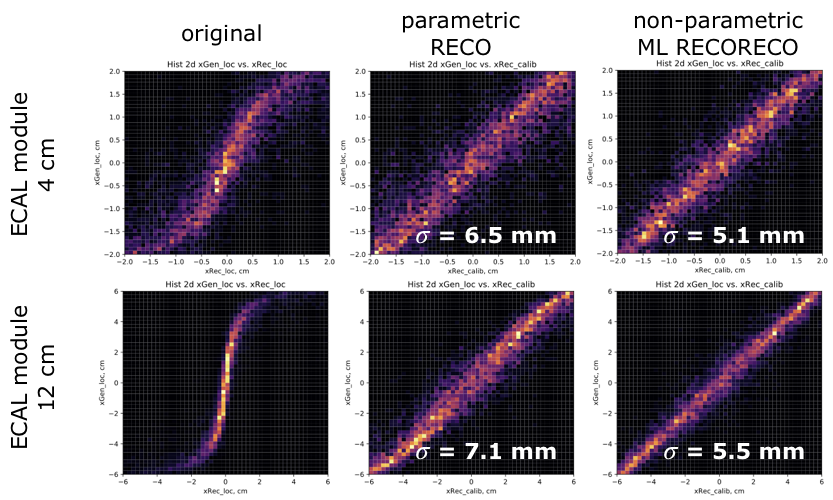}
\caption{\label{fig:scurve} ML-based reconstruction of the calorimeter
  cluster position provides spatial resolution similar to the customised
  reconstruction procedure, but without {\em a priori} knowledge about
  the particular spatial properties of the calorimeter under study. Left - correlation between cluster center and the true track position; middle - correlation corrected using parameterized correction; 
right - correlation using ML trained regressor.  }
\end{figure}

Figure~\ref{fig:scurve} demonstrates the quality of the spatial
reconstruction of the calorimeter cluster for the case of the LHCb
4~cm and 12~cm modules~\cite{Boldyrev:2020ydy}.
An automatically trained ML model (based on XGBoost~\cite{xgboost}, in this case),
agnostic to particular calorimeter details, produces a slightly better performance than the manually
selected parametric model. Importantly, the automatically trained
generic regressor provides a performance comparable with that of the manually
tuned one. This justifies the use of this surrogate regressor in the optimization cycle in place of a well-tuned reconstruction algorithm for extracting physics observables from the calorimeter response.

\paragraph{Figures of merit}

The local figures of merit (FOM) for the calorimeter optimization are the
characteristics mentioned above: energy, spatial and timing
resolutions, sustainability to high pile-up and huge radiation doses, etc.
However, the ultimate figure of merit for the optimization process is
the physics performance achievable using a given configuration of
the detector. Moreover, the FOM is actually not a single number, but
rather a dependency of the physics performance on the cost of the
configuration of the detector under study. 

\paragraph{Examples for optimizations}
To give an idea of the optimization procedure, let us consider
the future calorimeter of the LHCb experiment. Different regions of the
detector are very different in terms of requirements to the precision
of the signal measurement and background conditions.
We can use different technologies for calorimeter modules (Shahlik,
SpaCal), different materials (lead, tungsten) with different
granularities in different areas to optimize the overall performance.

Let us consider one of the target physics processes for the optimization, namely the
decay $B_s\rightarrow J/\psi (\mu^+\mu^-) \pi^0 (\gamma\gamma)$ with
emphasis on the $B_s$ signal reconstruction performance, which is driven by the photon
reconstruction in the calorimeter. 
Using simulated signal and background events, signal photons may be
propagated directly from the $\pi^0$ decay point to the calorimeter,
and the expected calorimeter response for a given module technology
and configuration may be extracted. This response needs to be merged
with the estimated contribution from the contributions of in-time and out-of-time pileup collisions.

\begin{figure}[htbp]
\centering 
\includegraphics[width=0.8\textwidth]{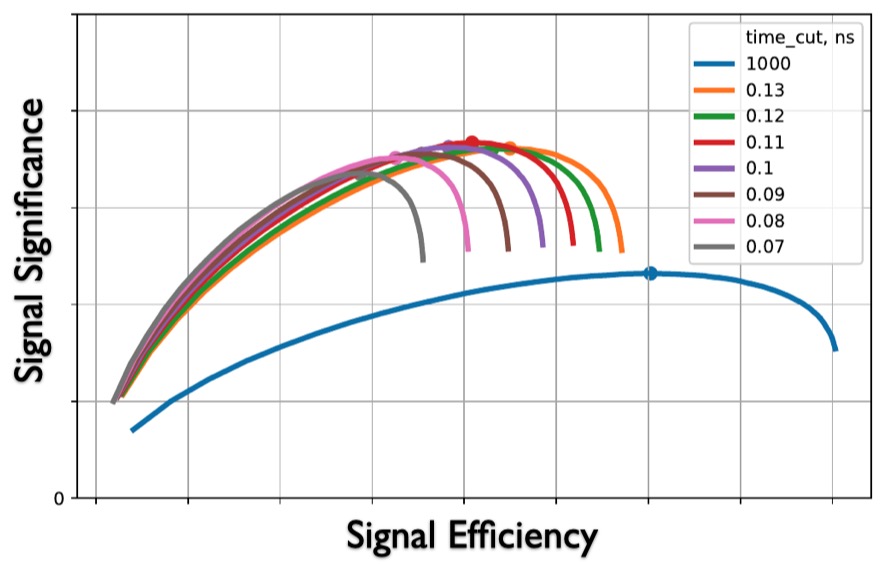}
\caption{\label{fig:calosignificance} Illustration of the physics
  signal significance for different efficiencies of the signal
  selections, evaluated using dynamically trained, ML-based
  calorimeter reconstruction algorithms.  Different curves correspond
  to different timing discrimination windows in this example.}
\end{figure}

\begin{figure}[htbp]
\centering 
\includegraphics[width=0.8\textwidth]{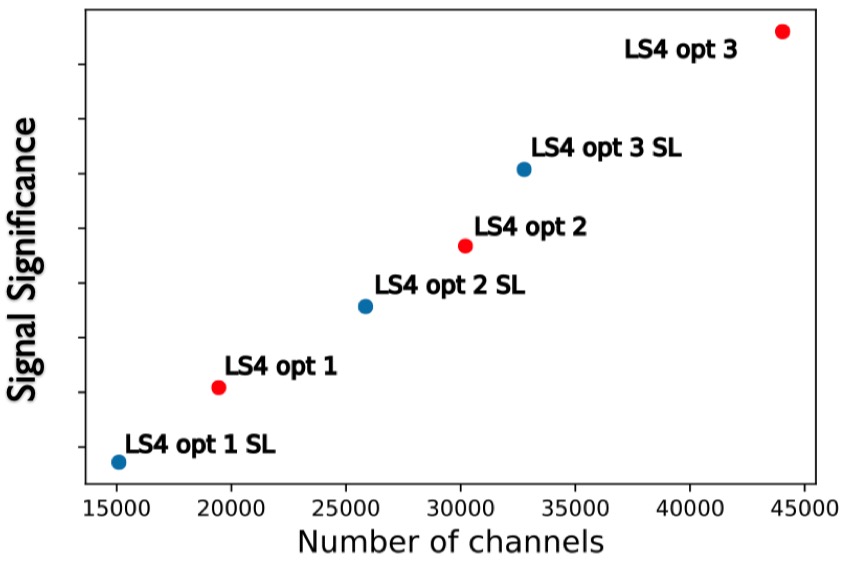}
\caption{\label{fig:calocosts} Dependency of the optimized Physics
  signal significance for different calorimeter configurations on the
  total number of readout channels used in those
  configuration. Dependencies like this help to make an educated
  decision about the optimal detector configuration which provides the
best balance between the ultimate physics performance and construction
costs.}
\end{figure}

To evaluate a realistic performance for reconstructing signal photon
energies, position, and timing, we trained ML-based regressors similar to those 
presented in Fig.~\ref{fig:scurve}. Using
reconstructed photon parameters we then evaluated the ultimate
performance for the physics signal---estimated, in this particular case, as 
the significance of the reconstructed
$B_s$---for different physics
selections. Figure~\ref{fig:calosignificance} show the 
significance curves obtained for different choices of the physics
reconstruction parameters, from which we can select the best  $B_s$
selection algorithm for a given calorimeter configuration, 
resulting in an optimal physics performance for the configuration under exam.

From another perspective, for each calorimeter
configuration one can evaluate the money cost of building such a detector from hardware.
In our case the overall detector cost is mostly driven by
the total number of readout channels. Thus, the physics
performance obtained above for each configuration is associated  with the corresponding
cost of building the detector using the configuration under exam.
Figure~\ref{fig:calocosts} illustrates the benefits brought to the physics performance
by the number of readout channels used for different configuration of calorimeter
cells, their granularities, and readout schemes. This kind of information 
vastly simplifies the problem of choosing the optimal balance between physics performance of the designed detector and the cost one has to pay to achieve that performance. 

\subsubsection {Hybrid Calorimeter for a Future Particle Collider}
\label{sec:calo-fcc}
The optimization of a hybrid calorimeter that may be used in a future particle collider
requires a reliable description of particle showers from electromagnetic and hadronic cascades in the material of the calorimeter.
If the detector layout is to be optimized based on the calorimeter shower structure, the corresponding differentiable pipeline must be interfaced with a differentiable surrogate of the calorimeter shower simulation.

To achieve a differentiable description of calorimeter showers, all three types of generative networks explained in Sec.~\ref{sec:gen-model} (GANs, VAEs, and Normalizing Flows) can be used. There are, however, differences in sampling time and quality that depend on the used architecture. Given this trade off, current applications of deep generative models at the LHC and beyond usually focus on a higher sampling speed, since a suboptimal performance in shower quality tends to matter less in downstream tasks~\cite{Buhmann:2021caf}. For the purpose of optimizing a calorimeter, however, a better sampling quality, \ie having more realistic showers, is more important. Therefore, we will be focusing on the generative model with the best sampling quality. Though, as already stated in Sec.~\ref{sec:gen-model}, the quantitative evaluation of the quality of generative models is a difficult task~\cite{2018arXiv180203446B}. This is especially true if models based on different architectures (VAEs, GANs, and normalizing flows) are compared to each other. One metric for comparison is given by a binary classifier to distinguish ``real'' (meaning based on GEANT4 training data) from ``fake'' (meaning generated by the generative model) samples~\cite{2016arXiv161006545L}. By using the Neyman-Pearson-Lemma, we can argue that $p_{\text{real}} = p_{\text{fake}}$ if a powerful classifier cannot distinguish the two sets. This test can therefore be seen as the ``ultimate'' test, as it looks not only at all voxels, but also their correlations~\cite{Krause:2021ilc}.

One of the first applications of deep generative modeling to calorimeter simulation was the CaloGAN~\cite{Paganini:2017hrr,Paganini:2017dwg}. There, a three-layer deep, simplified version of the ATLAS LAr detector was considered. The 3 layers were segmented into 288 + 144 + 72 voxels respectively, yielding a feature space of 504 dimensions. Showers initiated by $e^+, \gamma, $ and $\pi^+$, shot perpendicularly to the detector surface and of uniform incident energy in the $[1, 100]$ GeV range, could be generated by CaloGAN at a factor 20,000 faster than with GEANT4. Histograms of high-level features of these showers were close to their GEANT4 counterpart. However, it was shown in Ref.~\cite{Krause:2021ilc} that these showers still failed the classifier test and could be separated from the GEANT4 produced showers with almost 100\% accuracy. Samples of more advanced setups, like the Bib-AE~\cite{Buhmann:2020pmy}, were also shown to be separable by classifiers, see Ref.~\cite{2009.03796}.\\
The first generative model capable of generating samples that would confuse a classifier were based on normalizing flows~\cite{Krause:2021ilc,Krause:2021wez}. This approach, called CaloFlow, used the same detector geometry as the CaloGAN. The authors used a multi-step approach to generate high-quality samples. The first step uses a small NF to learn the probability $p_1(E_i|E_\text{inc})$, the distribution of the deposited energy into the three layers, $E_i$, conditioned on the incident energy $E_\text{inc}$. The second step uses a larger NF to learn the normalized energy depositions into all 504 voxel, $\mathcal{I}$, conditioned on the layer and incident energy: $p_2(\mathcal{I}| E_i,E_\text{inc})$. The normalization of energy depositions in each calorimeter layer helped to capture the different energy scales of the $E_i$, coming from the large size of layer 1 compared to layers 0 and 2. After generation, showers in each calorimeter layer were renormalized to have the correct energies $E_i$ that were used in generation. Such a step is necessary since generative models, even when trained on normalized showers, generate samples that are not perfectly normalized. 
There are two different architectural choices for the NF in the second step that were explored in Ref.~\cite{Krause:2021ilc,Krause:2021wez}. A MAF-based (see Sec.~\ref{sec:gen-model}) flow could be trained with the log-likelihood objective, resulting in a stable training without artefacts in generation and an optimal model selection based on the log-likelihood of a held-out test set. While this MAF was a factor 500 faster than GEANT4, it was still rather slow compared to other deep generative models. An IAF-based (see Sec.~\ref{sec:gen-model}) flow improved the sampling speed by a further factor 500, making CaloFlow as fast as CaloGAN. Due to memory constraints, such an IAF cannot be trained using the log-likelihood anymore. Instead a method called probability density distillation, originally developed for speech synthesis in Ref.~\cite{2017arXiv171110433V}, had to be used.

\begin{figure}[!ht]
    \centering
    \includegraphics[width=\textwidth, trim=50 0 75 50, clip]{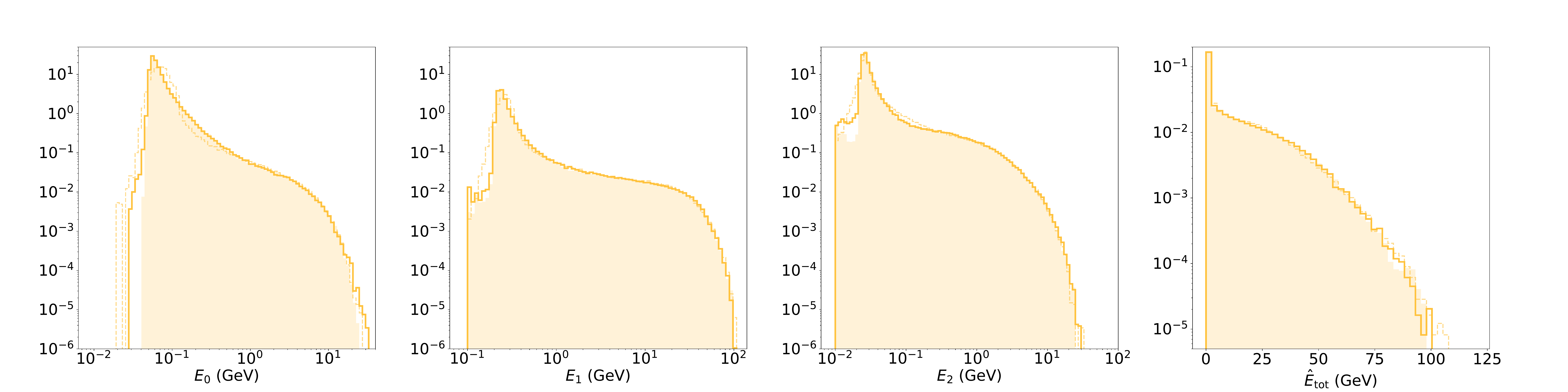}
    \includegraphics[width=\textwidth, trim=75 700 100 50, clip]{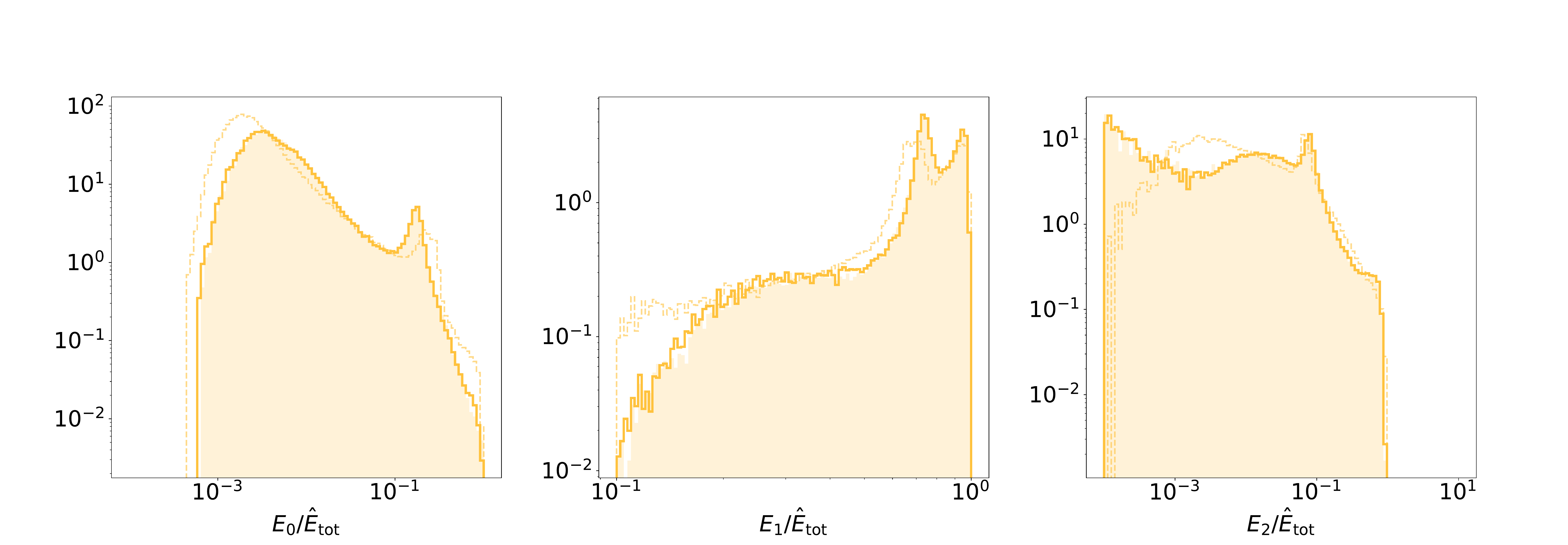}
    \includegraphics[width=0.75\textwidth]{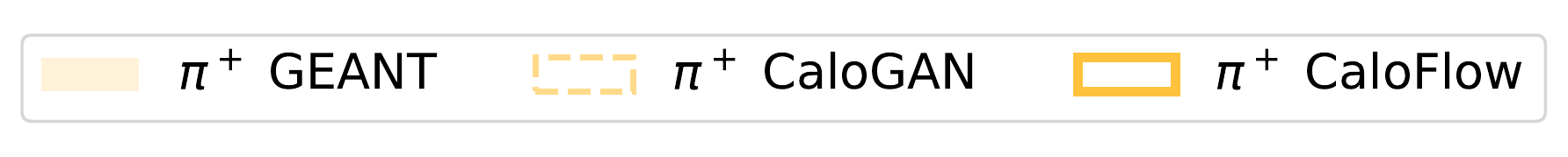}
    \caption{Distributions of energies in the 3 Calorimeter layers and total deposited energy (top) and ratio of layer energies to total deposited energy (bottom) for incident $\pi^+$ particles, comparing GEANT4 to CaloGAN~\cite{Paganini:2017hrr,Paganini:2017dwg} to CaloFlow~\cite{Krause:2021ilc,Krause:2021wez}.} 
    \label{fig:caloflow.histos.piplus}
\end{figure}

\begin{figure}[!ht]
  \centering
  \includegraphics[width=0.75\textwidth]{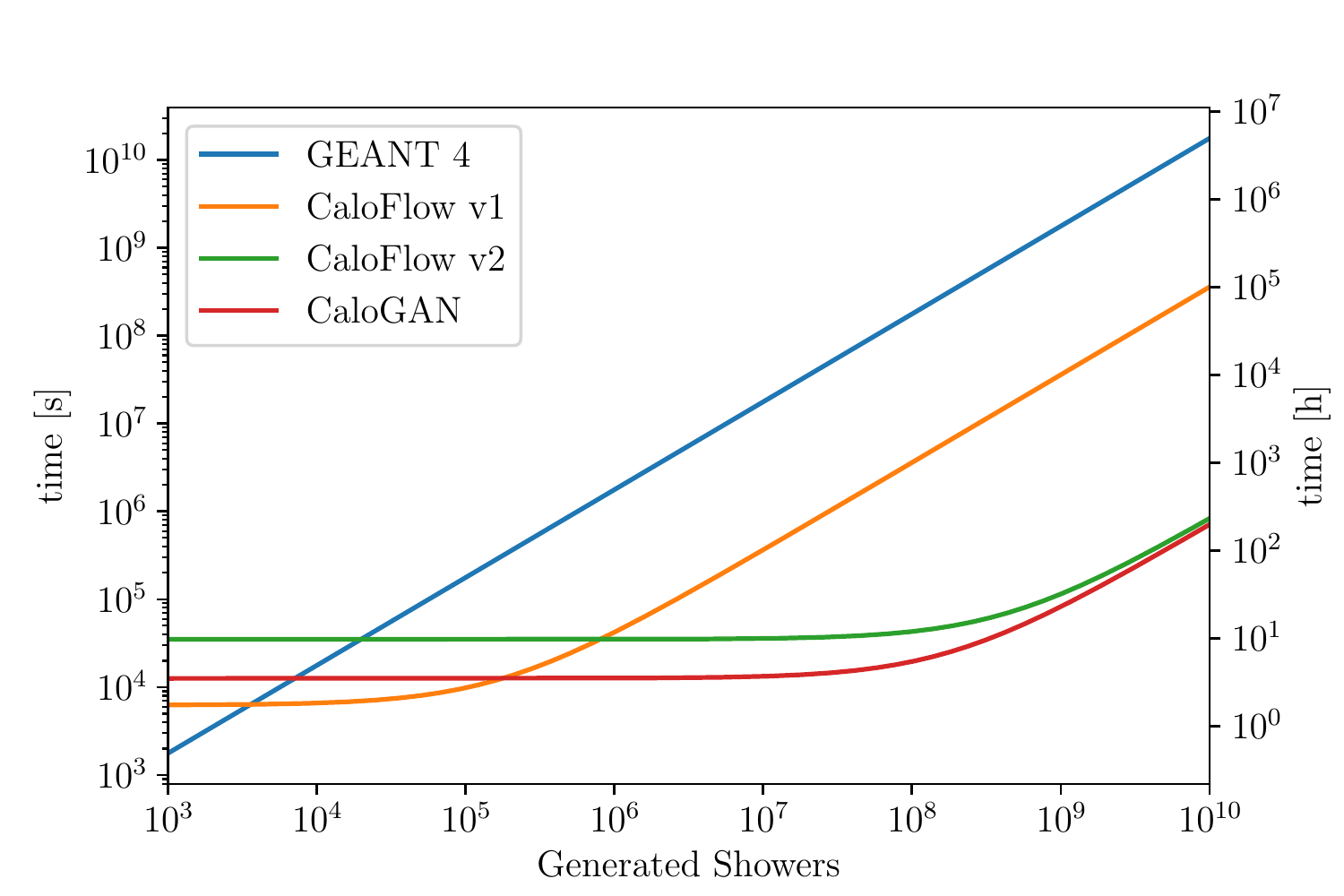}
  \caption{Comparison of shower generation times of GEANT4, CaloGAN~\cite{Paganini:2017hrr,Paganini:2017dwg}, CaloFlow v1~\cite{Krause:2021ilc}, and CaloFlow v2~\cite{Krause:2021wez}.}
  \label{fig:caloflow.timing}
\end{figure}

Figure~\ref{fig:caloflow.histos.piplus} shows histograms comparing the energy depositions in the 3 calorimeter layers of showers from GEANT4, CaloGAN, and CaloFlow. Figure~\ref{fig:caloflow.timing} shows the time needed to generate showers as a function of the number of requested showers. All deep generative surrogates outperform GEANT4.

\subsubsection{Electromagnetic Calorimeter of a Muon Collider Experiment}
\label{sec:calo-muoncoll}
The next generation of experiments in particle physics finds its common goal in expanding the phase space of events by increasing the energy scale of collisions beyond the TeV. Within this context a new international collaboration has been forming to study the design, challenges and outreach potential of a Muon Collider, which could be envisioned in Europe by the late 2040s~\cite{Long:2020wfp}. 

\begin{figure}
    \centering
    \includegraphics[width=0.5\textwidth]{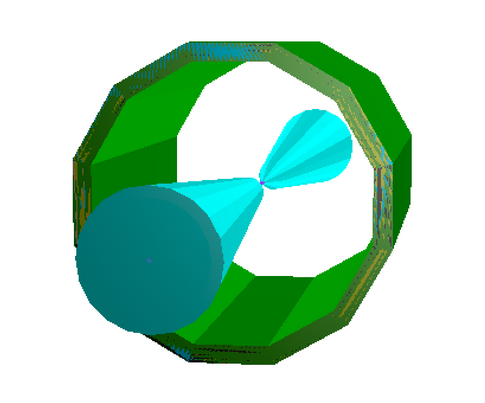}
    \caption{Muon Collider detector - nozzle and Crilin barrel design}
    \label{fig:nozzle}
\end{figure}

A Muon Collider comes with three main avantages, due to the properties of muons themselves: (I) a higher center-mass energy for each collision, since the total energy gets divided in less components with respect to a hadron collider; (II) little design limitation due to radiative effects, allowing us to use today's accelerator technology for a circular lepton collider; (III) the highest luminosity per energy used, which translates into highest energy efficiency for a collider experiment. It does however come with a set of challenges that need to be considered, especially due to muons not being stable particles: this generates a cloud of decay products that runs alongside the main beam line and interferes with detectors and instrumentation (Beam-Induced Background - BIB). The design of the detector, mainly mutuated from the ILC~\cite{Bartosik_2020}, does already include a double-cone shaped nozzle for shielding, as in Fig.~\ref{fig:nozzle}, which leaves a low-energy background inside the detector volume.

A new design for the electromagnetic calorimeter (ECAL) has been proposed (Crilin - crystal calorimeter with longitudinal information)~\cite{Cemmi:2021uum}, which implements an array of PbF2 crystals reducing the cost of instrumentation without loss in resolution from the original design. This solution is still in development phase and no definitive design has been proposed yet, which gives MODE a priceless chance to perform an optimization study in a systematic way and find the best possible setup to maximize outreach potential when considering BIB effects. 

\begin{figure}
    \centering
    \includegraphics[width=0.8\textwidth]{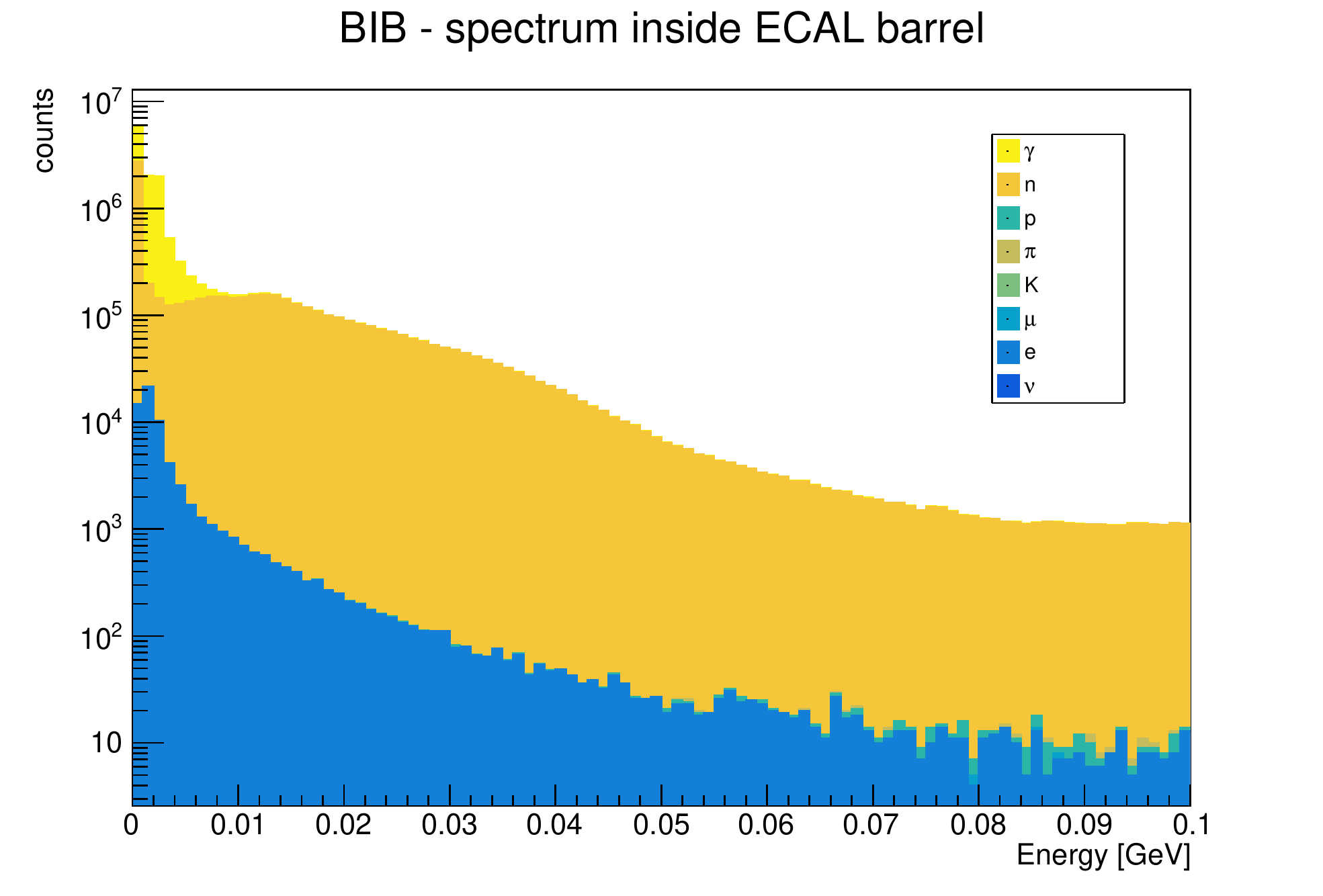}
    \caption{BIB spectrum components inside ECAL}
    \label{fig:spectrum}
\end{figure}

Figure~\ref{fig:spectrum} shows the BIB broken into its main components, obtained with a MARS15~\cite{Mokhov:2017klc} simulation of its interaction with the nozzle. The main contributions come from photons and neutrons and for a preliminary study we chose to focus on the former. 
We used Geant4 to study how the bulk material reacts to BIB photons, and obtain a continuous model for the detector response to be used for differentiation later. This has to be implemented in the geometric configuration of the detector, to get a toy model that can be dealt with for optimization studies. This toy model will also allow us to perform other studies, like event reconstruction using techniques such a Object Condensation~\cite{Kieseler:2020wcq}. 
At this point we will be able to set up a pipeline and define a cost function which allow us to study systematically the detector geometry. Various parameters such as crystal granularity an minimum resolution can be analyzed, as well as how we can best use the information at our disposal (for example where is the ideal placement for timing layers). 
The study can then be extended to more particle species for a more complete treatment of the BIB and possibly be beneficial to determine the final detector design.

\subsubsection{Optimization of the MUonE Detector}
\label{sec:muone}
The MUonE detector~\cite{Abbiendi:2677471} is an experiment proposed to be constructed and operated at the CERN North area, where it would intercept an intense 150-GeV muon beam directly upstream of the COMPASS experiment. The experiment aims at precisely measuring the $q^2$-differential cross section for elastic muon-electron scattering, which would allow to estimate the next-to-leading order hadronic contribution to the scattering process. That quantity would directly constrain one of the leading sources of uncertainty in the determination of the anomalous magnetic moment of the muon, $a_{\mu}=\frac{g-2}{2}$, improving the precision of comparisons of theoretical predictions with the experimental determination of that quantity, currently underway at Fermilab. Due to a long-standing disagreement of theory and measurement on this quantity, which might be produced by new physical processes contributing to the magnetic moment with quantum loop diagrams, the interest of improving the experimental precision is significant.

In its proposed setup, MUonE is composed of a set of 40 identical tracking stations, each 1-meter long, followed by an electromagnetic calorimeter. The stations measure the tracks of incoming muon and outgoing muon and electron in a set of silicon strip planes, and include beryllium targets for the scattering reaction. Given the well-defined goal of the experiment, the simple layout of the detector, and the very straightforward reconstruction of the event kinematics from measured particle trajectories, MUonE lends itself quite well to an exercise in optimization that may consider the full detector geometry, including positioning of active and passive layers as well as detailed characteristics of the detection elements, in the maximization of an objective that fully describes the final goal of the experiment. In fact, given the fully constrained kinematics of elastic scattering reactions, the stochastic elements of the problem are confined to charge collection processes in the silicon strips, the production of soft photons emitted in semi-elastic events, and the description of the detection of electron showers in the calorimeter; however, the details of the physics of the calorimeter, its performance and response may be neglected in a first study of the tracker optimization alone, as that detector provides redundancy in the measurement and a disentanglement of the symmetrical scattering situations where muon and electron emerge with the same divergence from the incoming muon direction. The information may be encoded as a probabilistic function that describes the correct identification of the electron shower, and a roughly Gaussian density function describing the estimate of the electron energy. 

\begin{center}
\begin{figure}[ht!]
\includegraphics[width=14cm]{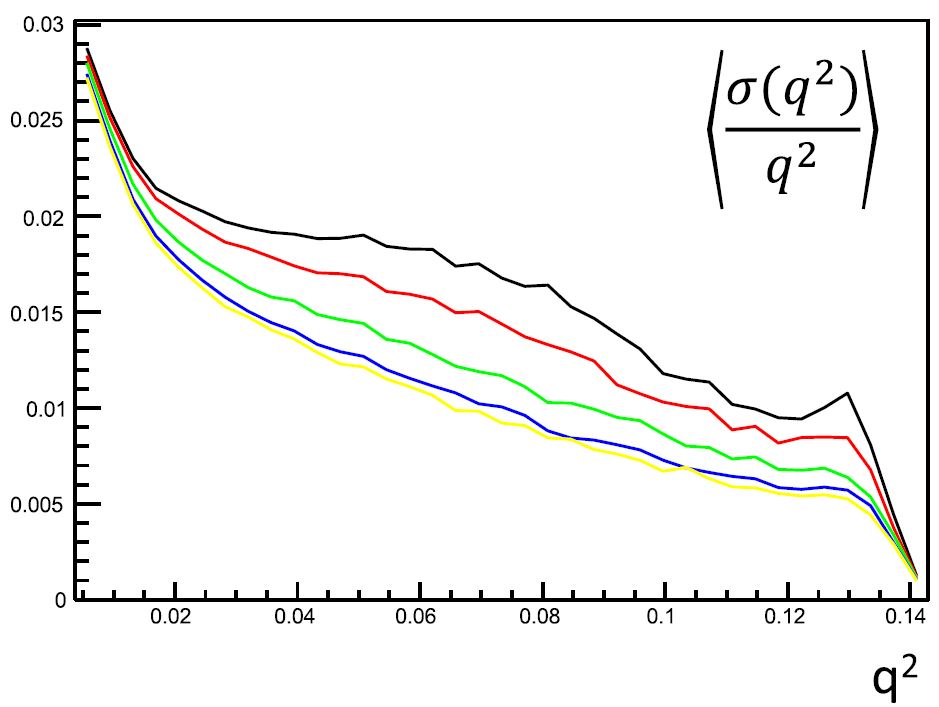}
\caption{Relative resolution in $q^2$ as a function of $q^2$ of muon-electron elastic scattering. The curves, coloured from black to yellow, show the improvements in resolution brought by separate optimization steps. The black curve corresponds to the original design, the yellow one to the result of a full optimization. A gain of a factor of 2 is achievable in relative resolution, which is the relevant figure of merit especially in the high-$q^2$ region.}
\label{f:muone_figure}
\end{figure}
\end{center}

More detail can be found in Ref.~\cite{dorigo2020geometry}, where a simple grid-scan approach has been followed, demonstrating that large gains are reachable (see Fig.~\ref{f:muone_figure}.  Although this brute-force procedure was found to be sustainable for the parameters considered, a more refined approach (as advocated by this document) would allow a more complex and precise multidimensional optimization with smaller computational cost.

As already noted in Sec.~\ref{sec:problem-statement}, the creation of a differentiable model of the geometry of silicon strips and target layers is straightforward. The other elements of the problem are listed below:\par

\begin{itemize}
    \item A differentiable model of the charge deposition in silicon strips by the three charged particles involved in the reactions. In a first approximation, this may be produced with a simple charge transport model as discussed in Ref.~\cite{dorigo2020geometry}; a more careful treatment may involve local generative surrogates. 
    \item A model of the self-calibration procedures that enable the constraining of the position uncertainty of detection and passive layers, as described in the above mentioned source.
    \item A model of the information-extraction procedures that allow to derive the quantity of interest and its uncertainty $\delta \alpha_{had} \pm \sigma_{\delta \alpha_{had}}$ from a given amount of identified elastic collision events.
\end{itemize}

\noindent
At the time of writing this document, the MUonE detector geometry is already defined, and construction of layers and stations has begun \footnote {Of the two solutions advocated in Ref.~\cite{dorigo2020geometry} to improve single-hit resolutions, the collaboration chose the tilting of sensors as in the right panel of Fig.~\ref{f:twosidedsensors}, which at the price of some complication in the geometry allows to reuse the construction jigs for the assembling of double-sided sensors, which do not provide a staggering of strips on the two sides.}. 
The above discussion is therefore mainly of academic value, yet this simple use case is illustrative of a number of possible similar problems involving silicon tracking detectors, to which many of the points made are relevant.

\subsubsection{Searches for Milli-charged Particles}
\label{sec:exotics}
It has always been intriguing why elementary particles in the standard model are having electric charges that are fraction 
($\pm 1, \pm 1/3, \pm 2/3$) of the charge of the electron? 
For instance, one can add a new gauge field $A_\mu^\prime$ that couples
to a new fermion $\chi$  with coupling $e^\prime$ and kinetically mixing (with coefficient $\kappa$) with the hypercharge 
field $B_\mu$. Then, by choosing a basis that will eliminate this 
kinetic mixing ($ A^\prime_\mu$ and $B_\mu$) to the standard model will lead to an electric charge for the new fermion $\chi$ that can be tuned by tuning $\kappa e^\prime$. Having small values of $\kappa e^\prime$ could be interpreted as a centicharged or even a millicharged particle. 

\begin{figure}[ht]
  \begin{center}
    \includegraphics[width=0.425\textwidth]{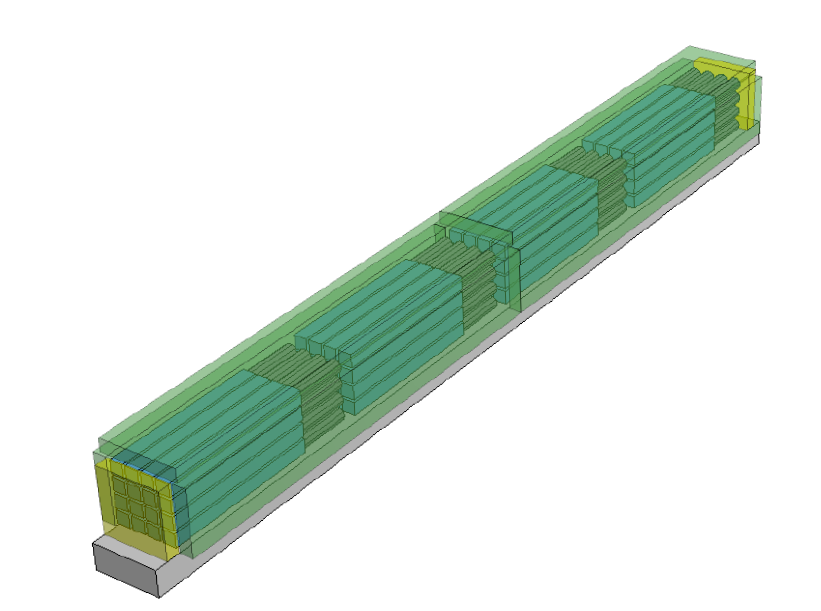}\hspace{2 cm}\includegraphics[width=0.425\textwidth]{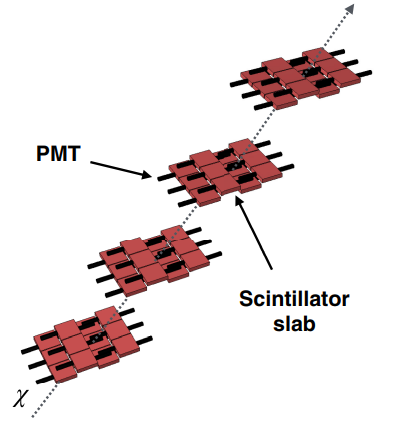}
    \caption{ Bar detector for the MilliQan detector, consisting of four layers each is composed of 4X4 scinitillator rectangular bars 5×5×60 cm$^3$ each with PMT on its end (left) and the slab detector for the MilliQan upgraded detector, consisting of four layers each has 4 slabs each of dimensions 5×5×60 cm$^3$ (right).(Reproduced from Ref.~\cite{milliqan2021}.)\label{f:milliqslab}}
\end{center}
\end{figure}

Search for millicharged particles ranges from Astrophysical observations 
to fixed targets experiments (SLAC-mQ~\cite{SLACmQ1998}) and finally to accelerator facilities searches like the LHC. In astrophysical proposals millicharged particles 
can be produced either in the atmosphere or in propagating through 
the Earth. Given the weak interaction of the millicharged particles with matter, neutrino detectors like Super-Kamiokande were proposed to detect them. Taking the advantage of the existing facilities is a cost effective solution if sensitivity is good enough for a viable detection/predictions of millicharged particles. The common denominator of most proposed neutrino dedicated experiment is the liquid detector structure. 
In Ref.~\cite{ArgoNeuT:2019ckq} the data from ArgoNeuT (a liquid Argon Time projection chamber) Experiment at 
Fermilab was reanalysed to check for mCP. The authors look for mCP in events triggered with triggering data acquisition set in coincidence with the NuMI beam spill signal. The majority of events does not signal neutrino propagation, due to the low neutrino interaction cross section and the spatial limit of the detector. Such “neutrino-empty” signals are explored for mCP presence.  In a recent mCP detection proposal 
using neutrino purposed experiment~\cite{Millicharged2021}, 
the Super-Kamiokande (liquid cherenkov detector, 50 kton) and the Jiangmen Underground Neutrino Observatory (JUNO, a 20 kton liquid scintillator detector) were proposed to detect mCP coming from/through  the atmosphere. In such scenario the detectors are already optimized/configured and there is less room for detector optimization. 
Dedicated mCP detectors like MilliQan~\cite{milliqan2020} has been designed installed in CMS drainage gallery. The initial detector 
design was a simple scintillator bars staked on top of each other 
and oriented toward the interaction point of CMS. Geant4 simulation were made to understand the behaviour of the detector and the layout 
of the different scintillator layers. Lots of lessons were 
learnt from the installed demonstrator  in 2018. A new design has been 
proposed and is under construction for the LHC Run~3~\cite{milliqan2021}. Besides the upgraded bar detector  design shown in Fig.~\ref{f:milliqslab} (left), a  
new slab detector has been added to increase angular acceptance, as illustrated in Fig.~\ref{f:milliqslab} (right).  
Muons constitute a major background for mCP detection. 
The scale of this kind of detectors is sufficiently contained, and the R\&D turnaround is sufficiently fast, that similar considerations may apply as in muon tomography (Sec.~\ref{sec:muography}), and an adaptation of TomOpt (Sec.~\ref{sec:tomopt}) is being considered for further optimization of the detector.  

\subsection{Astro-particle Physics and Neutrino Experiments}
\label{sec:astro}
In this section we sketch a common optimization problem which the astrophysics field will face in the near future. Multi-messenger observations of transient astrophysical events~\cite{Abbott_2017multimessenger, albert2019mm_o1search} aim to involve all available observatories. This common effort, started with announcing transients coming from gamma ray bursts~\cite{barthelmy2008gcn} which was expanded by announcements from GW detectors. Arrival of a trigger must be quickly and precisely evaluated because multi-messenger observations are incredibly fruitful when performed synchronously. For certain types of observatories, following a trigger may be costly thus we find important the development of common optimal strategy of observation.
Development of differentiable models including existing observatories and their characteristics, together with a suitable cost function would help in a quick calculation of optimal observation strategy for each event. Publication of such a strategy could be included in the GCN/TAN alert mechanism.

This section is devoted to the optimization of objective functions related with experiments.

\subsubsection{High-Energy Gamma-Ray Astronomy}

The field of very-high energy gamma-ray astrophysics studies the non-thermal universe at photon energies above tens of GeV to hundreds of TeV~\cite{2019scta.book.....C}. 
These gamma rays trace acceleration, propagation, and interaction of relativistic cosmic particles and provide insight into the most extreme environments around exploding stars and compact objects like black holes or neutron stars. 
Measurements of gamma rays from distance sources allow to characterise magnetic and photon background fields on intergalactic scales, and provide discovery space for fundamental physics topics such as the search for dark matter or for effects quantum gravity. 
Today's observatories have discovered about $10^3$ sources at energies above 1 GeV and roughly 150 sources at energies above 100 GeV.

\begin{figure}[!ht]
\includegraphics[width=0.35\textwidth]{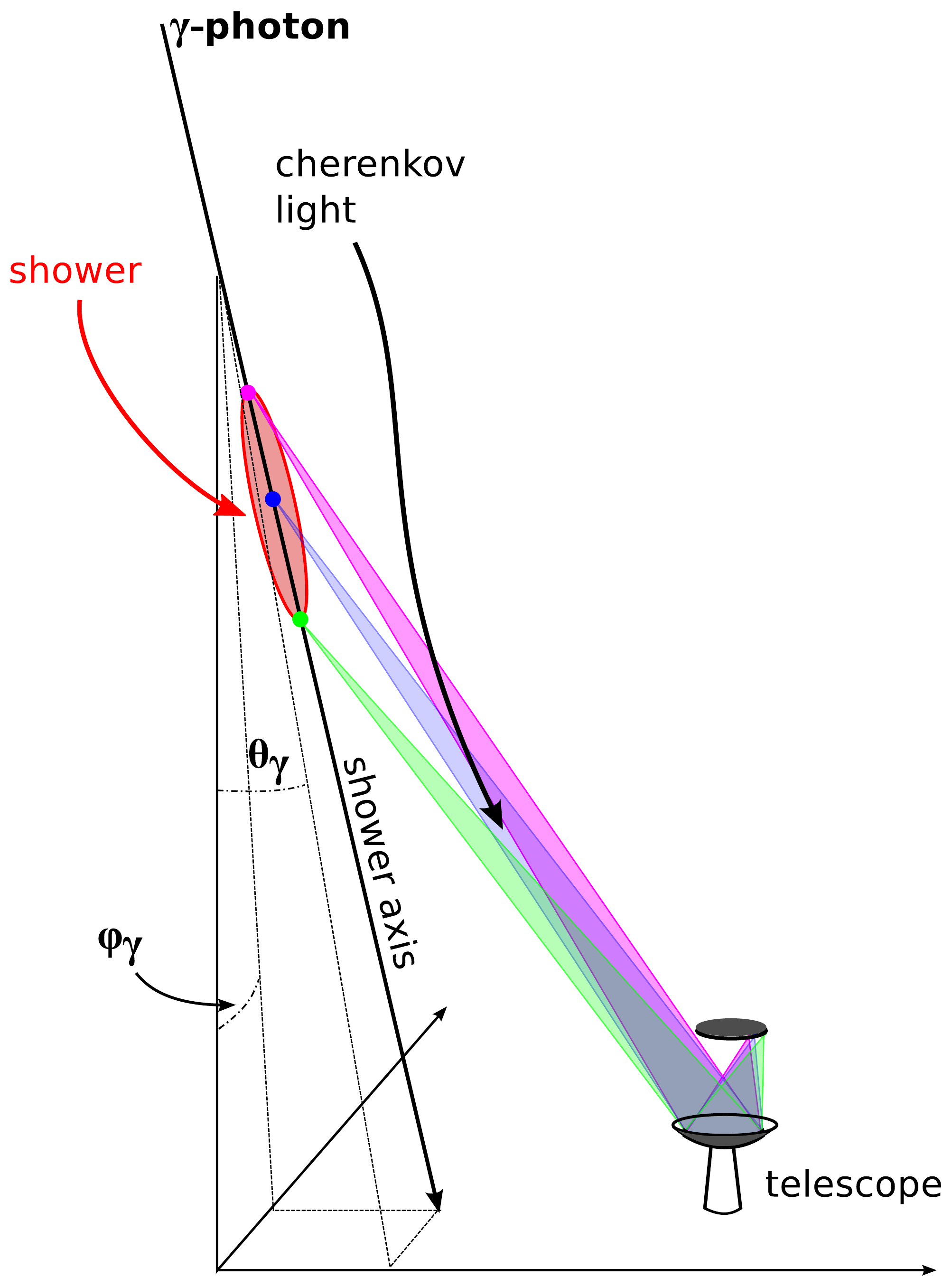}
\includegraphics[width=0.6\textwidth]{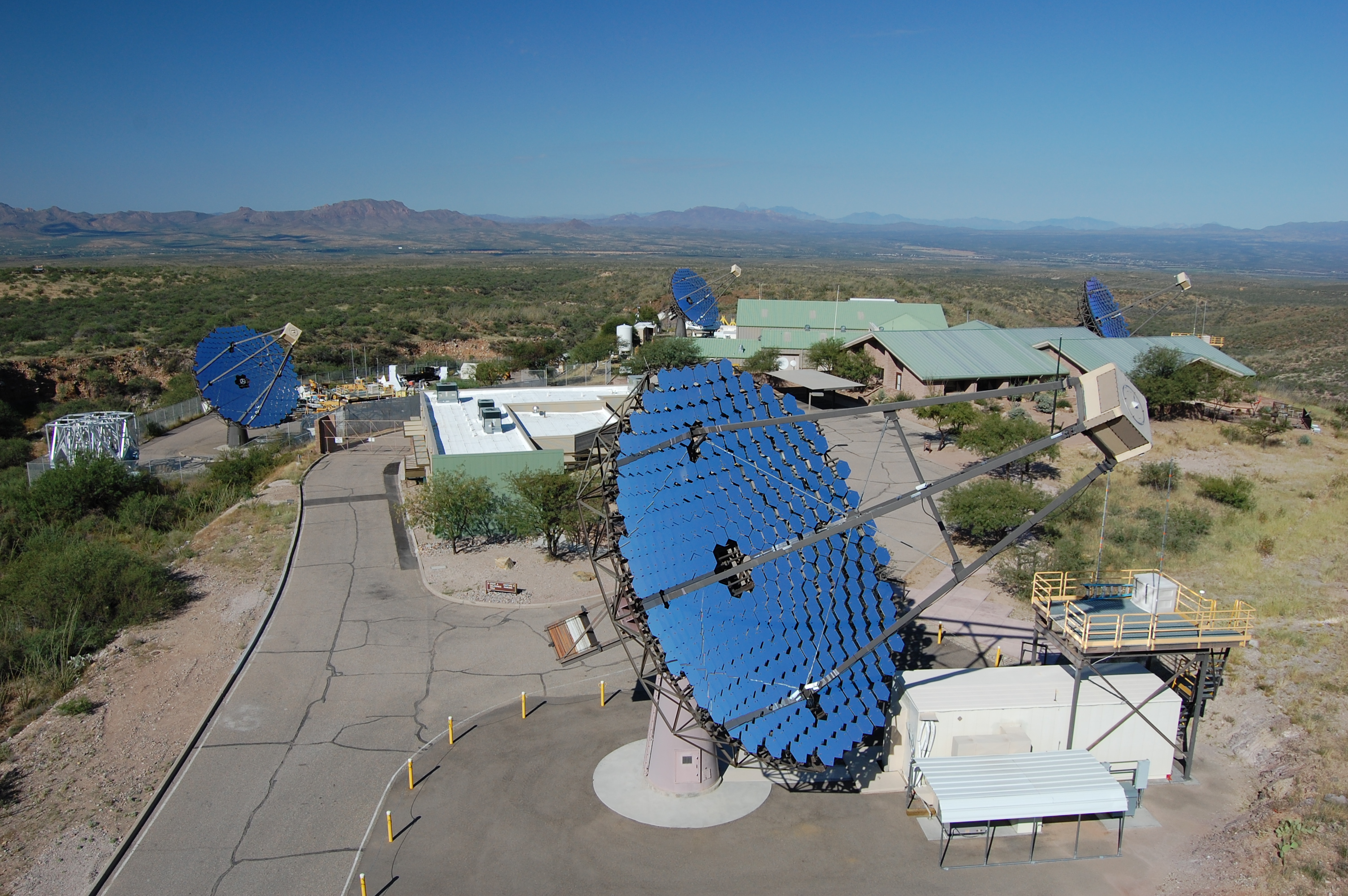}
\caption{\em Detection principle of imaging atmospheric Cherenkov telescopes (left). 
The sketch indicates the interaction of the primary gamma photon in the atmosphere, the development of the particle cascade, Cherenkov light emission and measurement by the telescope~\cite{2016PhDT.......565S}.
The VERITAS observatory consising of four 12-m diameter telescopes at the Fred Lawrence Whipple Observatory in Southern Arizona (right).}
\label{fig:iact-technique}
\end{figure}

The key characteristic of all high-energy astrophysical sources are low fluxes of less than $10^{-11}$ photons/cm$^{2}$/s and therefore the necessity of detectors with effective areas far beyond the scale of space-based instruments. 
Imaging atmospheric Cherenkov telescopes detect the Cherenkov light emitted by particle cascades initiated by the high-energetic photons when entering the atmosphere. 
This detection technique provides a calorimetric and imaging measurement, allowing to reconstruct the energy and direction of the primary photon and provide strong rejection power to the far more numerous background particle cascades initiated by charged cosmic rays.

The measurement principle and the main components are indicated in Fig.~\ref{fig:iact-technique}. 
All modern Cherenkov observatories consist of several telescopes. This is advantageous for two reasons: first to suppress background events by high-energy muons, and secondly for stereoscopic reconstruction of the main air shower axis providing improved direction and energy reconstruction.
The effective area of ground-based Cherenkov observatories is in the range of $10^{5}-10^{7}$ m$^{2}$.

Major operating instruments are the MAGIC observatory on the island of La Palma consisting of two 17-m telescopes, the H.E.S.S. array in Namibia with five telescopes, and the VERITAS observatory in southern Arizona with four 12-meter diameter telescopes (see Fig.~\ref{fig:iact-technique}). 
The next generation instrument is the Cherenkov Telescope Array (CTA), with detectors on sites in Chile and on La Palma with respectively 50 and 13 telescopes of different types. 
CTA is currently under construction, and it is expected to become operational in the mid 2020s.

\paragraph{State of the problem and current approach of optimization}

The complexity of the measurement process and the numerous background sources require detailed Monte Carlo simulations to understand the performance of an array of imaging Cherenkov telescopes. 
The Monte Carlo simulations include the full development of air showers in the atmosphere. 
All relevant hadronic and electromagnetic interactions, and the characterisation of atmospheric properties for Cherenkov photon generation and propagation (molecular density profile, aerosols) are part of the simulations. 
The telescope simulations comprises ray tracing of Cherenkov and background photons through the optical components, and the simulation of the photon detection and readout chain.

The current optimization approach relies on brute-force search covering a representative part of the parameter space by simulating large number of telescopes of different types (\egc different optical or camera designs) at different positions on the ground followed by a selection of a subset of telescopes at the analysis stage representing a candidate realisation for the observatory. 
This approach is inefficient, slow, and it involves a large amount of computing. 
As an example, a typical Monte Carlo production for CTA requires the simulation of $\approx 100$ billion gamma ray, protons, and electron events using about 1.5 PByte of storage for the detector simulation output and $100\times 10^{6}$ HS06 CPU hours~\cite{2021arXiv210804512G}.

It should be noted that this type of optimization is not only applied during construction, but also during operation: the optimal instrument configuration (\egc number of telescopes used, pointing mode, trigger settings) is different for the numerous types of astrophysical targets observed by gamma-ray observatories.

\paragraph{Figures of merits}

The figure of merits for the observatory observation are: 1.~detectability of an astrophysical signal for a given observation time. This is usually expressed in differential flux sensitivity, meaning the minimum flux per energy required to obtain a detection; 2.~sensitivity energy range and especially the required minimal energy threshold; 3.~precision of the direction and energy measurement; 4.~sky area covered by a single observation; 5. systematic uncertainty of flux, energy, and direction measurement.

\paragraph{Examples for optimizations}

Imaging atmospheric Cherenkov telescopes are complex instruments with a large number of parameters influencing the performance.  
A telescopes consist of one or two main mirrors with a collection area of the order of 100 m$^{2}$, a photo-detection plane consisting of an array of $10^{3}-10^{4}$ photo-multipliers or SiPM detectors, and a fast trigger and readout chain which allows to capture the nanosecond-long illumination from air showers. 
Fast trigger algorithms based on pre-defined patterns allow to suppress signals from night-sky background photons. 
Observatories can consist of a large number ($>50$ for CTA) of telescopes observing the same part of the sky simultaneously. 

Example 1 - optimization of telescope distances: assume in this example an array of $N$ imaging atmospheric Cherenkov telescopes ($4 < N < 100$). 
What is the optimal distance between those telescopes to achieve optimal performance for the figures of merits described before? 
A good starting point is the area illuminated by the Cherenkov light, which is roughly a circular area of 140 m radius. 
Increasing the telescope distance will enlarge the sensitive effective area and provide increase sensitivity at energies above several TeVs. 
Decreasing the telescope distance increases the telescope multiplicity per event and therefore improves the direction and energy reconstruction.

Example 2 - optimization of optical collection area with different telescope types: assume a gamma-ray observatory up to three different types of telescopes allowing to cover energy ranges from 10 GeV to 300 TeV. 
This setup is corresponds to the implementation of CTA, which will consist of three different telescope types (large-size telescopes with 23-m diameter mirrors; mid-sized telescopes with main dishes of 12 m diameter, and small-sized telescopes with about 4 m diameter dishes). 
How many of each telescope types are necessary for the optimum performances across the required energy range?

Example 3 - optimization of the observation mode for surveys, for which a large sky area is observed under consistent observation conditions. 
Parallel pointing of all telescopes restricts the observed area of the sky to the field-of view of each telescopes while maintaining large telescope multiplicities for precise reconstruction.
Divergent pointing, meaning that each telescope pointing is slightly offset from its neighbouring pointing allow to cover a larger area with reduced telescope multiplicity. 

\paragraph{Outlook and possible implementations}

The optimization of the configuration of ground-based arrays of imaging Cherenkov telescopes would benefit significantly from the suggested approach of differential programming. 
It would allow not only to find the optimum instrument, but also to optimize the observation mode of operating gamma-ray instruments. 
The simulation and analysis chains are implemented in modules and the new optimization approach can be implemented step-by-step (\egc by parameterizing the single telescope contribution and optimize the array configuration only).

\subsubsection{Interferometric Gravitational-Wave Detectors}

The field of experimental science, focused on the design of gravitational wave (GW) detectors~\cite{Willke20021377_geo600, Acernese2015_AdV}, has given spectacular results only recently when in 2015 the first direct detection of gravitational waves~\cite{Abbott2016_firstgw} was announced, but this field has much longer history. Since the first attempts to build a GW detector in 1960 by the construction of a Weber bar, in the field have risen competitive ideas which at the end turned into real technologies. Among them there are space missions like the Doppler space-craft tracking or the GW in-space antenna LISA~\cite{amaro2017laser} and at the end interferometric ground base detectors LIGO/Virgo/Kagra (LVK). The list of the latter can be extended to the next generation detector Einstein Telescope (ET)~\cite{maggiore2020etscience}. At present, these technologies are at different stage of development. Most of the Weber bar sites were closed several years ago after decades of development and measurement campaigns; Doppler space-craft tracking was implemented and tested in the past in several space-missions; LISA has proven its concept by completing all the tests of the LISA Pathfinder mission in 2017~\cite{PhysRevLett.120.061101_lisa_PF}. 
LVK has concluded the third observing run O3 which brought a set of new observations~\cite{abbott2021gwtc3} and permitted general studies on the Universe~\cite{lvk2021gwtc3_population}. LVK members are currently preparing for the O4 by upgrading their detectors and planing its activity in the future~\cite{abbott2020prospects} with the aim to continue the search for known coalescence signals and looking for yet undetected waveforms like continuous gravitational waves~\cite{abbott2017continouswaves} and waves originating from binary sources with extreme configuration~\cite{abbott2018searchrare}
In the following paragraph we will focus on the technology of ground-based interferometric antenna in order to find possible applications to differentiable programming (DP) in improving its design.

\begin{figure}[!ht]
\includegraphics[width=0.59\textwidth]{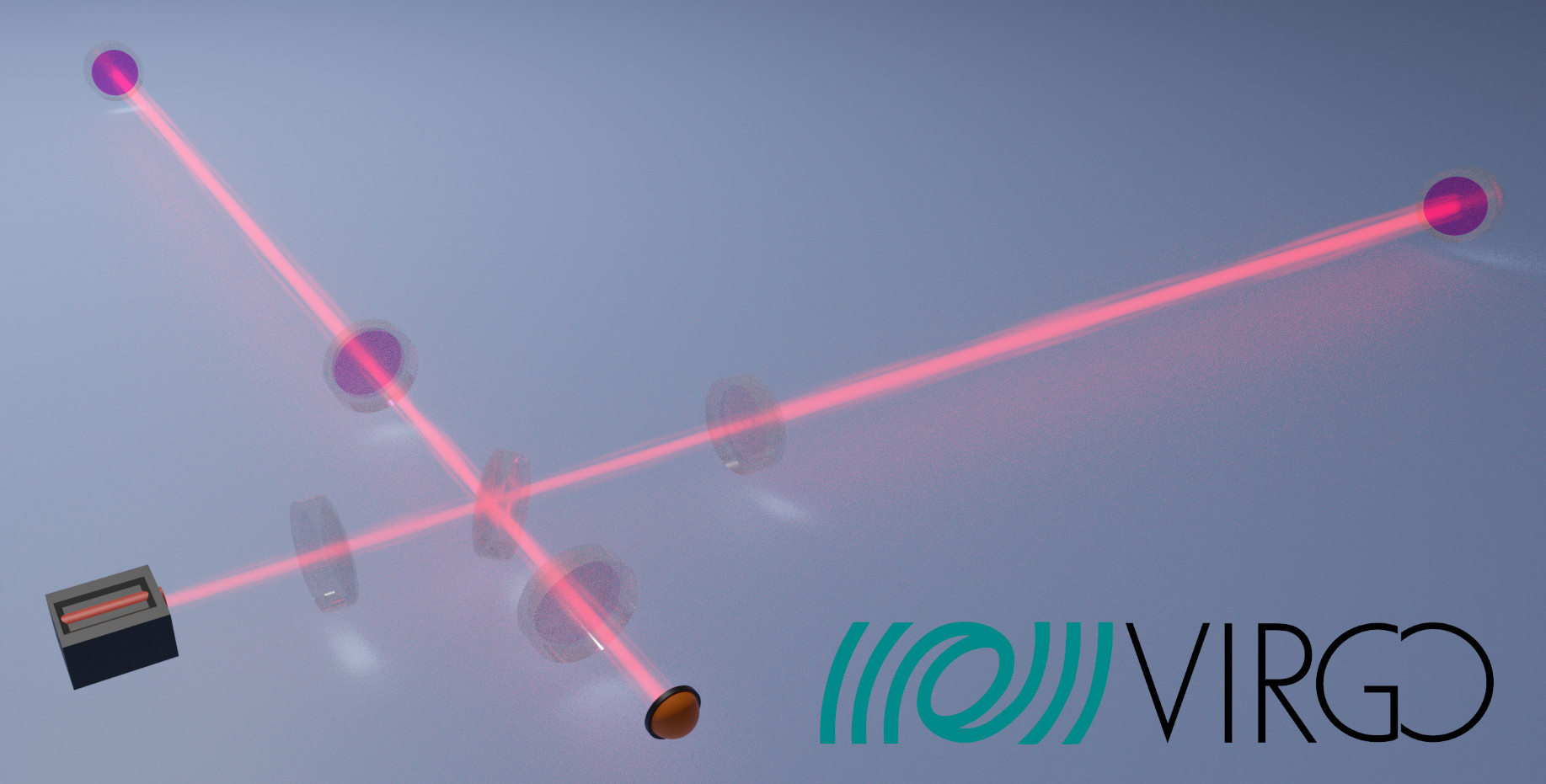}
\includegraphics[width=0.4\textwidth]{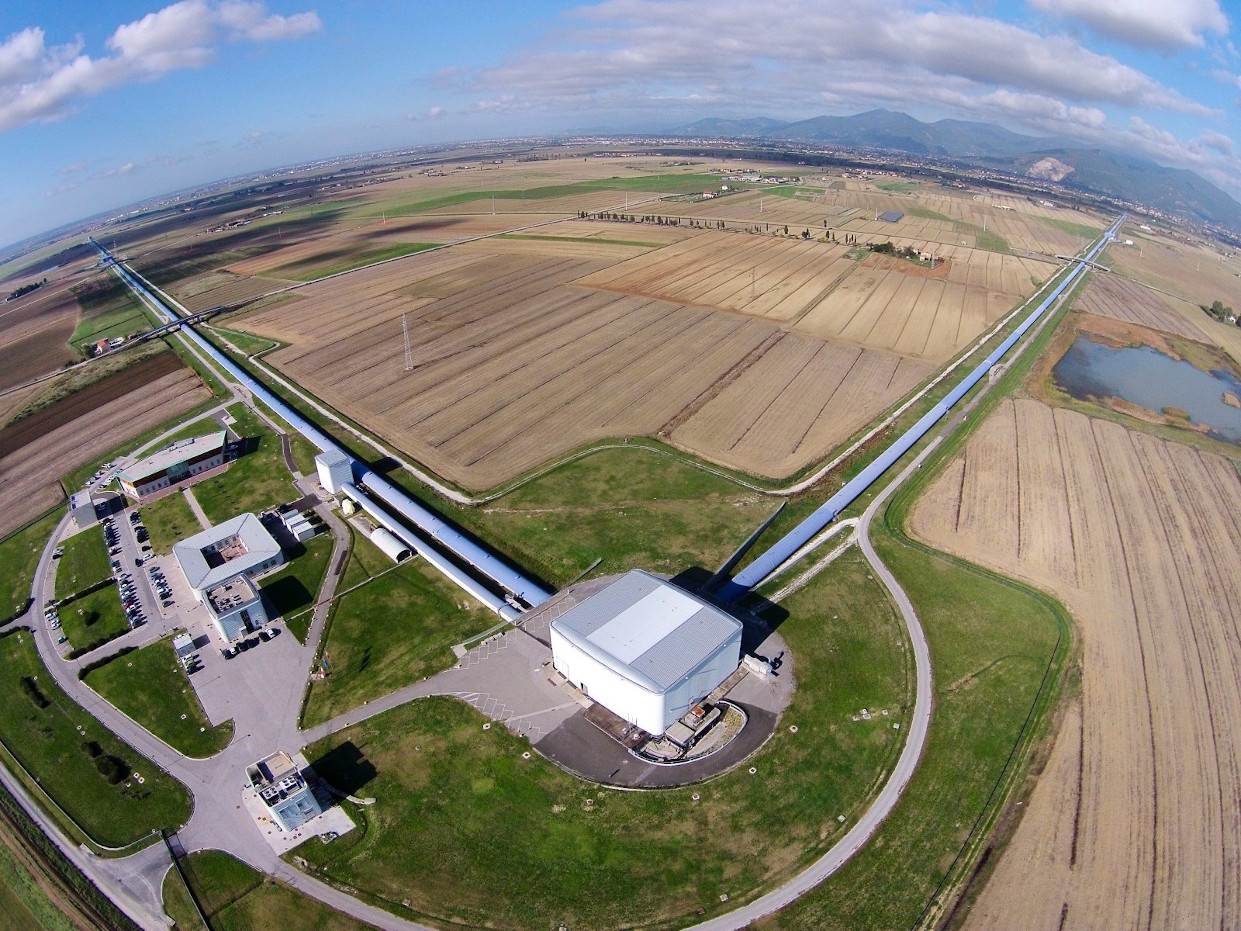}
\caption{\em Schematic diagram of the second generation interferometric gravitational wave detector. For simplicity, only the main optic elements are depicted. Laser source and detection are schematically depicted as black box and orange semi-circle (left). 
Virgo aerial view, the two 3-km arms of the interferometer meet the central building where the laser is generated, and five mirrors, including the beam-splitter mirror, are hosted. Optical benches used to acquire the interference pattern are also hosted in this building. On the left part, the other buildings host EGO offices. Credits: the Virgo Collaboration/N. Baldocchi (right).}
\label{fig:gw_itf}
\end{figure}

\paragraph{State of the problem and current approach of optimization}

More precisely, AdvancedVirgo+~\cite{acernese2017Virgo_status, acernese2018aVirgo_status, bersanetti2021aVirgo_status} is conceptually a quantum-enhanced doubly recycled large Michelson interferometer (ITF) with Fabry-Perot arm cavities, capable of detecting GWs in the frequency range between \SIrange{10}{20}{\kilo\hertz}. Figure~\ref{fig:gw_itf} depicts the configuration of seven mirrors which are the main optical components. Despite its apparent simplicity, the detector contains more than one thousand optical components divided in subsystems as follows: pre-stabilised laser, injection, detection, thermal compensation system and quantum noise reduction system. Auxiliary optics are placed on optical tables close to the main optics.
Gravitational wave detection is based on the sensing of the relative position of the main optical components (mirrors). Isolation of these components from vibrations is fundamental since the ambient where the detector is located is far from being quiet. Main mirrors hangs on fibres made of the same material as the substrate forming a monolithic block called a monolithic suspension~\cite{Amico20023318}. The entire set is subsequently suspended on the multi-stage vibration isolator called super attenuator~\cite{braccini1996seismic}. This sophisticated, actively controlled mechanical structures provide vibrations attenuation by the factor of $10^{13}$ in the ITF frequency operation range. Many of Virgo optical benches are also suspended on dedicated smaller suspensions. Suspended elements are encapsulated in vacuum tanks what assures cleanliness, additional layer of isolation both acoustic and electromagnetic and stabilization of the light rays path at the same time. The amount of components in the detector, the required precision of their relative position and the presence of suspended optical components requires implementation of active control loops. Many of the Virgo components are controlled actively or continuously monitored. For this purposes, Virgo uses a genuine, distributed data acquisition and signal processing system (DAQ)~\cite{bonnand2019acl, casanueva2019tools}.  During ITF operation DAQ assures the proper alignment, automatic lock and lock recovery and registers scientific data. Let us now consider a possible application of DP in the ITF design.

\paragraph{Figures of merits}

A direct measure of the ITF figure of merit is the distribution of noise in the noise budget and the detector observation range. LVK collaboration developed gwinc~\cite{gwinc}, a top-level simulation tool which includes and sums up all noise contributions. However, the important information provided by gwinc is not sufficient to optimize an individual subsystem and more detailed simulator are involved in the design process. We can distinguish three types of design in ITFs: mechanical, optical and electronic. In general, most of the design tasks belong to more than one class.
Considering the long period an ITF spends in the observing run, the second figure of merit must be attributed to the stability of the overall noise and to the detector duty-cycle.
In the next paragraph we describe possible cases of application of DP in this context.

\paragraph{Examples of optimizations}

At first glance the ITF optical system and ITF base topology are the main subjects of possible optimization. This is certainly the case for the ET design. In case of Virgo, its optical setup has been already heavily optimized for a wide spectrum of parameters. The optimization was performed manually but it was supported by the three main types of software: FFT simulators, ray-tracing simulators and modal simulators~\cite{degallaix2020oscar, freise2013finesse, brown2020pykat}. None of the mentioned tools is able to simulate all aspects of the detector and none of them was implemented in accordance with the DP idea. In this field, the first necessary effort is the development of DP-compliant simulators. The current Virgo base topology itself is a perfect reference for testing the future software. Unlike the ITF base topology, the optical subsystems surrounding the main optical elements certainly belong to the first class of design task, which can be approached by DP during their future upgrades. In particular, a possible candidate for DP optimization is the development bench for EPR squeezing~\cite{nguyen2021automated, giacoppo2021towards}.

The second class of challenging design tasks is constituted by mechanical structures. Among the most common parameters of mechanical systems, like resonant frequencies or breaking load, interferometric GW detectors rely on the characteristics of thermal noise. In Virgo, mechanical design is aided by analytical models specific for each design task, such as: super-attenuator, payload and monolithic suspension. In some cases, the finite-elements method was used to increase simulation completeness~\cite{Aisa2016644}. Careful analysis of existing models and their upgrade to a DP-compliant form is the first step to make. In our opinion, payload optimization is a good candidate to test the automatic optimization approach for mechanical design tasks.

The Einstein Telescope is a new generation GW detector, currently in a phase of conceptual design and funding. Its design is strongly inspired by the LIGO/Virgo/Kagra success. Certainly, ET will inherit many technologies from existing interferometric GW detectors. At variance with the Virgo case, ET can benefit from DP already at the conceptual design stage, where both base topology and specific subsystems can be optimized. This outcome strongly motivates the development of DP-compatible simulators for interferometric GW detectors.

\paragraph{Outlook and possible implementations}
In our opinion, the most efficient way to benefit from DP optimization is to follow the list of noise contributions and to provide a suitable software tool at the beginning of the design process. However, this approach may not be immediately implementable, hence the workflow may be conceived differently; specific simulators may be built and then applied to the most promising tasks of the design.

\subsubsection{Radio Detection of High-Energy Neutrinos}

Neutrinos are considered a perfect cosmic messenger as they traverse the universe unimpeded, and their flight direction points back to their sources. Ultra-high energy (UHE) neutrinos will provide insight into the inner processes of the most violent phenomena in our universe, those that happen in the vicinity of super-massive black holes (\egc in active galactic nuclei), in neutron star mergers, or gamma ray bursts. The detection of these ghostly, extremely energetic elementary particles would be one of the most important discoveries in astro-particle physics in the 21st century. 

To be able to measure the low flux of UHE neutrinos on Earth, a new detector technology has been developed, instrumenting polar ice sheets with radio antennas to search for neutrinos passing through the ice. A sparse array of radio-detector stations can instrument large volumes efficiently due to the large attenuation length (\SI{\approx 1}{km}) of radio signals in ice. Radio antennas measure the radio emission created by a neutrino-induced particle shower in ice via the Askaryan effect.

The technology to build and operate an array of radio detector stations that instrument the polar ice to catch the elusive UHE neutrino interactions has matured in small pilot arrays over the last decade~\cite{COSPAR2018, ARA2020-limit}. Now, for the first time, a sizeable detector is being constructed. The deployment of the Radio Neutrino Observatory in Greenland (RNO-G) started in June 2021~\cite{Aguilar:2020RNOG}. Detector completion is expected in 2024. Then, 35 autonomous radio-detector stations will instrument 300 Gigatons of ice. At the same time, an order of magnitude larger radio array is being planned as part of the IceCube-Gen2 efforts to build the next-generation neutrino observatory at the South Pole~\cite{IceCubeGen2-2020, HallmannICRC2021} with the start of production as early as 2025. 

\paragraph{Figures of merit}
The development of the NuRadioMC simulation code~\cite{NuRadioReco2019, NuRadioMC2019} enables a precise estimation of the figures of merits for arbitrary detector designs. The figures of merit are 1) the overall sensitivity of the detector, \iec the expected rate of detected neutrinos; 2) the achievable reconstruction resolution of the neutrino energy, direction and flavor; 3) the ability to reject rare backgrounds. These three quantities can be combined into high-level sensitivity estimates, \egc diffuse flux sensitivity or the discovery potential for point sources. At the same time, the monetary cost of the detector as well as deployment and engineering constraints need to be considered to formulate a sensible objective function; otherwise, a larger detector with more radio detector stations would always provide better performance and be preferred. Thus, the formulation of an objective function is challenging, but it is considered doable. 

\begin{figure}[t]
    \centering
    \includegraphics[width=0.49\textwidth]{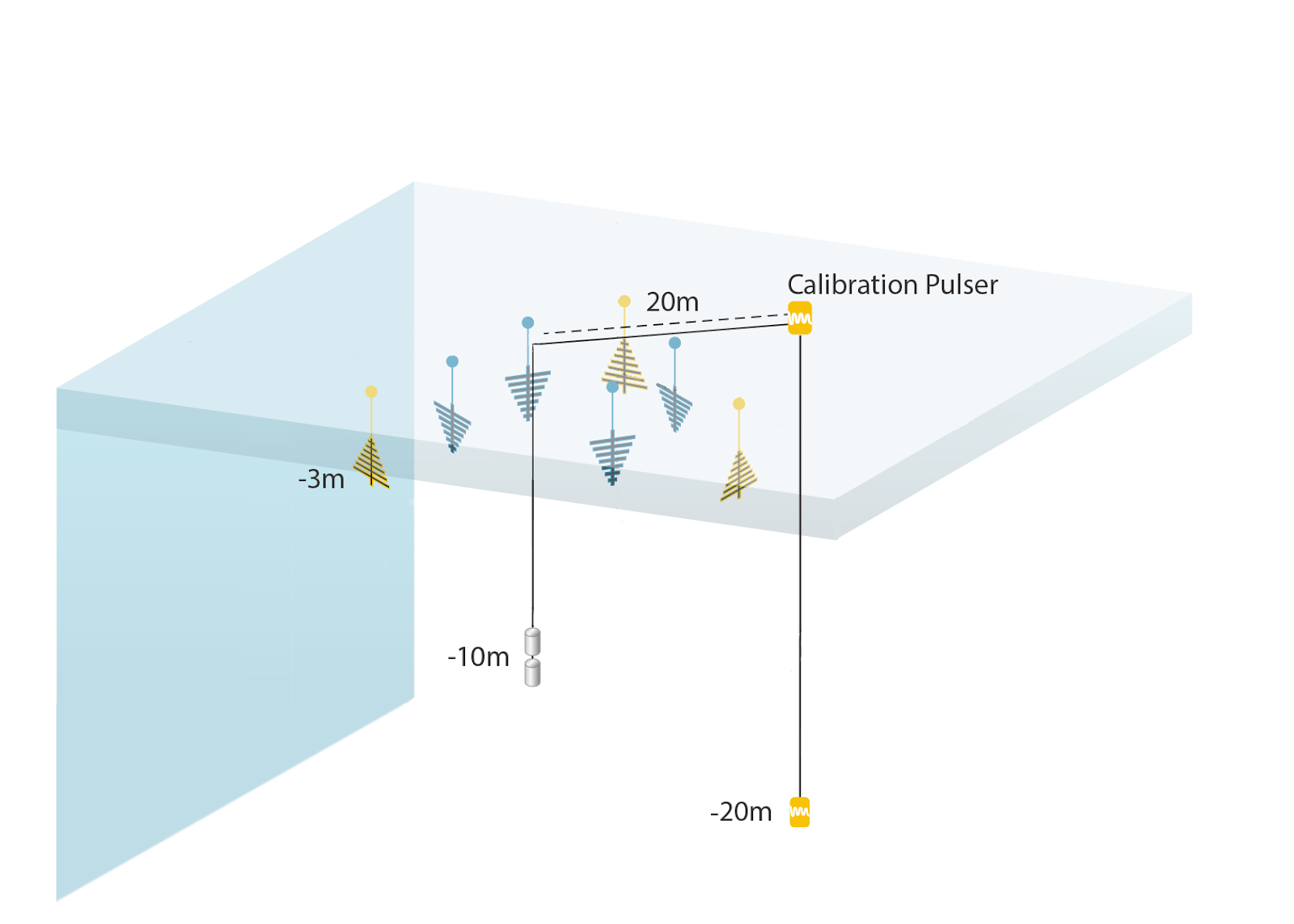}
    \includegraphics[width=0.49\textwidth]{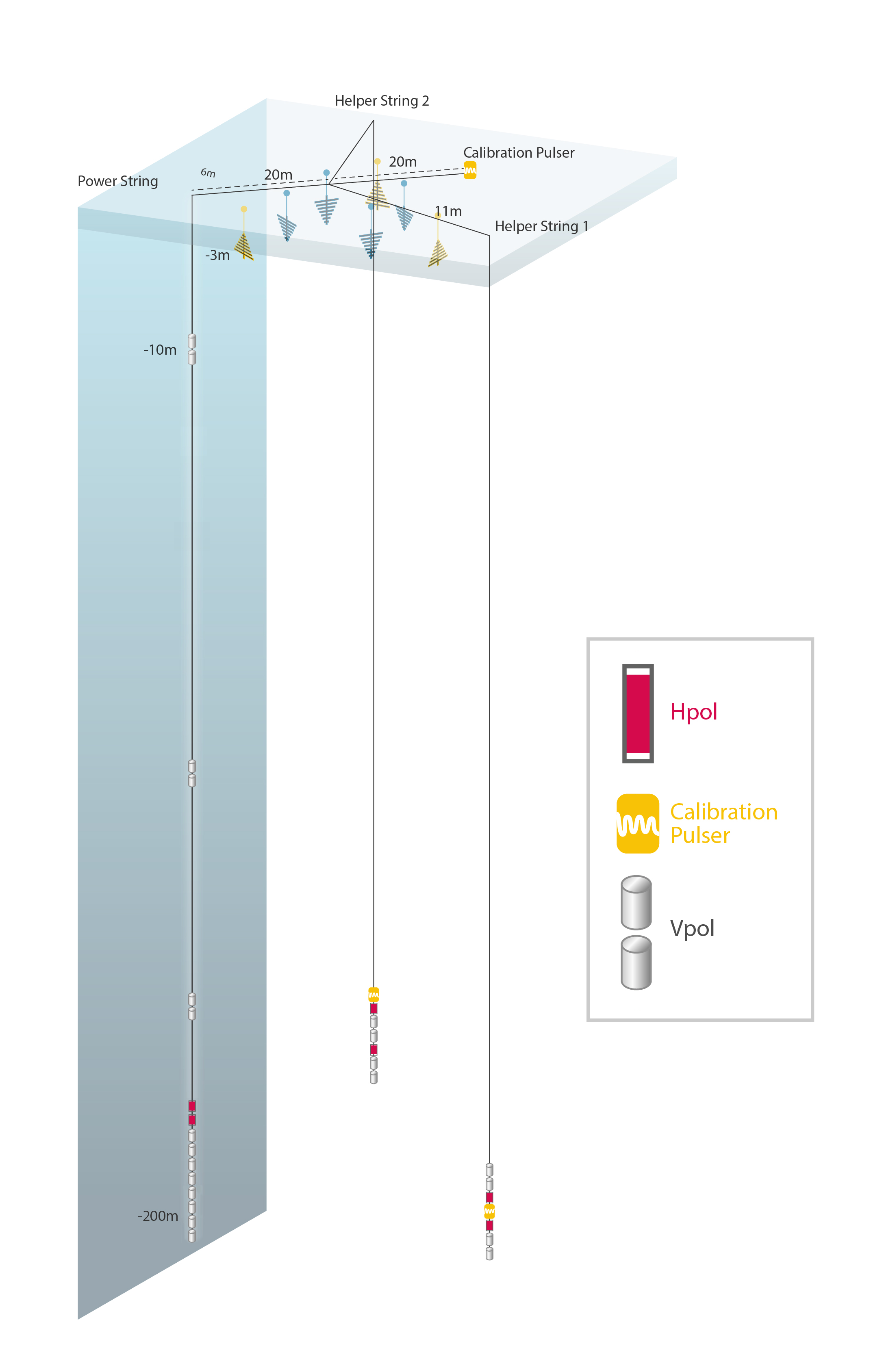}
    \caption{Two possible radio detector station designs considered for IceCube-Gen2. Figures reproduced from Ref.~\cite{HallmannICRC2021}.}
    \label{fig:Gen2station}
\end{figure}

\paragraph{State of the problem and current approach of optimization} A working radio detector that fulfils the main scientific objectives of the experiment can be built with a variety of different station designs. The main differences are in the number of antennas per station and their largest depth, as well as in the total number of radio stations and their separation. Two examples of station designs are shown in Fig.~\ref{fig:Gen2station}. For example, the sensitivity per station can be increased by installing the antennas deeper in ice, but at the same time the costs and deployment efforts increase. For similar monetary costs and deployment effort, a larger number of shallow detector stations (see Fig.~\ref{fig:Gen2station}, left) can be installed yielding the same overall sensitivity. Furthermore, the relative positions and orientations of antennas impact the reconstruction resolution. These two examples already show the large parameters space that needs to be considered.

An estimation of the figures of merit requires a time-consuming MC production with about 1 million core hours per design. The current optimization approach relies on estimating a good station and array layout using scaling relations, and only testing one or a few options with a full MC production. 

\paragraph{Outlook and plan towards an automated detector optimization} An automated detector optimization bears large potential to further optimize and fine-tune a radio neutrino detector. A first viable step would be the optimization of antenna positions within a single detector station with the objective of minimizing the reconstruction uncertainties of the neutrino energy and direction. This requires 1) a substantial speed up in the MC simulation and 2) an automated estimation of the reconstruction performance. Part 1) can be achieved using surrogate models (\eg through a generative adversarial neural network). Part 2) can be achieved through a deep-learning-based reconstruction method. The MC data set itself is used to learn to predict the neutrino direction and energy. The performance of the network is validated on an independent test dataset and used as a proxy for the reconstruction performance. Initial work (\eg Ref.~\cite{DLNuRadioICRC1, DLNuRadioICRC2}) shows that deep neural networks are capable of reconstructing the neutrino properties with good precision; however, large training data sets of $>$ 1 million events are required. The demand of training data size can be reduced by starting with a pre-trained network; \iec a large training data set only needs to be produced once, and using efficient data resampling. 

Optimizing the speed of estimating the reconstruction performance for different antenna positions is challenging but seems doable with the steps outlined above and within the framework of the MODE collaboration. 


\subsection{Cosmic-Ray Muon Imaging}
\label{sec:muography}
\begin{figure}[ht]
\includegraphics[width=\textwidth]{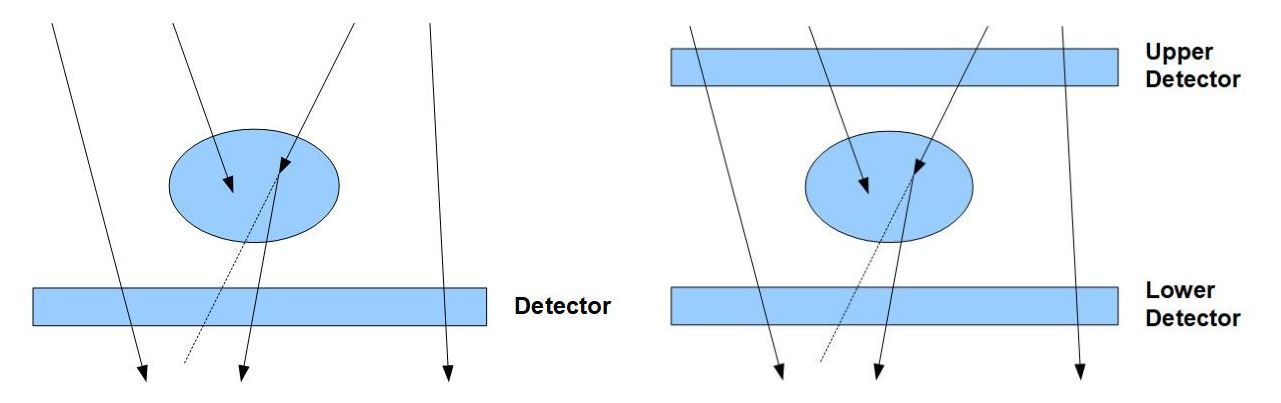}
\caption{\em Typical detector configurations in muon radiography, with respect to the object to be imaged, with the absorption method (left), where the fraction of muons surviving energy loss is measured, and with the scattering method (right), where the observable of interest is the root-mean-square of the deflection angle.}
\label{f:muography-methods}
\end{figure}

The abundant natural flux of atmospheric muons, and their large penetration power, have been exploited for the imaging of a large variety of objects spanning in size from O(1~m) to O(1~km), with applications including archaeology, volcanology, homeland security, nuclear safety, industrial process control, and many others~\cite{Bonechi:2019ckl}. 
In some applications, the volume of interest can be sandwiched between two trackers and one can measure the {\bf scattering} of the muons that pass the target volume, which is correlated with the atomic number $Z$ of the material. 
When the target volume is very large (\egc a mountain or an entire building), a single tracker is located downstream to measure the {\bf absorption} of the muon flux through the target, from which a density map can be derived. 
The two methods are illustrated in Fig.~\ref{f:muography-methods}. 

By optimizing the layout of the detection elements, possible large gains in the resolution and material identification potential of a muon tracking system are achievable. 
This is a domain of application where detector development can have a very fast turn-around, and where one may want to quickly react to changes in the practical constraints (\eg costs and budget may evolve during the lifetime of the project) or even in the goals of the project itself (\eg switching from a search for cavities to material discrimination, for which the figure of merit is not the same, or changing the object to be imaged). 
Therefore, an automatic optimization procedure is very appealing in the context of muon radiography, as it may have to be re-run several times along the lifetime of a single project.

The rest of this sub-section is organized as follows:
Section~\ref{sec:muography-fom} elaborates on the variety of figures of merit that depend on the specific use case; Section~\ref{sec:muography-parameters} lists the typical detector parameters to be optimized and distinguishes them by categories; Section~\ref{sec:tomopt} presents a new software project promoted by MODE, which has the goal to apply differentiable programming to a category of muon tomography applications; Section~\ref{sec:muography-industrial} provides examples of industrial applications that can profit from the MODE approach; finally, Sec.~\ref{sec:portable} illustrates a project for a portable and modular detection set-up for muography that would maximally profit from MODE. 


\subsubsection{Figures of Merit}
\label{sec:muography-fom}

The figure of merit to be maximized or minimized may be very different for different use cases, leading to different outcomes for the optimal detector setup. 

In many archaeological use cases the goal is to look for unknown voids, without assumptions on size, shape and location (see, \egc Ref.~\cite{ScanPyramids,MtEchia}). An appropriate figure of merit might thus be the sensitivity to localized excesses in muon flux with respect to a baseline (similarly to searches for new particles or resonances in particle and nuclear physics). 

In other cases, the goal is to identify the position, size, and shape of materials that are known or presumed to be present in the field of view of the detectors, \egc ice and rock in Ref.~\cite{SwissGlaciers}, or various special nuclear materials surrounded by shielding materials, in applications such as nuclear waste monitoring~\cite{Mahon2018}. 
In these cases, an appropriate figure of merit may be the sharpness of the images obtained. 
Depending on the atomic number $Z$ of the relevant materials, the size of the target volume, and other factors, the absorption or scattering methods may be more appropriate, and there are some grey areas where it is not obvious {\em a priori} which of the two detector configurations of Fig.~\ref{f:muography-methods} is the most convenient; the choice should be made after the parameters are optimized for each option. 

Identification of special nuclear materials is also the goal of the ``muon portals'' that exploit the scattering method for the inspection of cargoes at border controls~\cite{Riggi:2017izf} (see also Sec.~\ref{sec:tomopt} below). However, the relevant figure of merit is different in this case: more than forming an image, the aim is to fire an alarm when a forbidden material is present in the cargo. The constraints set by the border authorities include the time allowed for each inspection in the portal (typically of the order of minutes), and the false positive rate, because each alarm would cause a lengthy inspection, hence a financial compensation to the cargo company for the time lost in case of a false alarm. Given these constraints, therefore, a pertinent figure of merit would be the false negative rate given a fixed time and a false positive rate.

\subsubsection{Parameters of the Optimization Task} 
\label{sec:muography-parameters}
We distinguish here between local and global parameters. 
Local parameters are related to individual detector units, and are specific of the technology employed. Global parameters instead concern the geometry and spatial deployment of the overall detector setup.

The local parameters are the same as for any other tracking devices, such as the number of strips or pixels, and their pitch. 
Global parameters include the number of layers, their distance, and the surface area of a single layer (which can be composed of several adjacent detection units). Unless the location and orientation are entirely constrained by logistic considerations, the optimization can also take into account global parameters such as the distance from the target and the inclination of the apparatus with respect to the zenith angle (from where most of the muon flux is coming, with an approximate $\cos^2{\theta}$ dependence).

\begin{figure}[ht!]
\centering
\includegraphics[width=0.5\textwidth]{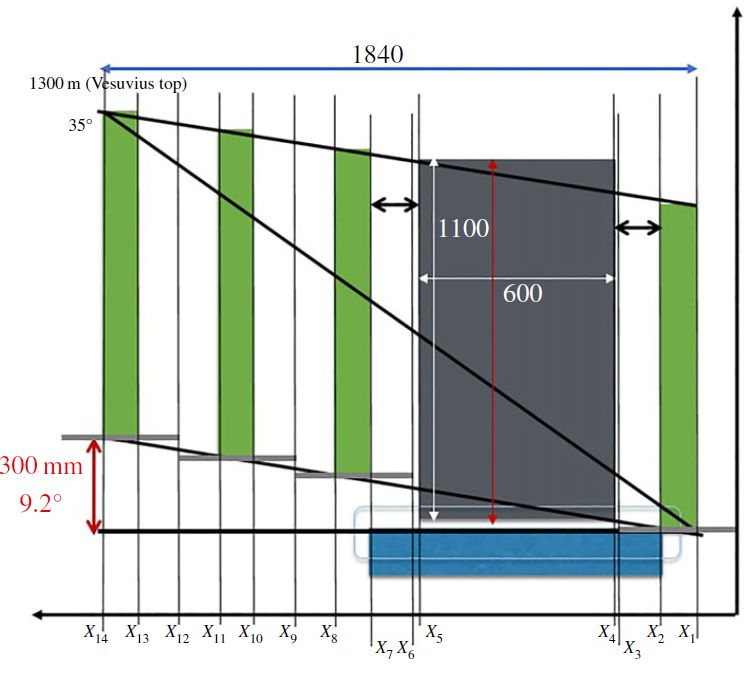} \\
\includegraphics[width=\textwidth]{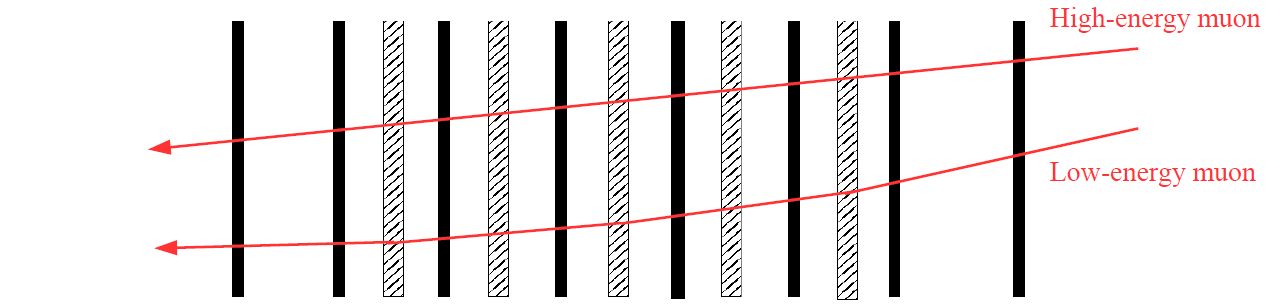}
\caption{\em Two ways to arrange passive material for momentum filtering or momentum estimation in a muon radiography apparatus. In the MURAVES experiment~\cite{MURAVES} (top) a thick lead wall (grey rectangle in the figure) is placed before the last active layer (the object of interest, Mt. Vesuvius, is on the left). In the muon telescopes of the Sakurajima Muography Observatory~\cite{Muography-SMO} (bottom) relatively thin slabs of passive material (grey) are alternated with MWPC detectors (black).}
\label{f:muography-passive-material}
\end{figure}

Multiple Coulomb scattering induces angular deflections whose root mean square depends on the inverse of momentum, implying that low-momentum muons are a nuisance to imaging as their arrival direction is less correlated with their trajectory within the target. This blurring can be reduced by either rejecting muons below a certain momentum threshold or by taking momentum into account as an input to the imaging algorithms. 
However, muon radiography detectors do not measure momentum directly, as that would demand the usage of large magnets, making the apparatus very expensive and cumbersome. 
Instead, the usage of slabs of passive material (\eg lead or steel) is a popular method to either estimate momentum indirectly (through the multiple scattering induced in a known volume of known material) or simply absorb, hence remove, the low-momentum muons up to a certain threshold (at the same time removing also some important backgrounds such as charged hadrons and $e^{\pm}$ by inducing their showering). 
Therefore, an additional set of global parameters, which is specific of muon radiography, relates to the presence of passive material, either for momentum filtering or as reference for scattering. Passive material can either be installed all in single place in the apparatus or diffused across the layers, as illustrated respectively in the top and bottom panels of Fig.~\ref{f:muography-passive-material}. In this sense the problem is similar to the deployment of the passive material in MUonE, as mentioned in Sec.~\ref{sec:muone}. 
The corresponding parameters for the optimization include the material, thickness, number, and position of the passive slabs.

Once an optimal set of parameters is identified for a given use case, and a set of muon detection planes is built, changing use case demands a re-optimization of the parameters, but a new optimization may have to be limited to the global parameters, as re-building all detector units from scratch may be unfeasible due to budget or time constraints (although the costs and timescales involved in typical muon radiography projects are much smaller than for many of the other examples in this document). 
For example, the angular resolution of a tracker depends on both the spatial resolution and the distance between planes; while the former depends on local parameters, the latter is a global one, and it is way cheaper to adjust. 

\subsubsection{\tomopt: Differential Muon Tomography optimization}\label{sec:tomopt}
    \tomopt is the first concrete step within the \mode collaboration to investigate the practicality and scalability of the optimization pipeline proposed in \autoref{sec:intro}. Muon tomography offers a comparatively simple optimization use case among detector systems. As discussed above, the optimization of such detectors can have a significant benefit, while allowing us to simultaneously bring up our own understanding of the requirements, feasibility, and potential limitations of physics-goal-oriented systems optimization.
    
    \paragraph{Package overview}
        While still in the development phase, the \tomopt package is planned to be a highly-modular, python-based, \pytorch~backed~\cite{pytorch}, framework providing users with a full suite for implementing and simulating all required aspects of the optimization pipeline in a flexible and extensible manner:
        \begin{itemize}
            \item Muon generation through random sampling of literature muon flux models, \eg Refs.~\cite{cosmic_muons1, cosmic_muons2};
            \item Definitions of initial detector-configurations;
            \item Passive volumes to be imaged (either pre-specified by the user, or randomly generated according to specified criteria), consisting of a range of materials;
            \item Muon propagation through volumes, and scattering within passive matter according to \geant-based models;
            \item Inference of properties of the passive volume (based either purely on classical methods, such as Point of Closest Approach (POCA)~\cite{POCA}, likelihood fitting, or contemporary methods such as deep-learning);
            \item Loss functions incorporating terms based on both the predictive performance of the detector system, and its cost (whether this be fiscal, heat generation, power requirement, exposure time, etc.);
            \item optimization of the detector system based on the loss, using a variety of standard gradient-based optimizers (SGD~\cite{gradient_descent}, Adam~\cite{adam}, etc.), over a training cycle backed by a stateful callback system.
        \end{itemize}
        
        Given \tomopt's emphasis on modularity, \pytorch was chosen as the back-end graph-computation library due to the \texttt{torch.nn.Module} class, which handles computation in the forwards pass and exposes parameters to the optimizer for the backwards pass. With its relatively abstract definition, but nonetheless generally useful functionality, it provides a convenient parent class for developing the various necessary modules in \tomopt. 
        
        In its current state, the full pipeline is in place, and development has now moved to focus on improvements to generalization, realism, and inference performance. Our intention is to publish demonstrations of optimization for a range of benchmark scenarios in 2022, along with the first public release of the package, at which point it will move in to the phase of open and continual, community-driven development under a free open-source licence.
    
    \paragraph{Muon propagation through volumes}
        Initial muon kinematics are sampled from literature models of muon flux, \eg Refs.~\cite{cosmic_muons1, cosmic_muons2}, and may undergo transport away from sea-level, as required. Muons then pass through the passive volume and detector systems, while undergoing multiple scattering. The passive volumes are modeled in terms of discrete voxels of varying material. As the muons pass through the passive volumes, they undergo multiple scattering according to the distance travelled in, and the radiation length (\xo) in the material of, each voxel. A parameterized scattering model is used, based on \geant simulation.
        
\begin{figure}[ht]
			\begin{center}
				\includegraphics[width=0.33\textwidth]{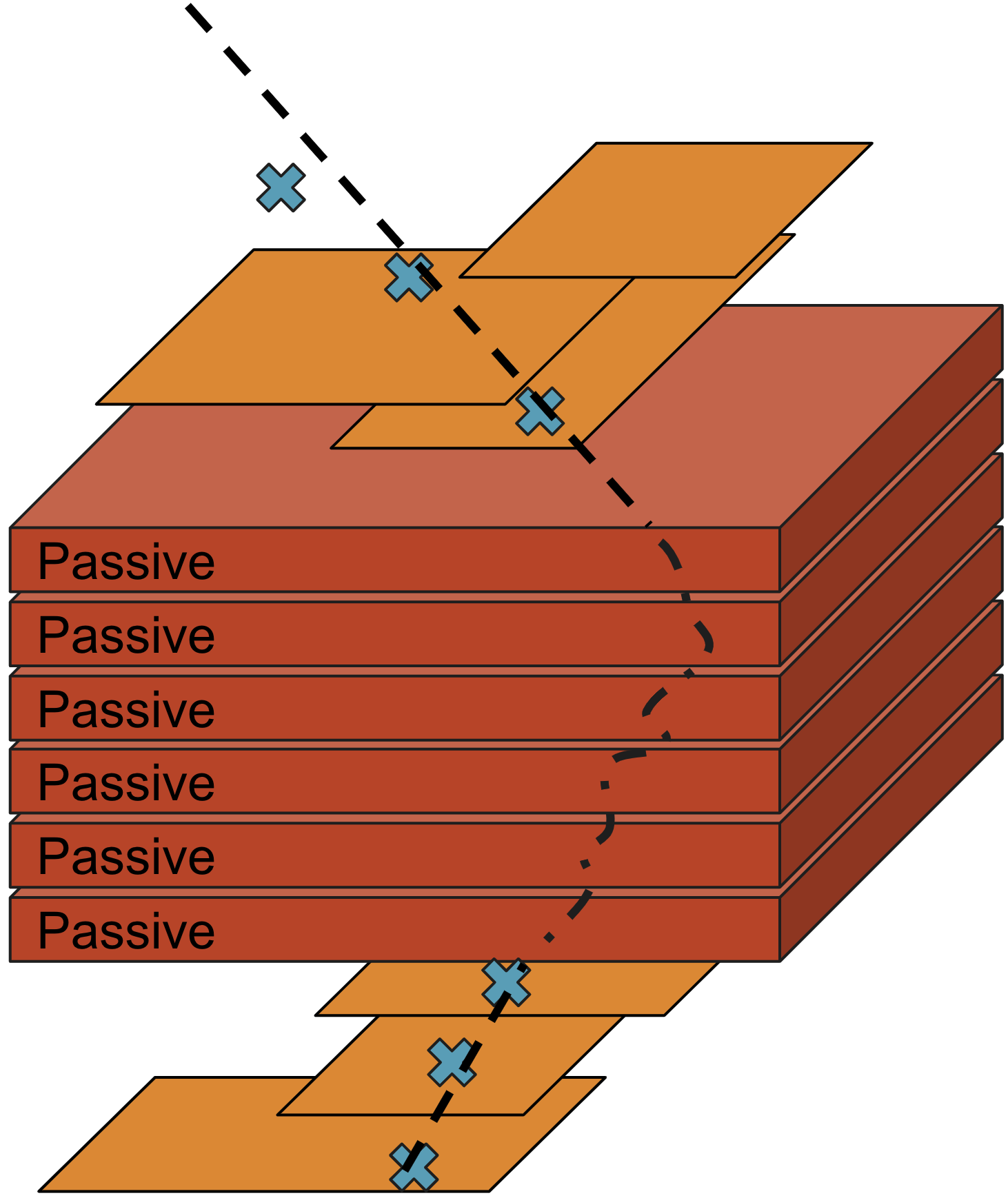}
				\caption{Example schematic for muon propagation (black dashed line) through a volume. The passive volume (red layers) consist of voxels of varying materials, and the detectors, indicated as orange panels, record hits (crosses). Note that although the muon does not pass through the top-most detector panel, a low-resolution hit is still recorded due to resolution and efficiency being currently modeled as a distribution which extends outside the panel.}
				\label{fig:tomopt:volume}
			\end{center}
	    \end{figure}

        Muon hits are recorded above and below the passive volume. To allow for a reasonably flexible detector system, detector panels are allowed to float within specified regions of space. Each panel has a fixed spatial-resolution for muon hit recording, with a fixed efficiency. The parameters to be optimized are their $(x,y,z)$ position and transverse spans ($x,y$). Currently, the number of panels is also fixed, but a simple extension will allow their variation at optimization stage. During optimization, the hit positions are made differential w.r.t. the panel parameters by replacing the resolution and efficiencies of the panels with parameters distributions, allowing hits to be recorded outside of the panels, but with a much lower precision than if the panel were centered on the muon. Figure~\ref{fig:tomopt:volume} shows a diagrammatic example of a muon propagating through both the detector and the passive volume.
        
    \paragraph{Inference}
        Inference is based on the POCA method, by fitting incoming and outgoing trajectories to the muon hits and extrapolating the trajectories inside the passive volume. Through inversion of the scattering model, an \xo may be predicted based on the differences between the trajectories. Since such a method effectively assigns the whole of the scattering to a single voxel, it is inherently biased. Additionally, the POCA caries an uncertainty due to the finite resolution of the detectors, which is computable via autograd-based uncertainty propagation.
        
        This uncertainty may be accounted for by replacing the POCA with a 3D Gaussian PDF which scales according to the POCA uncertainties. A weight per muon per voxel can be computed by integrating the PDF over each voxel. This weight can then be augmented by the uncertainty in the \xo per muon (again computable via autograd), and their hit efficiency. The final predictions for the each voxel in the passive volume is then a weighted average of the predictions of every muon.
        
        While this method accounts for the uncertainty on the POCA location, it does not completely address the inherent bias in the predictions. An alternate approach is to use a frozen graph neural network (GNN)~\cite{gnns} that has been pre-trained for the specific type of passive volumes expected. Such a network can take in features for each muon (including the POCA predictions) and output predictions for the volume (whether this be some overall prediction for the volume, or predictions per voxel). Since the GNN acts as a flexible and differentiable map from muon-level predictions to target space, it can learn to avoid biases in a way that the fixed-computation model-inversion approach cannot.
    
    \paragraph{Loss definitions and optimization}
        In general, the detector should aim to be as performant as possible, while being within budget. As previously discussed, the budget may contain multiple factors, not just pecuniary ones; such as size, imaging time, and power requirements. These will depend on the exact scenario for which the detector is being optimized. Similarly, the definition of ``performance" will also vary according to task. Provided these aspects can be expressed as analytic and differentiable functions of the detector predictions and parameters, then the detector can be optimized with respect to them via gradient descent.
        
\begin{figure}[ht]
			\begin{center}
				\includegraphics[width=0.63\textwidth]{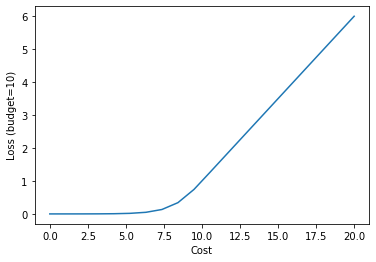}
				\caption{Cost component of the loss as a function of detector cost, for a target budget of 10 A.U.. A sigmoid component below the budget provides a slowly increasing gradient as the cost approaches the target budget, beyond which the loss increases linearly. The function is fully differentiable.}
				\label{fig:tomopt:budget}
			\end{center}
	    \end{figure}

        An example loss function might be: the mean squared-error of \xo predictions over every voxel in each volume, for the performance component; and the total cost of the detector, which scales according to the $xy$ span of each panel. The cost component, however, does not include any notion of a budget. Instead, the functional form can be adjusted to slowly turn on a cost penalisation as the actual cost approaches a pre-specified budget, beyond which the penalisation rapidly increases, such as the form shown in \autoref{fig:tomopt:budget}. The performance and cost components of the loss will also need to be scaled appropriately to provide a good optimization, but the scaling component can be approximated by first capturing the initial performance of the detector in a frozen state over a few warm-up epochs at the beginning of the optimization process.
        
	    While the loss can be computed for the voxels in a passive volume, to ensure complete generalization the loss should be averaged over a sufficient number of representative examples of passive volumes. This can be achieved by specifying the type of passive volumes expected, and generating examples on demand. Comparing to traditional DNN training over mini-batches of data, optimization here then happens over batches of passive volumes. Once a loss has been computed, it may be back-propagated to each of the detector parameters, which may then be updated using gradient descent.
	    
\begin{figure}[ht]
			\begin{center}
				\includegraphics[width=0.75\textwidth]{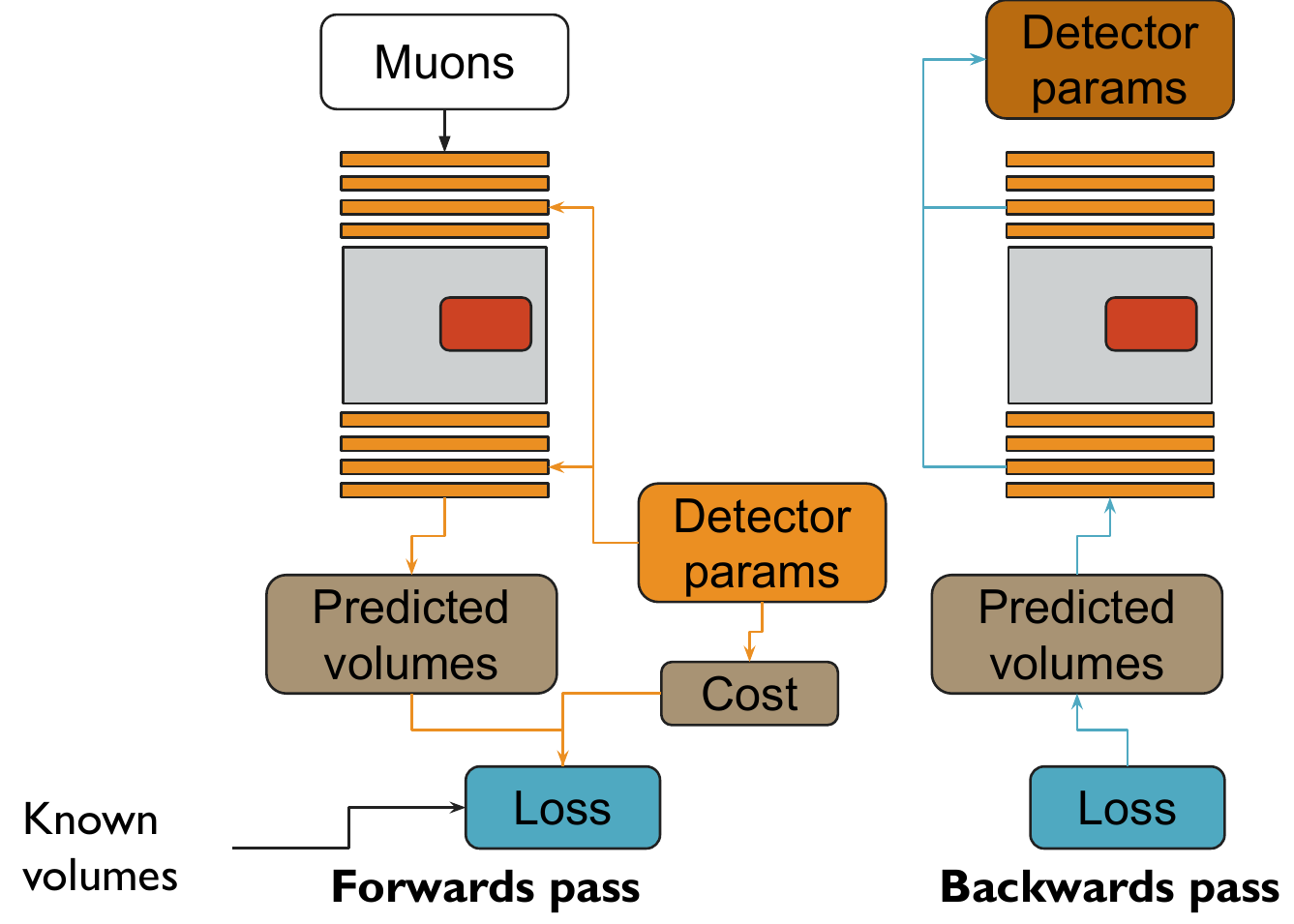}
				\caption{Example schematic for the parameter update loop in \tomopt. In the forwards pass, predicted volumes (or target features of the volumes) are computed based on inference of muons reconstructed by the parameterized detector. In the backwards pass, the derivative of loss is passed through the inference algorithm back to the detector parameters.}
				\label{fig:tomopt:update-loop}
			\end{center}
	    \end{figure}

    \paragraph{Update cycle}
	    Figure~\ref{fig:tomopt:update-loop} illustrates the forwards-backwards update cycle in \tomopt. In the forwards pass, batches of muons are used to scan known (and possible randomly generated) passive volumes. A fully differentiable inference algorithm is used to predict target values for the volumes, based on the detector readouts. Finally, the loss of these predictions is combined with the cost of the current detector system. In the backwards pass, the gradients of the loss with respect to the detector parameters is computed, and used to adjust detector parameters.
	    
\subsubsection{Industrial Applications}
\label{sec:muography-industrial}
Muography can offer solutions to many problems in the industrial sector working as a Non-Destructive Testing (NDT) technique to perform preventive maintenance of equipment, quality control of the production process and risk assessment and evaluation. There is a large variety of problems in the industry although most of them share the following characteristics: relatively large and dense objects, impossibility to have access to the object while the equipment is in production, and the presence of a harsh environment in terms of dust, temperature and space or time restrictions. Muography emerges as an interesting NDT due to its large power of penetration, the fact that its application does not require any physical contact with the target objects allowing inspection while the factory is in production and the possibility to perform a continuous monitoring.

Most of the industrial applications have another distinguishing feature with respect to other common applications of muography: the nominal geometries and densities of the objects are known with high precision and can be extracted from the engineering drawings. This implies that muon imaging algorithms are not required to reconstruct an unknown object but just to find variations on top of a well known, predefined design. A consequence of this is the dramatic reduction of complexity of the problem from an algorithmic point of view. Geometry variations can be encoded in a relatively small set of parameters that can be estimated using methods such as Deep Neural Networks working in regression mode or likelihood-based estimation methods. The following lines show a few examples of this kind of applications.

\begin{figure}[ht]
    \begin{center}
	\includegraphics[width=0.33\textwidth]{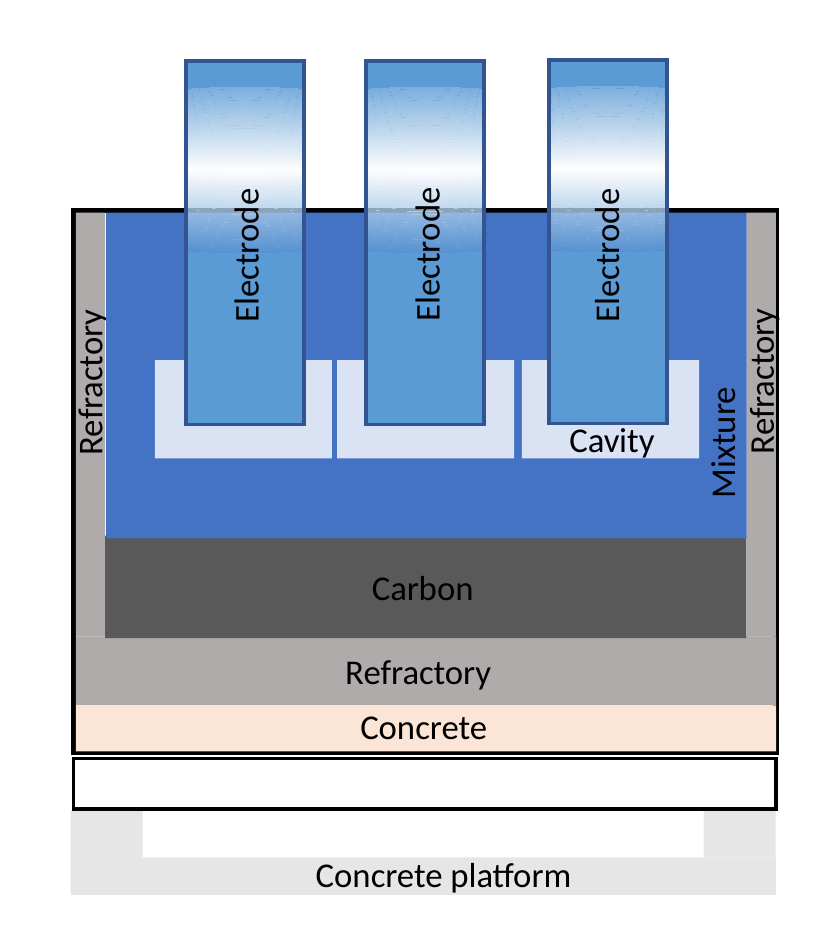}
	\includegraphics[width=0.33\textwidth]{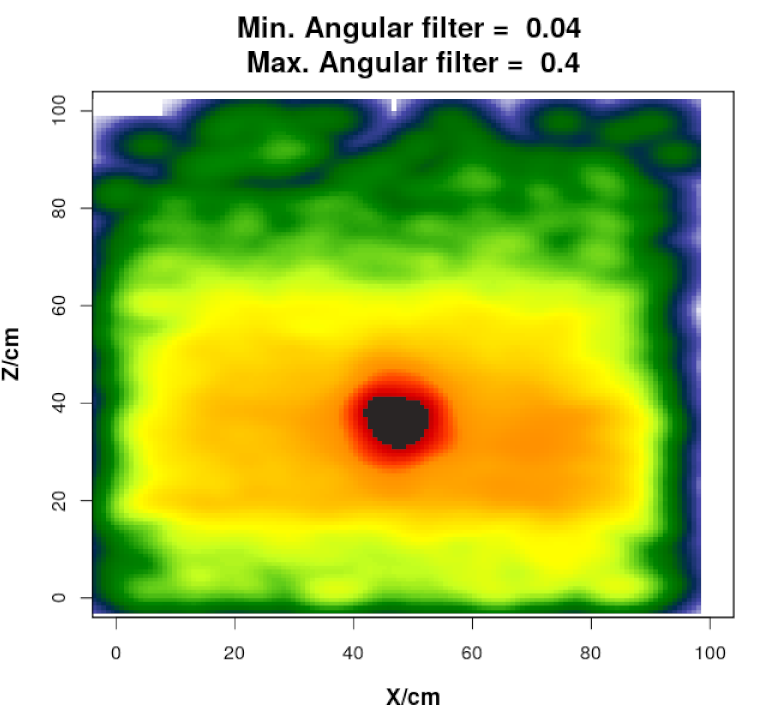}
	\caption{Schematic view of the structure of an electric furnace including three electrodes used to heat the mixture (left). POCA-based image of a steel pipe used to feed a convolutional neural network working in regression mode to infer the inner radius (right).}
	\label{fig:industry}
	\end{center}
\end{figure}

Electric arc furnaces are used in the processing of several raw minerals such as silicon or manganese. These furnaces are equipped with three large electrodes inserted vertically in the mineral mixture to produce electric discharges and heat the material. A schematic view of a generic electric furnace can be seen in Fig.~\ref{fig:industry} (left). The knowledge of the exact position of the electrodes is an interesting parameter that could help to understand the efficiency variations observed in this kind of equipment. Scattering muography can be applied to this problem by simply considering the angular and spatial deviation distributions characterized using their n-quantiles. This information can be used to train a Deep Neural Network using simulated datasets with different electrode locations and regressing over the height of the electrode with respect to the bottom of the furnace.

Pipes are used in almost every industrial plant to transport all kind of liquids and gases. The wear of the inner wall of the pipes is a general problem that forces the companies to perform inspections regularly in order to keep the integrity and safety of the facility. In cases where the pipes are thermally insulated, the application of other NDTs is complicated due to the properties of the low-density insulation layer. Scattering muography can be applied in order to measure the inner radius of the pipe. A first attempt uses a likelihood-based algorithm in which the distributions of the angular and spatial deviations are estimated for every muon by simulating its propagation through a pipe with a given radius. The distributions are interpreted as probability density functions and are combined in a likelihood function for all the muons. This function is then minimized against the radius in order to find the best match. A second approach was used by estimating the POCA observables of the muons and using the corresponding images to train a convolutional neural network (CNN) performing a regression to the inner radius. Figure~\ref{fig:industry} (right) shows an example of a real POCA image of a steel pipe used as input for the CNN algorithm.    

The industrial sector offers a large and heterogeneous class of problems where muography could provide cost-effective solutions. The estimation and imaging algorithms used in this context have been application-specific so far, complicating the exportation of the solutions to different problems. The MODE collaboration is aiming to device a general purpose procedure to solve the inverse problem associated to muography. This approach will allow to unify muography applications in the industry from the algorithmic point of view, allowing to quickly apply the technique in different cases and problems.

\subsubsection{Portable Modular Detectors for Flexible Muography}
\label{sec:portable}

\begin{figure}[ht!]
\centering
\includegraphics[width=0.9\textwidth]{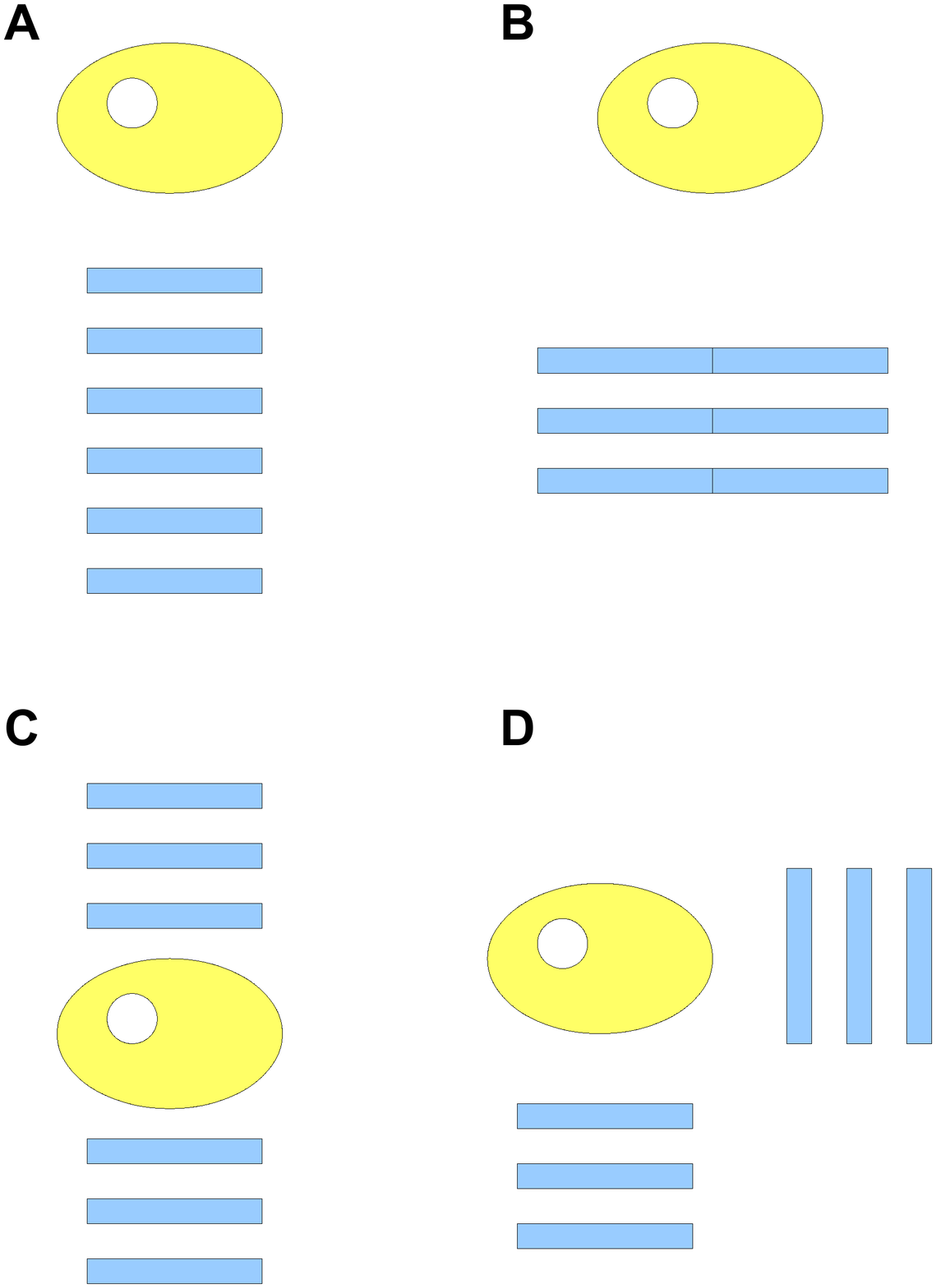}
\caption{\em Sketch of various detector configurations among which one could switch with fast turn-around in the context of a modular muography detector project such as the one described in Ref.~\cite{Gamage:2021dqd}. The same six detector planes may be deployed in order to maximize angular resolution and trigger redundancy (A) or to maximize the acceptance and therefore the statistics (B); to perform a 3D tomography instead of a simple 2D projection, where appropriate the same planes can be used to ``sandwich'' the volume of interest and exploit the scattering method (C) or to take 2D projections from orthogonal points of view (D). Reproduced from Ref.~\cite{Moussawi2021}.}
\label{f:muography-modularity}
\end{figure}

One of the ongoing trends in muography, especially for applications in geosciences, archaeology, and civil engineering, is the development of portable and inexpensive muon detectors, such as Ref.~\cite{Baccani:2018nrn,Kyushu2018}.
A recent project~\cite{Wuyckens2018,Basnet2020,Gamage:2021dqd,Moussawi2021} aims at the development of compact, autonomous, portable and modular muon radiography setups based on position-sensitive layers based on small-area resistive plate chambers (RPC), a technology chosen because of its good trade-off between position and time resolution, efficiency, cost, and ease of construction~\cite{AGUMuographyBook}. 
In this project, the goal is to allow a high degree of modularity of the geometry of the complete setup: ideally, the already mounted individual detector planes would be produced in large numbers and deployed {\it in situ} in the arrangement that most fits the specific use case while respecting the local constraints (\egc the optimal location may be in a narrow tunnel), as illustrated in Fig.~\ref{f:muography-modularity}. The same detection layers may be arranged to form two or one tracker depending on the relative importance of scattering or absorption on the final discrimination power. In either case, it is not always obvious {\em a priori} whether it is more convenient to have few layers with large area to collect more statistics (arranging several detector units side by side) or to maximize the number of layers crossed by the muons to improve resolution and redundancy of the setup. 
An automatic optimization algorithm based on machine learning, therefore, would allow for quick redesign of the geometry at every new measurement of a different target.

\subsection{Proton Computed Tomography}
\label{sec:protonct}
Radiation therapy using energetic protons for the treatment of cancer instead of X-rays is becoming more and more available on a world-wide scale. As the energy deposition of protons is concentrated around the \emph{Bragg peak}, it promises a better ratio between the doses deposited inside and outside the tumor~\cite{wilson_radiological_1946,cormack_representation_1963}. Despite the differences in how protons and photons interact with matter, the planning of proton treatment is still based on computed tomography images, \ie the spatial distribution of X-ray attenuation inside the patient (``phantom''), obtained from X-ray measurements from various angles. A direct measurement of the \emph{relative stopping power} (RSP, the ratio between the stopping power of a certain material and that of water) of protons inside the phantom can inform treatment planning in a better way. 

To this end, many prototypical \emph{proton computed tomography} (pCT) scanner designs have been reported in the literature~\cite{poludniowski_proton_2015,johnson_review_2018}. The processing of the raw measurements can be logically split into two parts:

First, a hardware-specific sub-procedure determines the positions and directions of protons both before entering and after exiting the phantom, and the energy loss in terms of a \emph{water-equivalent path length} (WEPL). 

Second, a reconstruction algorithm is applied to find the three-dimensional RSP image. In this respect, applying a \emph{model-based iterative reconstruction} (MBIR) algorithm means that a linear system
\begin{equation}\label{eq:pct-linsys}
A \cdot \text{RSP} = \text{WEPL}
\end{equation}
is approximately solved in a least-squares sense, possibly in addition to image quality objectives. The entries of the matrix $A$ reflect how much the voxels of the RSP image overlap with the proton paths. The paths must be estimated according to their entry and exit positions and directions~\cite{schulte_maximum_2008}, possibly taking the uncertainties in the measurements into account~\cite{krah_comprehensive_2018}. 

These two sub-procedures are visualised by the central and right arrow in Fig.~\ref{fig:pct-pipeline-coarse}. For optimization and other computational purposes, the raw measurement data can be manufactured by a randomized particle physics simulation, often called \emph{Monte Carlo} (MC) simulation, based on a model of the hardware and the RSP distribution inside the phantom. This is the leftmost sub-procedure listed in Fig.~\ref{fig:pct-pipeline-coarse}.

\paragraph*{Figures of merit.} The accuracy achievable with the utilized hardware and software design can be assessed by the following comparisons~\cite{meyer_optimization_2020}:
\begin{itemize}
\item Comparing the reconstructed proton paths with the proton paths found in the MC simulation.
\item Comparing the estimated WEPL with the true energy loss found in the MC simulation.
\item Comparing the reconstructed RSP solution of \eqref{eq:pct-linsys} with the original RSP that the MC simulation was run on.
\item Finally, when the reconstructed RSP image informs a proton therapy treatment plan (via a fourth sub-procedure on the right in Fig.~\ref{fig:pct-pipeline-coarse}), this plan can be assessed \eg in terms of doses inside and outside the tumor.
\end{itemize}

\paragraph*{Parameters for Optimization.}

On the hardware side, design parameters
\begin{itemize}
    \item of the proton beam, such as its energy, intensity, spot size and divergence, and
    \item the geometrical setup, including the relative positions, material budgets, and granularity of detector layers,
\end{itemize} 
can be optimized.  
The software also involves constants that could be tuned:
\begin{itemize}
\item Sophisticated path estimation via the (extended) most likely path~\cite{schulte_maximum_2008,collins-fekete_theoretical_2017,krah_comprehensive_2018} involves  coefficients of a certain polynomial fit to simulation data~\cite{williams_most_2004} or the assumed uncertainties of the measured positions and directions, among others.
\item If a neural network is used somewhere in the software pipeline, its weights can be considered as parameters of the software pipeline.
\end{itemize}

\paragraph*{Quantification of Uncertainties} Apart from gradient-based optimization, linearization can also be used to estimate the probability distribution of output variables given the probability distribution of input variables, and to identify sub-procedures that amplify uncertainty.

\paragraph{The Digital Tracking Calorimeter of the Bergen pCT collaboration} In the remainder of this section, we summarize a recent survey~\cite{aehle-derivatives-2022} on the applicability of differentiable programming to the software pipeline of the Bergen pCT collaboration~\cite{alme_high-granularity_2020}, displayed in Fig.~\ref{fig:pct-pipeline-coarse}.

The digital tracking calorimeter (DTC), which is designed and currently built by the collaboration, consists of 43 parallel layers of 108 ALPIDE chips each and is to be placed behind the phantom. Protons leaving the phantom activate several pixels in each layer until they are stopped, as visualized in Fig.~\ref{fig:pct-schematic-scanning-process}. 
The open-source package GATE~\cite{GATE}, based on the GEANT4 toolkit~\cite{GEANT,allison_geant4_2006,allison_recent_2016}, is used to simulate the passage of protons through both a model phantom with known RSP, and the DTC. Gate outputs the positions where protons hit ALPIDE chips, and the corresponding energy depositions, as floating-point numbers.
These data are further processed into the discrete detector response, that only tells which ALPIDE pixels have been activated in each read-out cycle. A library of sample activation clusters depending on the deposited energy can be used here.

A numerical study showed that GATE as a part of the \emph{MC simulation} sub-procedure is only piece-wise differentiable and has many discontinuities. It is as of yet unclear whether or not this is just a fixable implementation problem. In addition, the subsequent procedures in the pipeline could make it smooth by combining the MC results of many protons. As applying differentiable programming to GATE and its dependencies is a massive technical effort, either way a surrogate model should be used here first.

In the \emph{proton history reconstruction} sub-procedure, the discrete detector output is mapped back to floating-point hit positions and energy depositions. This procedure involves many ``discrete'' operations such as:
\begin{itemize}
    \item grouping neighboring pixels activated by the same proton into a ``cluster'';
    \item assigning an energy deposition to this cluster according to its size, which is an integer;
    \item matching clusters likely related to the same proton into a ``track''. 
\end{itemize}
As the derivative of such discrete functions is either 0 or non-existent, we cannot benefit from differentiable programming here and need a surrogate model. The residual WEPL of the proton is estimated by a fit of the Bragg-Kleeman equation~\cite{pettersen_design_2019} to the estimated energy loss in each layer.

In the \emph{model-based iterative reconstruction} sub-procedure, the matrix $A$ in \eqref{eq:pct-linsys} is generated and the linear system is solved iteratively. For typical setups, $A$ is too large to be stored in memory~\cite{biguri_tigre_2016}, so its elements are regenerated whenever required. 

Differentiating through the iterative solver in a ``black-box'' fashion is infeasible in reverse mode because of memory limits, and might suffer from a bad accuracy. This issue can probably be solved by exploiting the simple mathematical structure that allows to compute analytical derivatives.

\tikzstyle{varblock}=[align=center,draw=black,rounded corners=5,font=\scriptsize]
\tikzstyle{funarrowA}=[rounded corners=10pt]
\tikzstyle{funarrowB}=[above,midway,font=\scriptsize,align=center]

\begin{figure}
\centering
\begin{tikzpicture}

\node (detectoroutput) at (0,0) [varblock] {detector \\ output };

\coordinate (geantinput) at ($(detectoroutput)-(2,0)$) ;
\node (originalrsp) at ($(geantinput)+(-1.3,0)$) [varblock] {orig.\ RSP};
\node (detectorparameters) at ($(geantinput)+(-0.5,1.4)$) [varblock] {detector parameters};
\node (randompath) at ($(geantinput)+(-0.7,-1)$) [varblock] {random number generator};
\node (crosssections) at ($(geantinput)+(-1.6,0.7)$) [varblock] {physics param.};
\draw [funarrowA] (detectorparameters.south) -- (geantinput) -- (detectoroutput.west);
\draw [funarrowA] (originalrsp.east) -- (geantinput) -- (detectoroutput.west);
\draw [funarrowA] (randompath.north) -- (geantinput) -- (detectoroutput.west);
\draw [funarrowA] (crosssections.east) -- (geantinput) -- (detectoroutput.west);
\draw [->] (geantinput) -- (detectoroutput.west) node [funarrowB] {MC sim.\ };

\node (posanglewepl) at ($(detectoroutput)+(4.5,0)$) [varblock] {proton  positions, \\ directions,  WEPL};
\draw [->] (detectoroutput.east) -- (posanglewepl.west) node [funarrowB] {proton history \\ reconstruction };

\node (reconstructedrsp) at ($(posanglewepl)+(4.5,0)$) [varblock] {recon.\ RSP};
\draw [->] (posanglewepl.east) -- (reconstructedrsp.west) node [funarrowB] {MBIR \\ subprocedure};

\end{tikzpicture}
\caption{Overview of the software pipeline of the Bergen pCT collaboration. Reproduced from Ref.~\cite{aehle-derivatives-2022}.}
\label{fig:pct-pipeline-coarse}
\end{figure}
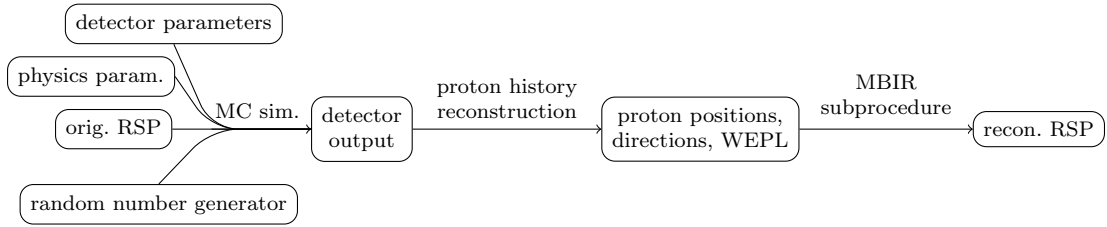

\newcommand{\minisq}[2]{ \fill (#1-0.05, #2) -- +(0.1,0) -- +(0.1,0.1) -- +(0,0.1) -- cycle; }
\newcommand{\sqrow}[1]{
  \draw[very thin] (#1-0.05,-2) -- +(0,4);
  \draw[very thin] (#1+0.05,-2) -- +(0,4);
  \foreach \yl in {0,...,40}
    \draw[very thin] (#1-0.05,\yl*0.1-2) -- +(0.1,0);
}

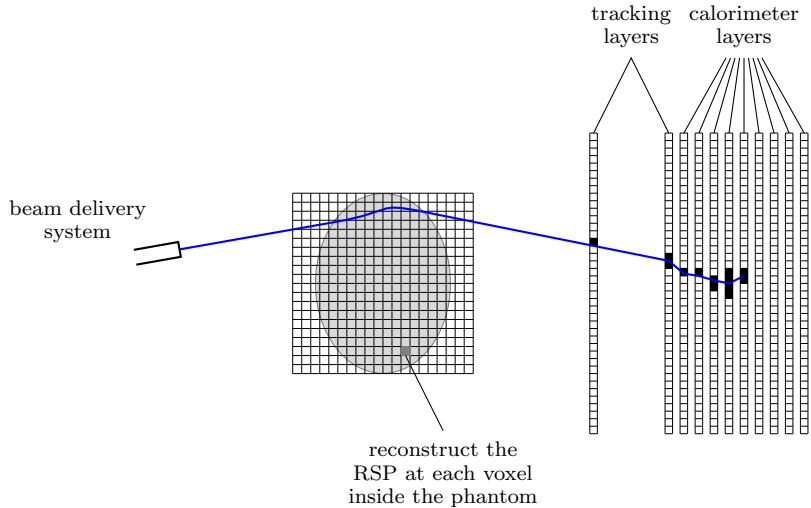
\begin{figure}[!h]
    \centering
    \begin{tikzpicture}

\sqrow{0}
\sqrow{1}
\foreach \xl in {1,...,9}{
  \sqrow{1+\xl*0.2}
}
\minisq{0}{0.5}
\minisq{1}{0.2}
\minisq{1}{0.3}
\minisq{1.2}{0.1}
\minisq{1.4}{0.1}
\minisq{1.6}{-0.1}
\minisq{1.6}{0}
\minisq{1.8}{-0.2}
\minisq{1.8}{-0.1}
\minisq{1.8}{0}
\minisq{1.8}{0.1}
\minisq{2}{0.1}
\minisq{2}{0.0}
\draw[blue,thick] (0,0.5) -- (1,0.3) -- (1.2,0.13) -- (1.4,0.1) -- (1.6,0.04) -- (1.8,0.00) -- (2.0,0.1);

\draw (0,2) -- (0.5,3) node[anchor=south,align=center]{\scriptsize tracking \\[-0.2cm] \scriptsize layers} -- (1,2);
\draw (2,3) node[anchor=south,align=center] (calolabel) {\scriptsize calorimeter \\[-0.2cm] \scriptsize layers} ;
\foreach \xl in {1,...,9}
  \draw (1+\xl*0.2,2) -- (calolabel);

\draw[gray,fill=gray!30] (-2.8,0) ellipse(0.69*1.3 and 0.92*1.3);

\foreach \xl in {0,...,20} {
  \draw[very thin] (\xl*0.12-4.0,-1.2) -- (\xl*0.12-4.0,1.2);
}
\foreach \yl in {0,...,20} {
  \draw[very thin] (-4.0,\yl*0.12-1.2) -- (-1.6,\yl*0.12-1.2);
}

\fill[gray] (12*0.12-4.0, 2*0.12-1.2) -- +(0.12,0) -- +(0.12,0.12) -- +(0,0.12) -- cycle;
\draw ($(12*0.12-4.0, 2*0.12-1.2) + (0.06,0.0)$) -- +(0.5,-1) node[anchor=north,align=center]{\scriptsize reconstruct the \\[-0.2cm] \scriptsize RSP at each voxel \\[-0.2cm] \scriptsize inside the phantom };

\draw (-5.8,0.4) node [above left,align=center] {\scriptsize beam delivery \\[-0.2cm] \scriptsize system};
\draw[thick, rotate around={10:(-5.8,0.4)}] (-6.1,0.3) -- (-5.5,0.3) -- (-5.5,0.5) -- (-6.1,0.5);

\begin{scope}[draw=blue,fill=blue,  thick]
\draw ($(-5.8,0.4)+({cos(10)*0.3},{sin(10)*0.3})$) -- +($({cos(10)*2.1},{sin(10)*2.1})$) coordinate (entrypoint) ;
\draw (0,0.5) -- +($2.3*(-1,0.2)$) coordinate (exitpoint) ;
\draw (entrypoint) .. controls ($(entrypoint)+({cos(10)*0.8},{sin(10)*0.8}) $)   and ($(exitpoint)+0.6*(-1,0.2)$)   .. (exitpoint);
\end{scope}
    
\end{tikzpicture}
    \caption{Schematic figure of the scanning process with the digital tracking calorimeter of the Bergen pCT collaboration~\cite{alme_high-granularity_2020}. Reproduced from Ref.~\cite{aehle-derivatives-2022}.}
    \label{fig:pct-schematic-scanning-process}
\end{figure}

\subsection{Low-Energy Particle Physics}
\label{sec:neutronpp}
Low-energy particle physics provides many unique research opportunities to search for exotic particle candidates or beyond Standard Model physics.
They range from decay~\cite{Gonz2, dubbers2021precise} to electric dipole moment (EDM)~\cite{chupp2019electric} and other measurements~\cite{jaeckel2010low}.
Specifically, tritium decay~\cite{aker2019improved}, neutron decay~\cite{saul2020limit, UCNA_Ab, wang2019design}, neutron lifetime~\cite{gonzalez2021improved}, and neutron EDM~\cite{abel2020measurement} produce key results.
These experiments are high-precision measurements designed for specific purposes, leading to complex designs.
These complex experiments have many tunable parameters, and we must design and operate them optimally to maintain continuous improvements of experimental results.
This quality requirement highlights the importance of advanced methods such as differential programming for the field. \\

Parameters for the optimization of such experiments are case-specific. 
However, we aim to set global and local parameters to improve measurement precision by reducing uncertainties. An example application is optimizing the geometry of a simplified scintillation energy detector.
Consider an incoming particle with kinetic energy $E$, which is converted in a scintillator to $n_{\gamma}$ photons proportional to the energy.
The photon sources are spread deeper in the scintillator depending on the energy of the incoming particle.
The photons travel through the scintillator or light guides to a photon detector, which creates the measurable signal of the event.
Due to attenuation in the material, a fraction of the photons is lost, leading to less precise results. 
Therefore, due to photon paths alone, we must optimize the geometry of the detector to increase energy resolution and minimize systematic corrections like energy-dependent detection efficiencies.
Such a detector is often used for low-energy particle physics, and we propose using a differentiable programming optimization pipeline as described in Sec.~\ref{sec:problem-description} with a surrogate model from simulations to achieve optimal detector performance. \\

Furthermore, we generalize from a simplified detector to other applications, such as optimizing the magnetic field configuration by changing coil dimensions, tuning operational coil parameters, or optimizing beam-line aperture placements for desired beam characteristics. 
We may choose whether to do end-to-end optimization for local or global experimental parameters, depending on the complexity and effective dimensionality of the experiment. In some cases, differentiable programming is not ideal, as other methods achieve  better convergence, such as Bayesian optimization~\cite{Shahriari2016Taking} as mentioned in Sec.~\ref{sec:accelerators} and Ref.~\cite{duris2020bayesian, roussel2021multiobj}. \\

To further illustrate the capabilities of differentiable programming for low-energy particle physics, we choose the optimization of a neutron beam-line as found in Ref.~\cite{saul2020limit,wang2019design}. 
We highlight how we can customize and adapt the optimization pipeline of Sec.~\ref{sec:problem-description} to the problem.
Consider a source of cold neutrons whose neutrons reach an experiment through neutron guides and a velocity selector. 
We position a set of $n$ quadratic neutron apertures between the velocity selector and the experiment to shape the resulting beam distribution.
The beam distribution can be calculated analytically by trigonometry and a set of integrals, requiring no surrogate model.
Optimizing the beam distribution is essential to reduce systematic effects, maintain experiment confinements, or other constraints like costs. The two most significant systematic effects are beam homogeneity and created background signals by the beam-line through neutron absorption. We may encode the desired beam homogeneity or shape in a target distribution $P(x)$ with $x$ being the distance perpendicular to the beam center.
Therefore, we set the optimization objective as KL divergence or \textit{relative entropy} $D_{\text{KL}}$~\cite{kullback1951information} of $P(x)$ and the resulting beam distribution of the current setup $Q(x)$ for a fixed detector position on the beam axis.
\begin{figure}[t]
    \centering
    \subfloat[\centering]{{\includegraphics[width=0.45\textwidth]{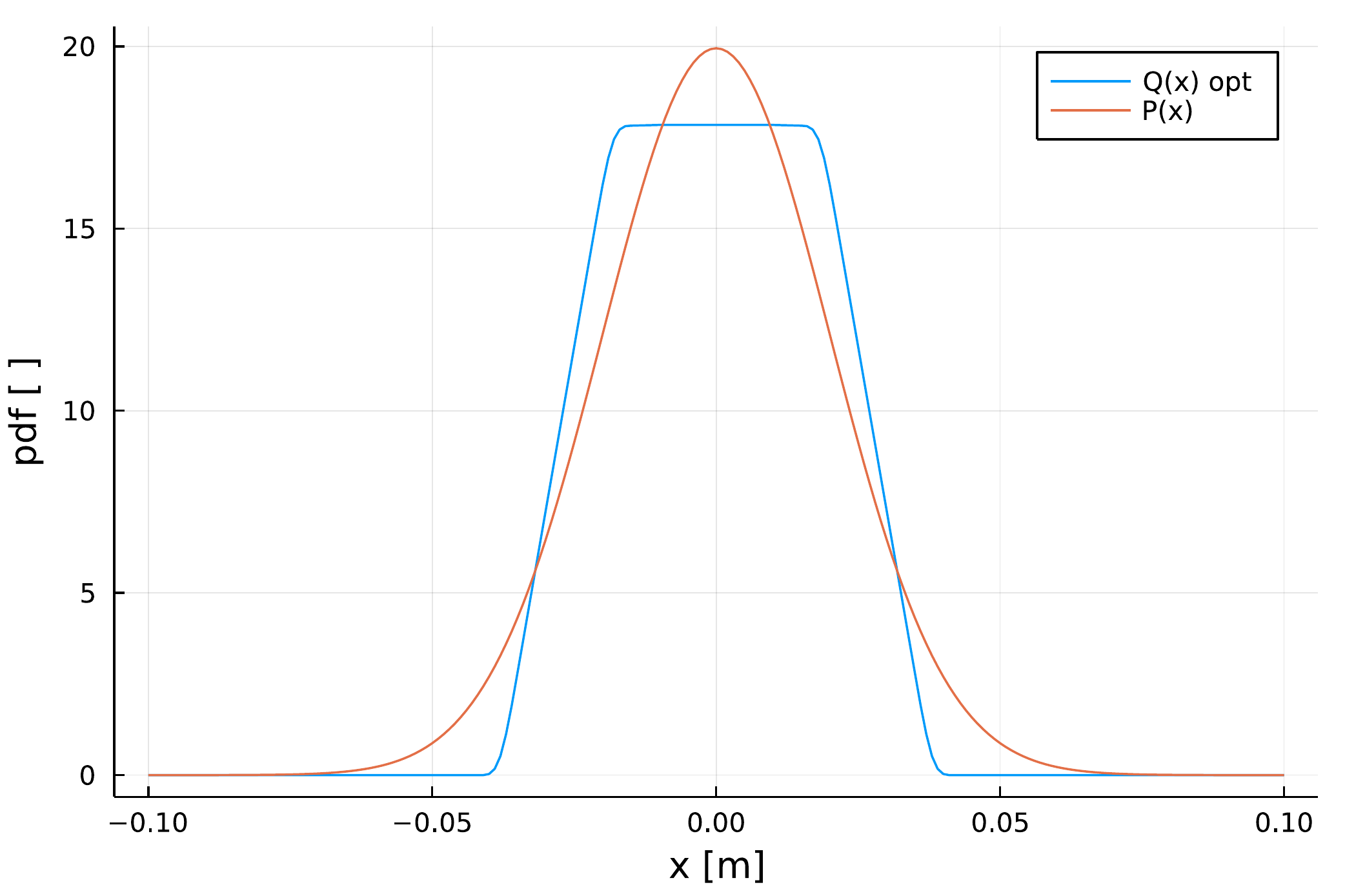} }}
    \qquad
    \subfloat[\centering]{{\includegraphics[width=0.45\textwidth]{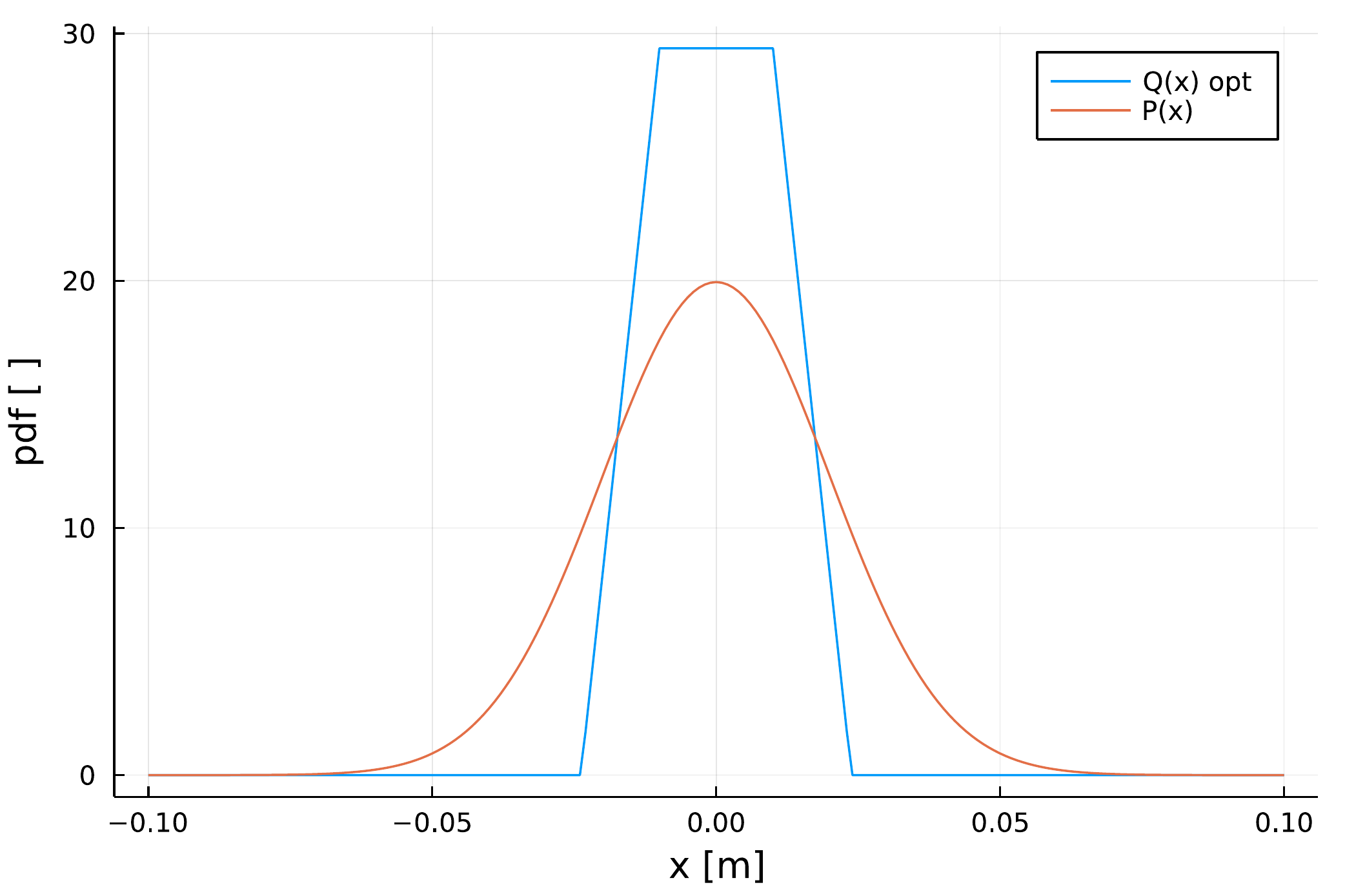} }}
    \caption{Two different optimized beam-line distributions $Q(x)$ for $n = 3$ apertures. (a) uses only $D_{\text{KL}}$ and $\mathcal{L}_2$, enforcing a good distribution approximation of $P(x)$ with equally sized apertures; (b) utilizes $\mathcal{L}$ as in \equationautorefname~\ref{equ:noble_loss}, sacrificing function approximation quality for lower background levels. }
    \label{fig:noble_Res}
\end{figure}
\begin{align*}
    D_{\text{KL}} (Q \mid\mid P) = \sum_{x \in \mathcal{X}} Q(x) \log \left(  \frac{Q(x)}{P(x)}  \right) .
\end{align*}
Furthermore, we can expand the objective value with additional terms addressing different systematic effects.
It is beneficial to place apertures further away from the experiment to minimize beam-line-induced background. 
We add the distance $p_i$ of aperture $i$ to the beam-line start as the first objective function adaption as
\begin{align*}
    \mathcal{L}_1 = \frac{\alpha_1}{n} \sum_{i = 1}^{n} p_n^2 ,
\end{align*}
with weighting parameter $\alpha_1$. We also add aperture width $w_i$ of aperture $i$ as regularization
\begin{align*}
    \mathcal{L}_2 = \frac{\alpha_2}{n} \sum_{i = 1}^{n} w_n^2 .
\end{align*}
Therefore, the total optimization objective $\mathcal{L}$ for the differential programming pipeline is
\begin{align}
    \mathcal{L} = D_{\text{KL}} (Q \mid\mid P) + \mathcal{L}_1 + \mathcal{L}_2 .
    \label{equ:noble_loss}
\end{align}
We keep the model general, so that it can be adapted to specific experiments. We present example results in Fig.~\ref{fig:noble_Res} with different regularization results for a Gaussian target distribution. The \textit{Julia} code and pipeline is available on \href{https://github.com/maxlampe/NobleAD}{\textit{GitHub}} and uses the \textit{ForwardDiff} package~\cite{RevelsLubinPapamarkou2016}. \\

\subsection{Error Analysis of Monte Carlo Data in Lattice QCD}
\label{sec:erran}
Lattice QCD is a computational framework based on discretizing space
and time on a hypercubic lattice of spacing $a$. This distance plays
the role of a cutoff, providing a regularization of the field theory. 
Although not very useful from a purely analytic point of view, its
appeal come from the fact that such discretized versions of QCD can be
\emph{simulated} on a computer using Monte Carlo methods, even in the
notoriously difficult non-perturbative regime of QCD. In the last
years Lattice QCD has matured significantly and is able to determine
many key quantities for high energy physics phenomenology
(see Ref.~\cite{Aoki:2021kgd} for a summary).  

Here we will focus on the applications of automatic differentiation to data analysis in lattice QCD along the lines described
in Ref.~\cite{Ramos:2018vgu}. While this topic straggles away from the strict boundaries of detector design optimization, we believe that a discussion the employed tools may be useful in the context discussed in this publication.
\begin{figure}[!h]
  \centering
  \includegraphics[width=\textwidth]{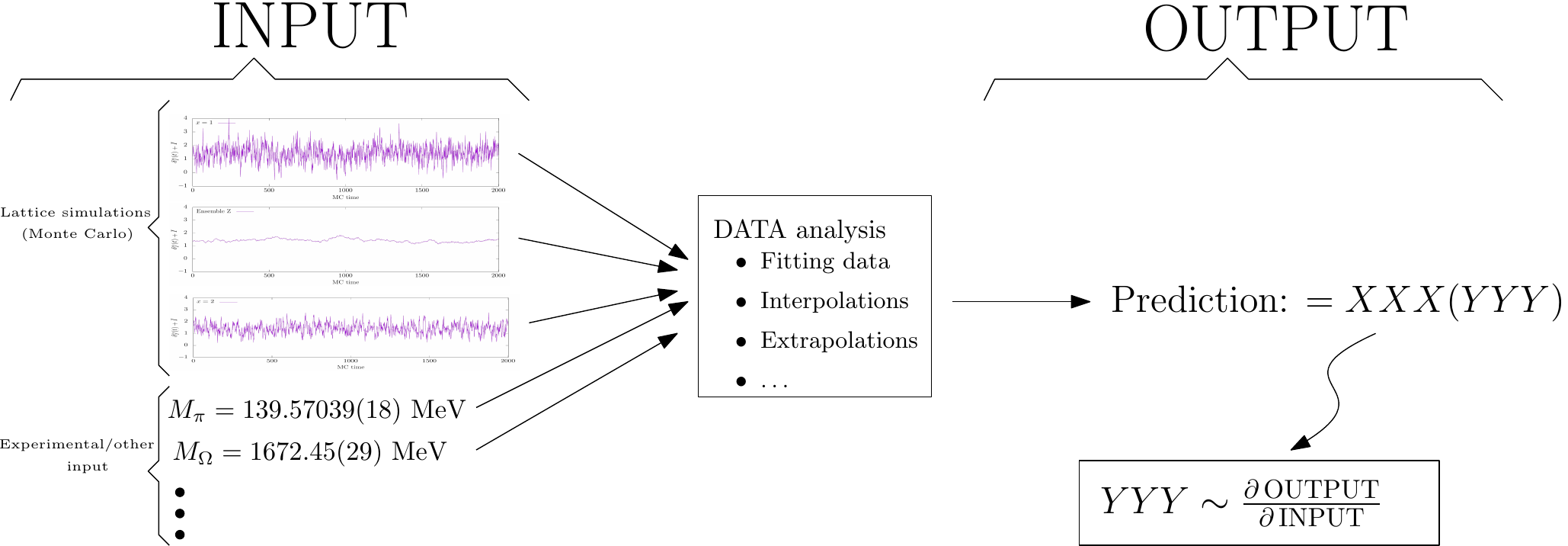}
  \caption{In lattice QCD several inputs (both experimental and the
    result of large scale Monte Carlo simulations) are used to
    determine several quantities of interest. 
  In Ref.~\cite{Ramos:2018vgu} it is proposed to determine the
  uncertainties by computing the derivatives of the results with
  respect to the inputs using automatic differentiation.}
  \label{fig:latt_diag}
\end{figure}

Data analysis in lattice QCD consists in processing the \emph{input}:
simulation data in the form of Monte Carlo averages over ensembles,
experimental inputs, etc... We aim at producing some prediction (\ie
the \emph{output} of the analysis), together with its
uncertainty (see figure). It is clear that the prediction is just a function of the \emph{input}, but in general this function is very complicated: it involves several non linear fits, interpolations, and other iterative processes like root finding. The challenges are the following: 
\begin{description}
\item[Correlations between inputs] Since producing Monte Carlo
  ensembles is numerically very expensive, several quantities are measured on each ensemble. This produces statistical correlations between different \emph{inputs} that affect the error estimates. 

\item[Auto-correlations of the data] Because of the very nature of
  Markov Chain Monte Carlo simulations, subsequent measurements of a
  quantity are not statistically independent. 
\end{description}

We use the $\Gamma$-method~\cite{Madras:1988ei,Wolff:2003sm} to quantify the
effect of these auto-correlations on the error estimates. It remains to
be decided how uncertainties are propagated from the inputs to the
predictions taking into account correctly the different
correlations. 

Our prediction $F$ is a generic function of the inputs $F \equiv
f(A_i^\alpha)$.  The index 
$i$ labels the input and the index $\alpha$ labels the source of
uncertainty. Correlations between different inputs
arise from the fact that several input (values of the index $i$) can
share the same source of uncertainty (value of the index
$\alpha$). 
The final value of the prediction is estimated via
\begin{equation}
  \bar F = f(\bar a_i^\alpha)\,,
\end{equation}
where $\bar a_i^\alpha$ are the central values of the inputs, or Monte
Carlo averages
\begin{equation}
  \bar a_i^\alpha = \frac{1}{N}\sum_{t=1}^{N_{\rm MC}^\alpha} a_i^\alpha(t)
\end{equation}
in case that the input consists on $N_{\rm MC}^\alpha$ Monte Carlo
measurements. In order to estimate the error in $\bar F$ we use linear
error propagation
\begin{equation}
  f(A_i^\alpha + \epsilon_i^\alpha) = F + f_i^\alpha \epsilon_i^\alpha + \mathcal O(\epsilon_i^2)\,. 
\end{equation}
where the derivatives $f_i^\alpha = {\partial f}/{\partial A_i^\alpha}$
are in practice evaluated at $\bar a_i^\alpha$
\begin{equation}
  \label{eq:dad}
  \bar f_i^\alpha = \frac{\partial f}{\partial A_i^\alpha}\Big|_{\bar a_i^\alpha}\,.
\end{equation}
The error in our prediction is estimated using the \emph{per-ensemble}
auto-correlation function 
\begin{equation}
  \Gamma_F^\alpha (t) = \sum_{ij} \bar f_i^\alpha \bar f_j^\alpha \Gamma_{ij}^{\alpha \alpha}(t)\,,
\end{equation}
with $\Gamma_{ij}^{\alpha\alpha}$ being the product of the
uncertainties of the inputs $i$ and $j$, or the
auto-correlation function in case of Monte Carlo input
(see Ref.~\cite{Ramos:2018vgu} for more details).

The proposal~\cite{Ramos:2018vgu} consists in using automatic
differentiation to compute the derivatives appearing in
\equationautorefname~\eqref{eq:dad}. A key issue is how to efficiently compute these derivatives when the function $f$
consists on an iterative process. 
The typical case is fitting the data to some model, that we will
examine in detail. 
The optimal value of the fit parameters $\bar p_i\, (i=1,\dots,N_{\rm
  parm})$ are given by the minimum of the function
\begin{equation}
  \chi^2(p_i;d_a)\,,\qquad p_i\, (i=1,\dots,N_{\rm parm})\,, \quad
  d_a\, (a=1,\dots,N_{\rm data})\,.
\end{equation}
where $d_a$ is the data (\ie
input) that is fitted.  Error propagation requires the derivatives of
the fit parameters with respect to the 
data with the condition that the $\chi^2$ function has to stay at the minimum. 
This derivatives can be exactly computed by the following equation~\cite{Ramos:2018vgu}:
\begin{equation}
  \frac{\delta p_i}{\delta d_a} = -\sum_{j=1}^{N_{\rm
      parm}}(H^{-1})_{ij}\partial_j \partial_a \chi^2\Big|_{(\bar p_i;\bar d_a)}\,.
\end{equation}
where $H_{ij} = \partial_j \partial_i \chi^2\Big|_{(\bar p_i;\bar
  d_a)}$ is the Hessian of the $\chi^2$ at the minimum. 

This Hessian can also be computed using automatic differentiation techniques.  The advantage of this approach are that one can rely on external efficient libraries for the minimization of the $\chi^2$ function (\ie to determine the central values of the fit parameters). 
Error propagation is performed \emph{once the minimum is found}, by
explicitly computing the Hessian of the $\chi^2$ function. 
Similar expressions can be easily obtained for any problem that can be
formulated as a ``root finding'' problem (the minimization of the
$\chi^2$ is nothing but finding the root of the gradient). 

We offer a few concluding comments for this section. First, using automatic differentiation for error analysis results in
\emph{robust} error estimates: as long as the central values of the
output are correctly computed, the exact nature of automatic
differentiation guarantees that errors will be correctly propagated. 
Second, Automatic differentiation techniques produce computationally
efficient analysis codes. 
They are significantly cheaper than resampling techniques
(see Ref.~\cite{Ramos:2018vgu} for an explicit comparison). 
Finally, some of the ideas discussed here like how to deal with Monte
Carlo data, and how to determine efficiently derivatives of
complicated iterative processes, can potentially be applied to other problems. 

Numerical implementations of these analysis techniques are freely
available:
\begin{description}
\item[Fortran] \url{https://gitlab.ift.uam-csic.es/alberto/aderrors}
\item[Julia] \url{https://gitlab.ift.uam-csic.es/alberto/aderrors.jl}
\item[Python] \url{https://github.com/fjosw/pyerrors} 
\end{description}

\clearpage
\section {System Architecture and Requirements}
\label{sec:hardware}
A complete optimization of a model of any one of the example use cases discussed in Sec.~\ref{sec:exemplary_use_cases}, properly implemented in terms of a pipeline connecting all elements of the problem, would require a substantial investment in computing resources. However, such investment would render dedicated resources idle or obsolete once the optimization is complete. A good way to face such a time-peaking computing demand is to rely on a cloud infrastructure providing scalable and manageable computational resources. Before we discuss the details of such infrastructure, let us outline the main principles we consider relevant for the optimization system to run successfully. 

\subsection{Guiding Principles}
First of all, the developed pipeline should support a variety of simulation packages that are commonly used in various physics branches: \eg GEANT4~\cite{GEANT}, Pythia~\cite{Sjostrand:2006za,Sjostrand:2007gs,Sjostrand:2014zea}, Genie~\cite{genie}, to mention just a few. Each of the cited packages relies on other software components that should be pre-installed on the system to allow for its execution. Also, to support the execution of the main differential engine, the system would require different sets of software components depending on the task at hand. Secondly, the system should support different scales of devices under optimization: as we have seen {\em supra} these range from a coarse, high-level comparison of design options for a simple problem (\eg the one of the MUonE detector, see {\em infra}, Sec.~\ref{sec:muone}) to a detailed estimation of the performance of an LHC-scale detector entailing the simultaneous consideration of several hundred design parameters; in other words, given the breadth of the possible application and their very different demands and scale, the usage of computing resources is going to be relatively small in some cases, while it can be enormous in others. Yet the scale of the selected optimization task should not be an issue implying specific attention to an end-user. 

An integrated end-to-end optimization system should run on many connected computing nodes in an automated fashion. Also, it is important to stress that running on regular cloud nodes might be quite expensive; the main commercial cloud providers offer so-called \textit{spot} nodes, which may be interrupted at a very short notice at any time. We suggest our system should be capable of running on such nodes as well. 
Then, in order to meet computing power peak demands the system should be able to exploit ``opportunistic'' resources and volunteer computing if matching hardware requirements. Opportunistic resources are resources not dedicated to the project that can be temporarily made available to it, for example from a site of the collaboration. Volunteer computing indicates 
a type of distributed computing in which computer owners can make available their computing resources to research projects (see, for example, \url{https://boinc.berkeley.edu/}).

Finally, the system should support handling many user tasks simultaneously, by providing users with independent computational resources while allowing them to share their results. In a realistic scenario, a large optimization problem would be studied by several researchers who would test different parts of the developed pipeline independently, by freezing some optimization parameters in turn. The capability to run in parallel several of these partial tasks would enable a quicker convergence of the global task.

Below we provide a synthetic outline of the described principles:
\begin{itemize}
    \item flexibility;
    \item distributed execution and scalability;
    \item interruption-friendly execution;
    \item user task independence \& result sharing.
\end{itemize}

\noindent
Of course, the above outline constitutes the final goal of a long development task, which must be approached in an incremental fashion.

\subsection{System Architecture Outline}

Following the guiding principles defined {\em supra}, the system should be merged into a cloud infrastructure ideally supporting all the major cloud providers (AWS, GCP, Azure, ...). Today's state-of-the-art virtual nodes management system is Kubernetes~\cite{kubernetes}, which allows flexible setup, execution, and monitoring of different tasks on various virtual nodes comprising an elastic cluster. Such a cluster may be populated by a few nodes (\egc a single node). Depending on the computing demands, it extends to a maximum number of nodes configured by the administrator. Also, it enables the creation of software environments (containers) that include all the packages required for specific software to run. In order to support both independent task execution and spot instances handling, the system should have storage, compute and code decoupled at the design time. Storage should incorporate both persistent structured data like a relational database for different run results and unstructured data volumes for handling temporary files and intermediate results. Computing resources will be provided with the help of Kubernetes and nodes of the cluster. The code for the computations of individual tasks should be aware of the storage capacities available to it as well as of the computational resources it can use. Part of the system responsible for the overall execution should ensure the availability of storage components, individual task tracking, and result sharing components.

\subsection{Infrastructure Requirements}

The optimization system relies on a few functionalities from the cloud provider. We list them below.

\begin{itemize}
    \item API for Kubernetes cluster access/management/container execution;
    \item Ability for Kubernetes cluster to run on spot instances;
    \item Kubernetes cluster monitoring;
    \item Support for structured DB setup and configuration;
    \item Support for unstructured persistent volumes setup and management;
    \item Billing API for monitoring and controlling of computational budget.
\end{itemize}

\subsection{Hardware Requirements}

Depending on the optimization task and its software computations packages different hardware resources should be provided. Thus, the system should run on relatively capable virtual nodes (24 CPUs, 128GB of RAM). However, some of the computational tasks might require additional hardware resources like GPU/TPU cards or extremely large RAM volumes. For such dedicated tasks the system should be able to instantiate a separate computational Kubernetes clusters.

\subsection{Main Software Components}

The optimization system should include the following main software components:

\begin{itemize}
    \item structural storage (database) management;
    \item volume storage management;
    \item cloud compute management;
    \item simulation package connection and software environment configuration;
    \item user task management;
    \item optimization monitoring/benchmarking interface;
    \item black-box optimization runtime (Python, Julia, ...);
    \item differentiable optimization runtime (PyTorch~\cite{pytorch}, Tensorflow~\cite{abadi2016tensorflow}, JAX~\cite{jax2018github}).
\end{itemize}

\noindent
The above is a non-exhaustive list, and is only meant to offer a view of the typical deployment needs of the system we imagine.

\subsection{Integration Requirements}

Finally, we believe that the system should also provide interfaces to ease the integration of the following:

\begin{itemize}
    \item cloud provider-specific Kubernetes interfaces, storage resources and computing resources configuration;
    \item new/custom physics simulators;
    \item external benchmarking services like Weights-and-biases or Comet.ml.
\end{itemize}

\section{Conclusions}
\label{sec:conclusions}

The history of particle detection is over 100 years old, and it is full of breakthroughs and paradigm-changing inventions which often ingeniously exploited technological advancements conceived for other applications to improve the performance of apparatus. In this document we argue that the most relevant recent advancement for particle detector development is the rise of differentiable programming. Coupled to the large computing power available today, the automatic calculation of derivatives of complex functions and computer code offers the possibility of scanning the high-dimensional space of design solutions in a systematic search of the global maximum of a carefully defined utility function. 
The potential gains---in terms of performance, decreased spending, or others---of a global model of the system under design are very large, and they come from the possibility of the optimization task to re-align the design goal to the specific choices that a designer is faced with when picking a value of the many free parameters of the system, once all external constraints (\egc total cost, time, space) are established. 

In this document we discuss a plan of investigations which, by considering and seeking a solution to a wide range of use cases of complexity varying from easy to very hard, may empower our community with the technology needed to attack detector design problems still harder and of larger scale. While automated optimization systems cannot at the present time have the ambition to substitute the hand and the intuition of the expert, they cannot any longer be ignored as assistants to the design task. Ultimately, we believe that their careful specification and systematic use may allow us to discover entirely new, groundbreaking solutions to our century-old problems, furthering the happy partnership of progress in technology and pure research we have witnessed until today. 

\clearpage
\section* {Acknowledgements}
We wish to thank all the participants of the first MODE workshop on differentiable programming that took place in Louvain-la-Neuve (Belgium) from 6 to 8 September 2021, for the fruitful discussions that influenced the content of this document. 

We gratefully acknowledge support by IRIS-HEP (Institute for Research and Innovation in Software for High Energy Physics, National Science Foundation grant OAC-1836650, \url{https://iris-hep.org/}) and JENAA (Joint ECFA-NuPECC-APPEC Activities, \url{http://www.nupecc.org/jenaa/}). 
A. Giammanco's work was partially supported by the EU Horizon 2020 Research and Innovation Programme under the Marie Sklodowska-Curie Grant Agreement No. 822185 (``INTENSE'') and the Research and Innovation Action for Security Grant Agreement No. 101021812 (``SilentBorder''), and by the Fonds de la Recherche Scientifique - FNRS under Grants No. T.0099.19 and J.0070.21.  P. Vischia’s work was supported by the FNRS under the Grant No. 40000963. C. Krause's work is supported by DOE grant DOE-SC0010008. M. Aehle, N. Gauger, and R. Keidel gratefully acknowledge the funding of the
research training group SIVERT by the German federal state of Rhineland-Palatinate.
The work of T. Dorigo, L. Layer and N. Simpson is supported by a Marie Sklodowska-Curie Innovative Training Network Fellowship of the European Commissions Horizon 2020 Programme under Contract Number 765710 INSIGHTS. L. Heinrich is supported by the Excellence Cluster ORIGINS, which is funded by the Deutsche Forschungsgemeinschaft (DFG, German Research Foundation) under Germany’s Excellence Strategy - EXC-2094-390783311. 
Alberto Ramos acknowledges financial support from the Generalitat Valenciana 
(genT program CIDEGENT/2019/040) and the Ministerio de Ciencia e
Innovacion (PID2020-113644GB-I00). H. Zaraket would like to thank the 
Erasmus Plus mobility program.  
Figure~\ref{f:muography-passive-material} (bottom) has been adapted from Ref.~\cite{Muography-SMO} thanks to the courtesy of L\'aszl\'o Ol\'ah.
We would also like to thank the International Muon Collider Collaboration 
for providing us with the data to produce Figs.~\ref{fig:nozzle} and~\ref{fig:spectrum}.

\begin{figure}[!ht]
\begin{center}
\includegraphics[width=10cm]{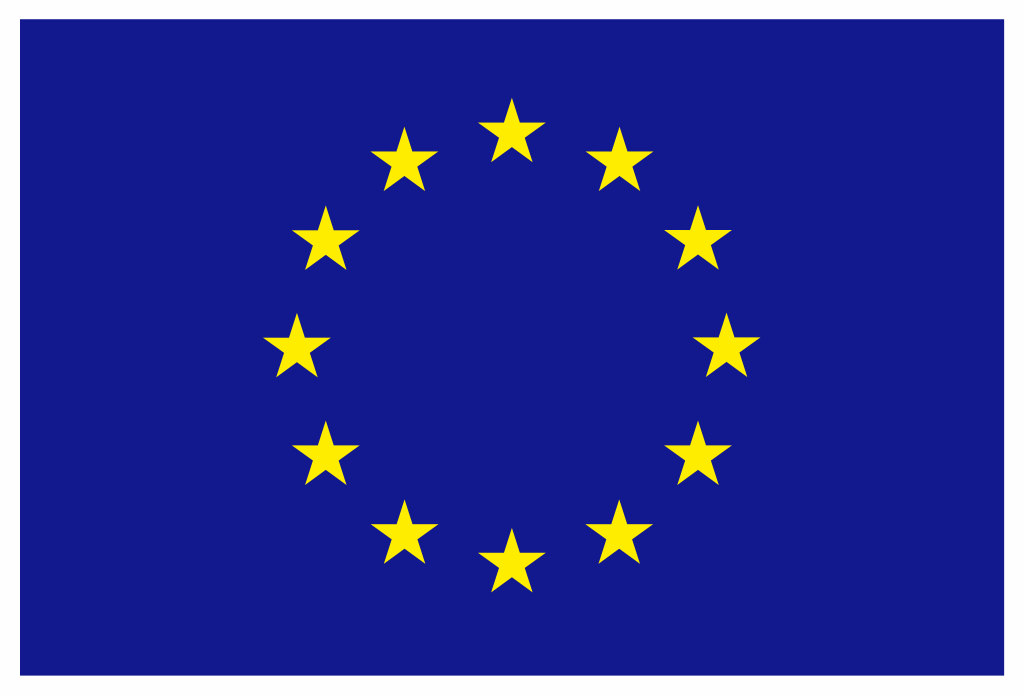}
\end{center}
\end{figure}

\clearpage

\bibliographystyle{unsrturl}
\bibliography{main.bib}

\begin{thebibliography}{100}

\bibitem{cranmer2020frontier}
K.~Cranmer, J.~Brehmer, and G.~Louppe.
\newblock The frontier of simulation-based inference.
\newblock {\em Proceedings of the National Academy of Sciences},
  117(48):30055--30062, 2020.
\newblock \href {http://arxiv.org/abs/1911.01429} {\path{arXiv:1911.01429}}.

\bibitem{npnipaper}
A.G. Baydin, K.~Cranmer, P.~de~Castro~Manzano, C.~Delaere, D.~Derkach,
  J.~Donini, T.~Dorigo, A.~Giammanco, J.~Kieseler, L.~Layer, G.~Louppe,
  F.~Ratnikov, G.C. Strong, M.~Tosi, A.~Ustyuzhanin, P.~Vischia, and H.~Yarar.
\newblock Toward machine learning optimization of experimental design.
\newblock {\em Nuclear Physics News}, 31(1):25--28, 2021.
\newblock \href {https://doi.org/10.1080/10619127.2021.1881364}
  {\path{doi:10.1080/10619127.2021.1881364}}.

\bibitem{Mishnayot:2014nya}
Y.~Mishnayot, M.~Layani, I.~Cooperstein, S.~Magdassi, and G.~Ron.
\newblock {Three-dimensional printing of scintillating materials}.
\newblock {\em Rev. Sci. Instrum.}, 85:085102, 2014.
\newblock \href {http://arxiv.org/abs/1406.4817} {\path{arXiv:1406.4817}},
  \href {https://doi.org/10.1063/1.4891703} {\path{doi:10.1063/1.4891703}}.

\bibitem{Giacomini:2019kqz}
G.~Giacomini, W.~Chen, G.~D'Amen, and A.~Tricoli.
\newblock {Fabrication and performance of AC-coupled LGADs}.
\newblock {\em JINST}, 14(09):P09004, 2019.
\newblock \href {http://arxiv.org/abs/1906.11542} {\path{arXiv:1906.11542}},
  \href {https://doi.org/10.1088/1748-0221/14/09/p09004}
  {\path{doi:10.1088/1748-0221/14/09/p09004}}.

\bibitem{Evans_2008}
L.~Evans and P.~Bryant.
\newblock {LHC} machine.
\newblock {\em Journal of Instrumentation}, 3(08):S08001--S08001, aug 2008.
\newblock \href {https://doi.org/10.1088/1748-0221/3/08/s08001}
  {\path{doi:10.1088/1748-0221/3/08/s08001}}.

\bibitem{Farrell:DLPS2017}
{\em {Particle Track Reconstruction with Deep Learning}}, 2017.

\bibitem{Farrell:2018cjr}
S.~Farrell et~al.
\newblock {Novel deep learning methods for track reconstruction}.
\newblock {\em {4th International Workshop Connecting The Dots 2018}}, 2018.
\newblock \href {http://arxiv.org/abs/1810.06111} {\path{arXiv:1810.06111}}.

\bibitem{Amrouche:2019wmx}
S.~Amrouche et~al.
\newblock {The Tracking Machine Learning Challenge: Accuracy Phase}.
\newblock In {\em The NeurIPS '18 Competition}, page 231. Springer
  International Publishing, 2020.
\newblock \href {https://doi.org/10.1007/978-3-030-29135-8\_9}
  {\path{doi:10.1007/978-3-030-29135-8\_9}}.

\bibitem{Ju:2020xty}
X.~Ju et~al.
\newblock {Graph Neural Networks for Particle Reconstruction in High Energy
  Physics detectors}.
\newblock {\em {33rd Annual Conference on Neural Information Processing
  Systems}}, 2020.
\newblock \href {http://arxiv.org/abs/2003.11603} {\path{arXiv:2003.11603}}.

\bibitem{Akar:2020jti}
S.~Akar et~al.
\newblock {An updated hybrid deep learning algorithm for identifying and
  locating primary vertices}, 2020.
\newblock \href {http://arxiv.org/abs/2007.01023} {\path{arXiv:2007.01023}}.

\bibitem{Shlomi:2020ufi}
J.~Shlomi et~al.
\newblock {Secondary Vertex Finding in Jets with Neural Networks}.
\newblock {\em {Eur. Phys. J. C}}, 81:540, 2021.
\newblock \href {https://doi.org/10.1140/epjc/s10052-021-09342-y}
  {\path{doi:10.1140/epjc/s10052-021-09342-y}}.

\bibitem{Choma:2020cry}
N.~Choma et~al.
\newblock {Track Seeding and Labelling with Embedded-space Graph Neural
  Networks}, 2020.
\newblock \href {http://arxiv.org/abs/2007.00149} {\path{arXiv:2007.00149}}.

\bibitem{Siviero:2020tim}
F.~Siviero et~al.
\newblock First application of machine learning algorithms to the position
  reconstruction in resistive silicon detectors.
\newblock {\em Journal of Instrumentation}, 16:P03019, 2021.
\newblock \href {https://doi.org/10.1088/1748-0221/16/03/p03019}
  {\path{doi:10.1088/1748-0221/16/03/p03019}}.

\bibitem{Fox:2020hfm}
P.J. Fox, S.~Huang, J.~Isaacson, X.~Ju, and B.~Nachman.
\newblock {Beyond 4D Tracking: Using Cluster Shapes for Track Seeding}.
\newblock {\em JINST}, 16(05):P05001, 2021.
\newblock \href {http://arxiv.org/abs/2012.04533} {\path{arXiv:2012.04533}},
  \href {https://doi.org/10.1088/1748-0221/16/05/P05001}
  {\path{doi:10.1088/1748-0221/16/05/P05001}}.

\bibitem{Amrouche:2021tlm}
S.~Amrouche, M.~Kiehn, T.~Golling, and A.~Salzburger.
\newblock {Hashing and metric learning for charged particle tracking}.
\newblock {\em {33rd Annual Conference on Neural Information Processing
  Systems}}, 2021.
\newblock \href {http://arxiv.org/abs/2101.06428} {\path{arXiv:2101.06428}}.

\bibitem{goto2021development}
Kiichi G. et~al.
\newblock {Development of a Vertex Finding Algorithm using Recurrent Neural
  Network}, 2021.
\newblock \href {http://arxiv.org/abs/{2101.11906}}
  {\path{arXiv:{2101.11906}}}.

\bibitem{Biscarat:2021dlj}
C.~Biscarat, S.~Caillou, C~Rougier, J.~Stark, and J.~Zahreddine.
\newblock {Towards a realistic track reconstruction algorithm based on graph
  neural networks for the HL-LHC}.
\newblock In {\em {25th International Conference on Computing in High-Energy
  and Nuclear Physics}}, 2021.
\newblock \href {http://arxiv.org/abs/2103.00916} {\path{arXiv:2103.00916}}.

\bibitem{Akar:2021gns}
{Akar, S.} et~al.
\newblock Progress in developing a hybrid deep learning algorithm for
  identifying and locating primary vertices.
\newblock {\em EPJ Web Conf.}, 251:04012, 2021.
\newblock \href {https://doi.org/10.1051/epjconf/202125104012}
  {\path{doi:10.1051/epjconf/202125104012}}.

\bibitem{Thais:2021qcb}
S.~Thais and G.~DeZoort.
\newblock {Instance Segmentation GNNs for One-Shot Conformal Tracking at the
  LHC}, 2021.
\newblock \href {http://arxiv.org/abs/2103.06509} {\path{arXiv:2103.06509}}.

\bibitem{Ju:2021ayy}
X.~Ju et~al.
\newblock {Performance of a geometric deep learning pipeline for HL-LHC
  particle tracking}.
\newblock {\em {Eur. Phys. J. C}}, 81:876, 2021.

\bibitem{Dezoort:2021kfk}
G.~Dezoort et~al.
\newblock {Charged particle tracking via edge-classifying interaction
  networks}, 2021.
\newblock \href {http://arxiv.org/abs/2103.16701} {\path{arXiv:2103.16701}}.

\bibitem{Edmonds:2021lzd}
A.~Edmonds, D.~Brown, L.~Vinas, and S.~Pagan.
\newblock Using machine learning to select high-quality measurements.
\newblock {\em Journal of Instrumentation}, 16:T08010, aug 2021.
\newblock \href {https://doi.org/10.1088/1748-0221/16/08/t08010}
  {\path{doi:10.1088/1748-0221/16/08/t08010}}.

\bibitem{Lavrik:2021zgt}
E.~Lavrik, M.~Shiroya, H.R. Schmidt, A.~Toia, and J.M. Heuser.
\newblock {Optical inspection of the silicon micro-strip sensors for the CBM
  experiment employing artificial intelligence}.
\newblock {\em Nucl. Instr. Meth. A}, 1021:165932, 2022.
\newblock \href {https://doi.org/https://doi.org/10.1016/j.nima.2021.165932}
  {\path{doi:https://doi.org/10.1016/j.nima.2021.165932}}.

\bibitem{Huth:2021zcm}
B.~Huth, A.~Salzburger, and T.~Wettig.
\newblock {Machine learning for surface prediction in ACTS}.
\newblock In {\em {25th International Conference on Computing in High-Energy
  and Nuclear Physics}}, 2021.
\newblock \href {http://arxiv.org/abs/2108.03068} {\path{arXiv:2108.03068}}.

\bibitem{Goncharov:2021wvd}
P.~Goncharov et~al.
\newblock {Ariadne: PyTorch Library for Particle Track Reconstruction Using
  Deep Learning}.
\newblock In {\em {24th International Scientific Conference of Young Scientists
  and Specialists}}, 2021.
\newblock \href {http://arxiv.org/abs/2109.08982} {\path{arXiv:2109.08982}}.

\bibitem{Lazar:2022ixi}
A.~Lazar et~al.
\newblock {Accelerating the Inference of the Exa.TrkX Pipeline}, 2022.
\newblock \href {http://arxiv.org/abs/2202.06929} {\path{arXiv:2202.06929}}.

\bibitem{Benedikt:2020ejr}
M.~Benedikt, A.~Blondel, P.~Janot, M.~Mangano, and F.~Zimmermann.
\newblock {Future Circular Colliders succeeding the LHC}.
\newblock {\em Nature Phys.}, 16(4):402--407, 2020.
\newblock \href {https://doi.org/10.1038/s41567-020-0856-2}
  {\path{doi:10.1038/s41567-020-0856-2}}.

\bibitem{2020arXiv200504305H}
D.~{Hernandez} and T.B. {Brown}.
\newblock {Measuring the Algorithmic Efficiency of Neural Networks}.
\newblock {\em arXiv e-prints}, page arXiv:2005.04305, May 2020.
\newblock \href {http://arxiv.org/abs/2005.04305} {\path{arXiv:2005.04305}}.

\bibitem{Kieseler:2020wcq}
J.~Kieseler.
\newblock {Object condensation: one-stage grid-free multi-object reconstruction
  in physics detectors, graph and image data}.
\newblock {\em Eur. Phys. J. C}, 80(9):886, 2020.
\newblock \href {http://arxiv.org/abs/2002.03605} {\path{arXiv:2002.03605}},
  \href {https://doi.org/10.1140/epjc/s10052-020-08461-2}
  {\path{doi:10.1140/epjc/s10052-020-08461-2}}.

\bibitem{hicks1974assessment}
R.M. Hicks and G.N. Murman, E.M. nd~Vanderplaats.
\newblock An assessment of airfoil design by numerical optimization.
\newblock Technical report, NASA Ames Research Center, 1974.

\bibitem{hicks_wing_1978}
R.M. Hicks and P.A. Henne.
\newblock Wing design by numerical optimization.
\newblock {\em Journal of Aircraft}, 15(7):407--412, 1978.
\newblock \href {https://doi.org/10.2514/3.58379} {\path{doi:10.2514/3.58379}}.

\bibitem{pironneau_optimum_1974}
O.~Pironneau.
\newblock On optimum design in fluid mechanics.
\newblock {\em J. Fluid Mech.}, 64(1):97--110, 1974.
\newblock \href {https://doi.org/10.1017/S0022112074002023}
  {\path{doi:10.1017/S0022112074002023}}.

\bibitem{pironneau_optimum_1973}
O.~Pironneau.
\newblock On optimum profiles in stokes flow.
\newblock {\em J. Fluid Mech.}, 59(1):117--128, 1973.
\newblock \href {https://doi.org/10.1017/S002211207300145X}
  {\path{doi:10.1017/S002211207300145X}}.

\bibitem{jameson_aerodynamic_1988}
A.~Jameson.
\newblock Aerodynamic design via control theory.
\newblock {\em J. Sci. Comput.}, 3(3):233--260, 1988.
\newblock \href {https://doi.org/10.1007/BF01061285}
  {\path{doi:10.1007/BF01061285}}.

\bibitem{towara_discrete_2013}
M.~Towara and U.~Naumann.
\newblock A discrete adjoint model for {OpenFOAM}.
\newblock {\em Procedia Computer Science}, 18:429--438, 2013.
\newblock \href {https://doi.org/10.1016/j.procs.2013.05.206}
  {\path{doi:10.1016/j.procs.2013.05.206}}.

\bibitem{Albring_etal2016b}
T.~Albring, M.~Sagebaum, and N.R. Gauger.
\newblock Efficient aerodynamic design using the discrete adjoint method in
  {SU2}.
\newblock {\em AIAA 2016-3518}, 2016.

\bibitem{luers_adjoint-based_2018}
M.~Luers et~al.
\newblock Adjoint-based volumetric shape optimization of turbine blades.
\newblock In {\em 2018 Multidisciplinary Analysis and Optimization Conference}.
  American Institute of Aeronautics and Astronautics, 2018.
\newblock \href {https://doi.org/10.2514/6.2018-3638}
  {\path{doi:10.2514/6.2018-3638}}.

\bibitem{nemili_accurate_2017}
A.~Nemili, E.~\"Ozkaya, N.R. Gauger, F.~Kramer, and F.~Thiele.
\newblock Accurate discrete adjoint approach for optimal active separation
  control.
\newblock {\em {AIAA} Journal}, 55(9):3016--3026, 2017.
\newblock \href {https://doi.org/10.2514/1.J055009}
  {\path{doi:10.2514/1.J055009}}.

\bibitem{zhou_discrete_2018}
B.Y. Zhou, S.~Ryong~Koh, N.R. Gauger, M.~Meinke, and W.~Schöder.
\newblock A discrete adjoint framework for trailing-edge noise minimization via
  porous material.
\newblock {\em Computers \& Fluids}, 172:97--108, 2018.
\newblock \href {https://doi.org/10.1016/j.compfluid.2018.06.017}
  {\path{doi:10.1016/j.compfluid.2018.06.017}}.

\bibitem{bombardieri_aerostructural_2021}
R.~Bombardieri, R.~Cavallaro, R.~Sanchez, and N.R. Gauger.
\newblock Aerostructural wing shape optimization assisted by algorithmic
  differentiation.
\newblock {\em Struct Multidisc Optim}, 64(2):739--760, 2021.
\newblock \href {https://doi.org/10.1007/s00158-021-02884-5}
  {\path{doi:10.1007/s00158-021-02884-5}}.

\bibitem{MGDLS2021}
M.~Morlighem, D.~Goldberg, T.~Dias~dos Santos, J.~Lee, and M.~Sagebaum.
\newblock Mapping the sensitivity of the {Amundsen} sea embayment to changes in
  external forcings using automatic differentiation.
\newblock {\em Geophysical Research Letters}, 48(23), 2021.
\newblock \href {https://doi.org/https://doi.org/10.1029/2021GL095440}
  {\path{doi:https://doi.org/10.1029/2021GL095440}}.

\bibitem{andersson2012casadi}
J.~Andersson, J.~{\AA}kesson, and M.~Diehl.
\newblock Casadi: A symbolic package for automatic differentiation and optimal
  control.
\newblock In {\em Recent advances in algorithmic differentiation}, pages
  297--307. Springer, 2012.

\bibitem{achdou_computational_2005}
Y.~Achdou and O.~Pironneau.
\newblock {\em Computational Methods for Option Pricing}.
\newblock Society for Industrial and Applied Mathematics, 2005.
\newblock \href {https://doi.org/10.1137/1.9780898717495}
  {\path{doi:10.1137/1.9780898717495}}.

\bibitem{goodfellow2016deep}
I.~Goodfellow, Y.~Bengio, and A.~Courville.
\newblock {\em Deep learning}.
\newblock MIT press, 2016.

\bibitem{lecun2015deep}
Y.~LeCun, Y.~Bengio, and G.~Hinton.
\newblock Deep learning.
\newblock {\em Nature}, 521(7553):436--444, 2015.

\bibitem{janai2020computer}
J.~Janai et~al.
\newblock Computer vision for autonomous vehicles: Problems, datasets and state
  of the art.
\newblock {\em Foundations and Trends{\textregistered} in Computer Graphics and
  Vision}, 12(1--3):1--308, 2020.

\bibitem{goldberg2017neural}
Y.~Goldberg.
\newblock Neural network methods for natural language processing.
\newblock {\em Synthesis lectures on human language technologies},
  10(1):1--309, 2017.

\bibitem{sutton2018reinforcement}
R.S. Sutton and A.G. Barto.
\newblock {\em Reinforcement learning: An introduction}.
\newblock MIT press, 2018.

\bibitem{pytorch}
A.~Paszke et~al.
\newblock Pytorch: An imperative style, high-performance deep learning library.
\newblock In H.~Wallach, H.~Larochelle, A.~Beygelzimer, F.~d\textquotesingle
  Alch\'{e}-Buc, E.~Fox, and R.~Garnett, editors, {\em Advances in Neural
  Information Processing Systems 32}, pages 8024--8035. Curran Associates,
  Inc., 2019.

\bibitem{abadi2016tensorflow}
M.~Abadi et~al.
\newblock $\{$TensorFlow$\}$: A system for $\{$Large-Scale$\}$ machine
  learning.
\newblock In {\em 12th USENIX symposium on operating systems design and
  implementation (OSDI 16)}, pages 265--283, 2016.

\bibitem{baydin_automatic_2017}
A.G. Baydin, B.A. Pearlmutter, A.A. Radul, and J.M. Siskind.
\newblock Automatic differentiation in machine learning: A survey.
\newblock {\em J. Mach. Learn. Res.}, 18(1):5595–5637, 2017.

\bibitem{olahfuncprog}
C.~Olah.
\newblock Neural networks, types, and functional programming, 2015.
\newblock Retrieved on 17-03-2022.
\newblock URL: \url{http://colah.github.io/posts/2015-09-NN-Types-FP/}.

\bibitem{dalrymple}
D.~Dalrymple.
\newblock Differentiable programming as the 2016 most important recent
  scientific news, 2016.
\newblock Retrieved on 17-03-2022.
\newblock URL: \url{https://www.edge.org/response-detail/26794}.

\bibitem{lecunfacebookpost}
Y.~LeCun.
\newblock Facebook post on differentiable programming, 2018.
\newblock Retrieved on 17-03-2022.
\newblock URL:
  \url{https://www.facebook.com/yann.lecun/posts/10155003011462143}.

\bibitem{Shirobokov:2020tjt}
S.~Shirobokov, V.~Belavin, M.~Kagan, A.~Ustyuzhanin, and A.G. Baydin.
\newblock {Black-Box Optimization with Local Generative Surrogates}, 2020.
\newblock \href {http://arxiv.org/abs/2002.04632} {\path{arXiv:2002.04632}}.

\bibitem{dorigo2020geometry}
T.~Dorigo.
\newblock Geometry optimization of a muon-electron scattering detector.
\newblock {\em Physics Open}, 4:100022, 2020.

\bibitem{ratnikov2020}
F.~Ratnikov.
\newblock {Using machine learning to speed up and improve calorimeter R\&D}.
\newblock {\em Journal of Instrumentation}, 15(05):C05032, 2020.

\bibitem{cisbani2020}
E.~Cisbani et~al.
\newblock {AI-optimized detector design for the future Electron-Ion Collider:
  the dual-radiator RICH case}.
\newblock {\em Journal of Instrumentation}, 15(05):P05009, 2020.

\bibitem{edelen:ml2}
A.~Edelen et~al.
\newblock Machine learning for orders of magnitude speedup in multiobjective
  optimization of particle accelerator systems.
\newblock {\em Physical Review Accelerators and Beams}, 23(4):044601, April
  2020.
\newblock Publisher: American Physical Society.
\newblock \href {https://doi.org/10.1103/PhysRevAccelBeams.23.044601}
  {\path{doi:10.1103/PhysRevAccelBeams.23.044601}}.

\bibitem{Koser:ML1}
D.~Koser et~al.
\newblock Input beam matching and beam dynamics design optimization of the
  {IsoDAR} {RFQ} using statistical and machine learning techniques.
\newblock {\em {arXiv}:2112.02579 [physics]}, 2021.
\newblock {(Submitted to Frontiers in Physics)}.
\newblock \href {http://arxiv.org/abs/2112.02579} {\path{arXiv:2112.02579}}.

\bibitem{van_der_veken:ml1}
F.~Van Der~Veken et~al.
\newblock Machine learning in accelerator physics: applications at the {CERN}
  {Large} {Hadron} {Collider}.
\newblock In {\em Proceedings of {Artificial} {Intelligence} for {Science},
  {Industry} and {Society} {PoS}({AISIS2019})}, volume 372, page 044. SISSA
  Medialab, July 2020.

\bibitem{meyer_optimization_2020}
S.~Meyer et~al.
\newblock Optimization and performance study of a proton {CT} system for
  pre-clinical small animal imaging.
\newblock {\em Phys. Med. Biol.}, 65(15):155008, 2020.
\newblock \href {https://doi.org/10.1088/1361-6560/ab8afc}
  {\path{doi:10.1088/1361-6560/ab8afc}}.

\bibitem{2020SciPy-NMeth}
P.~Virtanen et~al.
\newblock {{SciPy} 1.0: Fundamental Algorithms for Scientific Computing in
  Python}.
\newblock {\em Nature Methods}, 17:261--272, 2020.
\newblock \href {https://doi.org/10.1038/s41592-019-0686-2}
  {\path{doi:10.1038/s41592-019-0686-2}}.

\bibitem{shanno_conditioning_1970}
D.F. Shanno.
\newblock Conditioning of quasi-newton methods for function minimization.
\newblock {\em Math. Comp.}, 24(111):647--656, 1970.
\newblock \href {https://doi.org/10.1090/S0025-5718-1970-0274029-X}
  {\path{doi:10.1090/S0025-5718-1970-0274029-X}}.

\bibitem{goldfarb_family_1970}
D.~Goldfarb.
\newblock A family of variable-metric methods derived by variational means.
\newblock {\em Math. Comp.}, 24(109):23--26, 1970.
\newblock \href {https://doi.org/10.1090/S0025-5718-1970-0258249-6}
  {\path{doi:10.1090/S0025-5718-1970-0258249-6}}.

\bibitem{fletcher_new_1970}
R.~Fletcher.
\newblock A new approach to variable metric algorithms.
\newblock {\em The Computer Journal}, 13(3):317--322, 1970.
\newblock \href {https://doi.org/10.1093/comjnl/13.3.317}
  {\path{doi:10.1093/comjnl/13.3.317}}.

\bibitem{byrd_limited_1995}
R.H. Byrd, P.~Lu, J.~Nocedal, and C.~Zhu.
\newblock A limited memory algorithm for bound constrained optimization.
\newblock {\em {SIAM} J. Sci. Comput.}, 16(5):1190--1208, 1995.
\newblock \href {https://doi.org/10.1137/0916069} {\path{doi:10.1137/0916069}}.

\bibitem{zhu_algorithm_1997}
C.~Zhu, R.H. Byrd, P.~Lu, and J.~Nocedal.
\newblock Algorithm 778: L-{BFGS}-b: Fortran subroutines for large-scale
  bound-constrained optimization.
\newblock {\em {ACM} Trans. Math. Softw.}, 23(4):550--560, 1997.
\newblock \href {https://doi.org/10.1145/279232.279236}
  {\path{doi:10.1145/279232.279236}}.

\bibitem{burden2015numerical}
R.L. Burden, J.D. Faires, and A.M. Burden.
\newblock {\em Numerical Analysis}.
\newblock Cengage Learning, 2015.

\bibitem{griewank_evaluating_2008}
A.~Griewank and A.~Walther.
\newblock {\em Evaluating Derivatives}.
\newblock Other Titles in Applied Mathematics. Society for Industrial and
  Applied Mathematics, 2008.
\newblock \href {https://doi.org/10.1137/1.9780898717761}
  {\path{doi:10.1137/1.9780898717761}}.

\bibitem{wengert1964simple}
R.E. Wengert.
\newblock A simple automatic derivative evaluation program.
\newblock {\em Communications of the ACM}, 7(8):463--464, 1964.

\bibitem{linnainmaa1970representation}
S.~Linnainmaa.
\newblock The representation of the cumulative rounding error of an algorithm
  as a taylor expansion of the local rounding errors.
\newblock Master’s Thesis (in Finnish), Univ. Helsinki, 1970.

\bibitem{speelpenning1980compiling}
B.~Speelpenning.
\newblock Compiling fast partial derivatives of functions given by algorithms.
\newblock University of Illinois at Urbana-Champaign, 1980.

\bibitem{rumelhart1985learning}
D.E. Rumelhart, G.E. Hinton, and R.J. Williams.
\newblock Learning internal representations by error propagation.
\newblock Technical report, California Univ San Diego La Jolla Inst for
  Cognitive Science, 1985.

\bibitem{Walther2012Gsw}
A.~Walther and A.~Griewank.
\newblock Getting started with {ADOL-C}.
\newblock In U.~Naumann and O.~Schenk, editors, {\em Combinatorial Scientific
  Computing}, chapter~7, pages 181--202. Chapman-Hall CRC Computational
  Science, 2012.

\bibitem{Hogan2014FRM}
R.J. Hogan.
\newblock Fast reverse-mode automatic differentiation using expression
  templates in {C++}.
\newblock {\em {ACM} Transactions on Mathematical Software}, 40(4):26:1--26:24,
  2014.
\newblock \href {https://doi.org/10.1145/2560359} {\path{doi:10.1145/2560359}}.

\bibitem{Lotz2016Hat}
J.~Lotz.
\newblock {Hybrid approaches to adjoint code generation with DCO/C++}.
\newblock Department of Computer Science, RWTH Aachen University, 2016.
\newblock URL: \url{http://publications.rwth-aachen.de/record/667318}.

\bibitem{SaAlGauTOMS2019}
M.~Sagebaum, T.~Albring, and N.R. Gauger.
\newblock High-performance derivative computations using codipack.
\newblock {\em ACM Transactions on Mathematical Software (TOMS)}, 45(4), 2019.
\newblock \href {https://doi.org/10.1145/3356900} {\path{doi:10.1145/3356900}}.

\bibitem{Vassilev_Clad}
V.~Vassilev, M.~Vassilev, A.~Penev, L.~Moneta, and V.~Ilieva.
\newblock Clad {\textemdash} automatic differentiation using clang and {LLVM}.
\newblock {\em Journal of Physics: Conference Series}, 608:012055, 2015.
\newblock \href {https://doi.org/10.1088/1742-6596/608/1/012055}
  {\path{doi:10.1088/1742-6596/608/1/012055}}.

\bibitem{NEURIPS2020_9332c513}
W.~Moses and V.~Churavy.
\newblock Instead of rewriting foreign code for machine learning, automatically
  synthesize fast gradients.
\newblock In H.~Larochelle, M.~Ranzato, R.~Hadsell, M.~F. Balcan, and H.~Lin,
  editors, {\em Advances in Neural Information Processing Systems}, volume~33,
  pages 12472--12485. Curran Associates, Inc., 2020.

\bibitem{Bischof1997AAE}
C.H. Bischof, L.~Roh, and A.~Mauer.
\newblock {ADIC} --- {An} extensible automatic differentiation tool for
  {ANSI-C}.
\newblock {\em Software--Practice and Experience}, 27(12):1427--1456, 1997.
\newblock \href
  {https://doi.org/10.1002/(SICI)1097-024X(199712)27:12<1427::AID-SPE138>3.0.CO;2-Q}
  {\path{doi:10.1002/(SICI)1097-024X(199712)27:12<1427::AID-SPE138>3.0.CO;2-Q}}.

\bibitem{Bischof1996AAD}
C.H. Bischof, A.~Carle, P.~Khademi, and A.~Mauer.
\newblock {ADIFOR} 2.0: Automatic differentiation of {F}ortran 77 programs.
\newblock {\em IEEE Computational Science \& Engineering}, 3(3):18--32, 1996.

\bibitem{utke_openad_2008}
J.~Utke et~al.
\newblock Openad/f: A modular open-source tool for automatic differentiation of
  fortran codes.
\newblock {\em ACM Trans. Math. Softw.}, 34(4), 2008.
\newblock \href {https://doi.org/10.1145/1377596.1377598}
  {\path{doi:10.1145/1377596.1377598}}.

\bibitem{hascoet_tapenade_2013}
L.~Hascoet and V.~Pascual.
\newblock The tapenade automatic differentiation tool: Principles, model, and
  specification.
\newblock {\em {ACM} Trans. Math. Softw.}, 39(3):1--43, 2013.
\newblock \href {https://doi.org/10.1145/2450153.2450158}
  {\path{doi:10.1145/2450153.2450158}}.

\bibitem{autodiffwebsite}
M.~B\"ucker and F.~(founders) Schiller.
\newblock Community portal for automatic differentiation, 2000.
\newblock Retrieved on 17-03-2022.
\newblock URL: \url{http://www.autodiff.org}.

\bibitem{madjax}
L.~Heinrich and M.~Kagan.
\newblock {Differentiable Matrix Elements with MadJax}.
\newblock In {\em {20th Intern. Workshop on Adv. Computing and Analysis
  Techniques in Phys. Res.}}, 2022.
\newblock \href {http://arxiv.org/abs/2203.00057} {\path{arXiv:2203.00057}}.

\bibitem{Alwall:2014hca}
J.~Alwall et~al.
\newblock {The automated computation of tree-level and next-to-leading order
  differential cross sections, and their matching to parton shower
  simulations}.
\newblock {\em JHEP}, 07:079, 2014.
\newblock \href {http://arxiv.org/abs/1405.0301} {\path{arXiv:1405.0301}},
  \href {https://doi.org/10.1007/JHEP07(2014)079}
  {\path{doi:10.1007/JHEP07(2014)079}}.

\bibitem{jax2018github}
J.~Bradbury et~al.
\newblock {JAX}: composable transformations of {P}ython+{N}um{P}y programs.
\newblock \url{http://github.com/google/jax}, 2018.
\newblock Version 0.2.5.

\bibitem{hutchison_data-flow_2006}
B.~Dauvergne and L.~Hasco\"et.
\newblock The data-flow equations of checkpointing in reverse automatic
  differentiation.
\newblock In V.N. Alexandrov, G.D. van Albada, P.M.A. Sloot, and J.~Dongarra,
  editors, {\em Computational Science – {ICCS} 2006}, volume 3994, pages
  566--573. Springer Berlin Heidelberg, 2006.
\newblock \href {https://doi.org/10.1007/11758549_78}
  {\path{doi:10.1007/11758549_78}}.

\bibitem{bluhdorn_event-based_2021}
J.~Bl\"uhdorn, M.~Sagebaum, and N.R. Gauger.
\newblock Event-based automatic differentiation of {OpenMP} with {OpDiLib}.
\newblock {\em {arXiv}:2102.11572 [cs]}, 2021.
\newblock \href {http://arxiv.org/abs/2102.11572} {\path{arXiv:2102.11572}}.

\bibitem{aehle-derivatives-2022}
M.~Aehle et~al.
\newblock Derivatives in proton ct.
\newblock arXiv: 2202.05551, 2022.

\bibitem{Adelmann:2022ozp}
A.~Adelmann et~al.
\newblock {New directions for surrogate models and differentiable programming
  for High Energy Physics detector simulation}.
\newblock In {\em {2022 Snowmass Summer Study}}, 3 2022.
\newblock \href {http://arxiv.org/abs/2203.08806} {\path{arXiv:2203.08806}}.

\bibitem{kasim2020up}
M.~Kasim, D.~Watson-Parris, L.~Deaconu, S.~Oliver, P.~Hatfield, D.~Froula,
  G.~Gregori, M.~Jarvis, S.~Khatiwala, J.~Korenaga, et~al.
\newblock Up to two billion times acceleration of scientific simulations with
  deep neural architecture search.
\newblock In {\em APS Division of Plasma Physics Meeting Abstracts}, volume
  2020, pages BO05--001, 2020.

\bibitem{GEANT}
S.~Agostinelli et~al.
\newblock {GEANT4} --- a simulation toolkit.
\newblock {\em Nucl. Inst. Meth. A}, 506:250, 2003.
\newblock \href {https://doi.org/10.1016/S0168-9002(03)01368-8}
  {\path{doi:10.1016/S0168-9002(03)01368-8}}.

\bibitem{allison_geant4_2006}
J.~Allison et~al.
\newblock Geant4 developments and applications.
\newblock {\em IEEE Transactions on Nuclear Science}, 53(1):270--278, 2006.
\newblock \href {https://doi.org/10.1109/TNS.2006.869826}
  {\path{doi:10.1109/TNS.2006.869826}}.

\bibitem{allison_recent_2016}
J.~Allison et~al.
\newblock Recent developments in {Geant4}.
\newblock {\em Nucl. Instr. Meth. A}, 835:186--225, 2016.
\newblock \href {https://doi.org/10.1016/j.nima.2016.06.125}
  {\path{doi:10.1016/j.nima.2016.06.125}}.

\bibitem{FLUKA}
T.T. B{\"o}hlen et~al.
\newblock {The FLUKA code: developments and challenges for high energy and
  medical applications}.
\newblock {\em Nucl. Data Sheets}, 120:211, 2014.
\newblock \href {https://doi.org/https://doi.org/10.1016/j.nds.2014.07.049}
  {\path{doi:https://doi.org/10.1016/j.nds.2014.07.049}}.

\bibitem{Delphes}
J.~de~Favereau et~al.
\newblock {DELPHES 3, A modular framework for fast simulation of a generic
  collider experiment}.
\newblock {\em JHEP}, 02:057, 2014.
\newblock \href {http://arxiv.org/abs/1307.6346} {\path{arXiv:1307.6346}},
  \href {https://doi.org/10.1007/JHEP02(2014)057}
  {\path{doi:10.1007/JHEP02(2014)057}}.

\bibitem{Sinha:2022ogd}
A.K. Sinha et~al.
\newblock {SUPA: A Lightweight Diagnostic Simulator for Machine Learning in
  Particle Physics}, 2022.
\newblock \href {http://arxiv.org/abs/2202.05012} {\path{arXiv:2202.05012}}.

\bibitem{Belayneh:2019vyx}
D.~Belayneh et~al.
\newblock {Calorimetry with deep learning: particle simulation and
  reconstruction for collider physics}.
\newblock {\em Eur. Phys. J. C}, 80(7):688, 2020.
\newblock \href {http://arxiv.org/abs/1912.06794} {\path{arXiv:1912.06794}},
  \href {https://doi.org/10.1140/epjc/s10052-020-8251-9}
  {\path{doi:10.1140/epjc/s10052-020-8251-9}}.

\bibitem{Buhmann:2020pmy}
E.~Buhmann et~al.
\newblock {Getting High: High Fidelity Simulation of High Granularity
  Calorimeters with High Speed}.
\newblock {\em Comput. Softw. Big Sci.}, 5(1):13, 2021.
\newblock \href {http://arxiv.org/abs/2005.05334} {\path{arXiv:2005.05334}},
  \href {https://doi.org/10.1007/s41781-021-00056-0}
  {\path{doi:10.1007/s41781-021-00056-0}}.

\bibitem{Baldi:2016fzo}
P.~Baldi, K.~Cranmer, T.~Faucett, P.~Sadowski, and D.~Whiteson.
\newblock {Parameterized neural networks for high-energy physics}.
\newblock {\em Eur. Phys. J. C}, 76(5):235, 2016.
\newblock \href {http://arxiv.org/abs/1601.07913} {\path{arXiv:1601.07913}},
  \href {https://doi.org/10.1140/epjc/s10052-016-4099-4}
  {\path{doi:10.1140/epjc/s10052-016-4099-4}}.

\bibitem{2013arXiv1312.6114K}
D.P. {Kingma} and M.~{Welling}.
\newblock {Auto-Encoding Variational Bayes}.
\newblock {\em arXiv e-prints}, page arXiv:1312.6114, 2013.
\newblock \href {http://arxiv.org/abs/1312.6114} {\path{arXiv:1312.6114}}.

\bibitem{2014arXiv1401.4082J}
D.~{Jimenez Rezende}, S.~{Mohamed}, and D.~{Wierstra}.
\newblock {Stochastic Backpropagation and Approximate Inference in Deep
  Generative Models}.
\newblock {\em arXiv e-prints}, page arXiv:1401.4082, 2014.
\newblock \href {http://arxiv.org/abs/1401.4082} {\path{arXiv:1401.4082}}.

\bibitem{2019arXiv190602691K}
D.P. {Kingma} and M.~{Welling}.
\newblock {An Introduction to Variational Autoencoders}.
\newblock {\em arXiv e-prints}, page arXiv:1906.02691, 2019.
\newblock \href {http://arxiv.org/abs/1906.02691} {\path{arXiv:1906.02691}}.

\bibitem{2014arXiv1406.2661G}
I.J. {Goodfellow} et~al.
\newblock {Generative Adversarial Networks}.
\newblock {\em arXiv e-prints}, page arXiv:1406.2661, 2014.
\newblock \href {http://arxiv.org/abs/1406.2661} {\path{arXiv:1406.2661}}.

\bibitem{2020arXiv200106937G}
J.~{Gui}, Z.~{Sun}, Y.~{Wen}, D.~{Tao}, and J.~{Ye}.
\newblock {A Review on Generative Adversarial Networks: Algorithms, Theory, and
  Applications}.
\newblock {\em arXiv e-prints}, page arXiv:2001.06937, 2020.
\newblock \href {http://arxiv.org/abs/2001.06937} {\path{arXiv:2001.06937}}.

\bibitem{2015arXiv150505770J}
D.~{Jimenez Rezende} and S.~{Mohamed}.
\newblock {Variational Inference with Normalizing Flows}.
\newblock {\em arXiv e-prints}, page arXiv:1505.05770, 2015.
\newblock \href {http://arxiv.org/abs/1505.05770} {\path{arXiv:1505.05770}}.

\bibitem{2019arXiv190809257K}
I.~{Kobyzev}, S.J.D. {Prince}, and M.A. {Brubaker}.
\newblock {Normalizing Flows: An Introduction and Review of Current Methods}.
\newblock {\em arXiv e-prints}, page arXiv:1908.09257, 2019.
\newblock \href {http://arxiv.org/abs/1908.09257} {\path{arXiv:1908.09257}}.

\bibitem{2019arXiv191202762P}
G.~{Papamakarios}, E.~{Nalisnick}, D.~{Jimenez Rezende}, S.~{Mohamed}, and
  B.~{Lakshminarayanan}.
\newblock {Normalizing Flows for Probabilistic Modeling and Inference}.
\newblock {\em arXiv e-prints}, page arXiv:1912.02762, 2019.
\newblock \href {http://arxiv.org/abs/1912.02762} {\path{arXiv:1912.02762}}.

\bibitem{kullback1951information}
S.~Kullback and R.A. Leibler.
\newblock On information and sufficiency.
\newblock {\em The annals of mathematical statistics}, 22(1):79--86, 1951.

\bibitem{ATL-SOFT-PUB-2018-001}
{ATLAS Collaboration}.
\newblock {Deep generative models for fast shower simulation in ATLAS}.
\newblock Technical Report ATL-SOFT-PUB-2018-001, CERN, 2018.
\newblock URL: \url{http://cds.cern.ch/record/2630433}.

\bibitem{2020arXiv200606704D}
K.~{Deja}, J.~{Dubi{\'n}ski}, P.~{Nowak}, S.~{Wenzel}, and T.~{Trzci{\'n}ski}.
\newblock {End-to-end Sinkhorn Autoencoder with Noise Generator}.
\newblock {\em arXiv e-prints}, page arXiv:2006.06704, 2020.
\newblock \href {http://arxiv.org/abs/2006.06704} {\path{arXiv:2006.06704}}.

\bibitem{Buhmann:2021lxj}
E.~Buhmann et~al.
\newblock {Decoding Photons: Physics in the Latent Space of a BIB-AE Generative
  Network}.
\newblock {\em EPJ Web Conf.}, 251:03003, 2021.
\newblock \href {http://arxiv.org/abs/2102.12491} {\path{arXiv:2102.12491}},
  \href {https://doi.org/10.1051/epjconf/202125103003}
  {\path{doi:10.1051/epjconf/202125103003}}.

\bibitem{Howard:2021pos}
J.N. Howard, S.~Mandt, D.~Whiteson, and Y.~Yang.
\newblock {Foundations of a Fast, Data-Driven, Machine-Learned Simulator},
  2021.
\newblock \href {http://arxiv.org/abs/2101.08944} {\path{arXiv:2101.08944}}.

\bibitem{Buhmann:2021caf}
E.~Buhmann et~al.
\newblock {Hadrons, Better, Faster, Stronger}, 2021.
\newblock \href {http://arxiv.org/abs/2112.09709} {\path{arXiv:2112.09709}}.

\bibitem{Hariri:2021clz}
A.~Hariri, D.~Dyachkova, and S.~Gleyzer.
\newblock {Graph Generative Models for Fast Detector Simulations in High Energy
  Physics}, 2021.
\newblock \href {http://arxiv.org/abs/2104.01725} {\path{arXiv:2104.01725}}.

\bibitem{2017arXiv170107875A}
M.~{Arjovsky}, S.~{Chintala}, and L.~{Bottou}.
\newblock {Wasserstein GAN}.
\newblock {\em arXiv e-prints}, page arXiv:1701.07875, 2017.
\newblock \href {http://arxiv.org/abs/1701.07875} {\path{arXiv:1701.07875}}.

\bibitem{2017arXiv170400028G}
I.~{Gulrajani}, F.~{Ahmed}, M.~{Arjovsky}, V.~{Dumoulin}, and A.~{Courville}.
\newblock {Improved Training of Wasserstein GANs}.
\newblock {\em arXiv e-prints}, page arXiv:1704.00028, 2017.
\newblock \href {http://arxiv.org/abs/1704.00028} {\path{arXiv:1704.00028}}.

\bibitem{2018arXiv180203446B}
A.~{Borji}.
\newblock {Pros and Cons of GAN Evaluation Measures}.
\newblock {\em arXiv e-prints}, page arXiv:1802.03446, 2018.
\newblock \href {http://arxiv.org/abs/1802.03446} {\path{arXiv:1802.03446}}.

\bibitem{deOliveira:2017pjk}
L.~de~Oliveira, M.~Paganini, and B.~Nachman.
\newblock {Learning Particle Physics by Example: Location-Aware Generative
  Adversarial Networks for Physics Synthesis}.
\newblock {\em Comput. Softw. Big Sci.}, 1(1):4, 2017.
\newblock \href {http://arxiv.org/abs/1701.05927} {\path{arXiv:1701.05927}},
  \href {https://doi.org/10.1007/s41781-017-0004-6}
  {\path{doi:10.1007/s41781-017-0004-6}}.

\bibitem{Paganini:2017hrr}
M.~Paganini, L.~de~Oliveira, and B.~Nachman.
\newblock {Accelerating Science with Generative Adversarial Networks: An
  Application to 3D Particle Showers in Multilayer Calorimeters}.
\newblock {\em Phys. Rev. Lett.}, 120(4):042003, 2018.
\newblock \href {http://arxiv.org/abs/1705.02355} {\path{arXiv:1705.02355}},
  \href {https://doi.org/10.1103/PhysRevLett.120.042003}
  {\path{doi:10.1103/PhysRevLett.120.042003}}.

\bibitem{Paganini:2017dwg}
M.~Paganini, L.~de~Oliveira, and B.~Nachman.
\newblock {CaloGAN : Simulating 3D high energy particle showers in multilayer
  electromagnetic calorimeters with generative adversarial networks}.
\newblock {\em Phys. Rev.}, D97(1):014021, 2018.
\newblock \href {http://arxiv.org/abs/1712.10321} {\path{arXiv:1712.10321}},
  \href {https://doi.org/10.1103/PhysRevD.97.014021}
  {\path{doi:10.1103/PhysRevD.97.014021}}.

\bibitem{Bellagente:2019uyp}
M.~Bellagente, A.~Butter, G.~Kasieczka, T.~Plehn, and R.~Winterhalder.
\newblock {How to GAN away Detector Effects}.
\newblock {\em SciPost Phys.}, 8(4):070, 2020.
\newblock \href {http://arxiv.org/abs/1912.00477} {\path{arXiv:1912.00477}},
  \href {https://doi.org/10.21468/SciPostPhys.8.4.070}
  {\path{doi:10.21468/SciPostPhys.8.4.070}}.

\bibitem{Vallecorsa:2019ked}
S.~Vallecorsa, F.~Carminati, and G.~Khattak.
\newblock {3D convolutional GAN for fast simulation}.
\newblock {\em EPJ Web Conf.}, 214:02010, 2019.
\newblock \href {https://doi.org/10.1051/epjconf/201921402010}
  {\path{doi:10.1051/epjconf/201921402010}}.

\bibitem{SHiP:2019gcl}
C.~Ahdida et~al.
\newblock {Fast simulation of muons produced at the SHiP experiment using
  Generative Adversarial Networks}.
\newblock {\em JINST}, 14:P11028, 2019.
\newblock \href {http://arxiv.org/abs/1909.04451} {\path{arXiv:1909.04451}},
  \href {https://doi.org/10.1088/1748-0221/14/11/P11028}
  {\path{doi:10.1088/1748-0221/14/11/P11028}}.

\bibitem{Chekalina:2018hxi}
V.~Chekalina et~al.
\newblock {Generative Models for Fast Calorimeter Simulation. LHCb case}.
\newblock {\em {CHEP 2018}}, 2018.
\newblock \href {http://arxiv.org/abs/1812.01319} {\path{arXiv:1812.01319}},
  \href {https://doi.org/10.1051/epjconf/201921402034}
  {\path{doi:10.1051/epjconf/201921402034}}.

\bibitem{Carminati:2018khv}
F.~Carminati, A.~Gheata, G.~Khattak, Mendez L.P., S.~Sharan, and S.~Vallecorsa.
\newblock {Three dimensional Generative Adversarial Networks for fast
  simulation}.
\newblock {\em J. Phys. Conf. Ser.}, 1085(3):032016, 2018.
\newblock \href {https://doi.org/10.1088/1742-6596/1085/3/032016}
  {\path{doi:10.1088/1742-6596/1085/3/032016}}.

\bibitem{Vallecorsa:2018zco}
S.~Vallecorsa.
\newblock {Generative models for fast simulation}.
\newblock {\em J. Phys. Conf. Ser.}, 1085(2):022005, 2018.
\newblock \href {https://doi.org/10.1088/1742-6596/1085/2/022005}
  {\path{doi:10.1088/1742-6596/1085/2/022005}}.

\bibitem{Musella:2018rdi}
P.~Musella and F.~Pandolfi.
\newblock {Fast and Accurate Simulation of Particle Detectors Using Generative
  Adversarial Networks}.
\newblock {\em Comput. Softw. Big Sci.}, 2(1):8, 2018.
\newblock \href {http://arxiv.org/abs/1805.00850} {\path{arXiv:1805.00850}},
  \href {https://doi.org/10.1007/s41781-018-0015-y}
  {\path{doi:10.1007/s41781-018-0015-y}}.

\bibitem{Erdmann:2018kuh}
M.~Erdmann, L.~Geiger, J.~Glombitza, and D.~Schmidt.
\newblock {Generating and refining particle detector simulations using the
  Wasserstein distance in adversarial networks}.
\newblock {\em Comput. Softw. Big Sci.}, 2(1):4, 2018.
\newblock \href {http://arxiv.org/abs/1802.03325} {\path{arXiv:1802.03325}},
  \href {https://doi.org/10.1007/s41781-018-0008-x}
  {\path{doi:10.1007/s41781-018-0008-x}}.

\bibitem{Deja:2019vcv}
K.~Deja, T.~Trzcinski, and L.~Graczykowski.
\newblock {Generative models for fast cluster simulations in the TPC for the
  ALICE experiment}.
\newblock {\em EPJ Web Conf.}, 214:06003, 2019.
\newblock \href {https://doi.org/10.1051/epjconf/201921406003}
  {\path{doi:10.1051/epjconf/201921406003}}.

\bibitem{Derkach:2019qfk}
D.~Derkach, N.~Kazeev, F.~Ratnikov, A.~Ustyuzhanin, and A.~Volokhova.
\newblock {Cherenkov Detectors Fast Simulation Using Neural Networks}.
\newblock {\em Nucl. Instrum. Meth. A}, 952:161804, 2020.
\newblock \href {http://arxiv.org/abs/1903.11788} {\path{arXiv:1903.11788}},
  \href {https://doi.org/10.1016/j.nima.2019.01.031}
  {\path{doi:10.1016/j.nima.2019.01.031}}.

\bibitem{Erdmann:2018jxd}
M.~Erdmann, J.~Glombitza, and T.~Quast.
\newblock {Precise simulation of electromagnetic calorimeter showers using a
  Wasserstein Generative Adversarial Network}.
\newblock {\em Comput. Softw. Big Sci.}, 3(1):4, 2019.
\newblock \href {http://arxiv.org/abs/1807.01954} {\path{arXiv:1807.01954}},
  \href {https://doi.org/10.1007/s41781-018-0019-7}
  {\path{doi:10.1007/s41781-018-0019-7}}.

\bibitem{Oliveira:DLPS2017}
L.~de~Oliveira, M.~Paganini, and B.~Nachman.
\newblock {Tips and Tricks for Training GANs with Physics Constraints}.
\newblock In {\em {Proceedings of the Deep Learning for Physical Sciences
  Workshop at NIPS (2017)}}, 2017.

\bibitem{deOliveira:2017rwa}
L.~de~Oliveira, M.~Paganini, and B.~Nachman.
\newblock {Controlling Physical Attributes in GAN-Accelerated Simulation of
  Electromagnetic Calorimeters}.
\newblock {\em J. Phys. Conf. Ser.}, 1085(4):042017, 2018.
\newblock \href {http://arxiv.org/abs/1711.08813} {\path{arXiv:1711.08813}},
  \href {https://doi.org/10.1088/1742-6596/1085/4/042017}
  {\path{doi:10.1088/1742-6596/1085/4/042017}}.

\bibitem{Hooberman:DLPS2017}
B.~Hooberman et~al.
\newblock {Calorimetry with Deep Learning: Particle Classification, Energy
  Regression, and Simulation for High-Energy Physics}.
\newblock In {\em {Proceedings of the Deep Learning for Physical Sciences
  Workshop at NIPS (2017)}}, 2017.

\bibitem{2009.03796}
S.~Diefenbacher et~al.
\newblock {DCTRGAN: Improving the Precision of Generative Models with
  Reweighting}.
\newblock {\em Journal of Instrumentation}, 15:P11004, 2020.
\newblock \href {http://arxiv.org/abs/2009.03796} {\path{arXiv:2009.03796}},
  \href {https://doi.org/10.1088/1748-0221/15/11/p11004}
  {\path{doi:10.1088/1748-0221/15/11/p11004}}.

\bibitem{Maevskiy:2020ank}
A.~Maevskiy, F.~Ratnikov, A.~Zinchenko, and V.~Riabov.
\newblock {Simulating the time projection chamber responses at the MPD detector
  using generative adversarial networks}.
\newblock {\em Eur. Phys. J. C}, 81(7):599, 2021.
\newblock \href {http://arxiv.org/abs/2012.04595} {\path{arXiv:2012.04595}},
  \href {https://doi.org/10.1140/epjc/s10052-021-09366-4}
  {\path{doi:10.1140/epjc/s10052-021-09366-4}}.

\bibitem{Rehm:2021zoz}
F.~Rehm, S.~Vallecorsa, K.~Borras, and D.~Kr\"ucker.
\newblock {Validation of Deep Convolutional Generative Adversarial Networks for
  High Energy Physics Calorimeter Simulations}.
\newblock In {\em {AAAI-MLPS 2021 Spring Symposium at Stanford University}},
  2021.
\newblock \href {http://arxiv.org/abs/2103.13698} {\path{arXiv:2103.13698}}.

\bibitem{Rehm:2021qwm}
F.~Rehm, S.~Vallecorsa, K.~Borras, and D.~Kr\"ucker.
\newblock {Physics Validation of Novel Convolutional 2D Architectures for
  Speeding Up High Energy Physics Simulations}.
\newblock {\em EPJ Web Conf.}, 251:03042, 2021.
\newblock \href {http://arxiv.org/abs/2105.08960} {\path{arXiv:2105.08960}},
  \href {https://doi.org/10.1051/epjconf/202125103042}
  {\path{doi:10.1051/epjconf/202125103042}}.

\bibitem{Kansal:2021cqp}
R.~Kansal et~al.
\newblock {Particle Cloud Generation with Message Passing Generative
  Adversarial Networks}.
\newblock In {\em {Thirty-fifth Conference on Neural Information Processing
  Systems}}, 2021.
\newblock \href {http://arxiv.org/abs/2106.11535} {\path{arXiv:2106.11535}}.

\bibitem{Khattak:2021ndw}
G.R. Khattak, S.~Vallecorsa, F.~Carminati, and G.M. Khan.
\newblock {Fast Simulation of a High Granularity Calorimeter by Generative
  Adversarial Networks}, 2021.
\newblock \href {http://arxiv.org/abs/2109.07388} {\path{arXiv:2109.07388}}.

\bibitem{Anderlini:2021qpm}
L.~Anderlini.
\newblock {Machine Learning for the LHCb Simulation}.
\newblock In {\em {Proceedings of the 2021 Artificial Intelligence for the
  Electron Ion Collider (experimental applications) workshop}}, 2021.
\newblock \href {http://arxiv.org/abs/2110.07925} {\path{arXiv:2110.07925}}.

\bibitem{2019arXiv190604032D}
C.~{Durkan}, A.~{Bekasov}, I.~{Murray}, and G.~{Papamakarios}.
\newblock {Neural Spline Flows}.
\newblock {\em arXiv e-prints}, page arXiv:1906.04032, 2019.
\newblock \href {http://arxiv.org/abs/1906.04032} {\path{arXiv:1906.04032}}.

\bibitem{2016arXiv160508803D}
L.~{Dinh}, J.~{Sohl-Dickstein}, and S.~{Bengio}.
\newblock {Density estimation using Real NVP}.
\newblock {\em arXiv e-prints}, page arXiv:1605.08803, 2016.
\newblock \href {http://arxiv.org/abs/1605.08803} {\path{arXiv:1605.08803}}.

\bibitem{2017arXiv170507057P}
G.~{Papamakarios}, T.~{Pavlakou}, and I.~{Murray}.
\newblock {Masked Autoregressive Flow for Density Estimation}.
\newblock {\em arXiv e-prints}, page arXiv:1705.07057, 2017.
\newblock \href {http://arxiv.org/abs/1705.07057} {\path{arXiv:1705.07057}}.

\bibitem{2016arXiv160604934K}
D.P. {Kingma} et~al.
\newblock {Improving Variational Inference with Inverse Autoregressive Flow}.
\newblock {\em arXiv e-prints}, page arXiv:1606.04934, 2016.
\newblock \href {http://arxiv.org/abs/1606.04934} {\path{arXiv:1606.04934}}.

\bibitem{Krause:2021ilc}
C.~Krause and D.~Shih.
\newblock {CaloFlow: Fast and Accurate Generation of Calorimeter Showers with
  Normalizing Flows}, 2021.
\newblock \href {http://arxiv.org/abs/2106.05285} {\path{arXiv:2106.05285}}.

\bibitem{Krause:2021wez}
C.~Krause and D.~Shih.
\newblock {CaloFlow II: Even Faster and Still Accurate Generation of
  Calorimeter Showers with Normalizing Flows}, 2021.
\newblock \href {http://arxiv.org/abs/2110.11377} {\path{arXiv:2110.11377}}.

\bibitem{CaloChallenge}
M.~Faucci~Gianelli et~al.
\newblock Fast calorimeter simulation challenge 2022, 2022.
\newblock URL: \url{https://calochallenge.github.io/homepage/}.

\bibitem{Arbor}
M.~Ruan and H.~Videau.
\newblock {Arbor, a new approach of the Particle Flow Algorithm}.
\newblock In {\em {Proceedings, International Conference on Calorimetry for the
  High Energy Frontier (CHEF 2013): Paris, France, April 22-25, 2013}}, pages
  316--324, 2013.
\newblock \href {http://arxiv.org/abs/1403.4784} {\path{arXiv:1403.4784}}.

\bibitem{Pandora_1}
M.A. Thomson.
\newblock Particle flow calorimetry and the pandorapfa algorithm.
\newblock {\em Nucl. Instr. Meth. A}, 611(1):25–40, Nov 2009.
\newblock \href {https://doi.org/10.1016/j.nima.2009.09.009}
  {\path{doi:10.1016/j.nima.2009.09.009}}.

\bibitem{CLICPF_1}
J.S. Marshall, A.~M{\"u}nnich, and M.A. Thomson.
\newblock Performance of particle flow calorimetry at clic.
\newblock {\em Nucl. Instr. Meth. A}, 700:153–162, Feb 2013.
\newblock \href {https://doi.org/10.1016/j.nima.2012.10.038}
  {\path{doi:10.1016/j.nima.2012.10.038}}.

\bibitem{Pandora}
J.~S. Marshall and M.~A. Thomson.
\newblock {Pandora Particle Flow Algorithm}.
\newblock In {\em {Proceedings, International Conference on Calorimetry for the
  High Energy Frontier (CHEF 2013): Paris, France, April 22-25, 2013}}, pages
  305--315, 2013.
\newblock \href {http://arxiv.org/abs/1308.4537} {\path{arXiv:1308.4537}}.

\bibitem{Pandora_kit}
J.~S. Marshall and M.~A. Thomson.
\newblock The pandora software development kit for pattern recognition.
\newblock {\em The European Physical Journal C}, 75(9), Sep 2015.
\newblock \href {https://doi.org/10.1140/epjc/s10052-015-3659-3}
  {\path{doi:10.1140/epjc/s10052-015-3659-3}}.

\bibitem{Sefkow_2016}
F.~Sefkow, A.~White, K.~Kawagoe, R.~P{\"o}schl, and J.~Repond.
\newblock Experimental tests of particle flow calorimetry.
\newblock {\em Reviews of Modern Physics}, 88(1), Feb 2016.
\newblock \href {https://doi.org/10.1103/revmodphys.88.015003}
  {\path{doi:10.1103/revmodphys.88.015003}}.

\bibitem{Tran_2017}
H.L. Tran et~al.
\newblock Software compensation in particle flow reconstruction.
\newblock {\em The European Physical Journal C}, 77(10), Oct 2017.
\newblock \href {https://doi.org/10.1140/epjc/s10052-017-5298-3}
  {\path{doi:10.1140/epjc/s10052-017-5298-3}}.

\bibitem{CMSPFPaper}
{CMS Collaboration}.
\newblock Particle-flow reconstruction and global event description with the
  cms detector.
\newblock {\em Journal of Instrumentation}, 12(10):P10003–P10003, Oct 2017.
\newblock \href {https://doi.org/10.1088/1748-0221/12/10/p10003}
  {\path{doi:10.1088/1748-0221/12/10/p10003}}.

\bibitem{ATLASPF}
{ATLAS Collaboration}.
\newblock {Jet reconstruction and performance using particle flow with the
  ATLAS Detector}.
\newblock {\em Eur. Phys. J.}, C77(7), 2017.
\newblock \href {http://arxiv.org/abs/1703.10485} {\path{arXiv:1703.10485}},
  \href {https://doi.org/10.1140/epjc/s10052-017-5031-2}
  {\path{doi:10.1140/epjc/s10052-017-5031-2}}.

\bibitem{lecun1998gradient}
Y.~LeCun, L.~Bottou, Y.~Bengio, and P.~Haffner.
\newblock Gradient-based learning applied to document recognition.
\newblock {\em Proc. IEEE}, 86:2278, 1998.
\newblock \href {https://doi.org/10.1109/5.726791}
  {\path{doi:10.1109/5.726791}}.

\bibitem{Bols:2020bkb}
E.~Bols, J.~Kieseler, M.~Verzetti, M.~Stoye, and A.~Stakia.
\newblock {Jet Flavour Classification Using DeepJet}.
\newblock {\em JINST}, 15(12):P12012, 2020.
\newblock \href {http://arxiv.org/abs/2008.10519} {\path{arXiv:2008.10519}},
  \href {https://doi.org/10.1088/1748-0221/15/12/P12012}
  {\path{doi:10.1088/1748-0221/15/12/P12012}}.

\bibitem{TopTaggers}
A.~Butter, K.~Cranmer, D.~Debnath, B.M. Dillon, et~al.
\newblock {The Machine Learning Landscape of Top Taggers}.
\newblock {\em SciPost Phys.}, 7:014, 2019.
\newblock \href {http://arxiv.org/abs/1902.09914} {\path{arXiv:1902.09914}},
  \href {https://doi.org/10.21468/SciPostPhys.7.1.014}
  {\path{doi:10.21468/SciPostPhys.7.1.014}}.

\bibitem{FCChh-CDR}
A.~Abada et~al.
\newblock Fcc-hh: The hadron collider.
\newblock {\em The European Physical Journal Special Topics}, 228(4):755--1107,
  Jul 2019.
\newblock \href {https://doi.org/10.1140/epjst/e2019-900087-0}
  {\path{doi:10.1140/epjst/e2019-900087-0}}.

\bibitem{aleksa2019calorimeters}
M.~Aleksa et~al.
\newblock {Calorimeters for the FCC-hh}.
\newblock CERN-FCC-PHYS-2019-0003, 2019.
\newblock \href {http://arxiv.org/abs/1912.09962} {\path{arXiv:1912.09962}}.

\bibitem{scarselli2009graph}
F.~Scarselli et~al.
\newblock The graph neural network model.
\newblock {\em IEEE Transactions on Neural Networks}, 20(1), 2009.

\bibitem{GravNet}
S.R. Qasim, J.~Kieseler, Y.~Iiyama, and M.~Pierini.
\newblock {Learning representations of irregular particle-detector geometry
  with distance-weighted graph networks}.
\newblock {\em Eur. Phys. J.}, C79(7):608, 2019.
\newblock \href {http://arxiv.org/abs/1902.07987} {\path{arXiv:1902.07987}},
  \href {https://doi.org/10.1140/epjc/s10052-019-7113-9}
  {\path{doi:10.1140/epjc/s10052-019-7113-9}}.

\bibitem{JEDINet}
E.~Moreno et~al.
\newblock {JEDI-net: a jet identification algorithm based on interaction
  networks}.
\newblock {\em Eur. Phys. J.}, C80(1):58, 2020.
\newblock \href {http://arxiv.org/abs/1908.05318} {\path{arXiv:1908.05318}},
  \href {https://doi.org/10.1140/epjc/s10052-020-7608-4}
  {\path{doi:10.1140/epjc/s10052-020-7608-4}}.

\bibitem{ParticleNet}
H.~Qu and L.~Gouskos.
\newblock {ParticleNet: Jet Tagging via Particle Clouds}.
\newblock {\em Phys. Rev. D}, 101(5):056019, 2020.
\newblock \href {http://arxiv.org/abs/1902.08570} {\path{arXiv:1902.08570}},
  \href {https://doi.org/10.1103/PhysRevD.101.056019}
  {\path{doi:10.1103/PhysRevD.101.056019}}.

\bibitem{HEPTrkX2}
S.~Farrel et~al.
\newblock The hep.trkx project: deep neural networks for hl-lhc online and
  offline tracking.
\newblock {\em EPJ Web Conf.}, 150:00003, 2017.
\newblock \href {https://doi.org/10.1051/epjconf/201715000003}
  {\path{doi:10.1051/epjconf/201715000003}}.

\bibitem{Qasim:2021hex}
S.R. Qasim, K.~Long, J.~Kieseler, M.~Pierini, and R.~Nawaz.
\newblock {Multi-particle reconstruction in the High Granularity Calorimeter
  using object condensation and graph neural networks}.
\newblock {\em EPJ Web Conf.}, 251:03072, 2021.
\newblock \href {http://arxiv.org/abs/2106.01832} {\path{arXiv:2106.01832}},
  \href {https://doi.org/10.1051/epjconf/202125103072}
  {\path{doi:10.1051/epjconf/202125103072}}.

\bibitem{Shlomi:2020gdn}
J.~Shlomi, P.~Battaglia, and J.~Vlimant.
\newblock {Graph Neural Networks in Particle Physics}.
\newblock {\em {Machine Learning: Science and Technology}}, 2(2), 7 2020.
\newblock \href {http://arxiv.org/abs/2007.13681} {\path{arXiv:2007.13681}},
  \href {https://doi.org/10.1088/2632-2153/abbf9a}
  {\path{doi:10.1088/2632-2153/abbf9a}}.

\bibitem{HGCAL-TDR}
{CMS Collaboration}.
\newblock {The Phase-2 Upgrade of the CMS Endcap Calorimeter}.
\newblock Technical Report CERN-LHCC-2017-023. CMS-TDR-019, CERN, 2017.
\newblock URL: \url{https://cds.cern.ch/record/2293646}.

\bibitem{Pata:2021oez}
J.~Pata, J.~Duarte, J.~Vlimant, M.~Pierini, and M.~Spiropulu.
\newblock {MLPF: Efficient machine-learned particle-flow reconstruction using
  graph neural networks}.
\newblock {\em Eur. Phys. J. C}, 81(5):381, 2021.
\newblock \href {http://arxiv.org/abs/2101.08578} {\path{arXiv:2101.08578}},
  \href {https://doi.org/10.1140/epjc/s10052-021-09158-w}
  {\path{doi:10.1140/epjc/s10052-021-09158-w}}.

\bibitem{DeCastro:2018psv}
P.~de~Castro and T.~Dorigo.
\newblock {INFERNO: Inference-Aware Neural Optimisation}.
\newblock {\em Comput. Phys. Commun.}, 244:170--179, 2019.
\newblock \href {http://arxiv.org/abs/1806.04743} {\path{arXiv:1806.04743}},
  \href {https://doi.org/10.1016/j.cpc.2019.06.007}
  {\path{doi:10.1016/j.cpc.2019.06.007}}.

\bibitem{gilesstrong_2021_5040810}
G.C. Strong.
\newblock pytorch\_inferno, June 2021.
\newblock {Please check
  https://github.com/GilesStrong/pytorch\_inferno/graphs/contributors for the
  full list of contributors}.
\newblock \href {https://doi.org/10.5281/zenodo.4597140}
  {\path{doi:10.5281/zenodo.4597140}}.

\bibitem{asymptotics}
G.~Cowan, K.~Cranmer, E.~Gross, and O.~Vitells.
\newblock Asymptotic formulae for likelihood-based tests of new physics.
\newblock {\em The European Physical Journal C}, 71(2), 2011.
\newblock \href {https://doi.org/10.1140/epjc/s10052-011-1554-0}
  {\path{doi:10.1140/epjc/s10052-011-1554-0}}.

\bibitem{pyhf}
L.~Heinrich, M.~Feickert, and G.~Stark.
\newblock {pyhf}.
\newblock \url{https://github.com/scikit-hep/pyhf/releases/tag/v0.6.3}.
\newblock Version 0.6.3.
\newblock \href {https://doi.org/10.5281/zenodo.1169739}
  {\path{doi:10.5281/zenodo.1169739}}.

\bibitem{pyhf2}
L.~Heinrich, M.~Feickert, G.~Stark, and K.~Cranmer.
\newblock pyhf: pure-python implementation of histfactory statistical models.
\newblock {\em Journal of Open Source Software}, 6(58):2823, 2021.
\newblock \href {https://doi.org/10.21105/joss.02823}
  {\path{doi:10.21105/joss.02823}}.

\bibitem{relaxed}
N.~Simpson.
\newblock {relaxed: version 0.1.3}, 3 2022.
\newblock URL: \url{https://github.com/gradhep/relaxed}, \href
  {https://doi.org/10.5281/zenodo.6330891} {\path{doi:10.5281/zenodo.6330891}}.

\bibitem{Ginter:737473}
P.~Ginter.
\newblock {Transport from Spain to CERN of a vacuum vessel for the ATLAS barrel
  toroid magnet-system.}
\newblock ATLAS Collection., Nov 2003.
\newblock URL: \url{https://cds.cern.ch/record/737473}.

\bibitem{thomson_2013}
Mark Thomson.
\newblock {\em Modern Particle Physics}.
\newblock Cambridge University Press, 2013.
\newblock \href {https://doi.org/10.1017/CBO9781139525367}
  {\path{doi:10.1017/CBO9781139525367}}.

\bibitem{Zyla:2020zbs}
P.A. Zyla et~al.
\newblock {Review of Particle Physics}.
\newblock {\em PTEP}, 2020(8):083C01, 2020.
\newblock \href {https://doi.org/10.1093/ptep/ptaa104}
  {\path{doi:10.1093/ptep/ptaa104}}.

\bibitem{SAGAN2006356}
D.~Sagan.
\newblock Bmad: A relativistic charged particle simulation library.
\newblock {\em Nucl. Instr. Meth. A}, 558(1):356--359, 2006.
\newblock \href {https://doi.org/https://doi.org/10.1016/j.nima.2005.11.001}
  {\path{doi:https://doi.org/10.1016/j.nima.2005.11.001}}.

\bibitem{twopulse}
A.~Marinelli et~al.
\newblock High-intensity double-pulse x-ray free-electron laser.
\newblock {\em Nat. Commun.}, 6, 2015.

\bibitem{deb2002afast}
K.~Deb, A.~Pratap, S.~Agarwal, and T.~Meyarivan.
\newblock A fast and elitist multiobjective genetic algorithm: Nsga-ii.
\newblock {\em IEEE Transactions on Evolutionary Computation}, 6(2):182--197,
  2002.
\newblock \href {https://doi.org/10.1109/4235.996017}
  {\path{doi:10.1109/4235.996017}}.

\bibitem{s2eopt}
J.~Qiang.
\newblock Start to end beam dynamics optimization of x-ray fel light source
  accelerators.
\newblock {\em NAPAC16}, WEA3IO, Oct. 2016.

\bibitem{Snowmass_WP_AM}
S.~Biedron et~al.
\newblock {Snowmass21 Accelerator Modeling Community White Paper}.
\newblock In {\em {2022 Snowmass Summer Study}}, 3 2022.
\newblock \href {http://arxiv.org/abs/2203.08335} {\path{arXiv:2203.08335}}.

\bibitem{roussel2021multiobj}
R.~Roussel, A.~Hanuka, and A.~Edelen.
\newblock Multiobjective bayesian optimization for online accelerator tuning.
\newblock {\em Phys. Rev. Accel. Beams}, 24:062801, Jun 2021.
\newblock \href {https://doi.org/10.1103/PhysRevAccelBeams.24.062801}
  {\path{doi:10.1103/PhysRevAccelBeams.24.062801}}.

\bibitem{roussel2021turnkey}
Ryan Roussel et~al.
\newblock Turn-key constrained parameter space exploration for particle
  accelerators using bayesian active learning.
\newblock {\em Nature Communications}, 12, 2021.
\newblock \href {https://doi.org/10.1038/s41467-021-25757-3}
  {\path{doi:10.1038/s41467-021-25757-3}}.

\bibitem{KirschnerSafeBO2019}
J.~Kirschner, M.~Mutn{\'{y}}, N.~Hiller, R.~Ischebeck, and A.~Krause.
\newblock Adaptive and safe bayesian optimization in high dimensions via
  one-dimensional subspaces.
\newblock {\em CoRR}, abs/1902.03229, 2019.
\newblock \href {http://arxiv.org/abs/1902.03229} {\path{arXiv:1902.03229}}.

\bibitem{duris2020bayesian}
J.~Duris et~al.
\newblock Bayesian optimization of a free-electron laser.
\newblock {\em Physical review letters}, 124(12):124801, 2020.

\bibitem{hanuka2021physics}
A.~Hanuka et~al.
\newblock Physics model-informed gaussian process for online optimization of
  particle accelerators.
\newblock {\em Physical Review Accelerators and Beams}, 24(7):072802, 2021.

\bibitem{edelen2016neural}
A.L. Edelen et~al.
\newblock Neural networks for modeling and control of particle accelerators.
\newblock {\em IEEE Transactions on Nuclear Science}, 63(2):878--897, 2016.
\newblock \href {https://doi.org/10.1109/TNS.2016.2543203}
  {\path{doi:10.1109/TNS.2016.2543203}}.

\bibitem{scheinker2018demonstration}
A.~Scheinker, A.~Edelen, D.~Bohler, C.~Emma, and A.~Lutman.
\newblock Demonstration of model-independent control of the longitudinal phase
  space of electron beams in the linac-coherent light source with femtosecond
  resolution.
\newblock {\em Physical review letters}, 121(4):044801, 2018.
\newblock \href {https://doi.org/10.1103/PhysRevLett.121.044801}
  {\path{doi:10.1103/PhysRevLett.121.044801}}.

\bibitem{edelen2016first}
A.L. Edelen, S.G. Biedron, S.V. Milton, and J.P. Edelen.
\newblock First steps toward incorporating image based diagnostics into
  particle accelerator control systems using convolutional neural networks.
\newblock {\em arXiv preprint arXiv:1612.05662}, 2016.

\bibitem{ogren2021surrogate}
J.~{\"O}gren, C.~Gohil, and D.~Schulte.
\newblock Surrogate modeling of the {CLIC} final-focus system using artificial
  neural networks.
\newblock {\em Journal of Instrumentation}, 16(05):P05012, may 2021.
\newblock \href {https://doi.org/10.1088/1748-0221/16/05/p05012}
  {\path{doi:10.1088/1748-0221/16/05/p05012}}.

\bibitem{emma2018machine}
C.~Emma et~al.
\newblock Machine learning-based longitudinal phase space prediction of
  particle accelerators.
\newblock {\em Phys. Rev. Accel. Beams}, 21:112802, Nov 2018.
\newblock \href {https://doi.org/10.1103/PhysRevAccelBeams.21.112802}
  {\path{doi:10.1103/PhysRevAccelBeams.21.112802}}.

\bibitem{EdelenDiss}
A.~Edelen.
\newblock Neural networks for modeling and control of particle accelerators.
\newblock \emph{dissertation}, Colorado State University. Available at:
  \url{https://www.leelinska.com/wp-content/uploads/2021/06/Auralee_Edelen_Dissertation.pdf}.

\bibitem{Gupta_2021}
L.~Gupta et~al.
\newblock Improving surrogate model accuracy for the {LCLS}-{II} injector
  frontend using convolutional neural networks and transfer learning.
\newblock {\em Machine Learning: Science and Technology}, 2(4):045025, oct
  2021.
\newblock \href {https://doi.org/10.1088/2632-2153/ac27ff}
  {\path{doi:10.1088/2632-2153/ac27ff}}.

\bibitem{scheinker2021adaptiveML}
A.~Scheinker.
\newblock Adaptive machine learning for time-varying systems: low dimensional
  latent space tuning.
\newblock {\em Journal of Instrumentation}, 16(10):P10008, 2021.
\newblock \href {https://doi.org/10.1088/1748-0221/16/10/P10008}
  {\path{doi:10.1088/1748-0221/16/10/P10008}}.

\bibitem{mishra2021uncertainty}
A.A. Mishra, A.~Edelen, A.~Hanuka, and C.~Mayes.
\newblock Uncertainty quantification for deep learning in particle accelerator
  applications.
\newblock {\em Physical Review Accelerators and Beams}, 24(11):114601, 2021.

\bibitem{convery2021uncertainty}
O.~Convery, L.~Smith, Y.~Gal, and A.~Hanuka.
\newblock Uncertainty quantification for virtual diagnostic of particle
  accelerators.
\newblock {\em Phys. Rev. Accel. Beams}, 24:074602, Jul 2021.
\newblock \href {https://doi.org/10.1103/PhysRevAccelBeams.24.074602}
  {\path{doi:10.1103/PhysRevAccelBeams.24.074602}}.

\bibitem{NN_CSR}
A.~Edelen and C.~Mayes.
\newblock {Neural Network Solver for Coherent Synchrotron Radiation Wakefield
  Calculations in Accelerator-based Charged Particle Beams}.
\newblock 3 2022.
\newblock \href {http://arxiv.org/abs/2203.07542} {\path{arXiv:2203.07542}}.

\bibitem{Berz:1988aj}
M.~Berz.
\newblock {Differential Algebraic Description of Beam Dynamics to Very High
  Orders}.
\newblock {\em Part. Accel.}, 24:109--124, 1989.

\bibitem{scheinker2021adaptive}
A.~Scheinker, F.~Cropp, S.~Paiagua, and D.~Filippetto.
\newblock An adaptive approach to machine learning for compact particle
  accelerators.
\newblock {\em Scientific reports}, 11(1):1--11, 2021.

\bibitem{EdelenNeurIPS2017}
A.~Edelen, J.~Edelen, S.~Milton, S.~Biedron, and P.J.M. Van~der Slot.
\newblock Using neural network control policies for rapid switching between
  beam parameters in a free electron laser.
\newblock In {\em NeurIPS 2017}, Long Beach, CA, 2017.

\bibitem{roussel2022differentiable}
R.~Roussel et~al.
\newblock Differentiable preisach modeling for characterization and
  optimization of accelerator systems with hysteresis.
\newblock {\em arXiv preprint arXiv:2202.07747}, 2022.

\bibitem{Ivanov2020}
A.~Ivanov and I.~Agapov.
\newblock {Physics-based deep neural networks for beam dynamics in charged
  particle accelerators}.
\newblock {\em Physical Review Accelerators and Beams}, 23(7):074601, 7 2020.
\newblock \href {https://doi.org/10.1103/PhysRevAccelBeams.23.074601}
  {\path{doi:10.1103/PhysRevAccelBeams.23.074601}}.

\bibitem{lume}
C.E. Mayes et~al.
\newblock Lightsource unified modeling environment (lume), a start-to-end
  simulation ecosystem.
\newblock In {\em Proc. of IPAC}, page THPAB217, 2021.

\bibitem{christopher_mayes_2021_5559141}
C.~Mayes, R.~Roussel, and H.~Slepicka.
\newblock Xopt v0.5.0, 2021.
\newblock \href {https://doi.org/10.5281/zenodo.5559141}
  {\path{doi:10.5281/zenodo.5559141}}.

\bibitem{openPMDapi}
A.~Huebl, F.~Poeschel, F.~Koller, and J.~Gu.
\newblock {openPMD-api: C++ \& Python API for Scientific I/O with openPMD},
  2018.
\newblock URL: \url{https://github.com/openPMD/openPMD-api}, \href
  {https://doi.org/10.14278/rodare.27} {\path{doi:10.14278/rodare.27}}.

\bibitem{openPMDprojects}
{Curated catalogue of projects supporting openPMD}.
\newblock URL: \url{https://github.com/openPMD/openPMD-projects}.

\bibitem{openPMD}
A.~Huebl et~al.
\newblock openpmd: A meta data standard for particle and mesh based data, 2015.
\newblock \href {https://doi.org/10.5281/zenodo.591699}
  {\path{doi:10.5281/zenodo.591699}}.

\bibitem{LHCb:2018roe}
R.~Aaij et~al.
\newblock {Physics case for an LHCb Upgrade II - Opportunities in flavour
  physics, and beyond, in the HL-LHC era}.
\newblock {\em {CERN LHCC}}, 2018.
\newblock \href {http://arxiv.org/abs/1808.08865} {\path{arXiv:1808.08865}}.

\bibitem{LHCb:2000vji}
{LHCb Collaboration}.
\newblock {LHCb calorimeters: Technical design report}, 2000.

\bibitem{LHCb:2008vvz}
A.~Alves et~al.
\newblock {The LHCb Detector at the LHC}.
\newblock {\em JINST}, 3:S08005, 2008.
\newblock \href {https://doi.org/10.1088/1748-0221/3/08/S08005}
  {\path{doi:10.1088/1748-0221/3/08/S08005}}.

\bibitem{Apollinari:2017lan}
G.~Apollinari et~al.
\newblock {\em {High-Luminosity Large Hadron Collider (HL-LHC): Technical
  Design Report V. 0.1}}.
\newblock CERN Yellow Reports: Monographs. CERN, Geneva, 2017.
\newblock URL: \url{https://cds.cern.ch/record/2284929}, \href
  {https://doi.org/10.23731/CYRM-2017-004} {\path{doi:10.23731/CYRM-2017-004}}.

\bibitem{Barsuk:2000vna}
S.~Barsuk et~al.
\newblock {Design and construction of electromagnetic calorimeter for LHCb
  experiment}.
\newblock Technical report, CERN, Geneva, 2000.
\newblock URL: \url{https://cds.cern.ch/record/691508}.

\bibitem{Jenni:1987cz}
P.~Jenni, P.~Sonderegger, H.~P. Paar, and R.~Wigmans.
\newblock {THE HIGH RESOLUTION SPAGHETTI HADRON CALORIMETER: PROPOSAL}.
\newblock Technical report, NIKHEF, 1987.

\bibitem{Lucchini:2013loa}
M.~Lucchini et~al.
\newblock {Test beam results with LuAG fibers for next-generation
  calorimeters}.
\newblock {\em JINST}, 8:P10017, 2013.
\newblock \href {https://doi.org/10.1088/1748-0221/8/10/P10017}
  {\path{doi:10.1088/1748-0221/8/10/P10017}}.

\bibitem{fabjan2003calorimetry}
C.W. Fabjan and F.~Gianotti.
\newblock Calorimetry for particle physics.
\newblock {\em Reviews of Modern Physics}, 75(4):1243, 2003.

\bibitem{fedor:chef2019}
F.~Ratnikov.
\newblock Using machine learning to speed up and improve calorimeter r{\&}d.
\newblock {\em Journal of Instrumentation}, 15(05):C05032--C05032, may 2020.
\newblock \href {https://doi.org/10.1088/1748-0221/15/05/c05032}
  {\path{doi:10.1088/1748-0221/15/05/c05032}}.

\bibitem{Boldyrev:2020ydy}
A.~Boldyrev, D.~Derkach, F.~Ratnikov, and A.~Shevelev.
\newblock {ML-assisted versatile approach to Calorimeter R\&D}.
\newblock {\em JINST}, 15(09):C09030, 2020.
\newblock \href {http://arxiv.org/abs/2005.07700} {\path{arXiv:2005.07700}},
  \href {https://doi.org/10.1088/1748-0221/15/09/C09030}
  {\path{doi:10.1088/1748-0221/15/09/C09030}}.

\bibitem{xgboost}
T.~Chen and C.~Guestrin.
\newblock {XGBoost}: A scalable tree boosting system.
\newblock In {\em Proceedings of the 22nd ACM SIGKDD International Conference
  on Knowledge Discovery and Data Mining}, KDD '16, pages 785--794, New York,
  NY, USA, 2016. ACM.
\newblock \href {https://doi.org/10.1145/2939672.2939785}
  {\path{doi:10.1145/2939672.2939785}}.

\bibitem{2016arXiv161006545L}
D.~{Lopez-Paz} and M.~{Oquab}.
\newblock {Revisiting Classifier Two-Sample Tests}.
\newblock {\em arXiv e-prints}, page arXiv:1610.06545, 2016.
\newblock \href {http://arxiv.org/abs/1610.06545} {\path{arXiv:1610.06545}}.

\bibitem{2017arXiv171110433V}
A.~{van den Oord} et~al.
\newblock {Parallel WaveNet: Fast High-Fidelity Speech Synthesis}.
\newblock {\em arXiv e-prints}, page arXiv:1711.10433, 2017.
\newblock \href {http://arxiv.org/abs/1711.10433} {\path{arXiv:1711.10433}}.

\bibitem{Long:2020wfp}
K.~Long et~al.
\newblock {Muon colliders to expand frontiers of particle physics}.
\newblock {\em Nature Phys.}, 17(3):289--292, 2021.
\newblock \href {http://arxiv.org/abs/2007.15684} {\path{arXiv:2007.15684}},
  \href {https://doi.org/10.1038/s41567-020-01130-x}
  {\path{doi:10.1038/s41567-020-01130-x}}.

\bibitem{Bartosik_2020}
N.~Bartosik et~al.
\newblock Detector and physics performance at a muon collider.
\newblock {\em Journal of Instrumentation}, 15(05):P05001--P05001, 2020.
\newblock \href {https://doi.org/10.1088/1748-0221/15/05/p05001}
  {\path{doi:10.1088/1748-0221/15/05/p05001}}.

\bibitem{Cemmi:2021uum}
A.~Cemmi et~al.
\newblock {Radiation study of Lead Fluoride crystals}, 2021.
\newblock \href {http://arxiv.org/abs/2107.12307} {\path{arXiv:2107.12307}}.

\bibitem{Mokhov:2017klc}
N.V. Mokhov and C.C. James.
\newblock The mars code system user’s guide version 15(2016).
\newblock {\em {FERMILAB-FN-1058-APC}}, 2 2017.
\newblock \href {https://doi.org/10.2172/1462233} {\path{doi:10.2172/1462233}}.

\bibitem{Abbiendi:2677471}
G.~Abbiendi et~al.
\newblock {Letter of Intent: the MUonE project}.
\newblock Technical Report CERN-SPSC-2019-026, SPSC-I-252, CERN, Geneva, 2019.

\bibitem{SLACmQ1998}
A.A. Prinz et~al.
\newblock Search for millicharged particles at {SLAC}.
\newblock {\em Physical Review Letters}, 81(6):1175–1178, 1998.
\newblock \href {https://doi.org/10.1103/physrevlett.81.1175}
  {\path{doi:10.1103/physrevlett.81.1175}}.

\bibitem{ArgoNeuT:2019ckq}
R.~Acciarri et~al.
\newblock {Improved Limits on Millicharged Particles Using the ArgoNeuT
  Experiment at Fermilab}.
\newblock {\em Phys. Rev. Lett.}, 124(13):131801, 2020.
\newblock \href {http://arxiv.org/abs/1911.07996} {\path{arXiv:1911.07996}},
  \href {https://doi.org/10.1103/PhysRevLett.124.131801}
  {\path{doi:10.1103/PhysRevLett.124.131801}}.

\bibitem{Millicharged2021}
C.A. Arg{\"u}elles, K.J. Kelly, and V.M. Mu{\~n}oz.
\newblock Millicharged particles from the heavens: single- and
  multiple-scattering signatures.
\newblock {\em Journal of High Energy Physics}, 2021(11), 2021.
\newblock \href {https://doi.org/10.1007/jhep11(2021)099}
  {\path{doi:10.1007/jhep11(2021)099}}.

\bibitem{milliqan2020}
A.~Ball et~al.
\newblock Search for millicharged particles in proton-proton collisions at
  s=13~tev.
\newblock {\em Physical Review D}, 102(3), 2020.
\newblock \href {https://doi.org/10.1103/physrevd.102.032002}
  {\path{doi:10.1103/physrevd.102.032002}}.

\bibitem{milliqan2021}
A.~Ball et~al.
\newblock Sensitivity to millicharged particles in future proton-proton
  collisions at the lhc with the milliqan detector.
\newblock {\em Physical Review D}, 104(3), 2021.
\newblock \href {https://doi.org/10.1103/physrevd.104.032002}
  {\path{doi:10.1103/physrevd.104.032002}}.

\bibitem{Abbott_2017multimessenger}
B.P. Abbott et~al.
\newblock Multi-messenger observations of a binary neutron star merger.
\newblock {\em The Astrophysical Journal}, 848(2):L12, oct 2017.
\newblock \href {https://doi.org/10.3847/2041-8213/aa91c9}
  {\path{doi:10.3847/2041-8213/aa91c9}}.

\bibitem{albert2019mm_o1search}
A.~Albert et~al.
\newblock Search for multimessenger sources of gravitational waves and
  high-energy neutrinos with advanced ligo during its first observing run,
  antares, and icecube.
\newblock {\em The Astrophysical Journal}, 870(2):134, 2019.

\bibitem{barthelmy2008gcn}
S.~Barthelmy.
\newblock Gcn and voevent: A status report.
\newblock {\em Astronomische Nachrichten: Astronomical Notes}, 329(3):340--342,
  2008.

\bibitem{2019scta.book.....C}
{Cherenkov Telescope Array Consortium}, B.S. {Acharya}, et~al.
\newblock {\em {Science with the Cherenkov Telescope Array}}.
\newblock World Scientific, 2019.
\newblock \href {https://doi.org/10.1142/10986} {\path{doi:10.1142/10986}}.

\bibitem{2016PhDT.......565S}
C.~{Skole}.
\newblock {Search for extremely short transient gamma-ray sources with the
  VERITAS observatory}.
\newblock Humboldt University of Berlin, Germany, 2016.

\bibitem{2021arXiv210804512G}
O.~{Gueta}.
\newblock {The Cherenkov Telescope Array: layout, design and performance}.
\newblock {\em arXiv e-prints}, page arXiv:2108.04512, 2021.
\newblock \href {http://arxiv.org/abs/2108.04512} {\path{arXiv:2108.04512}}.

\bibitem{Willke20021377_geo600}
B.~Willke et~al.
\newblock The geo 600 gravitational wave detector.
\newblock {\em Classical and Quantum Gravity}, 19(7):1377 – 1387, 2002.
\newblock \href {https://doi.org/10.1088/0264-9381/19/7/321}
  {\path{doi:10.1088/0264-9381/19/7/321}}.

\bibitem{Acernese2015_AdV}
F.~Acernese et~al.
\newblock {Advanced Virgo: A second-generation interferometric gravitational
  wave detector}.
\newblock {\em Classical and Quantum Gravity}, 32(2), 2015.
\newblock \href {https://doi.org/10.1088/0264-9381/32/2/024001}
  {\path{doi:10.1088/0264-9381/32/2/024001}}.

\bibitem{Abbott2016_firstgw}
B.P. Abbott et~al.
\newblock Observation of gravitational waves from a binary black hole merger.
\newblock {\em Phys. Rev. Lett.}, 116:061102, Feb 2016.
\newblock \href {https://doi.org/10.1103/PhysRevLett.116.061102}
  {\path{doi:10.1103/PhysRevLett.116.061102}}.

\bibitem{amaro2017laser}
P.~Amaro-Seoane et~al.
\newblock Laser interferometer space antenna.
\newblock {\em arXiv preprint arXiv:1702.00786}, 2017.

\bibitem{maggiore2020etscience}
M.~Maggiore et~al.
\newblock Science case for the einstein telescope.
\newblock {\em Journal of Cosmology and Astroparticle Physics}, 2020(03):050,
  2020.

\bibitem{PhysRevLett.120.061101_lisa_PF}
M.~Armano et~al.
\newblock {Beyond the Required LISA Free-Fall Performance: New LISA Pathfinder
  Results down to $20\text{ }\text{ }\ensuremath{\mu}\mathrm{Hz}$}.
\newblock {\em Phys. Rev. Lett.}, 120:061101, 2018.
\newblock \href {https://doi.org/10.1103/PhysRevLett.120.061101}
  {\path{doi:10.1103/PhysRevLett.120.061101}}.

\bibitem{abbott2021gwtc3}
R.~Abbott et~al.
\newblock Gwtc-3: Compact binary coalescences observed by ligo and virgo during
  the second part of the third observing run.
\newblock {\em arXiv preprint arXiv:2111.03606}, 2021.

\bibitem{lvk2021gwtc3_population}
LIGO~Scientific Collaboration, Virgo Collaboration, KAGRA~Scientific
  Collaboration, et~al.
\newblock The population of merging compact binaries inferred using
  gravitational waves through gwtc-3.
\newblock {\em arXiv preprint arXiv:2111.03634}, 2021.

\bibitem{abbott2020prospects}
B.P. Abbott et~al.
\newblock Prospects for observing and localizing gravitational-wave transients
  with advanced ligo, advanced virgo and kagra.
\newblock {\em Living reviews in relativity}, 23(1):1--69, 2020.

\bibitem{abbott2017continouswaves}
B.P. Abbott et~al.
\newblock First low-frequency einstein@ home all-sky search for continuous
  gravitational waves in advanced ligo data.
\newblock {\em Physical Review D}, 96(12):122004, 2017.

\bibitem{abbott2018searchrare}
B.P. Abbott et~al.
\newblock Search for subsolar-mass ultracompact binaries in advanced ligo’s
  first observing run.
\newblock {\em Physical review letters}, 121(23):231103, 2018.

\bibitem{acernese2017Virgo_status}
F.~Acernese et~al.
\newblock Status of the advanced virgo gravitational wave detector.
\newblock {\em International Journal of Modern Physics A}, 32(28n29):1744003,
  2017.

\bibitem{acernese2018aVirgo_status}
F.~Acernese et~al.
\newblock Status of advanced virgo.
\newblock In {\em EPJ Web of Conferences}, volume 182, page 02003. EDP
  Sciences, 2018.

\bibitem{bersanetti2021aVirgo_status}
D.~Bersanetti et~al.
\newblock Advanced virgo: Status of the detector, latest results and future
  prospects.
\newblock {\em Universe}, 7(9):322, 2021.

\bibitem{Amico20023318}
P.~Amico et~al.
\newblock {Monolithic fused silica suspension for the Virgo gravitational waves
  detector}.
\newblock {\em Review of Scientific Instruments}, 73(9):3318, 2002.
\newblock \href {https://doi.org/10.1063/1.1499540}
  {\path{doi:10.1063/1.1499540}}.

\bibitem{braccini1996seismic}
S.~Braccini et~al.
\newblock Seismic vibrations mechanical filters for the gravitational waves
  detector virgo.
\newblock {\em Review of scientific instruments}, 67(8):2899--2902, 1996.

\bibitem{bonnand2019acl}
R.~Bonnand, A.~Masserot, B.~Mours, L.~Rolland, E.~Pacaud, M.~Was, and
  D.~Passuello.
\newblock {\em {The Algorithms for Control and Locking (Acl)server
  Documentation. Technical Report VIR-00XX-16}}, 2019.

\bibitem{casanueva2019tools}
J.~Casanueva and ISC team.
\newblock {ISC tools: VPM, Acl and Data Display. Technical Report
  VIR-0129A-18}, 2018.

\bibitem{gwinc}
{LIGO Scientific Collaboration and Virgo Collaboration}.
\newblock gwinc, 2022.
\newblock Version 0.4.1.
\newblock URL: \url{https://git.ligo.org/gwinc/pygwinc}.

\bibitem{degallaix2020oscar}
J.~Degallaix.
\newblock {OSCAR: A MATLAB based package to simulate realistic optical
  cavities}.
\newblock {\em SoftwareX}, 12:100587, 2020.

\bibitem{freise2013finesse}
A.~Freise, D.~Brown, and C.~Bond.
\newblock {Finesse, frequency domain INterferomEter simulation softwarE}.
\newblock {\em arXiv preprint arXiv:1306.2973}, 2013.

\bibitem{brown2020pykat}
D.D. Brown et~al.
\newblock Pykat: Python package for modelling precision optical
  interferometers.
\newblock {\em SoftwareX}, 12:100613, 2020.

\bibitem{nguyen2021automated}
C.~Nguyen et~al.
\newblock Automated source of squeezed vacuum states driven by finite state
  machine based software.
\newblock {\em Review of Scientific Instruments}, 92(5):054504, 2021.

\bibitem{giacoppo2021towards}
L.~Giacoppo et~al.
\newblock Towards ponderomotive squeezing with sips experiment.
\newblock {\em Physica Scripta}, 96(11):114007, 2021.

\bibitem{Aisa2016644}
D.~Aisa et~al.
\newblock {The Advanced Virgo monolithic fused silica suspension}.
\newblock {\em Nucl. Instr. Meth. A}, 824:644 – 645, 2016.
\newblock \href {https://doi.org/10.1016/j.nima.2015.09.037}
  {\path{doi:10.1016/j.nima.2015.09.037}}.

\bibitem{COSPAR2018}
A.~Anker et~al.
\newblock Targeting ultra-high energy neutrinos with the {ARIANNA} experiment.
\newblock {\em Advances in Space Research (in press)}, 2019.
\newblock \href {http://arxiv.org/abs/1903.01609} {\path{arXiv:1903.01609}},
  \href {https://doi.org/10.1016/j.asr.2019.06.016}
  {\path{doi:10.1016/j.asr.2019.06.016}}.

\bibitem{ARA2020-limit}
P.~Allison et~al.
\newblock {Constraints on the diffuse flux of ultrahigh energy neutrinos from
  four years of Askaryan Radio Array data in two stations}.
\newblock {\em Phys. Rev. D}, 102(4):043021, 2020.
\newblock \href {http://arxiv.org/abs/1912.00987} {\path{arXiv:1912.00987}},
  \href {https://doi.org/10.1103/PhysRevD.102.043021}
  {\path{doi:10.1103/PhysRevD.102.043021}}.

\bibitem{Aguilar:2020RNOG}
J.~A. Aguilar et~al.
\newblock {Design and Sensitivity of the Radio Neutrino Observatory in
  Greenland (RNO-G)}.
\newblock {\em JINST}, 16(03):P03025, 2021.
\newblock \href {http://arxiv.org/abs/2010.12279} {\path{arXiv:2010.12279}},
  \href {https://doi.org/10.1088/1748-0221/16/03/P03025}
  {\path{doi:10.1088/1748-0221/16/03/P03025}}.

\bibitem{IceCubeGen2-2020}
{The IceCube-Gen2 Collaboration et al.}
\newblock {IceCube-Gen2: The Window to the Extreme Universe}.
\newblock {\em Journal of Physics G: Nuclear and Particle Physics},
  48(6):060501, 2021.
\newblock \href {http://arxiv.org/abs/2008.04323v1}
  {\path{arXiv:2008.04323v1}}, \href {https://doi.org/10.1088/1361-6471/abbd48}
  {\path{doi:10.1088/1361-6471/abbd48}}.

\bibitem{HallmannICRC2021}
{Hallmann , S. and Clark, B. and Glaser, C. and Smith, D. for the IceCube-Gen2
  collaboration}.
\newblock Sensitivity studies for the icecube-gen2 radio array.
\newblock {\em PoS(ICRC2021)1183}, 2021.
\newblock \href {https://doi.org/10.22323/1.395.1183}
  {\path{doi:10.22323/1.395.1183}}.

\bibitem{NuRadioReco2019}
C.~Glaser et~al.
\newblock Nuradioreco: A reconstruction framework for radio neutrino detectors.
\newblock {\em The European Physical Journal C}, 79(6), 2019.
\newblock \href {http://arxiv.org/abs/1903.07023} {\path{arXiv:1903.07023}},
  \href {https://doi.org/10.1140/epjc/s10052-019-6971-5}
  {\path{doi:10.1140/epjc/s10052-019-6971-5}}.

\bibitem{NuRadioMC2019}
C.~Glaser et~al.
\newblock {NuRadioMC}: simulating the radio emission of neutrinos from
  interaction to detector.
\newblock {\em The European Physical Journal C}, 80(77), 2020.
\newblock \href {http://arxiv.org/abs/1906.01670} {\path{arXiv:1906.01670}},
  \href {https://doi.org/10.1140/epjc/s10052-020-7612-8}
  {\path{doi:10.1140/epjc/s10052-020-7612-8}}.

\bibitem{DLNuRadioICRC1}
C.~Glaser, S.~McAleer, P.~Baldi, and S.W. Barwick.
\newblock {Deep learning reconstruction of the neutrino energy with a shallow
  Askaryan detector}.
\newblock {\em PoS(ICRC2021)1051}, 2021.
\newblock \href {https://doi.org/10.22323/1.395.1051}
  {\path{doi:10.22323/1.395.1051}}.

\bibitem{DLNuRadioICRC2}
S.~Stjärnholm, O.~Ericsson, and C.~Glaser.
\newblock Neutrino direction and flavor reconstruction from radio detector data
  using deep convolutional neural networks.
\newblock {\em PoS(ICRC2021)1055}, 2021.
\newblock \href {https://doi.org/10.22323/1.395.1055}
  {\path{doi:10.22323/1.395.1055}}.

\bibitem{Bonechi:2019ckl}
L.~Bonechi, R.~D'Alessandro, and A.~Giammanco.
\newblock {Atmospheric muons as an imaging tool}.
\newblock {\em Rev. Phys.}, 5:100038, 2020.
\newblock \href {http://arxiv.org/abs/1906.03934} {\path{arXiv:1906.03934}},
  \href {https://doi.org/10.1016/j.revip.2020.100038}
  {\path{doi:10.1016/j.revip.2020.100038}}.

\bibitem{ScanPyramids}
K.~Morishima et~al.
\newblock {Discovery of a big void in Khufu's Pyramid by observation of
  cosmic-ray muons}.
\newblock {\em Nature}, 552(7685):386, 2017.
\newblock \href {http://arxiv.org/abs/1711.01576} {\path{arXiv:1711.01576}},
  \href {https://doi.org/10.1038/nature24647} {\path{doi:10.1038/nature24647}}.

\bibitem{MtEchia}
G.~Saracino et~al.
\newblock Imaging of underground cavities with cosmic-ray muons from
  observations at {Mt}. {Echia} ({Naples}).
\newblock {\em Sci. Rep.}, 7(1):1181, 2017.
\newblock \href {https://doi.org/10.1038/s41598-017-01277-3}
  {\path{doi:10.1038/s41598-017-01277-3}}.

\bibitem{SwissGlaciers}
R.~Nishiyama et~al.
\newblock First measurement of ice-bedrock interface of alpine glaciers by
  cosmic muon radiography.
\newblock {\em Geophys. Res. Lett.}, 44(12):6244, 2017.
\newblock \href {https://doi.org/10.1002/2017GL073599}
  {\path{doi:10.1002/2017GL073599}}.

\bibitem{Mahon2018}
D.~Mahon et~al.
\newblock {First-of-a-kind muography for nuclear waste characterization}.
\newblock {\em Phil. Trans. R. Soc. A}, 377:0048, 2018.
\newblock \href {https://doi.org/10.1098/rsta.2018.0048}
  {\path{doi:10.1098/rsta.2018.0048}}.

\bibitem{Riggi:2017izf}
F.~Riggi et~al.
\newblock {The Muon Portal Project: Commissioning of the full detector and
  first results}.
\newblock In {\em {Proceedings, 8th International Conference on New
  Developments in Photodetection (NDIP17), Tours, France, July 3-7, 2017}},
  volume 912, page~16, 2018.
\newblock \href {https://doi.org/10.1016/j.nima.2017.10.006}
  {\path{doi:10.1016/j.nima.2017.10.006}}.

\bibitem{MURAVES}
M.~D{\textquotesingle}Errico et~al.
\newblock Muon radiography applied to volcanoes imaging: the {MURAVES}
  experiment at {Mt. Vesuvius}.
\newblock {\em JINST}, 15(03):C03014, 2020.
\newblock \href {https://doi.org/10.1088/1748-0221/15/03/c03014}
  {\path{doi:10.1088/1748-0221/15/03/c03014}}.

\bibitem{Muography-SMO}
L.~Ol\'ah et~al.
\newblock High-definition and low-noise muography of the {Sakurajima} volcano
  with gaseous tracking detectors.
\newblock {\em Sci. Rep.}, 8(1):3207, 2018.
\newblock \href {https://doi.org/10.1038/s41598-018-21423-9}
  {\path{doi:10.1038/s41598-018-21423-9}}.

\bibitem{cosmic_muons1}
G.~Mengyun, C.~Ming-Chung, C.~Jun, L.~Kam-Biu, and Y.~Changgen.
\newblock A parametrization of the cosmic-ray muon flux at sea-level, 2015.
\newblock \href {http://arxiv.org/abs/1509.06176} {\path{arXiv:1509.06176}}.

\bibitem{cosmic_muons2}
P.~Shukla and S.~Sankrith.
\newblock {Energy and angular distributions of atmospheric muons at the Earth}.
\newblock {\em Int. J. Mod. Phys. A}, 33(30):1850175, 2018.
\newblock \href {http://arxiv.org/abs/1606.06907} {\path{arXiv:1606.06907}},
  \href {https://doi.org/10.1142/S0217751X18501750}
  {\path{doi:10.1142/S0217751X18501750}}.

\bibitem{POCA}
L.J. Schultz et~al.
\newblock Image reconstruction and material z discrimination via cosmic ray
  muon radiography.
\newblock {\em Nucl. Instr. Meth. A}, 519(3):687--694, 2004.
\newblock \href {https://doi.org/https://doi.org/10.1016/j.nima.2003.11.035}
  {\path{doi:https://doi.org/10.1016/j.nima.2003.11.035}}.

\bibitem{gradient_descent}
J.~Hadamard.
\newblock {\em M{\'e}moire sur le probl{\`e}me d'analyse relatif {\`a}
  l'{\'e}quilibre des plaques {\'e}lastiques encastr{\'e}es}.
\newblock M{\'e}moires pr{\'e}sent{\'e}s par divers savants {\`a}
  l'Acad{\'e}mie des sciences de l'Institut de France: {\'E}xtrait. Imprimerie
  nationale, 1908.

\bibitem{adam}
D.P. Kingma and J.~Ba.
\newblock Adam: {A} method for stochastic optimization.
\newblock In Y.~Bengio and Y.~LeCun, editors, {\em 3rd International Conference
  on Learning Representations, {ICLR} 2015, San Diego, CA, USA, May 7-9, 2015,
  Conference Track Proceedings}, 2015.

\bibitem{gnns}
F.~Scarselli, M.~Gori, A.C. Tsoi, M.~Hagenbuchner, and G.~Monfardini.
\newblock The graph neural network model.
\newblock {\em IEEE Transactions on Neural Networks}, 20(1):61--80, 2009.
\newblock \href {https://doi.org/10.1109/TNN.2008.2005605}
  {\path{doi:10.1109/TNN.2008.2005605}}.

\bibitem{Baccani:2018nrn}
G.~Baccani et~al.
\newblock {The MIMA project. Design, construction and performances of a compact
  hodoscope for muon radiography applications in the context of Archaeology and
  geophysical prospections}.
\newblock {\em JINST}, 13(11):P11001, 2018.
\newblock \href {http://arxiv.org/abs/1806.11398} {\path{arXiv:1806.11398}},
  \href {https://doi.org/10.1088/1748-0221/13/11/P11001}
  {\path{doi:10.1088/1748-0221/13/11/P11001}}.

\bibitem{Kyushu2018}
K.~Chaiwongkhot et~al.
\newblock {Development of a Portable Muography Detector for Infrastructure
  Degradation Investigation}.
\newblock {\em IEEE Trans. Nucl. Sci.}, 65:2316, 2018.
\newblock \href {https://doi.org/10.1109/TNS.2018.2855737}
  {\path{doi:10.1109/TNS.2018.2855737}}.

\bibitem{Wuyckens2018}
S.~Wuyckens, A.~Giammanco, P.~Demin, and E.~{Cortina Gil}.
\newblock {A portable muon telescope based on small and gas-tight Resistive
  Plate Chambers}.
\newblock {\em Phil. Trans. R. Soc. A}, 377:0139, 2018.
\newblock \href {http://arxiv.org/abs/1806.06602} {\path{arXiv:1806.06602}},
  \href {https://doi.org/10.1098/rsta.2018.0139}
  {\path{doi:10.1098/rsta.2018.0139}}.

\bibitem{Basnet2020}
S.~Basnet et~al.
\newblock {Towards portable muography with small-area, gas-tight glass
  Resistive Plate Chambers}.
\newblock {\em JINST}, 15(10):C10032, 2020.
\newblock \href {http://arxiv.org/abs/2005.09589} {\path{arXiv:2005.09589}},
  \href {https://doi.org/10.1088/1748-0221/15/10/C10032}
  {\path{doi:10.1088/1748-0221/15/10/C10032}}.

\bibitem{Gamage:2021dqd}
R.M.I.D. Gamage et~al.
\newblock {A portable muon telescope for multidisciplinary applications}.
\newblock {\em JINST}, 17(01):C01051, 2022.
\newblock \href {http://arxiv.org/abs/2109.14489} {\path{arXiv:2109.14489}},
  \href {https://doi.org/10.1088/1748-0221/17/01/C01051}
  {\path{doi:10.1088/1748-0221/17/01/C01051}}.

\bibitem{Moussawi2021}
M.~Moussawi et~al.
\newblock {A portable muon telescope for exploration geophysics in confined
  environments}.
\newblock In {\em {First International Meeting for Applied Geoscience \&
  Energy, 26 September - 1 October 2021, Denver (USA)}}, SEG Technical Program
  Expanded Abstracts, First International Meeting for Applied Geoscience \&
  Energy Expanded Abstracts, 3034-3038, 2021.
\newblock \href {https://doi.org/10.1190/segam2021-3581267.1}
  {\path{doi:10.1190/segam2021-3581267.1}}.

\bibitem{AGUMuographyBook}
A.~Giammanco, E.~Cortina~Gil, S.~Andringa, and M.~Tytgat.
\newblock {Resistive Plate Chambers in Muography}.
\newblock In L.~Olah, H.~Tanaka, and D.~Varga, editors, {\em {Muography:
  Exploring Earth's Subsurface with Elementary Particles}}, Geophysical
  Monograph Series (ISSN: 0065-8448), chapter~18. AGU - Wiley, 2022.
\newblock \href {https://doi.org/10.1002/9781119722748.ch18}
  {\path{doi:10.1002/9781119722748.ch18}}.

\bibitem{wilson_radiological_1946}
R.R. Wilson.
\newblock Radiological use of fast protons.
\newblock {\em Radiology}, 47(5):487--491, 1946.
\newblock \href {https://doi.org/10.1148/47.5.487}
  {\path{doi:10.1148/47.5.487}}.

\bibitem{cormack_representation_1963}
A.M. Cormack.
\newblock Representation of a function by its line integrals, with some
  radiological applications.
\newblock {\em Journal of Applied Physics}, 34(9):2722--2727, 1963.
\newblock \href {https://doi.org/10.1063/1.1729798}
  {\path{doi:10.1063/1.1729798}}.

\bibitem{poludniowski_proton_2015}
G.~Poludniowski, N.M. Allinson, and P.M. Evans.
\newblock Proton radiography and tomography with application to proton therapy.
\newblock {\em The British Journal of Radiology}, 88(1053):20150134, 2015.
\newblock \href {https://doi.org/10.1259/bjr.20150134}
  {\path{doi:10.1259/bjr.20150134}}.

\bibitem{johnson_review_2018}
R.P. Johnson.
\newblock Review of medical radiography and tomography with proton beams.
\newblock {\em Rep. Prog. Phys.}, 81(1):016701, 2018.
\newblock \href {https://doi.org/10.1088/1361-6633/aa8b1d}
  {\path{doi:10.1088/1361-6633/aa8b1d}}.

\bibitem{schulte_maximum_2008}
R.W. Schulte, S.N. Penfold, J.T. Tafas, and K.E. Schubert.
\newblock A maximum likelihood proton path formalism for application in proton
  computed tomography: Maximum likelihood path formalism for proton {CT}.
\newblock {\em Med. Phys.}, 35(11):4849--4856, 2008.
\newblock \href {https://doi.org/10.1118/1.2986139}
  {\path{doi:10.1118/1.2986139}}.

\bibitem{krah_comprehensive_2018}
N.~Krah, F.~Khellaf, J.M. L{\'e}tang, S.~Rit, and I.~Rinaldi.
\newblock A comprehensive theoretical comparison of proton imaging set-ups in
  terms of spatial resolution.
\newblock {\em Phys. Med. Biol.}, 63(13):135013, 2018.
\newblock \href {https://doi.org/10.1088/1361-6560/aaca1f}
  {\path{doi:10.1088/1361-6560/aaca1f}}.

\bibitem{collins-fekete_theoretical_2017}
C.A. Collins-Fekete, L.~Volz, S.K.N. Portillo, L.~Beaulieu, and J.~Seco.
\newblock A theoretical framework to predict the most likely ion path in
  particle imaging.
\newblock {\em Phys. Med. Biol.}, 62(5):1777--1790, 2017.
\newblock \href {https://doi.org/10.1088/1361-6560/aa58ce}
  {\path{doi:10.1088/1361-6560/aa58ce}}.

\bibitem{williams_most_2004}
D.C. Williams.
\newblock The most likely path of an energetic charged particle through a
  uniform medium.
\newblock {\em Phys. Med. Biol.}, 49(13):2899--2911, 2004.
\newblock \href {https://doi.org/10.1088/0031-9155/49/13/010}
  {\path{doi:10.1088/0031-9155/49/13/010}}.

\bibitem{alme_high-granularity_2020}
J.~Alme et~al.
\newblock A high-granularity digital tracking calorimeter optimized for proton
  {CT}.
\newblock {\em Frontiers in Physics}, 8:460, 2020.
\newblock \href {https://doi.org/10.3389/fphy.2020.568243}
  {\path{doi:10.3389/fphy.2020.568243}}.

\bibitem{GATE}
S.~Jan et~al.
\newblock {GATE} - {Geant4} {Application} for {Tomographic} {Emission}: a
  simulation toolkit for {PET} and {SPECT}.
\newblock {\em Physics in Medicine and Biology}, 49(19):4543--4561, 2004.

\bibitem{pettersen_design_2019}
H.E.S. Pettersen et~al.
\newblock Design optimization of a pixel-based range telescope for proton
  computed tomography.
\newblock {\em Physica Medica}, 63:87--97, 2019.
\newblock \href {https://doi.org/10.1016/j.ejmp.2019.05.026}
  {\path{doi:10.1016/j.ejmp.2019.05.026}}.

\bibitem{biguri_tigre_2016}
A.~Biguri, M.~Dosanjh, S.~Hancock, and M.~Soleimani.
\newblock {TIGRE}: a {MATLAB}-{GPU} toolbox for {CBCT} image reconstruction.
\newblock {\em Biomed. Phys. Eng. Express}, 2(5):055010, 2016.
\newblock \href {https://doi.org/10.1088/2057-1976/2/5/055010}
  {\path{doi:10.1088/2057-1976/2/5/055010}}.

\bibitem{Gonz2}
M.~Gonz{\'a}lez-Alonso, O.~Naviliat-Cuncic, and N.~Severijns.
\newblock New physics searches in nuclear and neutron $\ensuremath{\beta}$
  decay.
\newblock {\em Progress in Particle and Nuclear Physics}, 104:165--223, 2019.
\newblock \href {https://doi.org/https://doi.org/10.1016/j.ppnp.2018.08.002}
  {\path{doi:https://doi.org/10.1016/j.ppnp.2018.08.002}}.

\bibitem{dubbers2021precise}
D.~Dubbers and B.~M{\"a}rkisch.
\newblock Precise measurements of the decay of free neutrons.
\newblock {\em Annual Review of Nuclear and Particle Science}, 71:139--163,
  2021.

\bibitem{chupp2019electric}
T.E. Chupp, P.~Fierlinger, M.J. Ramsey-Musolf, and J.T. Singh.
\newblock Electric dipole moments of atoms, molecules, nuclei, and particles.
\newblock {\em Reviews of Modern Physics}, 91(1):015001, 2019.

\bibitem{jaeckel2010low}
J.~Jaeckel and A.~Ringwald.
\newblock The low-energy frontier of particle physics.
\newblock {\em Annual Review of Nuclear and Particle Science}, 60:405--437,
  2010.

\bibitem{aker2019improved}
M.~Aker et~al.
\newblock Improved upper limit on the neutrino mass from a direct kinematic
  method by katrin.
\newblock {\em Physical review letters}, 123(22):221802, 2019.

\bibitem{saul2020limit}
H.~Saul et~al.
\newblock Limit on the fierz interference term b from a measurement of the beta
  asymmetry in neutron decay.
\newblock {\em Physical review letters}, 125(11):112501, 2020.

\bibitem{UCNA_Ab}
X.~Sun et~al.
\newblock Improved limits on {Fierz} interference using asymmetry measurements
  from the ultracold neutron asymmetry (ucna) experiment.
\newblock {\em Phys. Rev. C}, 101:035503, Mar 2020.
\newblock \href {https://doi.org/10.1103/PhysRevC.101.035503}
  {\path{doi:10.1103/PhysRevC.101.035503}}.

\bibitem{wang2019design}
X.~Wang et~al.
\newblock Design of the magnet system of the neutron decay facility perc.
\newblock In {\em EPJ Web of Conferences}, volume 219, page 04007. EDP
  Sciences, 2019.

\bibitem{gonzalez2021improved}
F.M. Gonzalez et~al.
\newblock Improved neutron lifetime measurement with ucn $\tau$.
\newblock {\em Physical review letters}, 127(16):162501, 2021.

\bibitem{abel2020measurement}
C.~Abel et~al.
\newblock Measurement of the permanent electric dipole moment of the neutron.
\newblock {\em Physical Review Letters}, 124(8):081803, 2020.

\bibitem{Shahriari2016Taking}
B.~Shahriari, K.~Swersky, Z.~Wang, R.P. Adams, and N.~de~Freitas.
\newblock Taking the human out of the loop: A review of bayesian optimization.
\newblock {\em Proceedings of the IEEE}, 104(1):148--175, 2016.
\newblock \href {https://doi.org/10.1109/JPROC.2015.2494218}
  {\path{doi:10.1109/JPROC.2015.2494218}}.

\bibitem{RevelsLubinPapamarkou2016}
J.~{Revels}, M.~{Lubin}, and T.~{Papamarkou}.
\newblock Forward-mode automatic differentiation in {J}ulia.
\newblock {\em arXiv:1607.07892 [cs.MS]}, 2016.

\bibitem{Aoki:2021kgd}
Y.~Aoki et~al.
\newblock {FLAG Review 2021}.
\newblock CERN-TH-2021-191, JLAB-THY-21-3528, FERMILAB-PUB-21-620-SCD-T, 2021.
\newblock \href {http://arxiv.org/abs/2111.09849} {\path{arXiv:2111.09849}}.

\bibitem{Ramos:2018vgu}
A.~Ramos.
\newblock {Automatic differentiation for error analysis of Monte Carlo data}.
\newblock {\em Comput. Phys. Commun.}, 238:19--35, 2019.
\newblock \href {http://arxiv.org/abs/1809.01289} {\path{arXiv:1809.01289}},
  \href {https://doi.org/10.1016/j.cpc.2018.12.020}
  {\path{doi:10.1016/j.cpc.2018.12.020}}.

\bibitem{Madras:1988ei}
N.~Madras and A.D. Sokal.
\newblock {The Pivot algorithm: a highly efficient Monte Carlo method for
  selfavoiding walk}.
\newblock {\em J. Statist. Phys.}, 50:109--186, 1988.
\newblock \href {https://doi.org/10.1007/BF01022990}
  {\path{doi:10.1007/BF01022990}}.

\bibitem{Wolff:2003sm}
U.~Wolff.
\newblock {Monte Carlo errors with less errors}.
\newblock {\em Comput. Phys. Commun.}, 156:143--153, 2004.
\newblock [Erratum: Comput.Phys.Commun. 176, 383 (2007)].
\newblock \href {http://arxiv.org/abs/hep-lat/0306017}
  {\path{arXiv:hep-lat/0306017}}, \href
  {https://doi.org/10.1016/S0010-4655(03)00467-3}
  {\path{doi:10.1016/S0010-4655(03)00467-3}}.

\bibitem{Sjostrand:2006za}
T.~Sj{\"o}strand, S.~Mrenna, and P.Z. Skands.
\newblock {PYTHIA 6.4 Physics and Manual}.
\newblock {\em JHEP}, 05:026, 2006.
\newblock \href {http://arxiv.org/abs/hep-ph/0603175}
  {\path{arXiv:hep-ph/0603175}}, \href
  {https://doi.org/10.1088/1126-6708/2006/05/026}
  {\path{doi:10.1088/1126-6708/2006/05/026}}.

\bibitem{Sjostrand:2007gs}
T.~Sj{\"o}strand, S.~Mrenna, and P.Z. Skands.
\newblock {A Brief Introduction to PYTHIA 8.1}.
\newblock {\em Comput. Phys. Commun.}, 178:852--867, 2008.
\newblock \href {http://arxiv.org/abs/0710.3820} {\path{arXiv:0710.3820}},
  \href {https://doi.org/10.1016/j.cpc.2008.01.036}
  {\path{doi:10.1016/j.cpc.2008.01.036}}.

\bibitem{Sjostrand:2014zea}
T.~Sj{\"o}strand et~al.
\newblock An introduction to {PYTHIA} 8.2.
\newblock {\em Comput. Phys. Commun.}, 191:159, 2015.
\newblock \href {http://arxiv.org/abs/1410.3012} {\path{arXiv:1410.3012}},
  \href {https://doi.org/10.1016/j.cpc.2015.01.024}
  {\path{doi:10.1016/j.cpc.2015.01.024}}.

\bibitem{genie}
A.I. Castillo et~al.
\newblock {Genie: an interactive real-time simulation for teaching genetic
  drift.}
\newblock {\em Evo Edu Outreach}, 15:3, 2022.
\newblock \href {http://arxiv.org/abs/10.1101/268672v3.full}
  {\path{arXiv:10.1101/268672v3.full}}, \href
  {https://doi.org/10.1186/s12052-022-00161-7}
  {\path{doi:10.1186/s12052-022-00161-7}}.

\bibitem{kubernetes}
The~Kubernetes Community.
\newblock Kubernetes: Production-grade container orchestration, 2014.
\newblock Retrieved on 19-03-2022.
\newblock URL: \url{https://kubernetes.io/}.

\end{thebibliography}
\end{document}